\pdfoutput=1 
\documentclass[11pt,twoside]{uwthesis}
\usepackage{graphicx}
\usepackage{subfigure}
\usepackage{amssymb}
\usepackage{amsmath}
\usepackage{slashed}
\usepackage{multirow}
\usepackage{array}
\usepackage{hyperref}
\usepackage[capitalize]{cleveref}
\usepackage[numbers]{natbib}
\newcommand{\be}{\begin{equation}}
\newcommand{\ee}{\end{equation}}
\newcommand{\bea}{\begin{eqnarray}}
\newcommand{\eea}{\end{eqnarray}}
\newcommand{\bs}{\begin{split}}
\newcommand{\es}{\end{split}}
\newcommand{\bpm}{\begin{pmatrix}}
\newcommand{\epm}{\end{pmatrix}}

\newcommand{\eq}[1]{Eq. (\ref{eq:#1})}

\newcommand{\la}{\langle}
\newcommand{\ra}{\rangle}
\newcommand{\bsig}{\mbox{\boldmath$\sigma$}}
\newcommand{\btau}{\mbox{\boldmath$\tau$}}
\newcommand{\bpi}{\mbox{\boldmath$\pi$}}
\newcommand{\bgamma}{\mbox{\boldmath$\gamma$}}
\newcommand{\bnabla}{\mbox{\boldmath$\nabla$}}
\newcommand{\bfE}{\mbox{\boldmath$\Sigma$}}

\newcommand{\bfp}{{\bf p}}

\newcommand{\bfq}{{\bf q}}
\newcommand{\bfk}{{\bf k}}
\newcommand{\bfx}{{\bf x}}
\newcommand{\bfl}{{\bf l}}
\newcommand{\bfS}{{\bf S}}
\newcommand{\bfr}{{\bf r}}
\newcommand{\ubar}{\overline{u}}
\newcommand{\dbar}{\overline{d}}
\newcommand{\qbar}{\overline{q}}
\newcommand{\Nbar}{\overline{N}}

\newcommand{\cgz}{\left\la1/2\ m_1,1/2\ m_2\mid0\ 0\right\ra}
\newcommand{\ddr}{\frac{\partial}{\partial r}}

\newcommand{\mo}{\mu(\omega)}
\newcommand{\orh}{(\hat{\bfr})}

\begin{document}
 
%
\prelimpages

\Title{Charge Symmetry Breaking and \\
  Nuclear Pion Production Reactions}
\Author{Daniel R. Bolton}
\Year{2011}
\Program{Department of Physics}
\titlepage
 
\Chair{Gerald A. Miller}{Professor}{Department of Physics}
\Signature{Gerald A. Miller}
\Signature{Martin J. Savage}
\Signature{Stephen R. Sharpe}
\signaturepage

\setcounter{page}{-1}
\doctoralquoteslip

\setcounter{page}{-1}
\abstract{%
Large momentum transfer reactions such as pion production represent the frontier of Chiral Perturbation Theory and must be understood before more complex reactions can be considered.  Pion production is also interesting in its own right, one application being the hadronic extraction of a charge symmetry breaking parameter: the contribution of the down-up quark mass difference to the neutron-proton mass difference.  This dissertation reports on two primary projects: (1) a calculation of the charge symmetry breaking forward-backward asymmetry of the differential cross section of the $np\to d\pi^0$ reaction, and (2) the development of a new theoretical framework addressing the issue of reducibility in the impulse approximation's contribution to pion production.  It is shown that the traditional one-body impulse approximation must be replaced by a two-body operator which makes a larger contribution to $s$-wave pion production.
}
 
\tableofcontents
\listoffigures
\listoftables
 
\chapter*{Glossary}
\addcontentsline{toc}{chapter}{Glossary}
\thispagestyle{plain}
\begin{glossary}
\item[ChPT] Chiral Perturbation Theory, an EFT for low energy hadronic iteractions.
\item[contact term] An interaction vertex which can be thought of as originating from the integrating out of an intermediate heavy field.
\item[CS(B)] Charge Symmetry (Breaking), the approximate invariance (small symmetry breaking) of nature under the interchange of up and down quarks.
\item[EFT] Effective Field Theory, a quantum field theory formulated on a larger length scale than the fundamental theory.
\item[FRA] Free Recoil Approximation, the use of the approximations $\bfp^2=m_\pi m_N$ and $\bfk^2\approx0$ in the pion production kernel when including the distorted wave initial states.
\item[(H)B$\chi$PT] (Heavy) Baryon Chiral Perturbation Theory, inclusion of (non-)relativistic baryons (neutron, proton, and sometimes Delta) to $\chi$PT.
\item[IA] Impulse Approximation, the pion production diagram in which one nucleon is a spectator and the other produces a pion directly.
\item[isospin] A quantum number, similar to spin, which is used to describe systems in which the light quarks are treated as identical particles.
\item[LEC] Low Energy Constant, a parameter/coupling constant for a term of an EFT, often a non-renormalizable contact term.
\item[MCS] Momentum Counting Scheme, a reorganized power counting for pion production reactions accounting for the requisite large three-momentum.
\item[(N)LO] (Next-to-)Leading Order, a diagram's location within the expansion of an EFT.
\item[OPE] One-Pion-Exchange, self-explanatory nucleon-nucleon interaction.
\item[RS] Rescattering, the pion production diagram in which one nucleon produces a pion which is then scattered on-shell by the other nucleon.
\item[seagull] A vertex with two nucleons (incoming and outgoing) and two pions.
\item[threshold] In the context of pion production, the kinematical situation where a pion can just barely be produced; that is, the reaction in which the final state pion has zero three-momentum.
\item[WT] Weinberg-Tomozawa, a $\pi\pi NN$ vertex which is fixed by chiral symmetry and scales as the sum of the incoming and outgoing pion momenta.
\end{glossary}
 
\acknowledgments{
   {\narrower\noindent
  First of all, I would like to thank my advisor Jerry Miller for all the time he invested in teaching me nuclear physics.  You have always supported me when I did not know what to do next in a calculation or was confused by a concept.  Even though I may have wanted a quick answer at the time, looking back now I see how much more I learned in the struggle and I appreciate that long view you took.  I have been blessed to benefit from your wide view of the field.  I want to also mention the UW Physics Department which I am proud to have been a part of.  The staff (most notably the NT secretary Katie) and faculty (primarily Nikolai Tolich and Steve Sharpe) have played a large role in my professional development and have been constant sources of encouragement.  I have no doubt that my education in theoretical medium and high energy physics has been top-quality, thanks in no small part to Martin Savage, Larry Yaffe, and Ann Nelson.  Finally, I would like to acknowledge the contributions that my fellow graduate students (primarily Raul, Amy R, Amy N, Jason, and Shane) have made, both in working together and in making work fun.
   \par}
}

\dedication{\begin{center} To the people who {\it really} enabled me to complete this work: my wife and best friend Laura, and my parents, Ron and Diana. \end{center}}

%

\textpages
 
\chapter{Introduction}
\label{chap:intro}
 
Pion production reactions occur when two nucleons (protons or neutrons) collide at sufficiently high energy to produce a pion.  When the collision is that of a single neutron with a single proton, it is possible for them to collapse into a bound final-state deuteron after producing the pion.  A variety of different pion production reactions have been realized in laboratories; this thesis focuses on experiments with proton or neutron beams and liquid or gas-jet hydrogen targets.  Physicists study pion production for many reasons.  For example, this thesis will describe how a concert of experimental and theoretical effort can result in the extraction of the fundamental parameter $\delta m_N$, the portion of the neutron-proton mass difference attributable to the down-up quark mass difference.

This thesis begins in \cref{chap:chpt} with a review of the strong interaction, which is the primary force determining the properties of pion production reactions (cross sections, angular dependencies, energy dependencies, etc.).  To achieve a perfect description of these observables, one should perform scattering theory calculations making use of the potential energy as determined from the fundamental theory of the strong interaction, Quantum Chromodynamics (QCD).  This fundamental theory is written in terms of quark and gluon degrees of freedom.  At the same time, it is well-known that QCD exhibits ``color confinement" \citep{Gross:1973id}.  Among other things, this means that objects with non-zero color charge are never observed as well-defined asymptotic states.  This is in contrast with Quantum Electrodynamics (QED) where a proton is a clear example of a well-defined asymptotic state with electric charge.

Another property of QCD that affects its applicability is its non-perturbative nature at low energies.\footnote{``Low energies" is meant to exclude the high energy processes which can be found in stars and some colliders, and instead refer to nuclei in, say, this piece of paper.}  This property is connected with confinement in an incompletely-understood manner.  Again, in contrast with QED, which has a coupling constant $\alpha_\text{EM}\approx1/137$ in terms of which one is able to perform perturbation theory, QCD has a coupling constant $\alpha_S={\cal O}(1)$ preventing such an expansion.  Thus, one must either perform nonperturbative calculations such as lattice-regularized QCD (LQCD) or else describe the interactions of color neutral {\it hadrons}, where a separate perturbation theory based on an {\it effective field theory} approach is possible.

In this thesis the latter approach is taken and applied to the calculation of pion production observables which are then compared with experiment.  After decades of model-dependent perturbative theories of the strong interaction, a theory based on the {\it chiral} symmetry of QCD was developed, called Chiral Perturbation Theory ($\chi$PT). Chiral Perturbation Theory is guaranteed to reproduce the physics of QCD in a model-independent way, provided one can identify a convergent {\it power counting} scheme.  \Cref{chap:chpt} provides a development of the aspects of $\chi$PT relevant for pion production.

The next topic, found in \cref{chap:review}, is a review of pion production.  Much effort (both experimental and theoretical) has been devoted to its study over the past half-century, and \cref{chap:review} provides motivation for this effort as well as an overview of the reaction's features.  The salient feature is the large collision momentum $\tilde{p}\equiv\sqrt{m_\pi m_N}$ that is required in order to produce a pion.  A variety of reaction channels (particle and angular momentum) are possible, ranging from relatively-well understood to quite poorly understood.  This work will focus on the near threshold $np\to d\pi^0$ reaction.

\Cref{chap:csb,chap:wfncor,chap:nrred}, each based on a published paper, apply $\chi$PT to the calculation of several pion production observables.  \Cref{chap:csb} \citep{Bolton:2009rq} is an update of a theoretical calculation \citep{vanKolck:2000ip} from 2000 which relates $\delta m_N$ to the asymmetry in the differential cross section of a 2003 experiment \citep{Opper:2003sb}.  The calculation depends heavily on a particular amplitude which was updated \citep{Lensky:2005jc} in 2006.  In \cref{chap:csb}, the updated calculation of that amplitude is extended off-threshold and, using a model estimate for $\delta m_N$, a new post-diction is made for the asymmetry.  The theoretical calculation of the asymmetry is found to be $\sim2-3$ experimental uncertainties\footnote{The theory is not yet well-defined enough to assign an uncertainty.} above the experimental value.

While performing this calculation, the authors became interested in a formalism issue which is similar in nature to Ref. \citep{Lensky:2005jc}, but concerns a different amplitude, the impulse approximation (IA).  In the IA, pions are produced directly from a single nucleon without interacting with the spectator nucleon.  A na\"ive attempt at a solution is introduced in \cref{chap:csb}, but it is clear that further investigation is required.

This investigation is performed in \cref{chap:wfncor} \citep{Bolton:2010qu}.  This chapter begins with a clearer identification of the problem with the traditional formalism: strikingly different results are obtained for the IA amplitude before and after application of the Lippmann-Schwinger equation to the external nucleon-nucleon wave functions.  The same solution as in \cref{chap:csb} is pursued but more subtle modifications are made in order to correctly account for the large scale $\tilde{p}$.  Unfortunately, the large number of theoretical complications prohibit a clear resolution of the issue in this na\"ive manner.

A definitive conclusion is reached in \cref{chap:nrred} \citep{Bolton:2010uj}, which attacks the problem from its root in the reduction from a covariant 4D problem to a traditional 3D one.  It is shown that the traditional formalism for the IA is itself an approximation that is only valid in the absence of initial-state interactions.  For the full distorted-wave calculation, the IA is correctly included by using the Lippmann-Schwinger equation to turn it into a two-body operator.  The reduction to such a simple interpretation involves several approximations which are shown to be controlled in the Appendices.  The final conclusion is that the IA amplitude is enhanced by a factor of approximately two.

Finally, \cref{chap:discussion} gives a review of the topics discussed in this thesis.  Problems encountered in the calculations of \cref{chap:csb,chap:wfncor,chap:nrred} are recalled and possible solutions listed.  Having successfully shown in \cref{chap:nrred} that the IA is larger than was previously thought, at least two calculations need to be updated: the $p$-wave $NN\to d\pi$ amplitudes (and therefore, the $np\to d\pi^0$ asymmetry) and the $pp\to pp\pi^0$ total cross section.
\chapter{Chiral Perturbation Theory}
\label{chap:chpt}
 
Because of the non-perturbative nature of QCD, the theory cannot be used directly\footnote{Lattice QCD does provide means to accomplish this, but computational limitations make pion production calculations presently impractical.} to calculate pion production observables.  One way to overcome this limitation is to use an effective field theory (EFT) formulated in terms of the hadronic bound states of quarks and gluons, i.e. pions and nucleons\footnote{``Nucleons" refers to protons and/or neutrons which are put into the same multiplet in $\chi$PT.} \citep{Weinberg:1978kz}.  This theory, called chiral perturbation theory ($\chi$PT), is guaranteed to include all of QCD's features through a perturbative expansion containing all terms consistent with the symmetries of QCD \citep{Gasser:1983yg,Gasser:1984gg,Weinberg:1991um,Bernard:1993nj,Bernard:1995dp}.  The truth behind the usefulness of EFT is a separation of scales; the long distance (low energy) physics of pions and nucleons does not depend greatly on the short distance (high energy) physics of quarks and gluons.  Taking advantage of this separation of scales, the expansion of $\chi$PT is organized in terms of the small external momenta divided by the large hadronic scale which happens to also be the scale associated with the spontaneous breaking of the chiral symmetry of QCD, $\Lambda_\chi$.  The usefulness of $\chi$PT is primarily the fact that (once convergence is shown), it provides model-independent predictions for strong interaction phenomena.

In Sec. \ref{sec:qcd}, an overview of QCD and its symmetries is given.  Mesonic $\chi$PT is developed and the ``power counting" scheme that provides predictive power is explained in Sec. \ref{sec:chpt}.  \Cref{sec:hbchpt} expands the theory to include nucleons and a difficulty therein is overcome.  Finally, in \cref{sec:csbops}, Charge Symmetry Breaking operators are expressed within the formalism of $\chi$PT.
 
\section{\label{sec:qcd}Quantum Chromodynamics}

\subsection{The Lagrangian of QCD}
In this section, a brief description is given for the experimentally observed spontaneous breakdown of the SU(2)$_L\otimes$SU(2)$_R$ symmetry of QCD in the limit of massless $u$ and $d$ quarks to SU(2)$_V$.  This material has been summarized often, and this summary borrows heavily from Ref. \citep{Scherer:2002tk}.

To begin, consider a quark field $q_f(x)$ with flavor index $f$ (color and Dirac indices are suppressed) that transforms under a local color gauge transformation as,
\begin{align}
q_f(x)\rightarrow\text{exp}\left(-i\sum_{i=1}^8\Theta^i(x)\frac{\lambda^i}{2}\right)q_f(x)\equiv {\cal U}(x)q_f(x),
\end{align}
where $\lambda^i$ represents the eight Gell-Mann matrices which generate SU(3) and $\Theta^i(x)$ are functions of spacetime representing a particular gauge transformation.  The Lagrange density for the matter fields is then,
\begin{align}
{\cal L}_\text{matter}=\sum_f\overline{q}_f(x)\left(i\slashed{D}-m_f\right)q_f(x),
\end{align}
where the flavor sum runs over the six different quarks (u, d, s, c, b, and t) and the covariant derivative $D$ is required by gauge symmetry to transform as,
\begin{align}
D_\mu q_f\rightarrow {\cal U}D_\mu q_f.
\end{align}
This requirement is met by introducing a gauge field ${\cal A}_\mu$ (via $D_\mu=\partial_\mu+ig_3{\cal A}_\mu$) that transforms as,
\begin{align}
{\cal A}_\mu\rightarrow {\cal U}{\cal A}_\mu {\cal U}^\dagger+\frac{i}{g_3}\partial_\mu {\cal U}{\cal U}^\dagger,
\end{align}
where $g_3$ is the running strong coupling constant.  Also note that the gauge fields are matrices in color-space,
\begin{align}
{\cal A}_\mu=\sum_i{\cal A}_\mu^i\lambda^i.
\end{align}
The gauge field is then made dynamical by including the term,
\begin{align}
{\cal L}_\text{gauge}=-\frac{1}{4}\sum_i{\cal G}_{\mu\nu}^i{\cal G}^{i\,\mu\nu},
\end{align}
where the field-strength tensor ${\cal G}_{\mu\nu}^i=\partial_\mu {\cal A}_\nu^i-\partial_\mu{\cal A}_\nu^i-g_3f^{ijk}{\cal A}_\mu^j{\cal A}_\nu^k$ has a third term that is unique to non-Abelian gauge theories.  This term is written in terms of the structure constants of SU(3),
\begin{align}
f^{ijk}=\frac{1}{4i}\text{Tr}\left(\left[\lambda^i,\lambda^j\right]\lambda^k\right).
\end{align}
Finally, note that another term, quadratic in the field-strength tensor and totally antisymmetric in spacetime indices, is allowed by gauge invariance.  This term is important because it violates $CP$ invariance; that is, it is not invariant under simultaneous charge conjugation and parity transformations.  If its coupling constant were non-zero, then the strong force would break $CP$ symmetry; however, experimentally, no evidence for this term has been found \citep{Kim:2008hd}.

\subsection{The Chiral Symmetry of QCD}
Apart from the coupling $g_3$, the only parameters of QCD are the quark masses.  The ``current" quark masses\footnote{Because quarks are never observed as asymptotic states, any statement of their masses must come with a description of what that mass means.  Referred to here are the ``current" quark masses (the parameters in the Lagrangian) in contrast with the model-inspired ``constituent" quark masses.} are listed in Table \ref{tab:quarks}.
\begin{table}
\caption[Current quark masses]{\label{tab:quarks}Current quark masses in the $\overline{\text{MS}}$ renormalization scheme taken from the PDG \citep{Nakamura:2010zzi}.  The values shown are the masses $m_q(\mu)$ as evaluated at a particular scale $\mu$.  For the $u$, $d$, and $s$ quarks, the value shown corresponds to $\mu=2$ GeV.  For the $c$, $b$, and $t$ quarks the pole mass [$m_q(\mu=m_q)$] is used.}
\begin{center}
\begin{tabular}{cc}
\hline \hline
Quark & Mass (MeV) \\ \hline
up & 1.7 to 3.3 \\
down & 4.1 to 5.8 \\
strange & 80 to 130 \\
charm & 1,180 to 1,340 \\
bottom & 4,130 to 4,370 \\
top & $172,000\pm1,600$ \\
\hline \hline
\end{tabular}
\end{center}
\end{table}
It is clear that the masses of the up and down (together, ``light") quarks are negligible compared with the hadronic scale of $\sim1\text{ GeV}$.  For this reason, it is beneficial to consider the massless, or chiral, limit of QCD as a starting point for calculation.  The asymptotic states of interest in pion production are composed entirely (in the valence sense) of light quarks.  Thus the effects of the four heavier quarks, with masses $m_Q$, are seen through either electroweak or virtual pair production processes.  In moving to an effective field theory, these processes are ``integrated out" and replaced with contact interactions that are suppressed by factors of $1/m_Q$.

The strange quark mass is such that it could feasibly be included as a third light quark, and indeed this is sometimes done.  However, the review article \citep{Bernard:1995dp} discusses the fact that the resulting EFT is relatively slowly convergent.  For example, Ref. \citep{Ishikawa:2009vc} showed that the one-loop expansion of the octet and decouplet baryon masses in SU(3) HB$\chi$PT does {\it not} converge.  This conclusion was reached by using LQCD to investigate the quark mass dependence of the baryon masses.  On the other hand, a relatively new, covariant version of SU(3) B$\chi$PT was used in Ref. \citep{Geng:2008mf} to calculate baryon magnetic moments with reasonable convergence.  The application of EFT to pion production is already expected to be slowly convergent due to the large external momentum $\tilde{p}$.  Therefore, to avoid further convergence issues, strangeness is not considered in the pion production calculations of this thesis.

When the masses of the light quarks are neglected, there is nothing in the Lagrangian to distinguish one from the other.  For this reason they are often grouped into an ``isospin" doublet $q=\begin{pmatrix}u\\d\end{pmatrix}$ on which one may perform isospin rotations with the Pauli matrices.  The terminology used is that the light quarks are ``isospin-1/2" particles.  As will be discussed in \cref{sec:hbchpt}, the proton and neutron are also isospin-1/2 particles due to their similar masses.  The symbol used to represent the quantum number of isospin is $T$ (many authors use $I$ instead).

As shown in this and the next paragraph, the Lagrangian for {\it chiral} QCD manifests not only Lorentz and SU(3)$_c$ gauge symmetries, but also a {\it global} ``chiral" symmetry.  Consider the projection operators,
\begin{align}
P_L=\frac{1}{2}\left(1-\gamma_5\right),\qquad P_R=\frac{1}{2}\left(1+\gamma_5\right)
\end{align}
which satisfy $P_L+P_R=1$, $P_L^2=P_L$, $P_R^2=P_R$, and $P_LP_R=P_RP_L=0$.  These projectors separate the helicity ($\hat{p}\cdot{\bf S}$) eigenstates of the massless, free Dirac Equation: $q_L=P_Lq$, $q_R=P_Rq$.  Utilizing these definitions and adopting vector flavor notation, $q_L=\begin{pmatrix}u_L\\d_L\end{pmatrix}$,
\begin{align}
{\cal L}^\text{chiral}_\text{matter}
=\left(\overline{q}_L\,i\slashed{\partial}\,q_L+\overline{q}_R\,i\slashed{\partial}\,q_R\right).\label{eq:massless}
\end{align}
Note that the left- and right-handed sectors have decoupled and that the neglected mass term provides the coupling,
\begin{align}
{\cal L}_\text{mass}=-\left(\overline{q}_L\,M\,q_R+\overline{q}_R\,M\,q_L\right)\label{eq:qcdmass}
\end{align}
with $M=\text{diag}(m_u,m_d)$.

Chiral symmetry can now be formalized with the definition of the global chiral transformation,
\begin{align}
q_L\rightarrow\text{exp}\left[-i\sum_{a=1}^3\theta_a^L\frac{\tau_a}{2}-i\theta^L\right]q_L,
\end{align}
and similar for the right-handed fields.  Note the appearance of the Pauli matrices $\tau_a$ that generate the SU(2) symmetry group.  The U(1) transformation is parametrized by $\theta^L$.  The Lagrangian of \eq{massless} is invariant under each of SU(2)$_\text{L}$, SU(2)$_\text{R}$, U(1)$_\text{L}$, and U(1)$_\text{R}$ separately.  Due to the symmetry breaking pattern observed in nature (discussed below), it is beneficial to combine the left and right transformations to form vector and axial vector transformations.  To accomplish this, consider the Noether \citep{Noether:1918zz} symmetry currents,
\begin{align}\bs
L^\mu_a&=\frac{\partial\delta{\cal L}}{\partial(\partial_\mu\theta_a^L)}=\overline{q}_L\gamma^\mu\frac{\tau_a}{2}q_L
\\
L^\mu&=\frac{\partial\delta{\cal L}}{\partial(\partial_\mu\theta^L)}=\overline{q}_L\gamma^\mu q_L,
\es\end{align}
and similarly for the right-handed currents.  In these equations, $\delta{\cal L}$ represents the variation in the Lagrange density due to a {\it local}, infinitesimal chiral transformation.  Now form linear combinations,
\begin{equation}
\begin{aligned}
V^\mu_a&=R^\mu_a+L^\mu_a=\overline{q}\gamma^\mu\frac{\tau_a}{2}q, &\qquad A^\mu_a&=R^\mu_a-L^\mu_a=\overline{q}\gamma^\mu\gamma_5\frac{\tau_a}{2}a
\\
V^\mu&=R^\mu+L^\mu=\overline{q}\gamma^\mu q, & A^\mu&=R^\mu-L^\mu=\overline{q}\gamma^\mu\gamma_5q.
\end{aligned}
\end{equation}
Except for $A^\mu$, each of these currents are conserved in the chiral limit of the strong force.  Quantum fluctuations destroy the classical conservation of $A^\mu$, and instead one finds \citep{Bell:1969ts},
\be
\partial_\mu A^\mu=\frac{3g_3^2}{32\pi^2}\epsilon_{\mu\nu\rho\sigma}\sum_i{\cal G}_i^{\mu\nu}{\cal G}_i^{\rho\sigma}.
\ee
Another way to understand this divergence is to notice that although the Lagrangian itself is invariant under the iso-singlet axial transformation, the measure of the path integral is not.

Because later chapters will discuss the symmetry-breaking role of the quark masses, they should be considered briefly now.  Again, using Noether's Theorem, one finds that the terms of \eq{qcdmass} cause three of the currents to develop a divergence,
\begin{align}\bs
\partial_\mu V^\mu_a&=i\overline{q}\left[M,\frac{\tau_a}{2}\right]q
\\
\partial_\mu A^\mu_a&=i\overline{q}\left\{M,\frac{\tau_a}{2}\right\}\gamma_5q\label{eq:divamua}
\\
\partial_\mu A^\mu&=2i\overline{q}M\gamma_5q.
\es\end{align}
Thus the extent to which the quark masses are non-zero causes explicit breaking of the axial symmetries, and the extent to which $m_d-m_u\neq0$ causes explicit breaking of the vector, iso-vector symmetry.  The fact that the divergence of $A^\mu$ is proportional to the pseudoscalar density is referred to as the Partially Conserved Axial Current relation, and has been understood (at least macroscopically) since Gell-Mann first described the ``Eightfold Way" in 1964 \citep{GellMann:1964tf}.

\subsection{Spontaneous Symmetry Breaking of QCD}
Upon first thought, one should expect to see a spectrum of hadrons corresponding to the irreducible representations of the symmetry group $G$, \citep{GellMann:1964nj}
\begin{align}
G=\text{SU}(2)_L\otimes\text{SU}(2)_R\otimes\text{U}(1)_V.\label{eq:symgrp}
\end{align}
The multiplets are classified as mesons or baryons based on how they transform under U(1)$_V$ and there should be small splittings within each multiplet resulting from the non-zero quark masses.  If the spectrum is invariant under {\it separate} left- and right-handed transformations, the multiplets should contain approximately degenerate positive and negative parity particles.  The experimentally observed spectrum does not appear to satisfy this rule.  For example, the negative-parity proton has a mass of 1535 MeV, almost 600 MeV above the proton.

The fact that the full symmetry group is not manifest in the hadronic spectrum implies spontaneous symmetry breaking.  It can be shown that a scalar quark condensate provides such a mechanism,
\begin{align}
\la\overline{q}_{j,L}(x)q_{i,R}(x)\ra\equiv\Lambda^3\delta_{ij},
\end{align}
where $\la\cdot\ra$ denotes the expectation value of a quantity in the interacting QCD vacuum, and $\Lambda$ is a quantity with dimension of mass.  Under a chiral transformation, one finds
\begin{align}
\la\overline{q}_{j,L}(x)q_{i,R}(x)\ra\rightarrow\Lambda^3R_{il}\delta_{lk}L_{kj}^\dagger\equiv\Lambda^3\Sigma_{ij}.
\end{align}
Now, it is clear that $\Sigma_{ij}=\delta_{ij}$ if and only if $L=R$, which implies that the scalar quark condensate is only invariant under the $V=L+R$ (called the ``isospin") symmetry transformation.  Another way to state this is that a non-zero value for the condensate implies that the symmetry group of the ground state\footnote{The scalar quark condensate has the same quantum numbers as the vacuum} is smaller than the symmetry group of the Lagrangian,
\begin{align}
\la\overline{q}(x)q(x)\ra\neq0\implies\text{SU}(2)_L\otimes\text{SU}(2)_R\rightarrow\text{SU}(2)_V.
\end{align}
Thus, the ground state of the strong force is only invariant under SU(2)$_V\otimes$U(1)$_V$.  Goldstone's theorem \citep{Nambu:1960xd,Goldstone:1961eq} then states that the broken generators manifest as pseudoscalar bosons with small (non-zero due to the explicit breaking of \eq{divamua}) mass.  These criterion are of course met by the pions.  Note that physically, a scalar quark condensate means that it is more energetically favorable for the vacuum to be constantly fluctuating into $\overline{q}q$ pairs.  As this document is not meant to be be a complete review of QCD, the interested reader is encouraged to see more group-theoretical proofs of these statements in the already-mentioned Ref. \citep{Scherer:2002tk}.

\section[Chiral Perturbation Theory for Mesons]{\label{sec:chpt}$\chi$PT for Mesons}

\subsection{Leading Order Kinetic Term}
We have seen that the pions arise as the Goldstone bosons of the spontaneous symmetry breaking of $\text{SU}(2)_L\otimes\text{SU}(2)_R$.  To be more specific, the pion fields are fluctuations of the $\overline{q}q$ operator about the vacuum.  Therefore, when including them in the EFT Lagrangian, one must ensure that the pions transform in the same way as $\overline{q}q$ under the full symmetry group $G$ of \cref{eq:symgrp}.  The vacuum is defined by the (non-zero) chiral condensate which remains invariant under the subgroup $\text{SU}(2)_V$.  The physical pion states fill out the baryon-number-zero, isospin-one representation of the $\text{SU}(2)_V\otimes\text{U}(1)_V$ symmetry group.

The pion fields can be written as a $2\times2$ special unitary matrix $U$ with the correct transformation properties,
\begin{align}
U(x)\rightarrow RU(x)L^\dagger.
\end{align}
Although there is freedom in how $U$ is parametrized, the common convention is chosen here,
\begin{align}
U(x)=\exp\left(i\frac{\pi(x)}{f_\pi}\right),
\end{align}
where $\pi(x)=\sum_a\pi_a(x)\tau_a$ contains the pion fields, and $f_\pi=93$ MeV is the pion decay constant.  Note that in terms of the charge states, the pion matrix is given by\footnote{The charge states are formed to satisfy
\begin{align}
\left[\begin{pmatrix}+2/3 & 0\\0 & -1/2\end{pmatrix},\begin{pmatrix}\pi^0 & \sqrt{2}\pi^+\\ \sqrt{2}\pi^- & -\pi^0\end{pmatrix}\right]=\begin{pmatrix}0\times\pi^0 & +1\times\sqrt{2}\pi^+\\-1\times\sqrt{2}\pi^- & 0\times-\pi^0\end{pmatrix}
\end{align}}
\begin{align}
\pi=\begin{pmatrix}\pi^0 & \sqrt{2}\pi^+\\ \sqrt{2}\pi^- & -\pi^0\end{pmatrix},
\end{align}
so that in Cartesian coordinates $\pi^\pm=(\pi_1\pm\pi_2)/\sqrt{2}$.  The value of $f_\pi$ is experimentally determined from the leptonic decay $\pi\rightarrow\mu\nu$,
\begin{align}\bs
\la\mu^-(k)\overline{\nu}_\mu(q)|H|\pi^-(p)\ra&=\la0|\overline{u}\gamma_\mu(1-\gamma_5)d|\pi^-\ra\times{\cal M}_\text{EW}^\mu
\\
&=if_\pi p_\mu\times{\cal M}_\text{EW}^\mu
\es\end{align}
where ${\cal M}_\text{EW}^\mu$ represents the perturbatively calculable electroweak contribution to the process.

In order to construct Chiral Perturbation Theory, one follows the principle that any term consistent with the symmetry group $G$ of \eq{symgrp} must appear in the Lagrangian.  All such terms will contain derivatives $\partial_\mu U$ which will scale like the {\it typical} (also referred to as {\it external}) momentum, $p_\text{ext}$.  Recall that an effective theory is only an approximation to a more complete high energy theory.  Here this means that $\chi$PT is only valid for low energy processes with $p_\text{ext}\ll\Lambda_\chi$, where $\Lambda_\chi\sim1$ GeV is the energy scale at which chiral symmetry is restored.  Therefore, one is able to order the terms in the Lagrangian according to powers of $\chi\equiv p_\text{ext}/\Lambda_\chi.$\footnote{Practically, one expands in powers of $p_\text{ext}/4\pi f_\pi$, since factors of $4\pi f_\pi\sim\Lambda_\chi$ commonly appear in loop calculations}  This organization scheme is called {\it power counting}, and is central to $\chi$PT.

Although terms with mass dimension greater than four most certainly appear in the Lagrangian, $\chi$PT remains renormalizable order-by-order in the expansion.  In other words, at next-to-leading order (NLO), loops built from leading-order (LO) interactions are renormalized by NLO tree-level diagrams.  This procedure was worked out in great detail by Gasser \& Leutweyler \citep{Gasser:1983yg}.  A well-thought-out formalism for power counting particular diagrams exists \citep{Weinberg:1978kz}, but will not be presented at this time because a modified power counting scheme is necessary for pion production reactions.

The leading order Lagrangian consists of a single term with two derivatives to which all other possible forms can be reduced (up to unphysical total derivatives),
\begin{align}
{\cal L}=\frac{f_\pi^2}{4}\text{Tr}\left(\partial_\mu U\partial^\mu U^\dagger\right)+...,\label{eq:l2kin}
\end{align}
where the prefactor was included to ensure consistency of the kinetic terms with the Lagrangian for free scalars: $\frac{1}{2}\partial_\mu\phi\partial^\mu\phi$.  To see this, expand out \eq{l2kin} and look at the quadratic terms,
\begin{align}\bs
{\cal L}&=\frac{f_\pi^2}{4}\text{Tr}\left(i\frac{\partial_\mu(\pi_a\tau_a)}{f_\pi}\times(-i)\frac{\partial^\mu(\pi_b\tau_b)}{f_\pi}+...\right)+...
\\
&=\frac{1}{4}\partial_\mu\pi_a\partial^\mu\pi_b\text{Tr}(\tau_a\tau_b)+...
\\
&=\frac{1}{2}\partial_\mu\pi_a\partial^\mu\pi_a+...
\es\end{align}
Nevertheless, this is not the complete leading order Lagrangian because the quark masses, which in reality are non-zero, have not yet been included.

\subsection{Inclusion of Quark Masses}
One is not able to use the same symmetry principle to include the quark mass term because the quark masses explicitly break chiral symmetry: $\overline{q}_RMq_L\rightarrow\overline{q}_RR^\dagger MLq_L$.  However, if one {\it pretended} that the mass matrix transformed like $M\rightarrow RML^\dagger$, then the mass term {\it would} be invariant.  Thus one constructs a corresponding chrially invariant term in $\chi$PT,
\begin{align}
{\cal L}\supset\frac{f_\pi^2B}{2}\text{Tr}(MU^\dagger+UM^\dagger),\label{eq:l2mass}
\end{align}
where $\supset$ is understood to mean ``includes", the $f_\pi^2/2$ was included for convenience (see below) and $B$ is a mass scale used to make the term have correct mass dimension.  Note that the antisymmetric combination of $M$ and $U$ is excluded by parity\footnote{Under parity $\pi(t,{\bf x})\rightarrow-\pi(t,-{\bf x})$, so $U(t,{\bf x})\rightarrow U^\dagger(t,-{\bf x})$.}.  Again, by expanding \eq{l2mass}, one is able to connect $B$ with the mass of the pion (here the isospin limit $m_u=m_d\equiv m_q$ is employed; isospin breaking will be considered in later chapters),
\begin{align}
\frac{f_\pi^2B}{2}\text{Tr}(MU^\dagger+UM^\dagger)=\text{const}-Bm_q\pi_a^2+...
\end{align}
Identifying the quadratic term with the mass, it is clear that $m_\pi^2/2=Bm_q$.  As might be expected, $B$ can also be shown to be proportional to the aforementioned scalar quark condensate \citep{Scherer:2002tk}.

As a brief aside, although flavor SU(3) has not been considered in this chapter as a good symmetry, one is free to do this by replacing the $2\times2$ pion matrix with the 3$\times$3 meson matrix,
\begin{align}
\phi=\sum_{a=1}^8\phi_a\lambda_a=\begin{pmatrix}\pi^0+\frac{\eta}{\sqrt{3}} & \sqrt{2}\pi^+ & \sqrt{2}K^+ \\ \sqrt{2}\pi^- & -\pi^0+\frac{\eta}{\sqrt{3}} & \sqrt{2}K^0 \\ \sqrt{2}K^- & \sqrt{2}\overline{K}^0 & -\frac{2\eta}{\sqrt{3}}\end{pmatrix}.
\end{align}
Performing an expansion and looking at the quadratic terms, one finds the following expressions for the meson masses,
\begin{equation}\label{eq:mesonmass}
\begin{aligned}
m_\pi^2&=B(m_u+m_d) &\qquad m_{K^\pm}^2&=B(m_s+m_u)
\\
m_{\eta}^2&=\frac{B}{3}(4m_s+m_u+m_d) & m_{K^0}^2&=B(m_s+m_d).
\end{aligned}
\end{equation}
These equations can be combined to obtain a rough estimate for the light quark mass $\hat{m}=(m_u+m_d)/2$ in terms of meson masses and the strange quark mass\footnote{The strange quark mass can, in turn, be determined by a variety of methods; perhaps the most common modern method is lattice hadron spectroscopy with non-perturbative renormalization.  For example, see Ref. \citep{Allton:2008pn}.},
\begin{align}
\frac{\hat{m}}{m_s}=\frac{m_\pi^2}{2m_K^2-m_\pi^2}.
\end{align}
Finally, it is interesting to look at the mass splitting between the charged and neutral mesons.  To first order, there is no quark mass contribution to this splitting for the pions: $m_{\pi^+}^2-m_{\pi^0}^2=\delta_\text{EM}+0$.  However, this electromagnetic splitting should be the same for the kaons, which {\it do} have a quark mass contribution,
\begin{align}
m_{K^+}^2-m_{K^0}^2=\delta_\text{EM}+B(m_u-m_d).
\end{align}
Thus, using \eq{mesonmass} one can derive the following relations,
\begin{align}\bs
\frac{m_u}{m_d}&=\frac{m_{K^+}^2-m_{K^0}^2-\left(m_{\pi^+}^2-2m_{\pi^0}^2\right)}{m_{K^0}^2-m_{K^+}^2+m_{\pi^+}^2}=0.56
\\
\frac{m_s}{m_d}&=\frac{m_{K^0}^2+m_{K^+}^2-m_{\pi^+}^2}{m_{K^0}^2-m_{K^+}^2+m_{\pi^+}^2}=20.1.
\es\end{align}
Note that these relations are not exact for two reasons: they were derived at leading order in the EFT, and the EFT itself is based on an inexact symmetry.

Let us now return to the construction of the $\chi$PT Lagrangian.  It is important to note that the square of the pion mass is proportional to the quark mass.  For low energy processes, one has $m_\pi\sim p_\text{ext}$ and thus $m_\pi/\Lambda_\chi$ is considered perturbatively small.  This means that a single insertion of the quark mass matrix is counted as being ``of the same order" as $p_\text{ext}^2$.  Therefore the mass term of \eq{l2mass} is included in the leading order Lagrangian,
\begin{align}
{\cal L}^{(2)}=\frac{f_\pi^2}{4}\text{Tr}\left(\partial_\mu U\partial^\mu U^\dagger\right)+\frac{f_\pi^2B}{2}\text{Tr}\left(MU^\dagger+UM^\dagger\right).\label{eq:chptl2}
\end{align}
where the superscript $(2)$ refers to the fact that each of the terms in this expression scale like $p_\text{ext}^2$.  Once again, note that this expression is {\it not} invariant under chiral symmetry (due to the mass terms), but that the breaking of chiral symmetry is motivated by the way in which QCD breaks chiral symmetry.

\subsection{Beyond Leading Order: Low Energy Constants}
The building blocks of the mesonic $\chi$PT Lagrangian are $U$, $\partial_\mu U$, and $M$, and all terms consistent with the symmetry group $G$ must appear at some order.  It is interesting to note that there are no terms of order $p_\text{ext}^3$ because all the Lorentz indices must contract and $M\sim p_\text{ext}^2$.  To further illustrate the EFT principle, the NLO mesonic Lagrangian \citep{Gasser:1984gg} is displayed,
\begin{align}\bs
{\cal L}^{(4)}={}&L_1\left\{\text{Tr}\left[D_\mu U(D^\mu U)^\dagger\right]\right\}^2+L_2\text{Tr}\left[D_\mu U(D_\nu U)^\dagger\right]\text{Tr}\left[D^\mu U(D^\nu U)^\dagger\right]
\\
{}&+L_3\text{Tr}\left[D_\mu U(D^\mu U)^\dagger D_\nu U(D^\nu U)^\dagger\right]+2BL_4\text{Tr}\left[D_\mu U(D^\mu U)^\dagger\right]\text{Tr}\left[MU^\dagger+UM^\dagger\right]
\\
{}&+2BL_5\text{Tr}\left[D_\mu U(D^\mu U)^\dagger\left(MU^\dagger+UM^\dagger\right)\right]+4B^2L_6\left\{\text{Tr}\left[MU^\dagger+UM^\dagger\right]\right\}^2
\\
{}&+4B^2L_7\left\{\text{Tr}\left[MU^\dagger-UM^\dagger\right]\right\}^2+4B^2L_8\text{Tr}\left[UM^\dagger UM^\dagger+MU^\dagger MU^\dagger\right],\label{eq:chptl4}
\es\end{align}
where terms without $U$ have been ignored because they are unphysical.  In \eq{chptl4}, the $L_i$ are referred to as Low Energy Constants (LECs).  It is in these LECs that $\chi$PT rises or falls as a predictive theory.  The idea is that one can calculate a quantity (say, the $\pi\pi$ scattering length) to one-loop order using the vertices of ${\cal L}^{(2)}$ and absorbing the divergences into the operable LECs from ${\cal L}^{(4)}$.  The renormalized values of the LECs are then either obtained from experiment or calculated with LQCD.

External fields (such as the electromagnetic field) can be included in the QCD Lagrangian by adding the terms\footnote{Iso-scalar vector, scalar, and pseudoscalar terms can be included similarly.} $v_\mu=\sum_iv^i_\mu\overline{q}\gamma^\mu\frac{\lambda^i}{2}q$ and $a_\mu=\sum_ia^i_\mu\overline{q}\gamma^\mu\gamma_5\frac{\lambda^i}{2}q$.  When working in terms of right- and left-handed quark fields, the currents are rewritten as $v_\mu=r_\mu+l_\mu$, $a_\mu=r_\mu-l_\mu$; where $r_\mu\to Rr_\mu R^\dagger-i\partial_\mu RR^\dagger$ , $l_\mu\to Ll_\mu L^\dagger-i\partial_\mu LL^\dagger$ under a {\it local} chiral symmetry transformation.  The derivatives in \cref{eq:chptl4} have been written with a capital $D$ to indicate that these external fields can be included in $\chi$PT by using a covariant derivative $\partial_\mu U\to D_\mu U\equiv\partial_\mu U-ir_\mu U+iUl_\mu$ which transforms exactly as $\partial_\mu U$ does.

In the sense that $\chi$PT is a low energy theory which does not consider heavy quarks (or even baryons!) explicitly, the LECs are sometime referred to as ``contact terms."  As the language implies, effects of heavy particles occur on short time scales and when one probes a system with a low energy, such time scales cannot be resolved but instead appear shrunk to a (contact) point. By analogy, in Fermi's electroweak EFT, an interaction of two electrons via a $W$ boson is expressed as a single 4-pt vertex with $G_F$ as the LEC.

Of course, the procedure described following \eq{chptl4} does nothing to validate the theory; further predictions must be made.  Indeed, distinct combinations of LECs contribute to several experimentally independent observables including the pion mass, polarizability, and decay constant; the $\gamma\gamma\to\pi^0\pi^0$ reaction; and low energy pion-pion scattering.  As long as the renormalized LECs not so large as to prohibit the use of perturbation theory, the theory is generally accepted as useful.  This is clearly found to be the case for two-flavor mesonic $\chi$PT.  A modern survey of the statements in this paragraph can be found in Ref. \citep{Bijnens:2006zp}.

\section[Chiral Perturbation Theory for Baryons]{\label{sec:hbchpt}$\chi$PT for Baryons}

\subsection{Na\"ive Leading Order Lagrangian}
The next task of this chapter is to introduce baryons to $\chi$PT (creating B$\chi$PT).   Because their masses are nearly identical, the nucleons are placed into a isospin doublet,
\begin{align}
\psi(x)=\begin{pmatrix}p(x)\\n(x)\end{pmatrix}.
\end{align}
It is known phenomenologically that pions mediate the long-range part of the nucleon-nucleon interaction \citep{Yukawa:1935xg}.  It is useful to include this behavior when adding nucleons to $\chi$PT.  This can be accomplished by gauging the nucleon kinetic term, promoting the chiral transformation from global to local.  The local nature of the transformation comes from the pion fields by including $U(x)$ in the manner described below.  Because one is free to perform field redefinitions, it is useful to choose a basis such that the nucleon doublet transforms like $\psi\rightarrow K(L,R,U(x))\psi$, where $K$ is some undefined function of the $L$ and $R$ chiral transformations, in addition to $U(x)$.  The correct pion-nucleon interaction will then make the covariant derivative, $D_\mu \psi$, transform like the nucleon itself: $D_\mu \psi\rightarrow KD_\mu\psi$.  The expressions which satisfy these constraints are,
\begin{align}
D_\mu \psi&=\left[\partial_\mu+\frac{1}{2}\left(u^\dagger\partial_\mu u+u\partial_\mu u^\dagger\right)\right]\psi
\\
u(x)&\rightarrow Ru(x)K^{-1}(x)\label{eq:utrans}
\\
u(x)&\equiv\sqrt{U(x)}=e^{i\pi(x)/2f_\pi}.\label{eq:sqrtdef}
\end{align}
From \cref{eq:utrans,eq:sqrtdef} one finds that $K=\sqrt{RUL^\dagger}^{-1}R\sqrt{U}$.  To reproduce this parametrization, use the field redefinition $\psi'_L=u\psi_L$, $\psi'_R=u^\dagger\psi_R$ in the free form $\slashed{\partial}\psi$.  Though this definition may seem esoteric, it permits a simple construction of chirally invariant interactions.  Additionally, note that this parametrization preserves the necessary transformation under the SU(2)$_V$ subgroup, $\psi\rightarrow V\psi$, since for $L=R=V$ one finds that $K=V$.  One final observation regarding the kinetic term is that by parity the pion-nucleon interaction only contains terms with even numbers of pion fields.

Next, one constructs another chirally invariant term built from the $u(x)$.  Define,
\begin{align}
u_\mu=i\left(u^\dagger\partial_\mu u-u\partial_\mu u^\dagger\right),
\end{align}
which transforms like $u_\mu\rightarrow Ku_\mu K^\dagger$.  This term is odd under parity so one must include a $\gamma_5$ in addition to contracting with $\gamma_\mu$ to make it a scalar.  Putting all the pieces together, the leading order Lagrangian is written,
\begin{align}
{\cal L}^{(1)}=\overline{\psi}\left(i\slashed{D}-m_N+\frac{g_A}{2}\gamma^\mu\gamma_5u_\mu\right)\psi,\label{eq:chptl1}
\end{align}
where $m_N\approx1$ GeV is the nucleon mass and $g_A\approx1.27$ is the axial-vector coupling.

\begin{figure}
\centering
\includegraphics[height=1in]{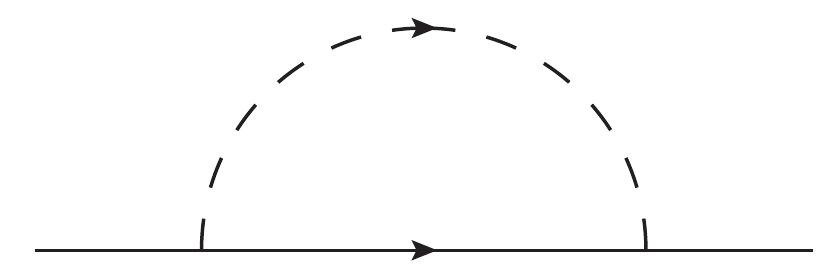}
\caption{\label{fig:sunset}One of the loop contributions to the nucleon mass.}
\end{figure}

The key difference between this construction and that of mesonic $\chi$PT is the introduction of a new mass scale $m_N$ which does not vanish in the chiral limit.  Schematically, this statement means that $\partial_0\psi\sim m_N\psi$.  This introduces a subtlety to the power counting scheme in which loops with nucleons contain terms that are effectively promoted to lower order.  As an example, it can be shown that in the $\overline{\text{MS}}$ scheme, the (subtracted) diagram of Fig. \ref{fig:sunset} (which contributes to the renormalized nucleon mass) contains both a term proportional to $g_A^2m_\pi^3/f_\pi^2$ and one proportional to $g_A^2m_\pi^2m_N/f_\pi^2$ \citep{Scherer:2002tk}.
This is a problem if one is to assign the diagram a particular order in the power counting scheme.

\subsection{\label{sec:HBChPT}Heavy Baryon Formalism}
One way to solve the power counting difficulty brought about by the nucleon mass is to remove it by performing the field redefinition $\psi\rightarrow e^{im_Nt}\psi$ and then integrating out the heavy components of the nucleon field.  The goal of this procedure, introduced by Ref. \citep{Jenkins:1990jv}, is to produce nucleons that obey $\partial_\mu N\sim p_\text{ext}N$ at the cost of new effective interactions that are suppressed by powers of $1/m_N$.  As shown below, the remaining dynamical field propagates (in the absence of interactions) with four-momentum $k_\mu=\left(E(\bfp)-m_N,\bfp\right)$.  Since B$\chi$PT is only to be used for low momentum transfer processes, it should then be a good approximation to assume that $k_\mu\ll m_N$ for all $\mu$.  The end result is a new EFT, Heavy Baryon Chiral Perturbation Theory (HB$\chi$PT), that is an expansion in both $p_\text{ext}/\Lambda_\chi$ and $p_\text{ext}/m_N$.

To develop HB$\chi$PT, first define projection operators,
\begin{align}
P_{v\pm}=\frac{1\pm\slashed{v}}{2},
\end{align}
which satisfy $P_{v+}+P_{v-}=1$, $P_{v\pm}^2=P_{v\pm}$, and $P_{v\pm}P_{v\mp}=0$ with the requirements that $v^2=1$ and $v_0\ge1$ satisfied by the choice $v=(1,{\bf 0})$ that is used in this work.  This choice makes it clear that one is separating out the heavy and light components the free nucleon spinor,
\begin{align}
P_{v+}\psi^{(+)}\stackrel{v=(1,{\bf 0})}{=}\begin{pmatrix}1 & 0 \\ 0 & 0 \end{pmatrix}\sqrt{E+m}\begin{pmatrix}\chi\\ \frac{\bsig\cdot\bfp}{E+m}\chi\end{pmatrix}e^{-ip\cdot x}.
\end{align}
Now one makes the field redefinition/split,
\begin{alignat}{2}
N(x)&=e^{im_Nv\cdot x}P_{v+}\psi(x), &\qquad H(x)&=e^{im_Nv\cdot x}P_{v-}\psi(x).\label{eq:psiredef}
\end{alignat}
Note that $\slashed{v}N(x)=N(x)$ and $\slashed{v}H(x)=-H(x)$.  The next step is to use the equations of motion to eliminate the heavy field $H(x)$.  The theory will then consist of solely the light field $N(x)$ which propagates with $\exp(-i(p-m_Nv)\cdot x)$.  This residual momentum will be referred to as $k_\mu$,
\begin{align}
p_\mu&\equiv m_Nv_\mu+k_\mu\label{eq:residualmom}
\\
v\cdot k&= -\frac{k^2}{2m_N},\label{eq:k0counting}
\end{align}
where the latter is found by squaring the former.

The Euler-Lagrange equations give the equation of motion for the leading order Lagrangian in \eq{chptl1},
\begin{align}
\frac{\partial{\cal L}^{(1)}}{\partial\psi}-\partial_\mu\frac{\partial{\cal L}^{(1)}}{\partial\left(\partial_\mu\psi\right)}=-\left(i\slashed{D}-m_N+\frac{g_A}{2}\gamma^\mu\gamma_5u_\mu\right)\psi(x)=0.\label{eq:eleqn}
\end{align}
Inverting \eq{psiredef}, inserting the resulting expression for $\psi$ into \eq{eleqn}, and multiplying by $-\exp(im_Nv\cdot x)$, one finds
\begin{align}
\left(m_N\slashed{v}+i\slashed{D}-m_N+\frac{g_A}{2}\gamma^\mu\gamma_5u_\mu\right)(N(x)+H(x))=0.\label{eq:neweom}
\end{align}
Next, left multiply \eq{neweom} first by $P_{v+}$ and second by $P_{v-}$ to create two differential equations.  After several steps of algebra, one obtains
\begin{align}\bs
\left(iv\cdot D+\frac{g_A}{2}\slashed{u}_\perp\gamma_5\right)N(x)+\left(i\slashed{D}_\perp+\frac{g_A}{2}v\cdot u\gamma_5\right)H(x)&=0
\\
\left(i\slashed{D}_\perp-\frac{g_A}{2}v\cdot u\gamma_5\right)N(x)+\left(-iv\cdot D-2m_N+\frac{g_A}{2}\slashed{u}_\perp\gamma_5\right)H(x)&=0,
\es\end{align}
where $V_{\perp\mu}=V_\mu-v\cdot Vv_\mu$ and various relations involving $v$ have been used; for example $v\cdot DH(x)=-v\cdot D\slashed{v}H(x)$.  Now the heavy field $H(x)$ is eliminated to find
\begin{align}\bs
{}&\left[\left(iv\cdot D+\frac{g_A}{2}\slashed{u}_\perp\gamma_5\right)\right.
\\
{}&+\left.\left(i\slashed{D}_\perp+\frac{g_A}{2}v\cdot u\gamma_5\right)\left(iv\cdot D+2m_N-\frac{g_A}{2}\slashed{u}_\perp\gamma_5\right)^{-1}\left(i\slashed{D}_\perp-\frac{g_A}{2}v\cdot u\gamma_5\right)\right]N(x)=0.
\es\label{eq:decoupled}\end{align}
Though the proof is technical (see Ref. \citep{Scherer:2002tk}), it can be shown that these nucleon spinors are normalized differently than the relativistic ones, such that one needs to multiply correlation functions by the normalization factor $\sqrt{E/m_Nv_0}$ for each external nucleon.

All that remains is to organize \eq{decoupled} into an expansion in $(2m_N)^{-1}$ and to manipulate the expression into a more commonly used form.  It is clear now that the derivatives in the denominator of \eq{decoupled} will produce small quantities and thus one is free to perform the expansion.  It is important to note here that this equation of motion can be generated from a Lagrangian identical to the LHS of \eq{decoupled} left multiplied by $\overline{N}$.  Nevertheless, this Lagrangian does not contain all the necessary terms for the EFT because it was obtained from simply the leading order relativistic equation of motion.  Other terms consistent with the symmetries must be included as well.  With that in mind, one proceeds by writing
\begin{align}
{\cal L}\supset\overline{N}\left[iv\cdot D+\frac{g_A}{2}\slashed{u}_\perp\gamma_5+\sum_{n=1}^{\infty}\frac{{\cal O}_n}{(2m_N)^n}\right]N.
\end{align}

The first $(2m_N)^{-1}$ correction (called a {\it recoil} correction) is given by,
\begin{align}
{\cal O}_1=-\slashed{D}_\perp^2-i\frac{g_A}{2}\left\{\slashed{D}_\perp\gamma_5,v\cdot u\right\}-\frac{g_A^2}{4}\left(v\cdot u\gamma_5\right)^2.
\end{align}
Note that there are two derivatives in each of these terms and regardless of whether they act on nucleons or pions, they will bring down factors of $p_\text{ext}$ making these terms ${\cal O}(p_\text{ext}^2)$.  These expressions can be dramatically simplified using the fact that $N=P_{v+}N$.  Because these NLO interactions will be heavily used throughout this thesis, a bit more time is devoted now to working on them.  First, consider the $\slashed{D}_\perp^2$ term.
\begin{align}\bs
\slashed{D}_\perp^2&=\left(g^{\mu\nu}-i\sigma^{\mu\nu}\right)D_{\perp\mu}D_{\perp\nu}
\\
&=D^2-(v\cdot D)^2-i\sigma^{\mu\nu}(D_\mu-v\cdot Dv_\mu)(D_\nu-v\cdot D_\nu)
\\
&\rightarrow D^2-(v\cdot D)^2-\frac{i\sigma^{\mu\nu}}{2}\left[D_\mu,D_\nu\right],
\es\end{align}
where $\sigma^{\mu\nu}=\frac{i}{2}\left[\gamma^\mu,\gamma^\nu\right]$.  The commutator of the covariant derivatives can be expressed in terms of the $u_\mu$ objects by use of $\partial_\mu(uu^\dagger)=0$ and deletion of terms symmetric in $\mu\leftrightarrow\nu$ (due to its contraction with $\sigma^{\mu\nu}$),
\begin{align}
[D_\mu,D_\nu]=\frac{1}{2}u_\mu u_\nu.
\end{align}
Next, use $v=(1,{\bf 0})$ to obtain,
\begin{align}\bs
{\cal L}\overset{v=(1,{\bf 0})}{=}\overline{N}{}&\left[iD_0-\frac{g_A}{2}\bgamma\cdot{\bf u}\gamma_5\right.
\\
{}&\left.+\frac{1}{2m_N}\left({\bf D}^2+\frac{i\sigma^{\mu\nu}}{4}u_\mu u_\nu+i\frac{g_A}{2}\left\{\bgamma\cdot{\bf D}\gamma_5,u_0\right\}-\frac{g_A^2}{4}u_0^2\right)+...\right]N.
\es\label{eq:l2gallilean}\end{align}
Finally, as the lower components of the nucleon spinors have been projected out (and employing the Dirac basis\footnote{The Dirac basis is defined by: $\gamma^0=\begin{pmatrix}1&0\\0&-1\end{pmatrix},\quad\bgamma^i=\begin{pmatrix}0&\bsig^i\\-\bsig^i&0\end{pmatrix},\quad\gamma_5=\begin{pmatrix}0&1\\1&0\end{pmatrix}$.}), the leading order HB$\chi$PT Lagrangian is given by,
\begin{align}
{\cal L}^{(1)}=N^\dagger\left[iD_0-\frac{g_A}{2}\bsig\cdot{\bf u}\right]N.\label{eq:hbchptl1}
\end{align}
The NLO expression includes the $(2m_N)^{-1}$ terms of \eq{l2gallilean} along with LEC terms originating from the projection of the NLO relativistic Lagrangian,
\begin{align}\bs
{\cal L}^{(2)}={}&N^\dagger\left[\frac{1}{2m_N}\left({\bf D}^2+\frac{i}{4}\epsilon^{ijk}\sigma_iu_ju_k+i\frac{g_A}{2}\left\{\bsig\cdot{\bf D},u_0\right\}-\frac{g_A^2}{4}u_0^2\right)\right.
\\
{}&\left.+2Bc_1\text{Tr}\left(M_+\right)+c_2u_0^2+c_3u\cdot u+c_4[\sigma^i,\sigma^j]u_iu_j+2Bc_5\left(M_+-\frac{1}{2}\text{Tr}(M_+)\right)\right]N,\label{eq:hbchptl2}
\es\end{align}
where $M_+=u^\dagger Mu^\dagger+uM^\dagger u$.  The $c_5$ term will later be shown to provide a charge symmetry breaking contribution to pion production.

\subsection{The $\Delta$ Resonance}

As outlined in Ref. \citep{Bernard:1995dp}, there are two reasons why it is important to include the $\Delta(1232)$ resonance as an explicit degree of freedom in HB$\chi$PT.  Firstly, the resonance lies a mere $\sim300$ MeV above the nucleon and, even for low momentum processes, should be expected to play an appreciable role in pion-nucleon dynamics.  Secondly, it is known that the $\pi N\Delta$ coupling is quite strong $g_{\Delta N\pi}\approx2g_{N\pi}$\footnote{In $p$-wave $\pi N$ scattering, the inclusion of the $\Delta$ as an intermediate state, in addition to the na\"{i}ve nucleon pole, dramatically improves the calculation of the $T=J=3/2$ scattering volume \citep{Ericson:1988gk}.}.  In this section a brief outline is given of the rather technical formalism for including the $S=3/2$, $T=3/2$ fields in HB$\chi$PT.

Spin-3/2 fields can be described in quantum field theory with the Rarita-Schwinger formalism \citep{Rarita:1941mf} that considers the tensor product of spin-1 and spin-1/2 fields, $\psi_\mu$.  The equation of motion is given by,
\begin{align}
\left(i\slashed{\partial}-m_\Delta\right)\psi_\mu(x)=0,
\end{align}
where the spin-1/2 component is removed with the constraint $\gamma^\mu\psi_\mu=0$.  Additionally, because the $\Delta$ is an isospin-3/2 particle, one considers the isospin doublets $\psi_\mu^i(x)$ for $i=1,2,3$ with the constraint $\tau^i\psi_\mu^i(x)=0$ projecting out the two unwanted degrees of freedom.  In this thesis the representation advocated by \citep{Hemmert:1997ye} is used,
\begin{align}\bs
\psi_\mu^1&=\frac{1}{\sqrt{2}}\begin{pmatrix}\Delta^{++}-\frac{1}{\sqrt{3}}\Delta^0 \\ \frac{1}{\sqrt{3}}\Delta^+-\Delta^-\end{pmatrix}_\mu
\\
\psi_\mu^2&=\frac{i}{\sqrt{2}}\begin{pmatrix}\Delta^{++}+\frac{1}{\sqrt{3}}\Delta^0 \\ \frac{1}{\sqrt{3}}\Delta^++\Delta^-\end{pmatrix}_\mu
\\
\psi_\mu^3&=\sqrt{\frac{2}{3}}\begin{pmatrix}\Delta^+ \\ \Delta^0\end{pmatrix}_\mu.
\es\end{align}

The general idea is the same as it was for the nucleons, one splits the $\psi_\mu^i$ into heavy and light fields and integrates out the light ones.  The additional difficulty is keeping track of the extra spin and isospin degrees of freedom.  To this end, define the projection operators,
\begin{align}\bs
P_{\mu\nu}^{3/2}&=g_{\mu\nu}-\frac{1}{3}\gamma_\mu\gamma_\nu-\frac{1}{3}\left(\slashed{v}\gamma_\mu v_\nu-v_\mu\gamma_\nu\slashed{v}\right)
\\
\xi_{ij}^{3/2}&=\delta^{ij}-\frac{1}{3}\tau^i\tau^j.
\es\end{align}
Because one wants to describe $N\Delta$ transitions, it is necessary to project out the only the nucleon mass (not the full Delta mass),
\begin{align}
T_\mu^i(x)=e^{im_Nv\cdot x}P_{v+}P_{\mu\nu}^{3/2}\xi_{ij}^{3/2}\psi_\nu^j(x).
\end{align}
Another important consideration is the ``point invariance" required of spin-3/2 fields.  Without going into full detail, this is the statement that the Lagrangian must be invariant under a transformation $\psi_\mu\rightarrow\psi_\mu+a\gamma_\mu\gamma_\nu\psi^\nu$ introducing an arbitrary admixture of spurious spin-1/2 fields.  This invariance is accomplished by the addition of terms into the Lagrangian which depend on the unphysical gauge parameter $A$.  This dependence can be absorbed into a field redefinition with the resulting generalized kinetic term,
\begin{align}
\Lambda_{\mu\nu}=(-i\slashed{\partial}+m_\Delta)g_{\mu\nu}-\frac{1}{4}\gamma_\mu\gamma_\lambda(-i\slashed{\partial}+m_\Delta)\gamma^\lambda\gamma_\nu.
\end{align}

Introducing the pions, one may write down a general expression with a covariant derivative and axial coupling, both of which are rank-2 tensors in spin and isospin \citep{Hemmert:1997ye}.  However, this would unnecessarily complicate matters for the purposes of this thesis.  All reactions considered here will have nucleons as asymptotic states and will not proceed beyond NLO.  Accordingly, diagrams containing $\pi\Delta\Delta$, $\pi\pi\Delta\Delta$, ..., vertices will not appear up to the order considered.  The purpose in describing the Delta lies simply in its propagator and the $\pi N\Delta$ vertex.  Therefore, it is sufficient to display these relevant terms only.

After performing the steps described in Sec. \ref{sec:HBChPT} to eliminate the spin-1/2, isospin-1/2, and heavy spin-3/2 isospin-3/2 components of the full $\psi_\mu^i$, the resulting Lagrangian is once again expressed as in expansion in $(2m_N)^{-1}$.  For the choice $v=(1,{\bf 0})$ one finds,
\begin{align}
{\cal L}_{\Delta\Delta}&=-T_i^{\dagger\mu}\left[i\partial_0-\delta-\frac{\bnabla^2}{2m_N}+...\right]g^{\mu\nu}\delta_{ij}T^\nu_j\label{eq:deltakin}
\\
{\cal L}_{\pi N\Delta}^{(1)}&=\frac{h_A}{2}\left[T_i^{\dagger\mu}g_{\mu\nu}w_i^\nu N+N^\dagger w_i^{\nu\dagger}g_{\nu\mu} T_i^\mu\right]\label{eq:lpiND1}
\\
{\cal L}_{\pi N\Delta}^{(2)}&=\frac{-1}{2m_N}\left[T_i^{\dagger\mu}h_Ai\partial_\mu w_j^0N+h.c\right]+...,\label{eq:lpiND2}
\end{align}
where $\delta=m_\Delta-m_N=293$ MeV, $w_i^\mu=\frac{1}{2}\text{Tr}\left(\tau_iu_\mu\right)$, $h_A$ is the analog of $g_A$ for the $N\Delta$ transition, and $h.c.$ represents the Hermitian conjugate of the preceding term.  In the calculations that follow, the value $h_A=2.1g_A$ is adopted based on $p$-wave scattering data summarized in Ref. \citep{Ericson:1988gk}.  Finally, note that in \eq{lpiND2}, terms with even numbers of pions and couterterms were left out, again due to their absence in the present calculations.

\section{\label{sec:csbops}Charge Symmetry Breaking}
In this section, Charge Symmetry (CS) and its breaking (CSB) are defined.  Examples are given of how CS(B) manifests in nature, as is a discussion of its inclusion in $\chi$PT.  A modern review of this topic can be found in Ref. \citep{Miller:2006tv}.  

\subsection{Definition of Charge Symmetry}
Charge symmetry is defined as the approximate invariance of a hadronic system or reaction under an isospin rotation of $\pi$ about the $y$-axis.\footnote{The $y$-axis is used because the $z$-axis of isospin is related to the electric charge, and rotating by $180^\circ$ about the $y$-axis modifies the charge.}   Formally, this means that the equation,
\begin{align}
[H_S,P_\text{CS}]=0,\label{eq:cs}
\end{align}
is satisfied, where $H_S$ is the strong Hamiltonian, and $P_\text{CS}=e^{i\pi\tau_2/2}$ is the charge symmetry operator.  A simple manifestation of CSB is the degeneracy of the nucleon doublet: $P_\text{CS}|p\ra=-|n\ra$, and $P_\text{CS}|n\ra=|p\ra$ with $m_n\approx m_p$.

\begin{figure}
\centering
\includegraphics[height=3in]{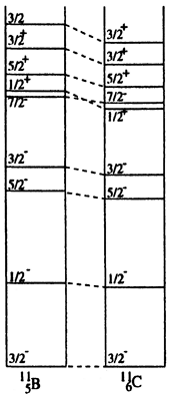}
\caption[Example of Charge Symmetry]{\label{fig:csspectra}Example of CS: spectra of two members of the $A=11$ multiplet.  The notation is $\text{Spin}^\text{Parity}$.  Taken from Ref. \citep{Wong:1998ex}.}
\end{figure}

The meaning of the name ``charge" symmetry becomes more clear when considering nuclei with large nucleon number, $A$.  The charge of a nucleus is proportional to the number of protons, $Z$; if $n$ of those protons were turned into neutrons, the charge would become proportional to $Z-n$.  However, if \cref{eq:cs} is satisfied, the nuclear properties (i.e. excitation energies, see \cref{fig:csspectra}) will remain the same.

A distinction which needs to be made is that charge symmetry is not the same as isospin invariance.  This confusion is understandable since the charge and isospin operators are related, $Q=e(1/2+T_3)$ (for nucleons).  However, charge symmetry refers to the specific situation described by \cref{eq:cs} while isospin invariance is more restrictive,
\begin{align}
[H_S,\btau/2]=0,
\end{align}
where $\btau$ is the total isospin vector of the system.  In the context of the nucleon-nucleon interaction (where isospin invariance is referred to as ``charge independence"), this distinction has been formalized with the definition of different ``classes" of forces \citep{Cheung:1979ma}.  A class II force, for example, maintains charge symmetry but breaks charge independence,
\begin{align}
V_{II}=c[\tau_3(i)\tau_3(j)-\frac{1}{3}\btau(i)\cdot\btau(j)],
\end{align}
where $i$ and $j$ are particle labels.  The Coulomb interaction leads to a class II force due to the fact that the electric charge operator is $Q\propto(1+\tau_3)$.

\subsection{Manifestations of CSB}
Charge Symmetry Breaking has two sources, the light quark mass difference and electromagnetism \citep{Slaus:1990nn}.  One example where the light quark mass difference dominates is the splitting of the neutron and proton masses,
\begin{align}
\frac{m_n-m_p}{(m_n+m_p)/2}=\frac{1.29\text{ MeV}}{938.92\text{ MeV}}=0.1\%.
\end{align}
If the source of this CSB were electromagnetism, one should expect the proton to be more massive.  Therefore, one attributes the $0.1\%$, {\it plus} the electromagnetic breaking, to the fact that $d$ quarks are more massive than $u$ quarks.  Another example of CSB is the $T=1$ $nn$ ($\sim-19\text{ MeV}$) and $pp$ ($\sim-17\text{ MeV}$) scattering lengths (after models are used to remove electromagnetic effects).  Reference \citep{Coon:1987kt} argues that the primary source of this difference is the vector meson $\rho-\omega$ mixing due to the non-zero value of $m_d-m_u$.

Isospin violation is typically dominated by electromagnetism \citep{Miller:1990iz}.  For example, consider the $T=1$ $nn$ ($\sim-19\text{ fm}$) and $np$ ($\sim-24\text{ fm}$) scattering lengths (again, after models are used to remove any residual electromagnetic effects), which at first glance would not seem to be affected by electromagnetism.  As shown in Ref. \citep{Ericson:1983vw}, the primary source of this difference is the the mass difference between the charged and neutral mesons\footnote{This mass difference, $m_{\pi^+}-m_{\pi^0}=[139.6-135.0]\text{ MeV}=4.6\text{ MeV}$, is clearly due primarily to electromagnetism since each particle has the same number of $u$ and $d$ valence quarks.} which mediate the nucleon-nucleon ($NN$) interaction.

Although there are a multitude of examples of CSB in hadronic physics (\cref{chap:csb} will discuss the angular distribution of the differential cross section of $np\to d\pi^0$ in great detail), the task of incorporating the light quark mass difference and electromagnetic CSB in $\chi$PT will be the focus of the remainder of this chapter.

\subsection{CSB in $\chi$PT}
The inclusion of CSB in $\chi$PT was worked out by Ref. \citep{vanKolck:1995cb}.  Let us begin by discussing CSB resulting from the light quark mass difference.  Reference \citep{vanKolck:1995cb} begins by noting that in the absence of strangeness, chiral symmetry is described by the group $\text{SU}(2)\otimes\text{SU}(2)\sim\text{SO}(4)$.  The quark mass difference term in the QCD Lagrangian is then identified as the third component of the $\text{SO}(4)$ vector $P=(-\overline{q}\btau q,\overline{q}i\gamma_5q)$,
\begin{align}\bs
{\cal L}^\text{QCD}&\supset-\bpm\ubar\\\dbar\epm\bpm m_u&0\\0&m_d\epm\bpm u\\d\epm
\\
&=-\frac{m_u+m_d}{2}\bpm\ubar\\\dbar\epm\bpm 1&0\\0&1\epm\bpm u\\d\epm-\frac{m_u-m_d}{2}\bpm\ubar\\\dbar\epm\bpm 1&0\\0&-1\epm\bpm u\\d\epm.
\es\end{align}
A term with the same transformation properties must appear in $\chi$PT, and the corresponding LEC must be proportional to $m_u-m_d$,
\begin{align}
{\cal L}^\text{$\chi$PT}\supset\frac{\delta m_N}{2}\left(\overline{N}\tau_3N-\frac{1}{2f_\pi^2}\overline{N}\pi_3\btau\cdot\bpi N\right),
\end{align}
where $\delta m_N\propto(m_u-m_d)$.  To derive this form, one need not discuss the group theory of $\text{SO}(4)$ since the LEC terms have already been written down in \cref{eq:hbchptl2}.  In that equation, one encounters the object $M_+$ which contains a term proportional to $m_u-m_d$,
\begin{align}\bs
M_+&=u^\dagger Mu^\dagger+h.c.
\\
&=\ldots+\frac{m_u-m_d}{2}\left(1-\frac{i}{2f_\pi}\btau\cdot\bpi-\frac{1}{8f_\pi^2}+\ldots\right)\tau_3\left(1-\frac{i}{2f_\pi}\btau\cdot\bpi-\frac{1}{8f_\pi^2}+\ldots\right)+h.c.
\\
&=\ldots+\frac{m_u-m_d}{2}2\left(\tau_3-\frac{1}{2f_\pi^2}\btau\cdot\bpi\pi_3+\ldots\right),
\es\end{align}
where the first $\ldots$ represents a term proportional to $(m_u+m_d)/2$.  The CSB term is traceless, so only the $c_5$ term of \cref{eq:hbchptl2} will contribute.  Truncating the expansion at two pions (diagrams with four or more pions will be parametrically suppressed) one finds,
\begin{align}
{\cal L}^{(2)}\supset 2Bc_5(m_u-m_d)\left(\tau_3-\frac{1}{2f_\pi^2}\btau\cdot\bpi\pi_3\right),
\end{align}
which tells us that $\delta m_N=4Bc_5(m_u-m_d)$.  Since $B(m_u+m_d)=m_\pi^2$, a final useful expression is
\begin{align}
\delta m_N=4c_5m_\pi^2\frac{m_u-m_d}{m_u+m_d}.
\end{align}
Sometimes the ratio that appears in this expression is referred to as a parameter $\epsilon=(m_u-m_d)/(m_u+m_d)$ which can be estimated from meson masses in the SU$(3)$ limit: $\epsilon\approx-1/3$.  Note that the $m_\pi^2$ behavior is required as this term is a member of the ${\cal L}^{(2)}$ Lagrangian.

There is also an electromagnetic CSB interaction which comes in at this order parametrized by the LEC $\overline{\delta}m_N$.  Reference \citep{vanKolck:1995cb} describes how hard photon exchanges between quarks break the $\text{SO}(4)$ symmetry in a manner described by the 34 component ($\qbar i\gamma^\mu\tau_3q$) of a rank-2 anti-symmetric tensor,
\begin{align}
T^\mu=\bpm\epsilon_{abc}\qbar i\gamma^\mu\gamma_5\tau_cq&\qbar i\gamma^\mu\tau_aq\\-\qbar i\gamma^\mu\tau_bq&0\epm,
\end{align}
so that the corresponding $\chi$PT term is,
\begin{align}
{\cal L}^{(2)}\supset\frac{\overline{\delta}m_N}{2}\left(\Nbar\tau_3N+\frac{1}{2f_\pi^2}\Nbar(\btau\cdot\bpi\pi_3-\tau_3\bpi^2)N\right).
\end{align}
Since the electromagnetic terms in $\chi$PT were not derived in the above sections, connection with the corresponding LEC, $f_2$, is not made at this time.
\chapter{Pion Production in Nucleon-Nucleon Reactions}
\label{chap:review}
This chapter presents a overview of pion production in nucleon-nucleon reactions.  First, \cref{sec:general} motivates the study of pion production and explains some features of the reverse reaction, deuteron breakup.  In \cref{sec:kinoverview} the thematic large threshold momentum is introduced and various kinematical relations are worked out.  Next, in \cref{sec:selrules} a discussion is devoted to the various reaction channels which are relevant for this work.  \Cref{sec:hybrid} presents the hybrid formalism that is used in subsequent chapters.  Finally, in \cref{sec:history}, a brief summary is given of the historical calculations which are being improved upon.

\section{\label{sec:general}General Remarks}
In this section, an outline is given of some general features of pion production reactions, which have been studied for many years \citep{Measday:1979if}.  Much of this information is drawn from the modern reviews found in Refs. \citep{Ericson:1988gk,Hanhart:2003pg,Moskal:2002jm}.  Since pions are the lightest hadrons, pion production is the inelasticity of lowest energy of the $NN$ interaction and is interesting for several reasons.  As explored in Ref. \citep{Hanhart:2000gp} and more recently in Ref. \citep{Baru:2009fm}, $p$-wave pion production can be used to study three-body nuclear forces.  This connection results from a $\pi NNNN$ contact term which contributes to a variety of quite different physical phenomena.  Secondly, pion production is a gateway to understanding production of more exotic mesons.  One example is the $pp\to pYK$ (where $Y$ is a hyperon and $K$ a kaon) reaction which can be used to extract information about the hyperon-nucleon potential.  Similarly, $NN\to NN\eta$, can be used to study the $NN\to N^*N$ transition due to the pronounced role of the $N^*(1535)$ as an intermediate state.  Other intermediate states can be studied; as will be discussed, the $\Delta(1232)$ makes a large contribution to the $p$-wave $NN\to d\pi$ amplitude.  Another motivation which is explored extensively in Ch. \ref{chap:csb} is the determination of Charge Symmetry Breaking observables, such as the light quark mass difference.  One final motivation that should be mentioned is the usefulness of pion production as a laboratory for the development of $\chi$PT as the EFT for the strong interaction.  The relative momentum of the initial-state nucleons required in order to produce a meson is large, and this fact provides a challenge to $\chi$PT (which is an expansion about small relative momenta); pion production could be considered the ``frontier" of $\chi$PT.  The continued development of $\chi$PT is important in order to better understand nuclear phenomena such as nucleosynthesis, fusion, etc.

\begin{figure}
\centering
\includegraphics[height=3in]{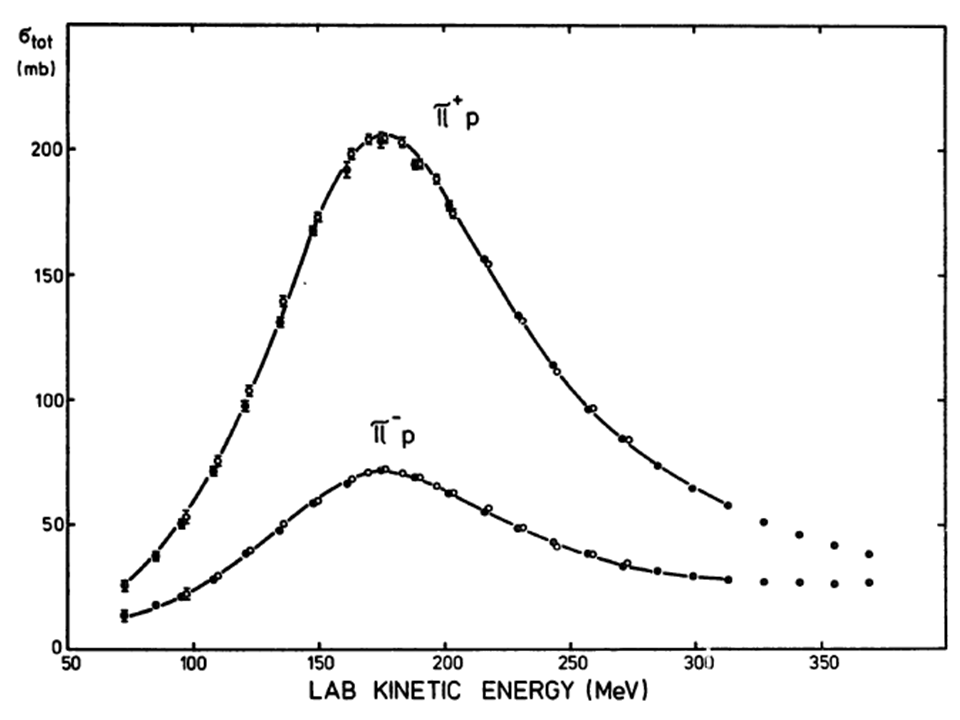}
\caption[Average total cross section of pion-deuteron scattering]{\label{fig:pidscatt}Average total cross section $\frac{1}{2}(\sigma_{\pi^+d}+\sigma_{\pi^-d})$ of pion-deuteron scattering (taken from Ref. \citep{Pedroni:1978it}).  The (un)filled circles show experimental data with two different target lengths and demonstrate the dominance of $\Delta$ intermediate state.  The dashed curve shows the na\"ive prediction obtained by summing the contributions of free $\pi p$ and $\pi n$ scattering, while the solid curve shows the multiple scattering calculation of Ref. \citep{Butterworth:1978kx}.}
\end{figure}

Before describing pion production reactions, a brief discussion is given of the reverse reaction, deuteron breakup $\pi d\to NN$.  The principle of detailed balance relates the two cross sections \citep{Brueckner:1951zzb},
\begin{align}
\sigma_{d\pi^+\to pp}=\frac{2}{3}\frac{p^2}{q^2}\sigma_{pp\to d\pi^+},\label{eq:detailedbalance}
\end{align}
where $\bfp$ is the nucleon momentum and $\bfq$ is the pion momentum in the center-of-mass system.  When considering $\pi d$ scattering, one sees that deuteron breakup is a primary contribution to the imaginary part of the scattering length, $a_{\pi d}$, a quantity which is described theoretically most recently by Ref. \citep{Baru:2007ca}.  Expressing the optical theorem in terms of the scattering length $a$, one finds
\begin{align}
\text{Im}\,a_{\pi^-d}=\lim_{q\to 0}\left(\frac{q\sigma_\text{tot}}{4\pi}\right).
\end{align}
By correcting for the other possible final-states ($NN\gamma,\ NNe^+e^-,\ \text{and }NN\pi$) one can at least approximate a relation between $\text{Im}\,a_{\pi d}$ and $\text{Im}\,a_{\pi d\to NN}$.  The most often quoted modern experiment \citep{Hauser:1998yd}, extracts the following value from the study of pionic deuterium,
\begin{align}
a_{\pi^-d}m_\pi=-0.0261(5)+i0.0063(7),
\end{align}
where $m_\pi=139.6$ MeV is the charged pion mass.  This connection with pionic deuterium will be revisited in Ch. \ref{chap:csb}, where a similar experiments' \citep{Strauch:2010rm} determination of the total cross section for $nn\to d\pi^-$ is used.  One final important feature of $\pi d$ scattering is the prominent contribution of the $\Delta(1232)$ resonance, clearly seen from the energy dependence of the total cross section shown in Fig. \ref{fig:pidscatt}.
This feature emphasizes the necessity of including the resonance as an explicit degree of freedom in the EFT.  Having gleaned useful information from deuteron breakup, let us now return to pion production reactions.

\section{\label{sec:kinoverview}Kinematics of \texorpdfstring{$NN\to d\pi$}{NN --> d pi}}
Several of the important features of $NN\to d\pi$ can be understood in terms of the large relative momentum required in the initial-state.  The deuteron binding energy (defined by $E_d\equiv m_n+m_p-E_b$) is only $E_b=2.2$ MeV, implying that virtually all of the energy of the produced pion is coming from the kinetic energy of the colliding nucleons ($m_N\approx940$ MeV).  Let the first nucleon have momentum $\bfp_1$ and the second $\bfp_2$, such that in the center of mass frame $\bfp_1=-\bfp_2$.  The total momentum of the $NN$ system is ${\bf P}=\bfp_1+\bfp_2=0$ and the relative momentum is $\bfp\equiv(\bfp_1-\bfp_2)/2=\bfp_1$.  Working in the non-relativistic limit and neglecting the small binding energy, one finds that even at threshold\footnote{``Threshold" refers to the reaction performed at just the right energy to create a pion with with no remaining energy for the pion to have a three-momentum.  Clearly, this reaction cannot be measured in the laboratory, but is instead studied via extrapolation.} the relative momentum is large,
\begin{align}
\frac{\tilde{p}^2}{m_N}=m_\pi\Rightarrow \tilde{p}=\sqrt{m_\pi m_N}=356\,\text{MeV}.\label{eq:pthresh}
\end{align}
This momentum sets the scale for pion production reactions.  Because this scale lies between the ``small" pion mass and the ``large" chiral symmetry breaking scale $\Lambda_\chi\approx m_N$, one can already see that the convergence of HB$\chi$PT is going to be slow.

In addition to the slow convergence of the EFT, there are at least three other ramifications of the large threshold momentum.  Firstly, when two particles collide at relatively large velocities, they are bound to approach each other closely.  This means that pion production reactions probe the short-range ($r\sim1/\tilde{p}\approx0.5$ fm) part of the nucleon-nucleon interaction, which is completely understood neither in HB$\chi$PT nor in nonperturbative QCD.  In fact, this difficulty requires one to use phenomenology for the initial-state interaction.  Of course, full coupled-channel calculations exist (for a modern example, see Ref. \citep{Blankleider:1999wu}); however, they suffer from separate difficulties, most notably the requirement of chiral symmetry.  More will be said about the use of phenomenology in Sec. \ref{sec:hybrid}, where the ``hybrid" formalism is described.

\begin{figure}
\centering
\includegraphics[height=1.5in]{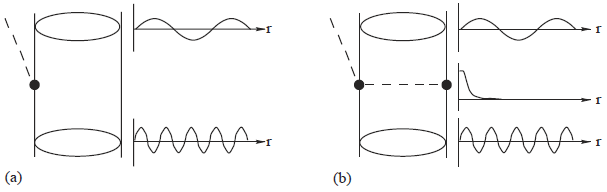}
\caption[Momentum mismatch for pion production operators]{\label{fig:mismatch}Momentum mismatch for (a) one-body and (b) two-body production operators, illustrated with the real-space wave functions (figure taken from Ref. \citep{Hanhart:2003pg}).  Note that this figure considers an unbound final-state with real, but small, momentum; the deuteron case manifests the same mismatch.}
\end{figure}

Secondly, and quite importantly, note that the wave function of the initial-state $NN$ pair is a somewhat distorted plane wave with momentum $\bfp$ while the wave function of the deuteron in momentum space contains much lower Fourier components.  This leads to a prominent momentum mismatch between the initial- and final-states for a one-body production operator.  In order to obtain a large overlap between these two states, a two-body production operator (primarily a meson exchange current, as will be seen) which transfers the large three-momentum is required.  This concept is illustrated in \cref{fig:mismatch}.

Thirdly, the fact that the initial-state contains the large scale, $\tilde{p}$, means that the energy $E$ dependence of the cross section will only depend mildly upon the dynamics of the initial-state, according to $(E/\tilde{p})^2$ \citep{Watson:1952ji}.  The same is true of the production operator, but not of the small-momentum final-state nucleons that interact quite strongly and therefore influence the cross section energy dependence \citep{Machner:1999ky}.  Since the focus of this work is near threshold kinematics, more will not be said about this point.

Let us work out two more kinematical relations before proceeding into a discussion of selection rules.  It is most straightforward for a theorist to work in the center of mass frame, but pion production experiments are performed in the lab frame, where a beam of neutrons\footnote{The $pp\to d\pi^+$ reaction is obviously easier to perform, but since the $np\to d\pi^0$ reaction is related by chiral symmetry and avoids Coulombic effects, it is chosen for this works' calculations.} with energy $E_n=m_n+T_L$ ($T_L$ is called the ``lab kinetic energy") is incident on a stationary proton target.  The invariants are written,
\begin{align}\bs
s_{L,i}&=(E_n(\bfp_1)+m_p,\bfp_1)^2=(m_n+m_p)^2+2m_pT_L \\
s_{C,i}&=(E_n(\bfp)+E_p(-\bfp),0)^2=m_n^2+m_p^2+2p^2+2\sqrt{m_n^2+p^2}\sqrt{m_p^2+p^2},
\es\end{align}
and equating them yields a relation between the center of mass relative momentum and the lab kinetic energy,
\begin{align}
p^2=\frac{m_p^2T_L(2m_n+T_L)}{(m_n+m_p)^2+2m_pT_L}\approx\frac{m_N T_L}{2}.
\end{align}

In the final state, in the center of mass frame, there is a pion with momentum $\bfq$ and a deuteron with momentum $-\bfq$, such that the invariant energy is $\sqrt{s_{C,f}}=E_\pi(\bfq)+E_d(\bfq)$.  It is common to refer to the pion momentum in terms of the dimensionless parameter $\eta=q/m_\pi$.\footnote{There are two potential pitfalls here: 1) the frame must be specified for $\eta$ to be meaningful (the CM frame is used unless stated otherwise), and 2) this definition is ambiguous for three-body final states like in $pp\to pp\pi^0$; the maximum pion momentum $\eta_\text{max}$ is a commonly used parameter in this situation.}  The algebra does not simplify nicely to allow an elucidating expression relating $q$ and $p$.  Nevertheless, one sees that at threshold ($\bfq=0$) when neglecting the nucleon mass difference ($m_n=m_p=m_N$),
\begin{align}
2\sqrt{m_N^2+p^2}=2m_N+m_\pi-E_b\quad\Rightarrow\quad p^2=(m_\pi-E_b)m_N+\frac{(m_\pi-E_b)^2}{4},
\end{align}
which is the relativistic version of \cref{eq:pthresh}.

\section{\label{sec:selrules}Selection Rules}
Though the arguments can be tricky to follow at times, it is crucial to understand the selection rules for pion production as well as the features of the various channels.  Using only angular momentum algebra and the Pauli Principle, one is able to make striking progress in understanding more features of the reaction.  The additional quantum number of isospin, and its conservation dictated by the chiral symmetry of QCD, already provides us with the relation $\sigma_{pp\to d\pi^+}=2\sigma_{np\to d\pi^0}$ due to the fact that an $np$ state can be either $T=0$ or $T=1$ with equal probability, $|np\ra=(|10\ra+|00\ra)/\sqrt{2}$.\footnote{Charge symmetry breaking in the cross section should not be expected to be visible on top of the much larger CS amplitudes.}  It is simplest to begin the discussion with a deuteron in the final-state, although a discussion is also given for the $pp\to pp\pi^0$ reaction due to its special place in historical calculations (see \cref{sec:history}).

The deuteron has the following quantum numbers: total angular momentum $J=1$, spin $S=1$, and isospin $T=0$.  Furthermore, it is a parity even particle, which would lead one to assume that it is a ${}^3S_1$ state.\footnote{The spectroscopic notation used is ${}^{2S+1}L_J$, where $L=0,1,2,\ldots$ is denoted $S,P,D,\ldots$}  Interestingly, the nuclear force has a tensor component which mixes in the other allowed state, ${}^3D_1$.  The pion is an $S=0$, $T=1$, odd parity state and it can be produced with arbitrary orbital angular momentum with respect to the deuteron $l_\pi$.

First consider $s$-wave pion production with $l_\pi=0$.  The final state is $J=1$ and odd in parity and, since the strong force conserves both these quantities, the initial state must be as well (so $L_i=1,3,5,\ldots$).  Although isospin breaking operators exist, the largest contributions to pion production come from isospin conserving transitions where the initial state is $T=1$.  According to the Pauli Principle, since it is anti-symmetric under $1\leftrightarrow2$ in space and symmetric in isospin, the initial state must be symmetric in spin with $S=1$.  To satisfy $J=1$ with $S=1$, the only possible state is ${}^3P_1$.  So one sees that only a single channel is able to contribute to s-wave pion production.

Next consider $p$-wave pion production with $l_\pi=1$.  The final state can now be $J=0,1,2$ and is even parity.  That the parity is the opposite of $s$-wave production tells us that the initial state must be $S=0$ in order to satisfy the Pauli Principle, thus one has $L=J$.  The possible initial states are then ${}^1S_0$ and ${}^1D_2$.  Higher $l_\pi$ will only contribute appreciably at higher energies than this work is concerned with.  A schematic argument for this point, which was formally shown by Ref. \citep{Watson:1952ji}, is now given.

The cross section is given by the transition rate per unit flux,
\begin{align}
\sigma=\frac{2\pi}{v}|T|^2\rho,
\end{align}
where $v$ is the beam velocity, $T$ is the transition matrix, and $\rho$ is the density of states.  An exposition of the details of how the cross section is calculated is given in \cref{sec:hybrid}, but for now let us just concern ourselves with the energy dependence.  The T-matrix is proportional to the pion wave function which, to leading order, is just a plane wave,
\begin{align}
e^{i\bfq\cdot\bfx}=4\pi\sum_{l,m}i^{l}j_{l}(qr)Y_{lm}(\hat{\bfq})Y^*_{lm}(\hat{\bfr}),
\end{align}
which for small $q$ in the $l_\pi$ channel leads to the dependence, $T_{l_\pi}\propto q^{l_\pi}$.
Then, using that the density of states $\rho\propto q$, one finds that near threshold,
\begin{align}
\sigma_{l_\pi}\propto q^{2l_\pi+1},\label{eq:qdep}
\end{align}
validating the claim that only the low $l_\pi$ channels will contribute appreciably to the cross section.  This also means that the deuteron breakup cross section diverges ($\sigma\sim1/q$) at threshold, according to \cref{eq:detailedbalance}.

Finally, consider the reaction $pp\to pp\pi^0$ in the $l_\pi=0$ channel, which is clearly $T=1$ in both the initial and final-states.  Again, since the final-state nucleons have minimal kinetic energy and are interacting strongly, $L_f=0$ will make the largest contribution.  Antisymmetry then forces the final-state into ${}^1S_0$; the initial-state has opposite parity with ${}^3P_0$.  The energy dependence of this reaction is more complicated for two reasons: 1) the three-body final-state kinematics are complicated and 2) the strong meson exchange current is forbidden by isospin conservation enhancing the effects of $l_\pi>0$ channels.  Further details about these issues can be found in Ref. \cite{Hanhart:2003pg}.

\begin{figure}
\centering
\includegraphics[height=2.5in]{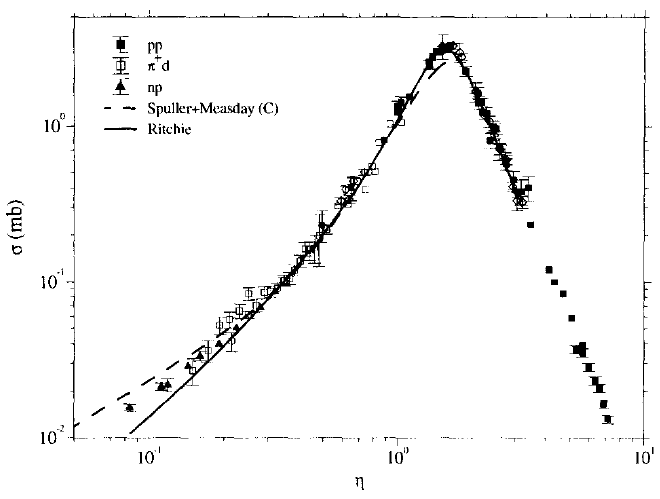}
\caption[Delta peak in $pp\to d\pi^+$ cross section]{\label{fig:drochner}$pp\to d\pi^+$ cross section up to $\eta\approx7$ (figure taken from Ref. \citep{Drochner:1998ja}).  The prominence of the $\Delta$ peak is clear.  Three different experiments are compared with two theories.}
\end{figure}

\begin{figure}
\centering
\includegraphics[height=2.5in]{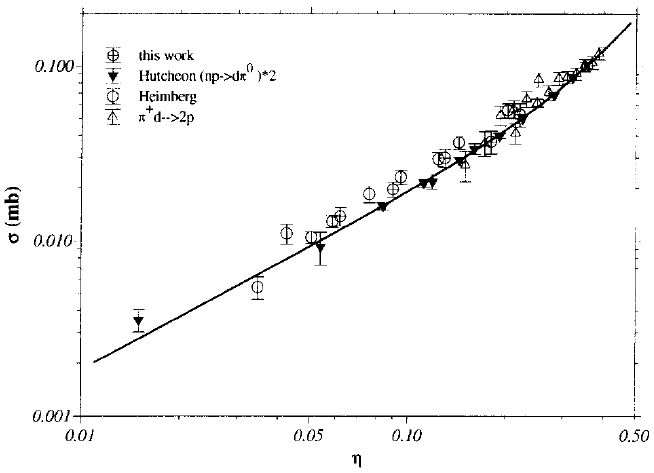}
\caption[COSY $pp\to d\pi^+$ cross section near threshold]{\label{fig:drochner2}COSY (labelled ``this work") $pp\to d\pi^+$ cross section up to $\eta\approx0.5$ (figure taken from Ref. \citep{Drochner:1998ja}).  A deviation from the neutral pion production experiment is highlighted with a fit to the latter data.  The linear dependence on the pion momentum near threshold predicted by \cref{eq:qdep} is visible.}
\end{figure}

\begin{figure}
\centering
\includegraphics[height=2.5in]{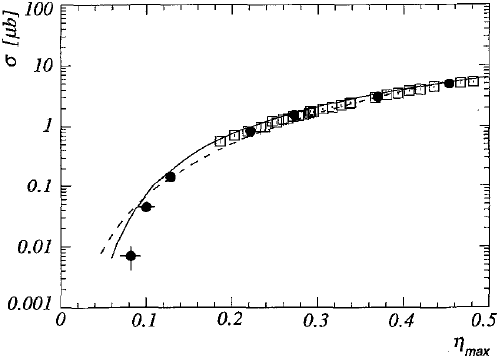}
\caption[CELSIUS $pp\to pp\pi^0$ cross section]{\label{fig:bondar}CELSIUS (filled circles) $pp\to pp\pi^0$ cross section as a function of maximum pion momentum $\eta_\text{max}$ (figure taken from Ref. \citep{Bondar:1995zv}).  The data is compared with the IUCF experiment in \cref{fig:meyer} along with theoretical (heavy meson exchange) predictions from Refs. \citep{Lee:1993xh,Horowitz:1993hk}.}
\end{figure}

\begin{figure}
\centering
\includegraphics[height=2.5in]{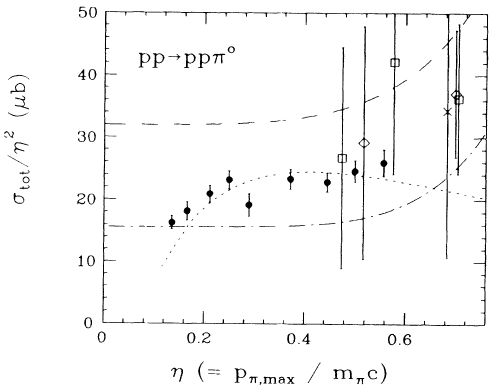}
\caption[IUCF $pp\to pp\pi^0$ cross section]{\label{fig:meyer}IUCF (filled circles) $pp\to pp\pi^0$ cross section as a function of maximum pion momentum $\eta_\text{max}$ (figure taken from Ref. \citep{Meyer:1990yf}).  The data is compared with three older experiments, fits (dashed and dot-dashed curves), and the theoretical energy dependence due to phase space and the final-state $pp$ interaction (dotted curve).}
\end{figure}

To develop an even better feel for these reactions, plots have been included of modern experimental data for $pp\to d\pi^+$ in \cref{fig:drochner,fig:drochner2} and for $pp\to pp\pi^0$ in \cref{fig:bondar,fig:meyer}.

\section{\label{sec:hybrid}The Hybrid Formalism}
Having now described the general features, kinematics, and selection rules of pion production reactions, the next topic to discuss is the methodology by which one includes the important initial- and final-state interactions.  It should be stressed that the ideal method would be to perform a non-perturbative calculation of these interactions using an elastic kernel derived from HB$\chi$PT.  Unfortunately, this is not yet possible since chiral potentials are only reliable for $p\lesssim250$ MeV \citep{Epelbaum:2004fk}.  Instead, one uses the hybrid approach in which one calculates the production operator perturbatively using HB$\chi$PT and convolves this operator with wave functions that are calculated non-perturbatively from phenomenological potentials that only obey chiral symmetry in their long-range pion tails.  This hybrid approach may seem ad-hoc, but it has been successfully applied to a variety of other processes in the two-nucleon sector (see Ref. \citep{Bedaque:2002mn} for a nice review, and Refs. \citep{Entem:2003ft,Bogner:2003wn} for other applications).  Note that in the remainder of this thesis, the ``operator" or ``kernel" refers to the part of pion production which is calculated perturbatively, while the procedure of combining this operator with the non-perturbative parts of the matrix element to obtain an observable is referred to as ``convolving" or ``folding".

To summarize, the operator is calculated as the sum of two-particle-irreducible diagrams involving four nucleon lines (two incoming and two outgoing) and one pion line.  Then, a phenomenological potential is used to calculate the $NN$ scattering wave functions and the deuteron bound state wave function (see \cref{sec:nnwfns}).  Finally, the operator is convolved with the wave functions to obtain the matrix element.  As pointed out in Ref. \citep{Park:2000ct}, the phenomenological potentials always involve some form of high-momentum cutoff.  Therefore, in order to retain consistency, it is important to implement this cutoff in the convolution integrals as well.  More will be said about this at various points throughout the remainder of this thesis.

Let us now be more precise about the implementation of the hybrid formalism by deriving an expression for the cross section in the center of mass frame where the pion momentum is $\bfq$.  First one must define expressions for the initial- and final-states.  To form an interacting $NN$ state with total momentum ${\bf P}=\bfp_1+\bfp_2$, one uses a superposition of free particle states
\begin{align}\bs
\left |\psi({\bf P})\right\ra&=\int\frac{d^3p_1}{(2\pi)^3}\frac{d^3p_2}{(2\pi)^3}\psi\left(\frac{|\bfp_1-\bfp_2|}{2}\right)\left |N(\bfp_1),N(\bfp_2)\right\ra\delta({\bf P}-\bfp_1-\bfp_2)
\\
&=\int\frac{d^3p}{(2\pi)^3}\psi(p)\left |N(\bfp+{\bf P}/2),N(-\bfp+{\bf P}/2)\right\ra,
\es\end{align}
where spin and isospin have been ignored for now.  The wave function $\psi(p)$ is obtained by solving the Schr\"{o}dinger equation with the appropriate $NN$ potential.  For an overview of the standard numerical techniques required to calculate such wave functions, see \cref{sec:nnwfns}.  In the center of mass frame, ${\bf P}=0$, and if the nucleons forming the deuteron have momentum $\bfk_{1,2}$, then $\bfk_1+\bfk_2\equiv{\bf K}=-\bfq$.  The invariant matrix element is then
\begin{align}
\mathcal{M}\left(N(\bfp_1),N(\bfp_2)\rightarrow \pi(\bfq),d({\bf K})\right)=\int\frac{d^3k}{(2\pi)^3}\frac{d^3p}{(2\pi)^3}\psi_d^*(k)\hat{\mathcal{M}}\left(p,k,q\right)\psi_{np}(p).
\label{eq:mxel}
\end{align}
Note that one treats the initial-state as two separate particles, but the deuteron as a single particle.  The sum of the two-particle-irreducible diagrams calculated in momentum space with external momenta $\bfq,\bfp_{1,2},\bfk_{1,2}$ (the operator) is denoted $\hat{\mathcal{M}}$, and is convolved with the external wave functions.  Also note that the wave functions will include spin and isospin kets on which the operator acts.

Let us now perform the calculation in position space by inserting Fourier representations of both wave functions
\begin{align}
\psi_{NN}(\bfr)=\int\frac{d^3p}{(2\pi)^3}e^{i\bfp\cdot\bfr}\psi_{NN}(\bfp).
\label{eq:psift}
\end{align}
A Fourier representation of the operator with respect to $\bfl\equiv\bfk-\bfp$ is also inserted
\begin{align}
\hat{\mathcal{M}}\left(\bfr\right)=\int\frac{d^3l}{(2\pi)^3}e^{i\bfl\cdot\bfr}\hat{\mathcal{M}}\left(\bfl,\bfq\right).
\label{eq:mft}
\end{align}
As described in \cref{sec:method}, $\bfl$ is the momentum that appears in pion production reactions: $\mathcal{M}(\bfp,\bfk,\bfq)\rightarrow\mathcal{M}(\bfl,\bfq)$.

Now one can rewrite Eq. (\ref{eq:mxel})
\begin{align}\bs
\mathcal{M}&=\int d^3rd^3r'd^3r''\psi_d^*(r'')\hat{\mathcal{M}}\left(\bfr\right)\psi_{np}(r')\int\frac{d^3k}{(2\pi)^3}\frac{d^3p}{(2\pi)^3}e^{i\bfk\cdot\bfr\,''}e^{-i(\bfk-\bfp)\cdot\bfr}e^{-i\bfp\cdot\bfr\,'}
\\
&=\int d^3r\psi_d^*(r)\hat{\mathcal{M}}\left(\bfr\right)\psi_{np}(r).
\label{eq:m}
\es\end{align}
With these choices of Fourier representations, the momentum integrals evaluate to delta functions which allow evaluation of the spatial integrals.  The result is an integral over a single spatial variable.

Next, one expresses the invariant S-matrix element in terms of $\mathcal{M}$
\begin{align}
\left\la\pi^0(q)d(K)\mid S\mid p(p_1),n(p_2)\right\ra=1-i(2\pi)^4\delta^4(q+K-p_1-p_2)\mathcal{M}(p_1,p_2\rightarrow q,K).
\label{eq:smatrix}
\end{align}
In the center of mass frame, the spin-averaged differential cross section is, \citep{Amsler:2008zzb}
\begin{align}
d\sigma=\frac{1}{4|\vec{p}\,|\sqrt{s}}\frac{1}{4}\sum_{m_d,m_1,m_2}\left|\mathcal{M}\right|^2(2\pi)^4\delta^4(q+K-p_1-p_2)\frac{d^3q}{(2\pi)^32\omega_q}\frac{d^3K}{(2\pi)^32E_d},
\end{align}
where the four $NN$ spin states have been averaged over and the three spin states of the deuteron summed over.  The vector part of the delta function gives $\bfq=-{\bf K}$ and the energy part gives $E_1+E_2=\omega_q+E_d$.  This allows one to perform all but the $d\Omega_K$ integral,
\begin{align}
\frac{d\sigma}{d\Omega}=\frac{|\bfq\,|}{64\,\pi^2\,s\,|\bfp\,|}\frac{1}{4}\sum_{m_d,m_1,m_2}|\mathcal{M}|^2.
\label{eq:dsdoab}
\end{align}
What remains is to derive expressions for the wave functions and the operator appearing in Eq. (\ref{eq:m}); this is done in \cref{sec:states}.

There is one remaining subtlety: in using phenomenological $NN$ potentials, one assumes that the $NN$ pair is in its center of mass frame.  This is not true in the reaction's center of mass frame because the (final-state) deuteron has a non-zero total momentum, $-\bfq$.  In the frame where the deuteron is at rest, the T matrix is written in terms of the stationary potential, $V$, and the two nucleon propagator, $G$, using the Lippmann-Schwinger Equation $T=V+VGT$.  However, since the T matrix is invariant, it can also be expressed it in terms of the boosted potential in the center of mass frame, $V'$, and the boosted two nucleon propagator, $G'$ as $T=V'+V'G'T\approx V'+V'G'V'$.  Inserting the boosted expression of T into the stationary one yields
\begin{align}
V'+V'G'V'=V+VG(V'+V'G'V').
\end{align}
Rearranging this equation, one obtains
\begin{align}\bs
V'&=V+V(G-G'')V'
\\
G''&=(\frac{1}{V}V'-GV')G'
\\
&\approx G'-GV'G'.
\es\end{align}
Thus the boosted potential up to next-to-leading order is
\begin{align}\bs
V'&=V+V(G-G')V'
\\
&\approx V+V(G-G')V.
\label{eq:boostedpot}
\es\end{align}
Corrections to the leading order approximation, $V'=V$, will come in at orders higher than will be considered in this work.

\section{\label{sec:history}History of Pion Production Calculations}
Let us begin this section with an overview of experimental results for the total cross section of $pp\to d\pi^+$ just above threshold.  In this regime, according to \cref{eq:qdep}, the total cross section takes the form $\sigma=Aq+Bq^3$.  Thus the total cross section is typically reported in terms of two parameters,
\begin{align}\bs
\sigma(pp\to d\pi^+)&=\alpha\eta+\beta\eta^3
\\
\sigma(np\to d\pi^0)&=\frac{1}{2}\left(\alpha\eta+\beta\eta^3\right)\label{eq:alphadefch3}
\es\end{align}
\Cref{tab:exptch3} displays the results of the most recent experiments.
\begin{table}
\caption[Experimental total cross section parameters for \texorpdfstring{$NN\to d\pi$}{NN --> d pi}]{\label{tab:exptch3}Experimental total cross section parameters for $pp\to d\pi^+$ (Coulomb corrected) and the isospin-equivalent $np\to d\pi^0$.  The parametrization used is $\sigma(pp\to d\pi^+)=\alpha\eta+\beta\eta^3$.}
\begin{center}
\begin{tabular}{lcc}
\hline\hline
Experiment & $\alpha\ (\mu\text{b})$ & $\beta\ (\mu\text{b})$\\ \hline
$np\rightarrow d\pi^0$ (TRIUMF, 1989) \citep{Hutcheon:1989bt} & $184\pm5$ & $781\pm79$\\
$\pi^+d\to pp$ (1991) \citep{Ritchie:1991rc} & $174\pm3$ & $982\pm38$\\
$\vec{p}p\rightarrow d\pi^+$ (IUCF, 1996) \citep{Heimberg:1996be} & $208\pm5$ & $1220\pm100$\\
$pp\rightarrow d\pi^+$ (COSY, 1998) \citep{Drochner:1998ja} & $205\pm9$ & $791\pm79$\\
Pionic deuterium (PSI, 2010) \citep{Strauch:2010rm} & $252^{+5}_{-11}$ & N/A\\ \hline\hline
\end{tabular}
\end{center}
\end{table}

\begin{figure}
\centering
\includegraphics[height=1.5in]{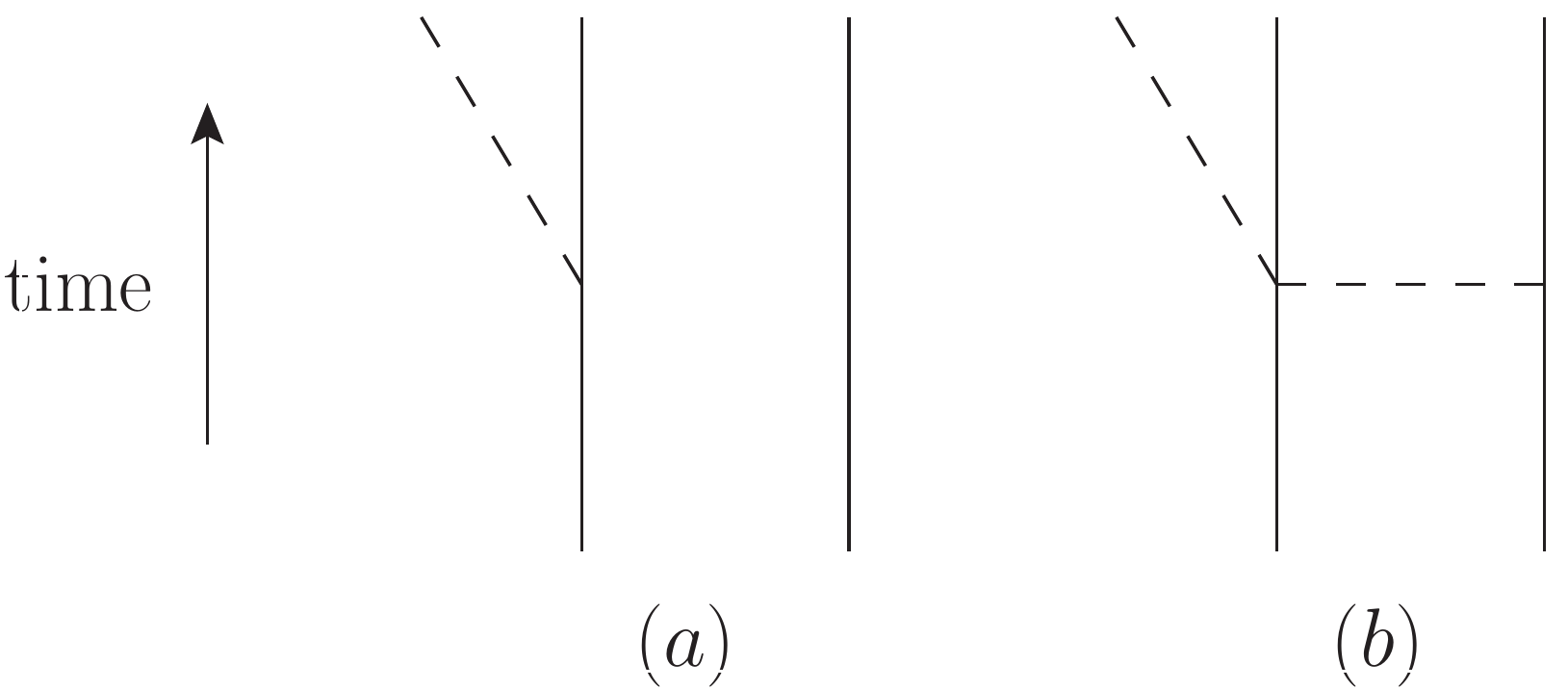}
\caption[Amplitudes included in Koltun and Reitan's work]{\label{fig:KR}Amplitudes included in Koltun and Reitan's work: (a) impulse approximation and (b) pion rescattering.  Solid lines represent nucleons and dashed lines pions.}
\end{figure}

Pion production has been studied theoretically since the 1960s when Koltun and Reitan published their seminal paper obtaining $\alpha=146\ \mu\text{b}$ \citep{Koltun:1965yk}.  This calculation came well before the rise of $\chi$PT and considered just two amplitudes\footnote{Of course these amplitudes also appear in $\chi$PT with additional constraints coming from chiral symmetry.}: the impulse approximation (IA, sometimes referred to as ``direct production") and pion rescattering (RS).  Diagrams representing these processes are shown in \cref{fig:KR}.
In the IA amplitude a pion is produced by a single nucleon and does not interact with the spectator nucleon.  The leading order $\pi NN$ vertex is proportional to the pion's momentum; therefore, at threshold (where the momentum is zero) the IA proceeds through a recoil correction which is suppressed by $p/m_N$.  In the RS amplitude, the $\pi\pi NN$ vertex is commonly referred to as the Weinberg-Tomozawa (WT) vertex (KR used the $\pi N$ scattering phase shifts to fix the strength of the WT vertex).  

Although Koltun and Reitan's results are as accurate as should be expected for $pp\to d\pi^+$, this was not the case for the $pp\to pp\pi^0$ reaction; models found threshold cross sections to be 5-10 times smaller than experiments \citep{Miller:1991pi}.  The clear reason for this situation is the fact that the large RS diagram which contributes to $pp\to d\pi^+$ is isospin-forbidden for $pp\to pp\pi^0$, leaving smaller and less understood diagrams to dominate.  This discrepancy with experiment prompted several more model-dependent solutions, among which were heavy-meson exchange (in the kernel) \citep{Lee:1993xh,Horowitz:1993hk} and off-shell effects in the RS diagram \citep{Hernandez:1995kj,Hanhart:1995ut}.  Although individually these solutions seemed to resolve the issue, adopting both simultaneously changed the situation from dramatic under-prediction to dramatic over-prediction.  Chiral Perturbation Theory should provide a final solution to this problem since it is model independent (provided convergence is achieved), but the first such calculations \citep{Park:1995ku,Hanhart:1997jd} only worsened agreement with the data.

The next development was the recognition by Ref. \citep{Cohen:1995cc} that the large momentum scale $\tilde{p}=\sqrt{m_\pi m_N}$ necessitated a reordering of the diagrams into a modified power counting scheme which expands in powers of $\chi\equiv \tilde{p}/m_N\approx m_\pi/\tilde{p}$.  Sometimes this new scheme is referred to as the ``momentum counting scheme" (MCS) in contrast with the ``Weinberg" scheme that simply counts all momenta as ${\cal O}(m_\pi)$.  In the MCS, one examines a diagram and deduces whether the momentum in a vertex factor will carry the small final-state momentum or the large initial-state momentum.  To be more explicit about this modified power counting scheme, let us consider $p$-wave pion production from the two diagrams (IA and RS) shown in \cref{fig:KR}.

\begin{figure}
\centering
\includegraphics[height=1.5in]{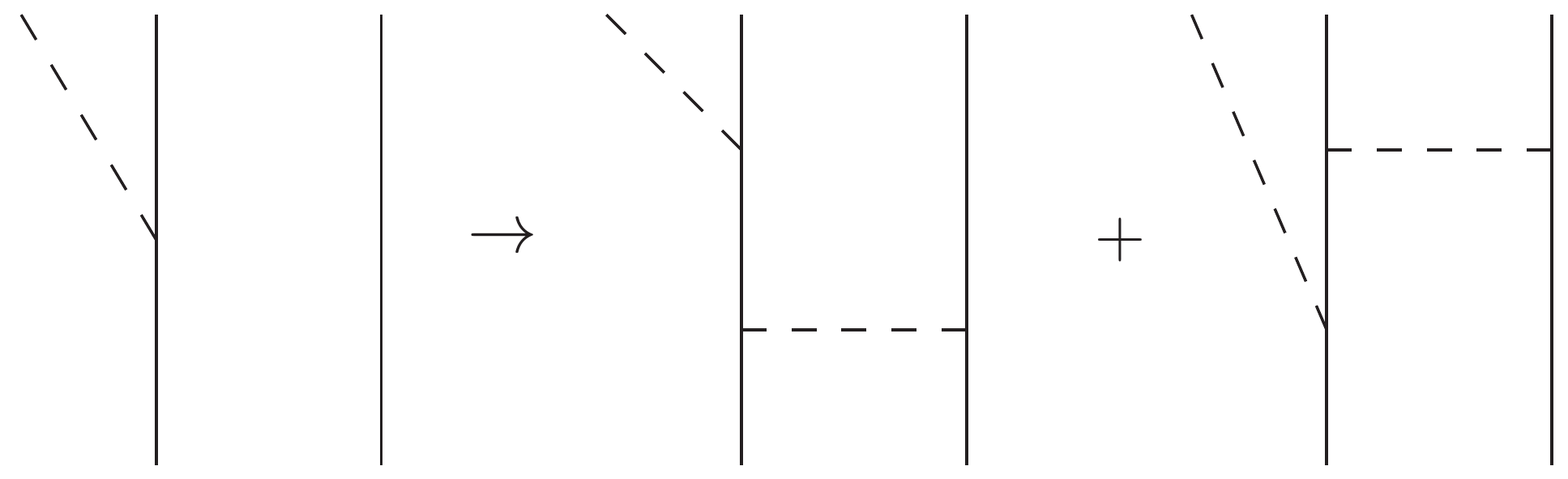}
\caption[Truly irreducible impulse approximation]{\label{fig:IAope}Irreducible diagrams which represent the true power counting for the impulse approximation}
\end{figure}

However, before doing this, let us take a brief aside to discuss a point that will become a major theme of this thesis.  Formally, the IA diagram requires a modification: it must also include a meson exchange due to the fact that a nucleon cannot emit a pion and remain on-shell.  The definition of an irreducible diagram in $A=2$ HB$\chi$PT, as laid out by Refs. \citep{Weinberg:1990rz,Weinberg:1991um}, is that no intermediate states are off shell by $\leq m_\pi^2/m_N$.  At best, the nucleon could be off-shell by $m_\pi/2$ before and after the pion emission; this led Ref. \citep{Cohen:1995cc} to claim that the power counting of the IA diagram should be determined by considering the diagrams of \cref{fig:IAope}.
The question of how to correctly include one pion exchange (OPE) in the IA is one that will be revisited several times throughout the following chapters.

Now let us return to the task of power counting and start by considering the first diagram on the RHS of \cref{fig:IAope}.  For $p$-wave production, the pion production vertex gives a $q$ while the intermediate nucleon propagator gives a $1/m_\pi$.  Counting $q\sim m_\pi$, as dictated by the Weinberg counting scheme, one sees that the IA contributes to the $\chi^0$ term in the expansion.  The RS diagram for $p$-wave production (smaller than it is for $s$-wave production) gets a $m_\pi$ from the WT vertex, a $1/\tilde{p}^2$ from the pion propagator, and a $q$ from the pion emission.  Thus the RS diagram should be counted as $m_\pi q/\tilde{p}^2$.  In the traditional Weinberg counting this ratio becomes $\sim\chi^0$ while in the MCS counting, it becomes $\sim\chi^2$.  This example illustrates the subtleties that pion production theorists have been attempting to work out in the past decade.

Unfortunately, the first attempts at applying the MCS \citep{daRocha:1999dm,Dmitrasinovic:1999cu,Ando:2000ema} seemed to indicate a possible divergence of the power counting expansion; the NLO contributions were not always suppressed as they should be.  It was found that the modified power counting scheme becomes still more subtle when loops are involved, and after the work Refs. \citep{Hanhart:2000gp,Hanhart:2002bu}, most questions seemed to be answered at least qualitatively.  It was shown that with correct counting, the first set of loops sum to zero for $pp\to pp\pi^0$.  On the other hand, the NLO $pp\to d\pi^+$ loops remained finite and the form of the kernel required an ad-hoc counterterm in order to render the convolution integral finite \citep{Gardestig:2005sn}.  This final issue was solved by Ref. \citep{Lensky:2005jc} which recognized another reducibility issue with the diagram involving the recoil correction to the WT vertex.  This work will be reviewed in \cref{sec:power counting}, but the summary is that the NLO loops modify the WT vertex by simply multiplying it by a factor of 4/3.  This factor brings the theoretical calculation into greater agreement with experiment.

This is the point where this thesis joins the task.  The theoretical model-dependent ambiguities have been largely resolved by the use of $\chi$PT with an appropriately modified power counting scheme.  The $pp\to pp\pi^0$ cross section within $\chi$PT appears to be free of theoretical inconsistencies, but due to its small size and the physics community's still-incomplete knowledge of the relevant LECs, agreement with experimental cross section data is still a relevant question.  The next task which one should consider for $pp\to d\pi^+$ is the calculation of other pion production observables.
\chapter[Charge Symmetry Breaking in the \texorpdfstring{$n p\to d\pi^0$}{np --> d pi0} Reaction]{Charge Symmetry Breaking in the $\lowercase{np}\rightarrow\lowercase{d}\pi^0$ reaction}
\label{chap:csb}

The asymmetry in the angular distribution of $np\rightarrow d\pi^0$ due to Charge Symmetry Breaking is calculated using Heavy Baryon Chiral Perturbation Theory.  Recent developments in power counting have proven successful in describing total cross sections, and we apply them to the asymmetry calculation.  Reducibility in one of the leading order diagrams is examined.  We compare the updated theory with experimental results for a set of physically reasonable CSB parameters and find that the theoretical asymmetry is larger than experiment by $\sim2-3$ experimental uncertainties.  This chapter is a modified version of our published paper \citep{Bolton:2009rq}; we have reorganized and rewritten parts in effort to make it flow better with the rest of this thesis.

\section{Introduction\label{sec:csbintro}}

In the preceding chapters, we have reviewed the development of (Heavy Baryon) Chiral Perturbation Theory and its application to pion production in nucleon-nucleon reactions.  We have motivated the investigation of pion production by noting how its special feature (the large threshold momentum $\tilde{p}=\sqrt{m_\pi m_N}=356$ MeV) provides a challenge to HB$\chi$PT.  Once theorists arrive at an ordering scheme for the expansion, error estimates will become more reliable and calculations of different and more exotic reactions possible.

One particularly interesting quantity that pion production can help determine is the magnitude of the CSB parameters discussed in \cref{sec:csbops}.  Charge Symmetry Breaking in the nuclear force is by no means a new topic (see Refs. \citep{Cheung:1979ma, Epelbaum:1999zn} for example), but pion production provides a unique opportunity to study CSB with an observable that vanishes in the CS limit.  Additionally, the algebraic combination of $\delta m_N$ and $\overline{\delta}m_N$ that appears in the $np\to d\pi^0$ asymmetry is distinct from the sum which appears in the nucleon mass difference.  This distinction allows for a clean connection between a hadronic experiment and the down-up mass difference.

Recall that CS is broken both by the mass difference between up and down quarks and also by electromagnetic effects.  Both of these effects enter HB$\chi$PT through interactions of neutral pions with nucleons \citep{Weinberg:1977hb}.  Unfortunately, the lack of a neutral pion beam prohibits a direct measurement of these operators.  The addition of a second nucleon opens new doors; the $np\rightarrow d\pi^0$ reaction circumvents the need for a neutral pion beam and also minimizes electromagnetic effects in the initial- and final-states.  To be more specific, the angular distribution of this reaction is symmetric about $90^\circ$ in the center of mass frame when CS is respected.  A recent experiment at TRIUMF \citep{Opper:2003sb} was able to observe that this distribution is asymmetric at the $\approx2\sigma$ level.

This report advances previous work in several ways.  The authors of Ref. \citep{Lensky:2005jc} showed that a vertex which was thought to be higher order in fact contributes at leading order to $s$-wave pion production.  This advance lead to an improved understanding of the total cross section of $pp\rightarrow d\pi^+$; we extend this calculation off threshold for $np\to d\pi^0$ and make a comparison with the experimental data.  Another ingredient to the CSB calculation is the CS $p$-wave amplitude.  We point out a problem with the standard $\Delta(1232)$ amplitude and implement a cutoff in order to match $p$-wave data.  Finally, we calculate the asymmetry of $np\rightarrow d\pi^0$, including both RS and IA CSB diagrams, bringing the calculations of Refs. \citep{Niskanen:1998yi,vanKolck:2000ip} up to date.

This chapter also contains a discussion (alluded to in \cref{sec:history}) about reducibility in the IA diagram.  It has been assumed that the energy transfer required for pion production occurs in the external wave function when one considers the IA diagram.  At the same time, it has been recognized that OPE must be included in order to perform the MCS power counting for the IA.  As a first attempt to resolve this discrepancy, we tried introducing ``wave function corrections" and found them to be negligible for most of the amplitudes.  The following two chapters will continue in this path toward an eventual solution which is quite different than the na\"ive wave function corrections.  The eventual solution is recognized by considering the way in which 3D wave functions are obtained by integrating over the relative energy in a 4D formalism.

In \cref{sec:reactionch4} we review the kinematics and selection rules specific to the $np\rightarrow d\pi^0$ process.  We also form expressions for the initial- and final-states in this section.  In \cref{sec:strong} we present the diagrams with vertices from the CS part of the Lagrangian (\cref{sec:lagrangian}).  In this section we also provide a review of the power counting/reducibility developed by Ref. \citep{Lensky:2005jc} and discuss its impact on neutral pion production.  Next, in \cref{sec:csb}, we present diagrams with vertices from the CSB part of the Lagrangian which contribute to the reaction.  Our results for the asymmetry are given in \cref{sec:csbresults}.  In \cref{sec:discussion}, we compare our work with another recent calculation of the asymmetry and discuss the problems we encountered along with possible solutions.

\section{The \texorpdfstring{$np\to d\pi^0$}{n p --> d pi0} Reaction\label{sec:reactionch4}}

\subsection{\label{sec:kinematics}Kinematics and selection rules}

In this section, we build upon \cref{chap:review} and discuss the specific kinematics used in the calculation of this chapter.  There are two frames to keep in mind: the center of mass frame (\textbf{C}), and the lab frame where the proton is at rest (\textbf{E} for experiment).  The experiment of interest \citep{Opper:2003sb} is $np\rightarrow d\pi^0$ performed in the \textbf{E} frame.  In this frame the invariant is expressed as $s_\mathbf{E}=(m_p+m_n)^2+2m_pT_L$ where $T_L$ is the kinetic energy defined by $E_n=m_n+T_L$.  The asymmetry experiment was performed at $T_L=279.5$ MeV, slightly above the threshold value of $T_L=275.1$ MeV.  To simplify the formalism, we use the \textbf{C} frame to do the calculation.  In terms of the pion momentum $\bfq$ and the deuteron mass $m_d$ we have $\sqrt{s_\mathbf{C}}=\sqrt{m_d^2+\bfq^2}+\sqrt{m_\pi^2+\bfq^2}$.  In this near-threshold regime, the parameter $\eta$ describes the excess energy, $\eta=|\bfq\,|/m_\pi$.  Equating the invariants we find $\eta_\mathbf{C}=0.169$ ($q_\mathbf{C}=22.86$ MeV) at the experimental energy. 

Also discussed in \cref{sec:selrules} were the selection rules for the isospin related $pp\to d\pi^+$ and $np\to d\pi^0$ reactions.  It was shown that to produce $s$-wave pions, a ${}^3P_1$ initial-state was required, and likewise for $p$-wave pions, a ${}^1S_0$ or ${}^1D_2$ state.  In the context of this calculation we need to distinguish these CS, or ``strong" amplitudes with $T_i=1$ from CSB amplitudes.  Charge Symmetry Breaking operators transform as vectors under isospin, thus the initial neutron-proton state must have $T_i=0$.  The same arguments that were used for the strong operators then show that $s$-wave pions are produced from $^1P_1$ $np$ pairs while $p$-wave pions are produced from the coupled channels $^3S_1$ and $^3D_1$ in addition to $^3D_2$.  These conclusions are summarized in \cref{tab:channels}.
\begin{table}
\caption{\label{tab:channels}Channels for the $np$ wave function in $np\rightarrow d\pi^0$}
\begin{center}
\begin{tabular}{lcc}
\hline\hline
& Strong & CSB \\ \hline
$l_\pi=0$ & $^3P_1$ & $^1P_1$ \\
$l_\pi=1$ & $^1S_0$, $^1D_2$ & $^3S_1$, $^3D_1$, $^3D_2$ \\ \hline\hline
\end{tabular}
\end{center}
\end{table}

The observable of interest in the experiment is the forward/backward asymmetry in the differential cross-section given by
\begin{align}
A_{fb}=\frac{\int_0^{\pi/2}d\Omega\ [\sigma(\theta)-\sigma(\pi-\theta)]}{\int_0^\pi d\Omega\ \sigma(\theta)}.
\label{eq:asymmetry}
\end{align}
A non-zero asymmetry will only be observed when initial-states of opposite parity interfere.  However, the interference can only occur for states with the same total spin because the spin z-components are summed,
\be
\sum_{m_1=-1/2}^{+1/2}\sum_{m_2=-1/2}^{+1/2}\left\la\frac{1}{2}\ m_1,\frac{1}{2}\ m_2\middle\vert 1\ m_1+m_2\right\ra\left\la \frac{1}{2}\ m_1,\frac{1}{2}\ m_2\middle\vert 0\ 0\right\ra=0.
\ee
Thus for calculating the asymmetry, we are concerned with two terms: ($s$-wave strong)$\cdot$($p$-wave CSB) and ($p$-wave strong)$\cdot$($s$-wave CSB).

\subsection{Initial and final states\label{sec:states}}

Recall the expressions derived in \cref{sec:hybrid} for the differential cross section in the context of the hybrid approach,
\begin{align}
\frac{d\sigma}{d\Omega}&=\frac{|\bfq\,|}{64\,\pi^2\,s\,|\bfp\,|}\frac{1}{4}\sum_{m_d,m_1,m_2}|\mathcal{M}|^2.
\\
{\cal M}&=\int d^3r\psi_d^*(r)\hat{\mathcal{M}}\left(\bfr\right)\psi_{np}(r)
\end{align}
For the $np\to d\pi^0$ reaction, it remains to find expressions for the external wave functions.

In the absence of interactions the wave function $\psi_{np}(\bfr)$ is expressed by performing a partial wave expansion on an anti-symmetrized wave function of a free proton and a free neutron with relative momentum $\bfp$.  First we consider the strong operators where the $np$ pair is in an isospin-1 state,
\begin{align}\bs
\left(\bfr\mid\psi_{np}\right\ra&=\mathcal{P}_{T=1}\frac{1}{\sqrt{2}}\left(e^{i\bfp\cdot\bfr}|m_1,m_2\ra\otimes|T_{z,1},T_{z,2}\ra-e^{-i\bfp\cdot\bfr}|m_2,m_1\ra\otimes|T_{z,2},T_{z,1}\ra\right)
\\
&=\frac{1}{\sqrt{2}}\left(e^{i\bfp\cdot\bfr}|m_1,m_2\ra-e^{-i\bfp\cdot\bfr}|m_2,m_1\ra\right)\otimes\frac{1}{\sqrt{2}}\left|T=1,T_z=0\right\ra,
\label{eq:npwfn}
\es\end{align}
where $\mathcal{P}_{T=1}$ is the isospin projector.  The bra $\left(\bfr\;\right|$ indicates that we are choosing a basis for space, but not for spin or isospin.  Implicit in the notation is the dependence of $(\bfr\mid\psi_{np}\ra$ on the momentum $\bfp$, the spin z-components of the two nucleons, $m_i$, and the isospin z-components of the two nucleons, $T_{z,i}$, with the requirement that $T_{z,1}+T_{z,2}=0$.

The exponentials are now expanded and the presence of the strong interaction is added by changing the spherical Bessel functions, $j_L(pr)\rightarrow e^{i\delta_{L}}u_{L,J}(r)/pr$.  The $u_{L,J}$ functions and the $\delta_{L}$ phase shifts are obtained by solving the Schr\"odinger equation with a phenomenological $NN$ potential (we use Argonne V18 \citep{Wiringa:1994wb}).  This technique is described in \cref{sec:nnwfns}.  Finally, the spherical harmonics are combined with the spin kets to form states with definite total angular momentum.  The notation for these states is $\left|(SL)J,m_J\right\ra\otimes\mid T,T_z\ra$.  For the allowed quantum numbers, we find
\begin{align}\bs
(\bfr\mid\psi_{np}(^{2S+1}L_J&,T=1)\ra=4\pi (i)^Le^{i\delta_{L}}\frac{u_{L,J}(r)}{pr}\left\la1/2\ m_1,1/2\ m_2\mid S\ m_s\right\ra
\\
&\times\sum_{m_i}\left\la S\ m_s,L\ m_i-m_s\mid J\ m_i\right\ra Y^{L\ *}_{m_i-m_s}(\hat{p})\left(\hat{r}\mid(SL)J,m_i\right\ra\otimes\left|1,0\right\ra,
\label{eq:npwfnT1}
\es\end{align}
where $m_s=m_1+m_2$ and the second Clebsch Gordan coefficient allows us to make the sum over $m_i=m_l+m_s$ rather than $m_l$.  For the CSB operators, we have $T=0$ np wave functions and find
\begin{align}\bs
(\bfr\mid\psi_{np}(^{2S+1}L_J&,T=0)\ra=\pm4\pi (i)^Le^{i\delta_{L}}\frac{u_{L,J}(r)}{pr}\left\la1/2\ m_1,1/2\ m_2\mid S\ m_s\right\ra
\\
&\times\sum_{m_i}\left\la S\ m_s,L\ m_i-m_s\mid J\ m_i\right\ra Y^{L\ *}_{m_i-m_s}(\hat{p})\left(\hat{r}\mid(SL)J,m_i\right\ra\otimes\left|0,0\right\ra,
\label{eq:npwfnT0}
\es\end{align}
where the $\pm$ refers to $T_{z,1}=\pm1/2$.  Similar analysis gives the final-state wave function of the deuteron as a function of its polarization, $m_f$,
\begin{align}
\left\la\psi_d(m_f)\mid\bfr\right)=\left\la0,0\right|\otimes\left(\frac{u(r)}{r}\left\la(10)1,m_f\mid\hat{r}\right)+\frac{w(r)}{r}\left\la(12)1,m_f\mid\hat{r}\right)\right).
\label{eq:dwfn}
\end{align}

\section{Strong Amplitudes\label{sec:strong}}

\subsection{Diagrammatic Expansion\label{sec:diagrams}}

Before we calculate the effects of CSB, we need to discuss the CS diagrams which lead to a calculation of the total cross section in agreement with experiment.  As is always the case in an EFT, there are a host of possible diagrams one could draw, and predictive power is only obtained by following a power counting scheme.  In HB$\chi$PT, one orders contributions in powers of the external momenta divided by the chiral symmetry breaking scale.  In the $np\to d\pi^0$ reaction, both $q$ and $\tilde{p}$ appear as external momenta and we need to keep track of both in the power counting.  Recall that because the expansion parameter $\chi\equiv\tilde{p}/m_N=\sqrt{m_\pi/m_N}=0.40$ is large, the expansion should not be expected to converge quickly.

The Lagrangian is given in \cref{sec:lagrangian}.  The index of a ``type $i$" vertex is given by
\begin{align}
\nu_i=d_i+\frac{f_i}{2}-2,
\end{align}
where $d_i$ is the sum of the number of derivatives, $m_\pi$'s, and $\delta$'s (the $\Delta$N mass difference), and $f_i$ is the number of fermion fields.  In standard power counting at tree level, the sum of the $\nu_i$ for each vertex in a diagram indicates the power of $\chi$ at which that diagram contributes.  This rule, however, requires modification due to the relatively large value of $\tilde{p}$ as discussed in \cref{sec:history}.  We will come back to this issue again in \cref{sec:power counting}.

\begin{figure}
\centering
\includegraphics[height=1.5in]{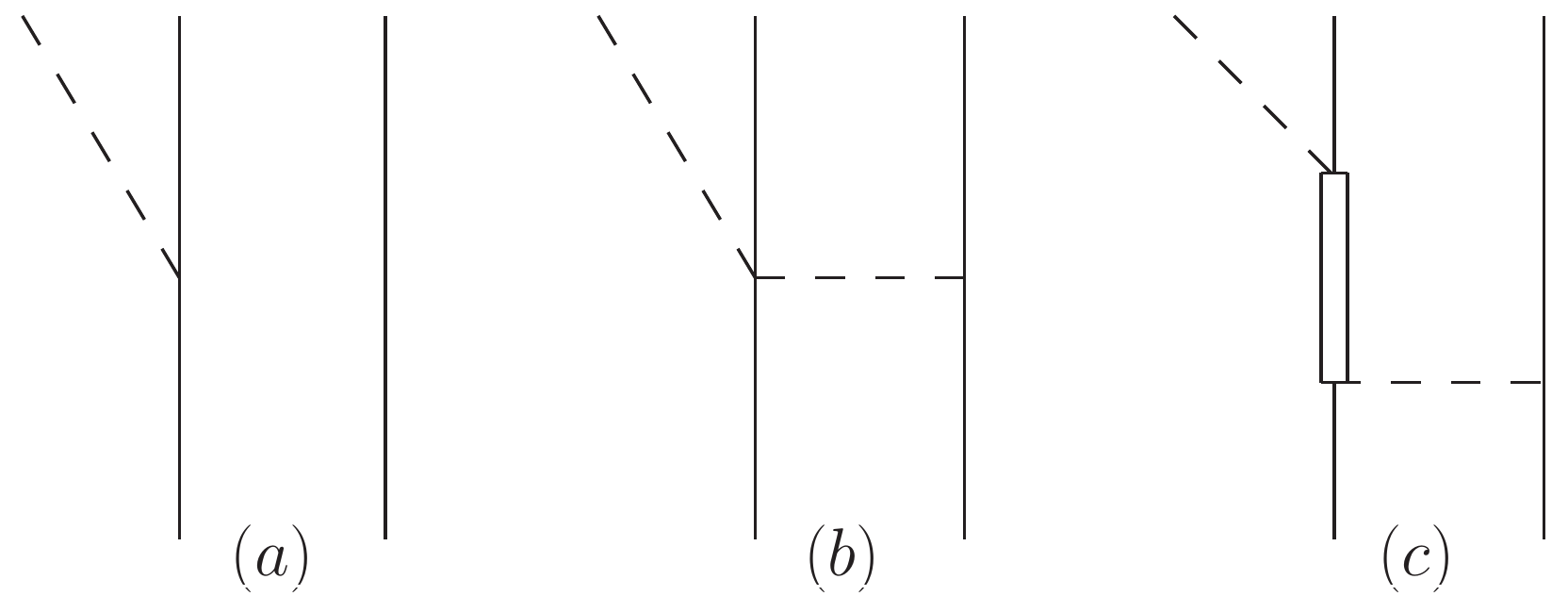}
\caption[Leading order contributions to $np\rightarrow d\pi^0$]{Leading order contributions to $np\rightarrow d\pi^0$.  Solid lines represent nucleons, the double solid line represents a $\Delta$, and dashed lines represent pions.\label{fig:stronglo}}
\end{figure}

There are three two-particle irreducible diagrams which can be drawn using the vertices from $\mathcal{L}^{(0)}$.  They will be referred to as the impulse approximation, or IA, (\cref{fig:stronglo}a); pion rescattering, or RS, (\cref{fig:stronglo}b); and Delta (\cref{fig:stronglo}c) diagrams.

Let us recall a few facts about these three diagrams.  We have seen in \cref{sec:kinoverview} that the RS diagram is enhanced relative to the IA diagram due to the ``momentum mismatch" being provided for in the RS but not the IA diagram.  Using the leading order vertices, the RS diagram contributes to the channel with $s$-wave pions.  Although the Delta diagram provides the momentum transfer required, the Delta resonance is at $1232$ MeV and the $\pi N$ energy is $\approx 1080$ MeV, so the Delta diagram is also somewhat suppressed for our situation of interest.  Finally, we note that both the IA and the Delta diagrams with leading order vertices contribute to the channels in which the pion is in a $p$-wave.

\begin{figure}
\centering
\includegraphics[height=1.5in]{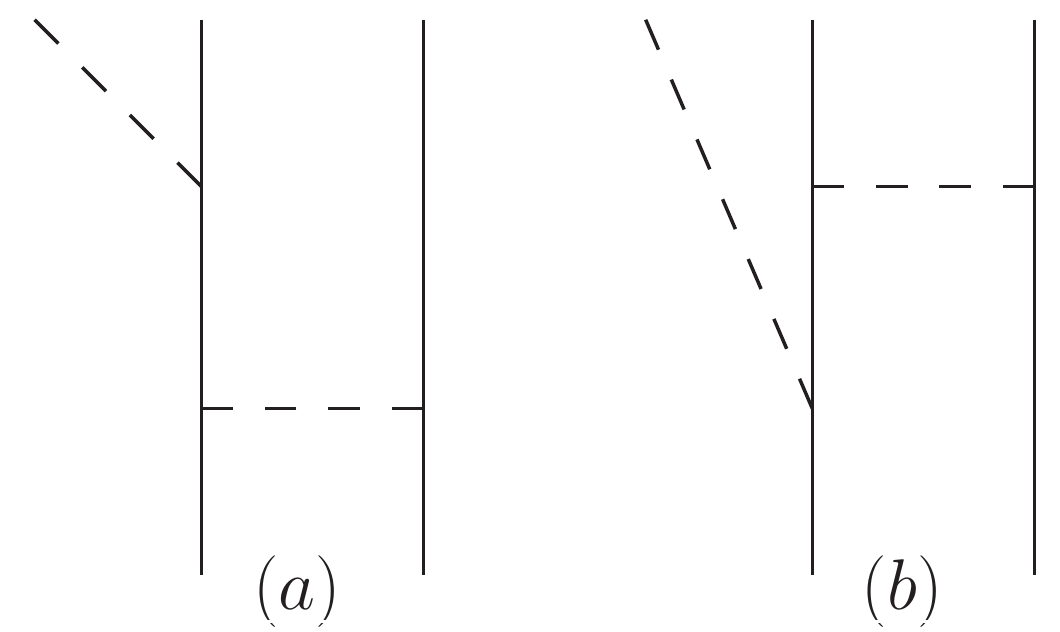}
\caption[Reducible impulse diagrams]{Kinematically consistent, but topologically reducible impulse contributions.  Solid lines represent nucleons and dashed lines represent pions.\label{fig:strongia}}
\end{figure}

We saw in \cref{sec:history} that there is ambiguity as to what is meant by \cref{fig:stronglo}a, and we must now make a decision on how it is to be included.  On the one hand, we know that a single nucleon cannot emit a pion and remain on-shell.  But on the other hand, the diagrams in \cref{fig:strongia} (the same diagrams that were used in \cref{fig:IAope} to power count the IA) which remedy this problem by including OPE appear to be two-particle-reducible, topologically.

\begin{figure}
\centering
\includegraphics[height=1.5in]{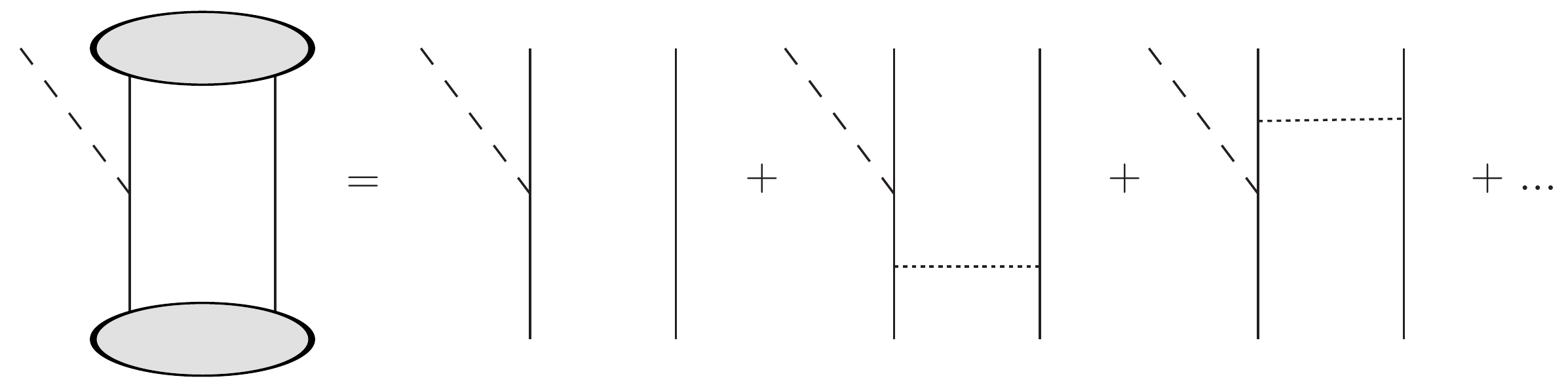}
\caption[``Hybrid" approach]{``Hybrid" approach.  Solid lines represent nucleons, dashed lines represent pions, dotted lines represent pions with $\omega=0$, and filled ovals represent $NN$ strong interactions.\label{fig:strongiawfn}}
\end{figure}

Near threshold the energies of the exchanged pions in \cref{fig:strongia} are $\omega\approx m_\pi/2$.  However, the diagram of \cref{fig:stronglo}a can be thought of as being evaluated as the sum of the diagrams shown in \cref{fig:strongia}.  This is due to hybrid nature of the calculation; once the operator is calculated, it is convolved with $NN$ wave functions.  One of the major terms of the strong interaction potential at low energy arises from static OPE ($\omega=0$).\footnote{We ignore the effects of the rest of the wave function for the moment.}  The effects of static OPE are schematically shown in \cref{fig:strongiawfn}.

\begin{figure}
\centering
\includegraphics[width=6in]{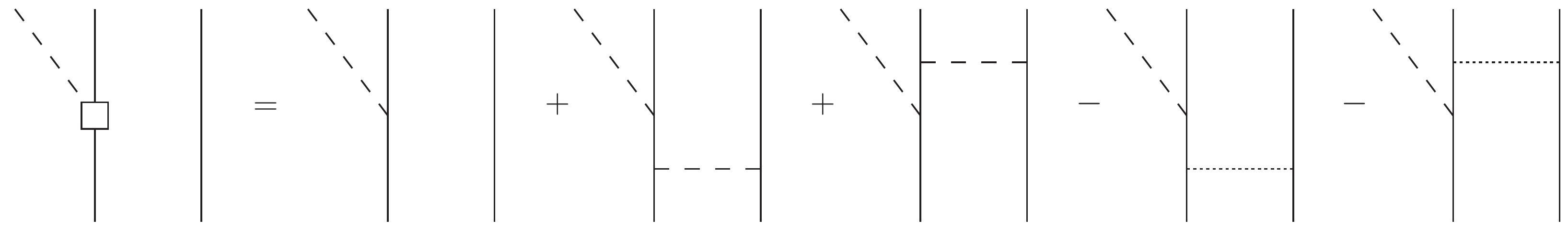}
\caption[Complete impulse amplitude]{Complete impulse contribution.  Solid lines represent nucleons, dashed lines represent pions, dotted lines represent ``wave function" pions with $\omega=0$, and the square represents the operator which is used for the full impulse approximation.\label{fig:strongiasum}}
\end{figure}

Thus we make the following na\"ive choice: to obtain the correct impulse contribution, we add up the contributions from \cref{fig:stronglo}a and \cref{fig:strongia} and then subtract what is already included in the wave functions (the last two diagrams of \cref{fig:strongiawfn}).  This calculation is schematically shown in \cref{fig:strongiasum}.

The OPE propagator is of Yukawa form,
\begin{align}
D_\pi(\omega,r)=-\frac{e^{-\mu(\omega) r}}{4\pi r},
\end{align}
where $\mu(\omega)=\sqrt{m_\pi^2-\omega^2}\approx\sqrt{3}/2m_\pi$.  Thus subtracting off the final two diagrams in \cref{fig:strongiasum} amounts to making the replacement
\begin{align}
\frac{e^{-\sqrt{3}m_\pi r/2}}{r}\rightarrow\frac{e^{-\sqrt{3}m_\pi r/2}}{r}-\frac{e^{-m_\pi r}}{r},
\end{align}
in the exchanged pion propagator.  The final four terms of \cref{fig:strongiasum} comprise a correction to the impulse diagram, which we will refer to as a ``wave function correction."  We find that this correction is $\sim4\%$ of the total impulse amplitude at the experimental energy, and we include it in our calculation.  However, it is important to note that a different method is developed in \cref{chap:nrred} to handle the IA within the hybrid formalism.  To give a complete picture, we leave this attempt as it was at the time this paper was published.

\subsection{Power Counting\label{sec:power counting}}

Now we will look more closely at the size of these diagrams in the MCS counting scheme, using the counting techniques summarized in the review article \citep{Hanhart:2003pg}.  In \cref{sec:history} we explained how to power count the IA and RS diagrams for $p$-wave production, and in this section, we expand to the rest of the diagrams and channels.  The first step is to consider how the external momenta and masses appear in the calculation of a particular diagram.  Then, MCS can be used to assign each diagram's status as LO, NLO, etc.

Let us examine the Feynman rules which we will use.  The propagators are calculated from the Lagrangian in \cref{eq:l0}
\begin{align}\bs
D_N(p)&=\frac{i}{p^0+i\epsilon}
\\
D_\Delta(p)&=\frac{i}{p^0-\delta+i\epsilon}
\\
D_\pi(p)&=\frac{-i}{\bfp^2+(m_\pi^2-(p^0)^2)-i\epsilon}.\label{eq:props}
\es\end{align}
Also from the leading order Lagrangian, we see that both the $\pi NN$ and the $\pi N\Delta$ vertices are proportional to $|\bfq|$, the pion momentum \textit{at that vertex}.  Note that this momentum is $\tilde{p}$ in the OPE vertices.  The only other vertex appearing thus far is the WT, which is proportional to $\omega_{q,\text{in}}+\omega_{q,\text{out}}$.

The external particles have the same momenta in each diagram.  The produced pion has $q=(\omega_q,\bfq)\approx(m_\pi,0)$, and the incoming nucleons have $p_{1,2}=(E_{1,2},\pm\bfp)\approx\left(m_\pi/2,\pm{\bf \tilde{p}}\right)$ in the Heavy Baryon formalism in which the nucleon mass is subtracted off of the energy component.

We saw in \cref{sec:history} that for the $p$-wave IA diagram of \cref{fig:strongia}a the final emission contributes $q$, the nucleon propagator $1/m_\pi$, and the OPE $\tilde{p}\cdot1/\tilde{p}^2\cdot\tilde{p}$ so that the whole diagram is $\sim q/m_\pi$.  These same arguments give that $s$-wave RS diagram is $\sim\frac{3m_\pi/2}{\tilde{p}}$, and the $p$-wave Delta diagram is $\sim\frac{q}{m_\pi-\delta}$.  Thus, according to the MCS scheme, the $p$-wave IA and Delta diagrams are order $\chi^0$ and the $s$-wave RS diagram is order $\chi^1$.  However, more practically, we note that $q/m_\pi=\eta\approx\chi^2$ and $\delta\approx 2m_\pi$ so that the IA and Delta diagrams are numerically $\sim\chi^2$.  This ordering comes in agreement with the fact that the $s$-wave diagrams should make a larger contribution to the cross section than $p$-wave diagrams.

Na\"ively, these three amplitudes from \cref{fig:stronglo} represent the complete LO calculation, $\chi^1$ for $s$-wave and $\chi^2$ for $p$-wave.  However, it is well documented that these three amplitudes alone do not correctly reproduce the experimental data for the reaction\footnote{Admittedly, this is not necessarily a serious problem since we have only considered the leading order Lagrangian and have already stated the convergence will be slow.  A more serious problem is expounded upon at the end of this section.}.  Near threshold ($\eta\approx0.05$), the most recent $np\to d\pi^0$ experiment \citep{Hutcheon:1989bt} found $\alpha=\sigma/\eta\approx90\ \mu$b,\footnote{Note that in this chapter we use a different definition for $\alpha$; its values in this chapter are two times smaller than in the other chapters.} while for these first three diagrams, we find $\alpha\approx55\ \mu$b.

The solution to this problem was discovered by Ref. \citep{Lensky:2005jc}, who noticed that the $\nu=1$ ``recoil" correction to the WT vertex, which is found in \cref{eq:l1} goes like $(\bfq_\text{in}+\bfq_\text{out})\cdot(\bfp_\text{in}+\bfp_\text{out})/(2m_N)$.
\begin{figure}
\centering
\includegraphics[height=1.5in]{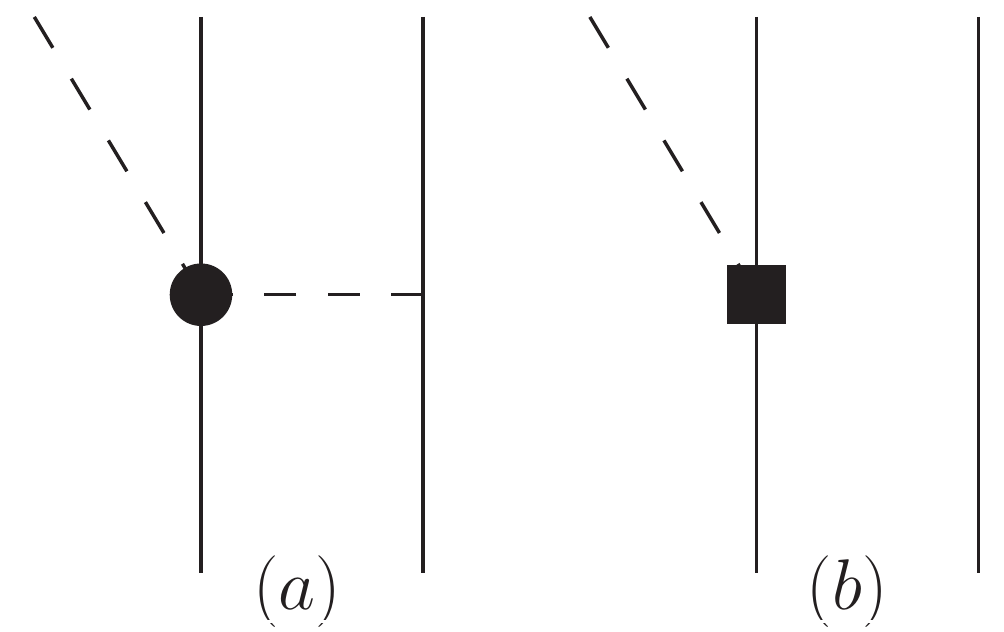}
\caption[Recoil corrections]{Recoil corrections.  Solid lines represent nucleons, dashed lines represent pions, and the filled circle and square represent $\nu=1$ vertices.\label{fig:strongrecoil}}
\end{figure}
For \cref{fig:strongrecoil}a, this vertex (a filled circle) $\sim{\bf\tilde{p}}^2/(2m_N)=m_\pi/2$ and thus this diagram is of order $\chi^1$, the same order as the $\nu=0$ rescattering diagram.  Similarly, one finds that the $s$-wave portion of the recoil correction to the impulse diagram (\cref{fig:strongrecoil}b) is of order $\chi^1$.  In this diagram, the filled square represents the sum analogous to \cref{fig:strongiasum} for the recoil diagram. We find that the wave function corrections are more important ($\sim20\%$) in this case.  Finally, the $s$-wave portion of the Delta diagram's recoil correction is found to be higher order and is therefore ignored.

The recoil corrections to the propagators have also been included in the calculation where applicable.  For this reaction, the only such diagram is \cref{fig:strongia}b where the 3-momentum in the nucleon propagator is large ($\sim\tilde{p}$).  For that propagator, we use the corrected version,
\begin{align}
D_N(p)=\frac{i}{p^0-\bfp^2/2m_N+i\epsilon}\approx-\frac{i}{m_\pi}.
\end{align}
Using this propagator rather than the $\nu=0$ version doubles the size of \cref{fig:strongia}b.  Nevertheless, this diagram (minus its $\omega_\pi=0$ analog) is already very small.  Thus the net effect of correcting the propagators is small for this reaction at this order.

\begin{figure}
\centering
\includegraphics[height=2in]{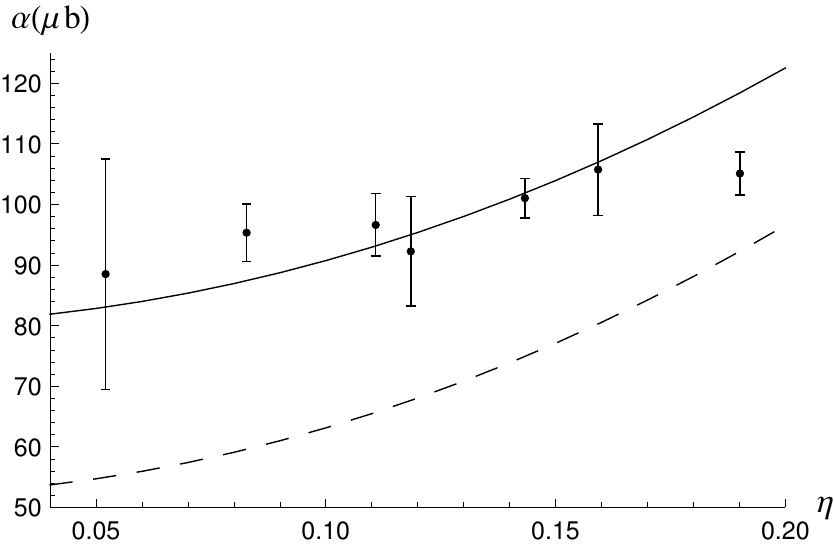}
\caption[Cross section for $np\rightarrow d\pi^0$]{Cross section for $np\rightarrow d\pi^0$ in terms of $\alpha=\sigma/\eta$ as a function of the pion center of mass momentum, $\eta=q/m_{\pi^0}$.  Circles with error bars display the experimental results of Ref. \citep{Hutcheon:1989bt}.  The dashed line displays the results of including the diagrams in \cref{fig:stronglo} and the solid line displays the results of also including the recoil terms discussed in the text.\label{fig:alpha}}
\end{figure}

Including all these recoil corrections (especially \cref{fig:strongrecoil}a) brings the theoretical cross section near the experimental results as shown by the solid curve of \cref{fig:alpha}.
Due to the relative scatter of the data shown in \cref{fig:alpha} it is difficult to tell how well the theory is reproducing the experiment.  Regardless, it is clear that theoretical improvement has been made.

\begin{figure}
\centering
\includegraphics[height=1.5in]{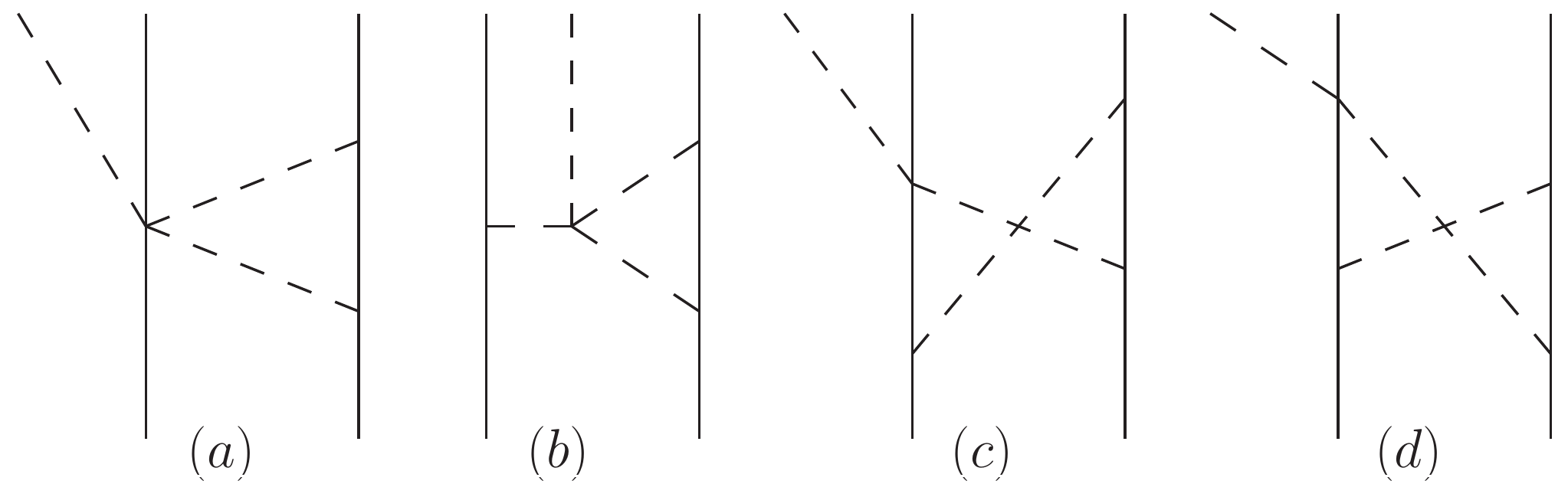}
\caption[Irreducible loops for $np\to d\pi^0$]{Irreducible loops.  Solid lines represent nucleons and dashed lines represent pions.\label{fig:irrloops}}
\end{figure}

It should also be noted that the subtlety of reducibility and recoil corrections in this reaction resolves questions about NLO loop diagrams discovered by Ref. \citep{Gardestig:2005sn}.  Namely, the sum of all the NLO irreducible loops in \cref{fig:irrloops} is found to be proportional to $\bfp$.  This is a problem because such sensitivity of the operator to the $NN$ wave function is not physical.

\begin{figure}
\centering
\includegraphics[height=1.5in]{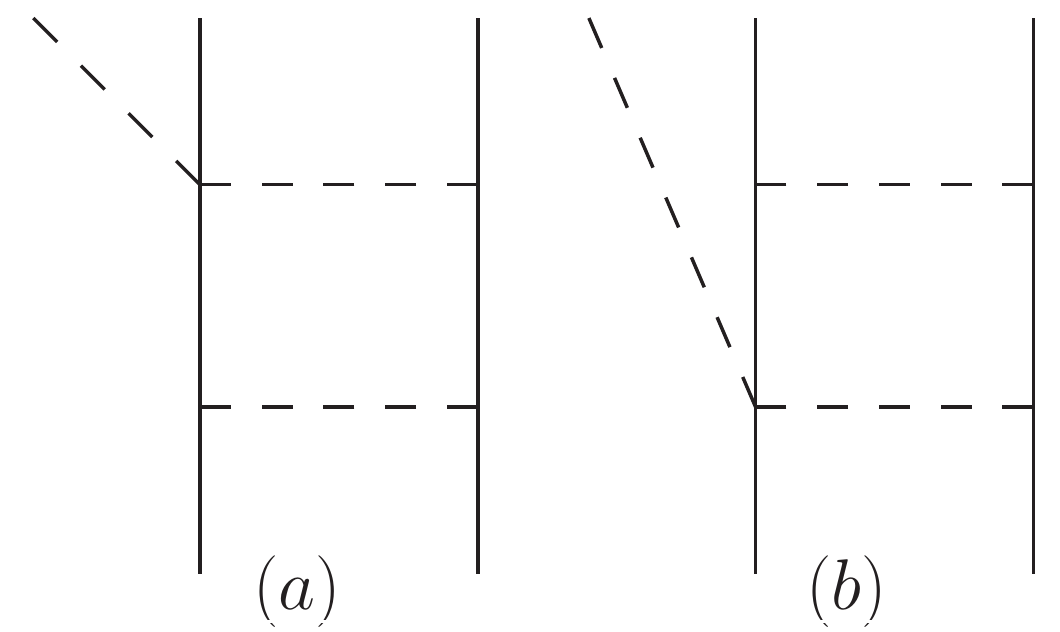}
\caption[Reducible loops for $np\to d\pi^0$]{Reducible loops.  Solid lines represent nucleons and dashed lines represent pions.\label{fig:redloops}}
\end{figure}

The solution to the problem is, once again, to consider including OPE in the operator, this time for the rescattering diagram.  There are two resulting diagrams shown in \cref{fig:redloops}.  
It was shown in Ref. \citep{Lensky:2005jc} that in these topologically reducible loops, the recoil corrections to the nucleon propagators need to be included in addition to the WT's recoil correction.  This key argument is an application of the MCS power counting scheme.  Reference \citep{Lensky:2005jc} then showed that (part of) the energy dependence of the WT vertex ``cancels" one of the nucleon propagators leaving a reducible diagram similar to \cref{fig:irrloops}a.  This diagram is equal in magnitude and opposite in sign to the aforementioned NLO sum, resolving the issue.

The other term that remains from the original loop integral after this manipulation is still of reducible form but now has an on-shell WT vertex $\sim2m_\pi$.  This term would already appear upon convolution of the rescattering diagrams discussed above (including recoil corrections) with external wave functions, i.e. this term is truly reducible.  This result can be stated another way: cancellation of the irreducible loops comes from a short range piece of including OPE in the rescattering operator.  Thus a complete NLO calculation must include this OPE.  However, its remaining reducible piece would only be included in the calculation if the recoil correction to the WT vertex is included in the rescattering operator.  Therefore it is both consistent and necessary to include the WT recoil correction in the pion production operator.

\subsection{P-Wave Observables\label{sec:pwave}}

Another important test of the theory is how well it describes $p$-wave pions \citep{Baru:2009fm}.  This is especially important for the asymmetry, which involves strong $p$-waves at leading order.  The differential cross section can be expanded in Legendre polynomials,
\begin{align}
\frac{d\sigma}{d\Omega}=\alpha_0+\alpha_1P_1(\cos(\theta))+\alpha_2P_2(\cos(\theta))+...,\label{eq:legendre}
\end{align}
where $\theta$ is the angle between $\bfp$ and $\bfq$.  Note that the total cross section plotted in \cref{fig:alpha} is $\alpha=4\pi\alpha_0/\eta$.  As discussed in \cref{sec:observables}, $\alpha_2$ receives contributions almost exclusively from $p$-wave pions.  The ratio $\alpha_2/\alpha_0$ is therefore used as a test for this part of the theory.  We find that the diagrams of \cref{fig:stronglo} along with their recoil corrections overestimate the data by approximately a factor of two.

Upon closer inspection we find that the $^1S_0$ amplitude (which is supposed to be small) is relatively large.  This amplitude is coming mainly from the Delta diagram, as can be seen in \cref{sec:observables} where the values of the reduced matrix elements are listed.  To remedy the situation in the simplest way possible, we implement a cutoff for the Delta diagram,
\begin{align}\bs
D_\pi=\frac{-i}{\bfp^2+\mu^2}\to D_\pi^c(\Lambda)&\equiv\frac{-i}{\bfp^2+\mu^2}\left(\frac{\Lambda^2}{\bfp^2+\Lambda^2}\right)
\\
&=\left(\frac{-i}{\bfp^2+\mu^2}-\frac{-i}{\bfp^2+\Lambda^2}\right)\frac{\Lambda^2}{\Lambda^2-\mu^2}.\label{eq:cutoff}
\es\end{align}

One can show that doing this essentially softens the OPE potential for $r<\log(\Lambda/\mu)/\Lambda$.  Note that one consequence of such a cutoff is that it modifies both the $^1S_0$ and the $^1D_2$ channels.  Clearly this cutoff is not an acceptable long-term solution for an effective field theory, but the fact that such a procedure is necessary is interesting given that the reaction occurs at an energy  $\sim150\ \text{MeV}$ below the Delta resonance.  That $p$-wave pion production is highly sensitive to the strength of its contact term,
\begin{align}
{\cal L}_\text{ct}=-\frac{d_1}{f_\pi}\left[N^\dagger\left(\btau\cdot\vec{\sigma}\cdot\vec{\nabla}\bpi\right)N\right]\left[N^\dagger N\right],\label{eq:d1ct}
\end{align}
was discussed by Ref. \citep{Hanhart:2000gp}.  Also note that if we were to instead use a dipole cutoff, we would find a larger value is needed, $\Lambda\to\sqrt{2}\Lambda$.  The $\pi N\Delta$ vertex should fall at least as fast as the $\gamma NN$ vertex which, according to ``quark counting", goes as $1/Q^4$.  In this chapter, we use the cutoff in \cref{eq:cutoff} for simplicity, but recognize that the corresponding $\Lambda$ will be small.

In a complete calculation, the cutoff dependence of the offending $^1S_0$ amplitude will be absorbed by the contact term, $d_1=d_1(\Lambda)$.  Naive dimensional analysis (NDA) of this term leads one to expect $d_1\sim1/\Lambda^3$.  Other tree-level diagrams will contribute to the $^1S_0$ amplitude at this order; however, for the purposes of this thesis, we will assume that the Delta diagram is large enough to justify ignoring these other diagrams.  In this simplified case, we can investigate the validity of the NDA scaling by looking at the $\Lambda$-dependence of the Delta amplitude.

\begin{figure}
\centering
\includegraphics[height=2in]{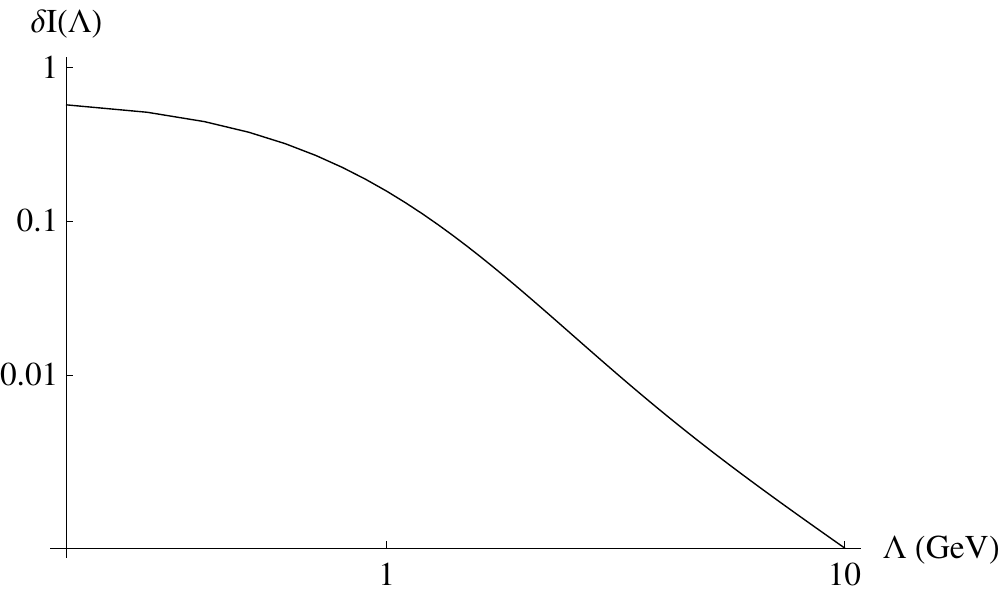}
\caption[Cutoff dependence of the Delta diagram's $^1S_0$ amplitude]{\label{fig:1S0cutoff}Cutoff dependence of the Delta diagram's $^1S_0$ amplitude.  The vertical axis shows the change in the dimensionless radial integral $I$ which depends on the cutoff according to \cref{eq:cutoff}.  This change is defined according to $\delta I(\Lambda)=I(\infty)-I(\Lambda)$.}
\end{figure}

The log-log plot of \cref{fig:1S0cutoff} shows the change in the ${\cal M}^\Delta_{^1S_0}$ amplitude (the corresponding radial integral is what appears in the plot) as a function of the cutoff.
The true value of $d_1$ is set by requiring that some experimental $p$-wave observable is exactly reproduced; for this simplified analysis we use the condition,
\begin{align}
{\cal M}_{^1S_0}={\cal M}^\Delta_{^1S_0}+{\cal M}^\text{ct}_{^1S_0}=0.\label{eq:1S0cond}
\end{align}
For large values of the cutoff, the amplitude is unmodified and the contact term takes on a large magnitude in order to satisfy \cref{eq:1S0cond}.  For small values of the cutoff, the amplitude is suppressed and the contact term is smaller in magnitude.  The calculated scaling behavior, given by the slope of the log-log plot, is $\sim-2$, indicating that $\delta{\cal M}^\Delta_{^1S_0}(\Lambda)\sim1/\Lambda^2$.

There are two ways to interpret the $1/\Lambda^2$ scaling.  Firstly, higher order contact terms are not required since the calculated scaling is not dependent on a larger negative power of the cutoff.  Secondly, it appears that the contact term contributes at lower order than NDA would lead one to believe.  This discrepancy is either due to the cutoff dependence of the other diagrams which contribute to the $^1S_0$ amplitude (and were ignored in \cref{fig:1S0cutoff}), or else a modified power counting scheme will be required.  In such a modified scheme, the $d_1$ contact term would be promoted to leading order.

In fact, there already exists a modified power counting scheme for spin-triplet $NN$ scattering at low energies.  The original identification of this breakdown of Weinberg counting was given in Ref. \citep{Beane:2001bc}.  As discussed recently in Ref. \citep{Birse:2007sx}, the relevant small scale is the strength of OPE in the triplet channel, specifically, the tensor interaction parametrized by,
\begin{align}
\lambda_\pi=\frac{16\pi f_\pi^2}{g_A^2m_N},
\end{align}
which is numerically $\sim2m_\pi$.  In pion production, the $NN\to N\Delta$ transition that participates in ${\cal M}^\Delta_{^1S_0}$ (of which the final state is spin-triplet) proceeds exclusively through this tensor interaction, so it is not surprising that we observe this scaling behavior.  Further discussion of this issue is beyond the scope of this thesis.

\begin{figure}
\centering
\includegraphics[height=2in]{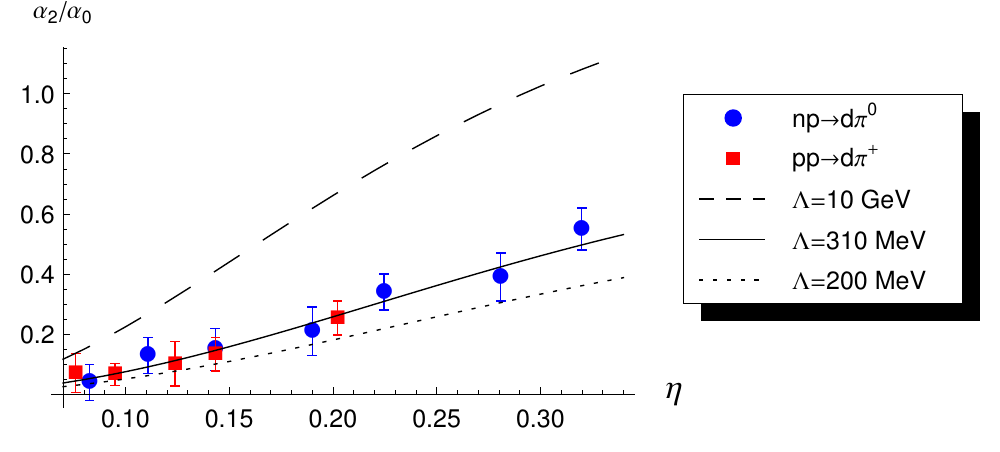}
\caption[Legendre coefficients of the differential cross section]{\label{fig:a2}Legendre coefficients of the differential cross section for different values of the cutoff.  Data is from an $np\to d\pi^0$ experiment (Ref. \citep{Hutcheon:1989bt}, circles) and an $\vec{p}p\to d\pi^+$ experiment (Ref. \citep{Heimberg:1996be}, squares) in which the data have not be corrected for Coulomb effects.}
\end{figure}

\Cref{fig:a2} shows the effects of this cutoff on the ratio $\alpha_2/\alpha_0$, where $\Lambda=10\ \text{GeV}$ represents the original theory (such a large cutoff has no significant effect).
The amplitudes for $np\to d\pi^0$ are related to those for $pp\to d\pi^+$ (which are bigger by $\sqrt{2}$) when charge independence is respected.  Thus, the ratio plotted should have the same value for both reactions.  By adjusting the cutoff to fit the data, we find $\Lambda=310$ MeV.

\begin{figure}
\centering
\includegraphics[height=2in]{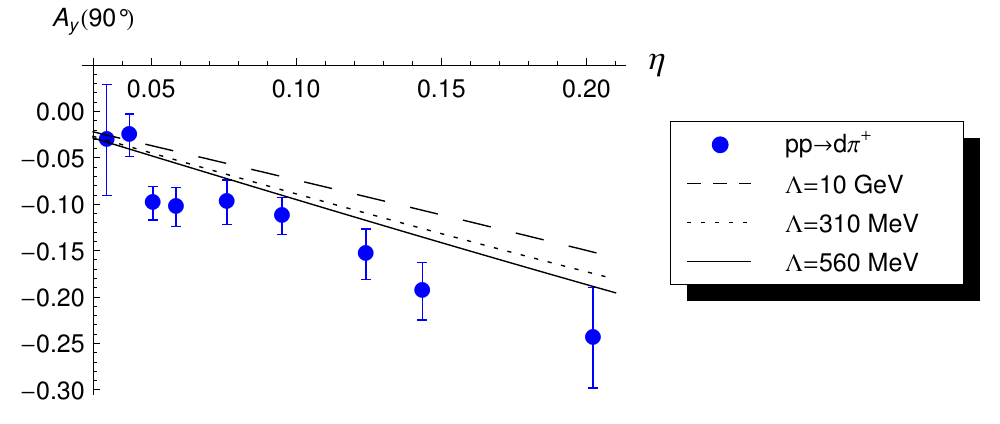}
\caption[Analyzing power for different values of the cutoff]{\label{fig:ay}Analyzing power for different values of the cutoff.  Data is from a $\vec{p}p\to d\pi^+$ experiment (Ref. \citep{Heimberg:1996be}, circles) in which the data have not be corrected for Coulomb effects.}
\end{figure}

Another useful observable for testing $p$-wave pion production is the analyzing power, $A_y$, which is defined 
\begin{align}
A_y(\theta)&\equiv\frac{d\sigma_\uparrow(\theta)-d\sigma_\downarrow(\theta)}{d\sigma_\uparrow(\theta)+d\sigma_\downarrow(\theta)}\label{eq:aydef}
\\
d\sigma_{\uparrow,\downarrow}(\theta)&\equiv\frac{|\bfq|}{64\pi^2s|\bfp|}\frac{1}{4}\sum_{m_d,m_2}\left|\mathcal{M}\left(m_{1,y}=\pm1/2,\theta\right)\right|^2,\label{eq:dsdoup}
\end{align}
where $m_{1,y}=\pm1/2$ refers to the fact that the beam is polarized perpendicular to the scattering plane.  In the z-basis, these states are
\begin{align}
\mathcal{M}\left(m_{1,y}=\pm1/2\right)=\frac{\mathcal{M}\left(m_1=1/2\right)\pm i\mathcal{M}\left(m_1=-1/2\right)}{\sqrt{2}}.
\end{align}
As shown in \cref{sec:observables}, $A_y$ is proportional to the product of $s$-wave and $p$-wave amplitudes.  \Cref{fig:ay} shows the effects of the cutoff on this observable.
Again, charge independence implies that $A_y$ should be the same for both neutral and charged pion production.  We find the best agreement with the analyzing power data for $\Lambda=560\ \text{MeV}$.  Below, we will display our results using both the original theory and a cutoff taken to be the geometric mean of two fits, $\Lambda=417\ \text{MeV}$.

\section{Charge Symmetry Breaking Amplitudes\label{sec:csb}}

\subsection{Leading-Order CSB Amplitudes}
\begin{figure}
\centering
\includegraphics[height=1.5in]{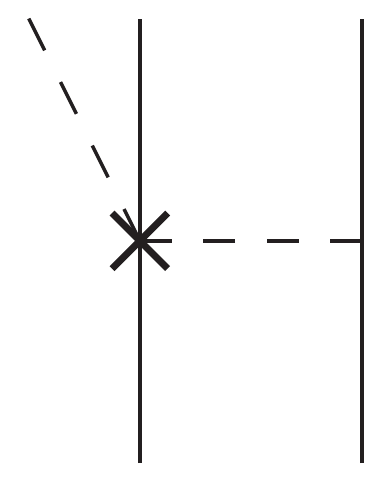}
\caption[Leading CSB diagrams]{Leading CSB contribution.  Solid lines represent nucleons, dashed lines represent pions, and crosses represent $\nu=1$ CSB vertices.\label{fig:csbrs}}
\end{figure}

The fact that the up and down quarks have different mass is reflected in the Lagrangian by including terms which break chiral symmetry \citep{Weinberg:1994tu}.  The leading such terms are given in \cref{eq:l1} and have coupling constants $\delta m_N$ and $\overline{\delta}m_n$ which are constrained by
\begin{align}
\delta m_N+\overline{\delta}m_N=m_n-m_p.
\label{eq:csbsize}
\end{align}
Recall from \cref{sec:csbops} that the $\delta m_N$ term has its origins in the quark mass difference and its size is $\sim (m_d-m_u)\equiv\epsilon(m_d+m_u)$ with $\epsilon\approx1/3$.  Chiral symmetry tells us that $(m_d+m_u)\propto m_\pi^2$, and so dimensional analysis along with the hadronic scale $m_N$ yields $\delta m_N\sim\epsilon m_\pi^2/m_N$.  The $\overline{\delta}m_N$ term is of electromagnetic origins, but is of the same order as $\delta m_N$.  These CSB operators appear in the rescattering diagram depicted in \cref{fig:csbrs}
where the CSB vertex is denoted with a cross.  The size of this diagram is $\delta m_N/\tilde{p}\approx\epsilon m_\pi^2/(m_N\tilde{p})=\epsilon\chi^3$.  Note that although the full nucleon mass difference appears explicitly in the Lagrangian at this order, the corresponding operator ($N^\dagger\tau_3N$) does not change the parity and thus does not contribute to an asymmetry.

The Cottingham formula \citep{Gasser:1982ap} (a model for the electromagnetic contribution to $m_n-m_p$) can be used to obtain $\overline{\delta}m_N=-(0.76\pm0.30)$ MeV.  The constraint of \cref{eq:csbsize} then fixes $\delta m_N=2.05\pm0.30$ MeV. In this chapter, we use this value of $\delta m_N$ for our final results, though other values have been suggested.  Reference \citep{vanKolck:2000ip} discusses models which predict values for $\overline{\delta}m_N$ leading to $1.83\leq\frac{\delta m_N}{\text{MeV}}\leq2.83$.  Additionally, a recent LQCD study \citep{Beane:2006fk} found $\delta m_N=2.26\pm0.72$, having used the MILC collaboration's determination of $m_u/m_d$ \citep{Aubin:2004fs}.

Though we do not present the full details of the calculation of this CSB diagram, let us briefly consider on the isospin algebra which one encounters,
\begin{align}
\hat{{\cal M}}\propto\left[\delta m_N(\btau(1)\cdot\btau(2)+\tau(1)_3\tau(2)_3)-\overline{\delta}m_N(\btau(1)\cdot\btau(2)-\tau(1)_3\tau(2)_3)\right].
\end{align}
Evaluating this between $T=0$ states yields,
\begin{align}
{\cal M}\propto\left[\delta m_N-\frac{\overline{\delta}m_N}{2}\right].
\end{align}
This is the distinct algebraic combination we referred to in the introduction.  Since this combination is different than that appearing in \cref{eq:csbsize}, one can use \cref{eq:csbsize} to eliminate $\overline{\delta}m_N$ in favor of the well-known nucleon mass difference.

Let us now discuss another LO source of CSB coming from a more detailed evaluation of the strong RS diagram shown in \cref{fig:stronglo}b which was made by Ref. \citep{Filin:2009yh}.  We became aware of this group's efforts when nearing the end of our study and worked with the authors of Ref. \citep{Filin:2009yh} to resolve differences between our calculations.  For clarity in this thesis, we include this new CSB amplitude into the main body of the calculation.

Charge symmetry breaking occurs in the strong RS diagram as a consequence of the time derivative in the WT vertex.  The diagram is typically considered by treating the external nucleons as identical.  The fact that they are not identical means that the energy transferred in the pion exchange will not only include the $m_\pi/2$ required by the kinematics, but also a term $\pm(m_n-m_p)/2$ (positive if the WT vertex is on the proton, negative if on the neutron).  The effect of this contribution is shown in Ref. \citep{Filin:2009yh} to be equivalent to a change in the CSB rescattering diagram,
\begin{align}
\left[\delta m_N-\frac{\overline{\delta}m_N}{2}\right]\to\left[\delta m_N-\frac{\overline{\delta}m_N}{2}\right]+\frac{\delta m_N+\overline{\delta}m_N}{2}=\frac{3\delta m_N}{2}.
\label{eq:newcsb}
\end{align}

\subsection{Next-To-Leading-Order CSB Amplitudes}
\begin{figure}
\centering
\includegraphics[height=1.5in]{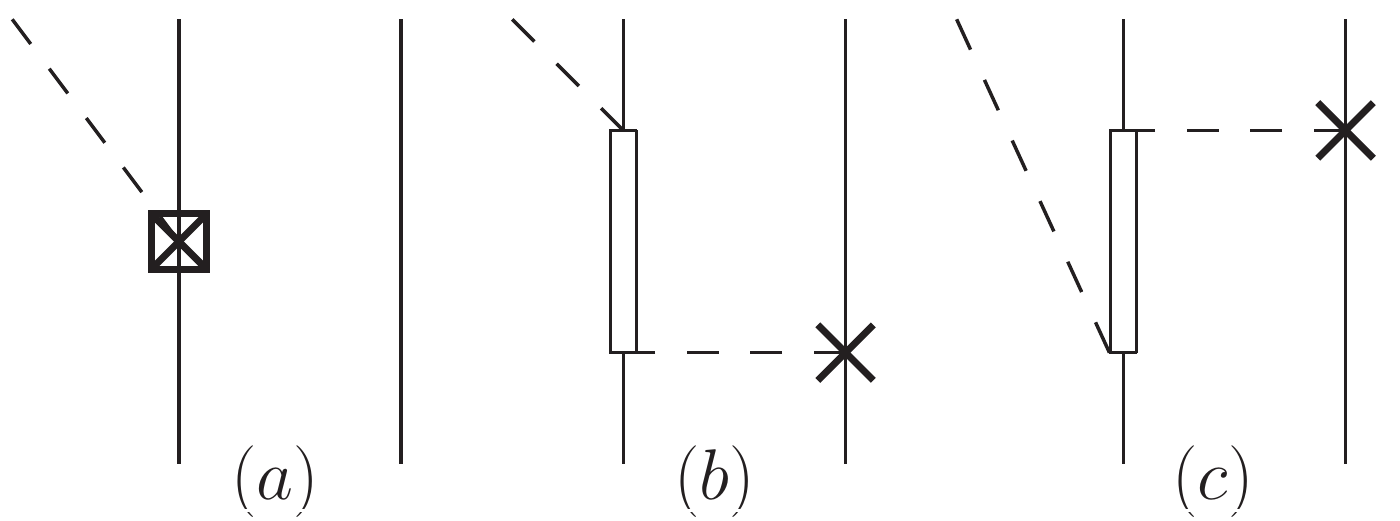}
\caption[$\nu=2$ CSB diagrams]{$\nu=2$ CSB contributions.  Solid lines represent nucleons, double solid lines represent $\Delta$'s, dashed lines represent pions, crosses represent $\nu=2$ CSB vertices, and the boxed cross represents the full impulse CSB diagram including OPE.\label{fig:csbpwave}}
\end{figure}

Another CSB term given in \cref{eq:l2} involves one derivative and one $m_\pi^2$ ($\beta_1\sim\epsilon m_\pi^2/m_N^2$) and is thus a $\nu=2$ vertex with momentum dependence $|\bfq|$.  This vertex appears in the diagrams of \cref{fig:csbpwave} whose sizes are $\beta_1q/m_\pi\approx\epsilon\eta\chi^4$.  In \cref{fig:csbpwave}a, the boxed cross represents the sum analogous to \cref{fig:strongiasum} for the CSB impulse diagram.  Again, the wave function corrections are small ($2\%$).

Little is known about $\beta_1$, the impulse CSB coupling.  As a starting point, Ref. \citep{vanKolck:1996rm} notes that this CSB operator can be viewed as arising from $\pi-\eta$ mixing.  The result shown is that
\begin{align}
\beta_1=\frac{g_\eta f_\pi}{m_Nm_\eta^2}\langle\pi^0|H|\eta\rangle=c_\eta\left(\frac{\epsilon\ m_\pi^2}{m_N^2}\right),
\label{eq:beta1}
\end{align}
with the idea that the value $c_\eta\sim{\cal O}(1)$ is ``natural".  As discussed in the review \citep{Miller:2006tv},
\begin{align}
0.10\leq\frac{g_\eta^2}{4\pi}\leq0.51.
\label{eq:etasize}
\end{align}
Also, Ref. \citep{Coon:1995xp} gives $\langle\pi^0|H|\eta\rangle=-0.0039\ \text{GeV}^2$, and we use $m_\eta=547.51$ MeV.  These values result in $-0.47\leq c_\eta\leq-0.21$.  Thus it is at least plausible that the $\beta_1$ term could originate naturally from $\eta-\pi$ mixing.  We note that the $\eta'$ could also give such a term, but do not consider it here.

Using \cref{eq:beta1,eq:etasize}, we obtain $-3.2\times10^{-3}\leq\beta_1\leq-1.4\times10^{-3}$.  Note that the value used in the original calculation of the asymmetry by Ref. \citep{Niskanen:1998yi} was $\beta_1=-8.7\times10^{-3}$, which we refer to as the ``extreme value".  However, according to the above discussion, the ``natural" size is $\beta_1\sim-\frac{\epsilon m_\pi^2}{m_N^2}\approx-6\times10^{-3}$.  Thus even though its origins may not lie exclusively with the $\eta$, the aforementioned ``extreme" value for $\beta_1$ is not extreme at all from the effective field theory's point of view.

\section{Asymmetry Results\label{sec:csbresults}}

Contributions to the asymmetry come from interference terms between the $p$-wave part of the strong amplitude and the $s$-wave part of the CSB amplitude, and vice versa.  The issue is somewhat complicated because, in contrast to threshold emission, each diagram can contribute in both the $s$-wave and the $p$-wave.  However, contributions to the sub-leading parity ($s$-wave for the impulse and Delta diagrams and $p$-wave for the rescattering diagrams) are formally higher order.  An example of this is the strong rescattering diagram for $p$-wave pions considered in \cref{sec:history} which was found to contribute according to $\sim\eta\chi^2$.

\begin{figure}
\centering
\includegraphics[width=\linewidth]{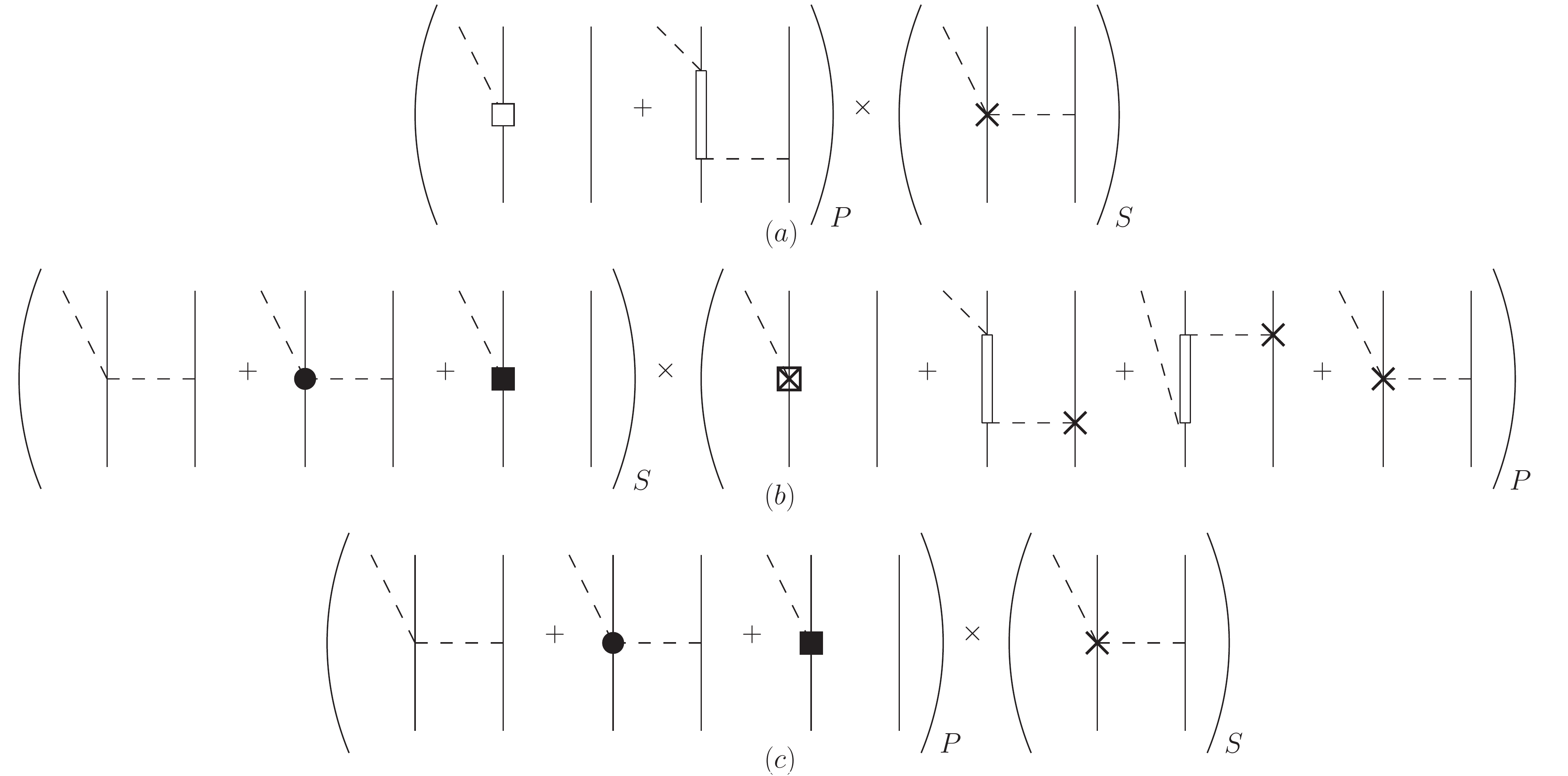}
\caption[Interference terms for the asymmetry in $np\rightarrow d\pi^0$]{Interference terms for the asymmetry in $np\rightarrow d\pi^0$.  Solid lines represent nucleons, double solid lines represent $\Delta$'s, and dashed lines represent pions.  The filled circle and square represent $\nu=1$  strong vertices and crosses represent CSB vertices.  The boxes represent full impulse diagrams in the sense of \cref{fig:strongiasum}.\label{fig:asymmetry}}
\end{figure}

The contributions to the asymmetry are depicted in \cref{fig:asymmetry}.
\Cref{fig:asymmetry}a includes strong $p$-waves and CSB $s$-waves and has size $(\eta)\times(\epsilon\chi^3)$.  \Cref{fig:asymmetry}b includes strong $s$-waves and CSB $p$-waves and has size $(\chi)\times(\epsilon\eta\chi^4)$.  \Cref{fig:asymmetry}c includes strong $p$-waves and CSB $s$-waves and has size $(\eta\chi^2)\times(\epsilon\chi^3)$.  Thus we find that in these kinematics \cref{fig:asymmetry}a ($\sim\epsilon\eta\chi^3$) is the LO contribution, and \cref{fig:asymmetry}b,c ($\sim\epsilon\eta\chi^5$) both come in at NLO.\footnote{Here we refer to the $\sim\epsilon\eta\chi^5$ terms as NLO even though they are suppressed by $\chi^2$ because no terms suppressed by a $\chi^1$ appear.}  Other interference terms involving these diagrams are higher order, $\sim\epsilon\eta^3\chi^5$ or smaller.  Finally we note that this work is not intended to be a complete NLO calculation since loops and higher order vertices may contribute at this order.  The identification of all NLO diagrams is beyond the scope of this work.

The details of the calculation are spelled out in \cref{chap:csbdetails}.  The coupling of the WT vertex (and its recoil correction) is determined by chiral symmetry, but the LECs $f_\pi$, $g_A$, and $h_A$ appear in the calculation and we use $f_\pi=91.9$ MeV, $g_A=1.267$, and $h_A=2.1g_A$.  We use the following masses: $m_N=(m_n+m_p)/2=938.919$ MeV, $m_n-m_p=1.293$ MeV, $m_\Delta=1232$ MeV, and $m_\pi=m_{\pi^0}=134.977$ MeV.  In \cref{sec:reduced}, the amplitudes are expressed in terms of reduced matrix elements based on the partial wave decomposition.  Next, in \cref{sec:method}, we present an example calculation: the reduced matrix elements corresponding to the RS diagram.  Finally, in \cref{sec:observables}, we connect the observables (cross section and $p$-wave observables) with the reduced matrix elements.  Also in this section one can find a table of our results for each individual reduced matrix element.

In \cref{tab:csbresults}, we display our asymmetry results.  Because the asymmetry is linear in the CSB amplitudes (and therefore the CSB parameters), we are able to present our results as a set of coefficients, $\{x,y\}$ defined by
\begin{align}
A_{fb}\times10^4=x\cdot\left(\frac{\delta m_N}{\text{MeV}}\right)+y\cdot(\beta_1\times10^3).
\label{eq:parametrization}
\end{align}
\begin{table}
\caption[Asymmetry in $np\rightarrow d\pi^0$]{Asymmetry in $np\rightarrow d\pi^0$ as a function of CSB parameters $\delta m_N$ and $\beta_1$.  ``LO" and ``NLO" represent the sums discussed in \cref{fig:asymmetry}.  The Delta cutoff is of monopole form; a dipole cutoff would be $\sqrt{2}$ bigger.\label{tab:csbresults}}
\begin{center}
\begin{tabular}{lcccc}
\hline\hline
\multicolumn{4}{c}{$A_{fb}\times10^4=x\cdot(\frac{\delta m_N}{\text{Mev}})+y\cdot(\beta_1\times10^3)$} & $A_{fb}(2.05\ \text{MeV},-0.0032)\times10^4$\\ \cline{1-4}
Order & Delta Cutoff & x & y & \\ \hline
LO (no recoil) & None & 33.7 & 0 & 69.1\\ 
LO (no recoil) & $\Lambda=417\ \text{MeV}$ & 27.6 & 0 & 56.6\\ 
NLO (no recoil) & None & 37.6 & 1.4 & 72.6\\ 
NLO (no recoil) & $\Lambda=417\ \text{MeV}$ & 32.5 & 1.8 & 61.0\\[.1in]
LO & None & 25.0 & 0 & 51.2\\ 
LO & $\Lambda=417\ \text{MeV}$ & 18.9 & 0 & 38.9\\ 
NLO & None & 28.1 & 1.4 & 53.1\\ 
NLO & $\Lambda=417\ \text{MeV}$ & 22.1 & 1.6 & 40.2\\[.1in]
NLO & $\Lambda=310\ \text{MeV}$ & 20.1 & 1.7 & 36.0\\ 
NLO & $\Lambda=560\ \text{MeV}$ & 24.0 & 1.6 & 44.3\\ \hline\hline
\end{tabular}
\end{center}
\end{table}

The primary advance made in this work over the previous calculation \citep{vanKolck:2000ip} is the inclusion of the rescattering and impulse recoil corrections; this improvement is shown in moving from the top four rows to the next four rows.  At LO in the asymmetry calculation this simply increases $\alpha_0$, but at NLO it affects both the numerator and the denominator of the asymmetry.  The final result, using the set of physically reasonable parameters from \cref{sec:csb}, is shown in the last column.  Our most accurate calculation (the third to last row) uses a cutoff to suppress the ${}^1S_0$ amplitude and includes the NLO interference terms to find,
\begin{align}
A^\text{thy}_\text{fb}=40.2\times10^{-4}.
\end{align}
The effects of using different values for the cutoff can be seen in the last two rows of the table.  Due to the large number of theoretical issues which still need to be addressed (primarily the cutoff in the strong $p$-waves and the model dependence of the CSB couplings), we believe that providing an uncertainty at this time would be false advertising.

The experiment of Ref. \citep{Opper:2003sb} found $A_{fb}=[17.2\pm9.7]\times10^{-4}$ and thus our calculation overestimates the data by approximately $2.5\,\sigma^\text{expt}$.  It is interesting to note that using the most ``extreme" set of parameters discussed in \cref{sec:csb} ($\delta m_N=1.83\ \text{MeV}$, $\beta_1=-8.7\times10^{-3}$) in the cutoff NLO calculation yields $A_{fb}=26.5\times10^{-4}$.  Clearly the current calculation cannot be fully judged until more accurate values for the CSB couplings are known.

\section{Discussion\label{sec:discussion}}

\subsection{Comparison with other calculations}

The first calculation of the $np\to d\pi^0$ asymmetry used a $N\Delta$ coupled channel formalism and included the CSB impulse vertex as well as other, smaller effects arising directly from the neutron-proton mass difference \citep{Niskanen:1998yi}.  This study reported $A_{fb}=-28\times10^{-4}$.  The second calculation included only the CSB rescattering vertex, and found $A_{fb}=60\times10^{-4}$ \citep{vanKolck:2000ip}.  Both these calculations were preformed before the work of Ref. \citep{Lensky:2005jc} which brought the total cross section into agreement with experiment.  Our work brings the asymmetry calculation up to date.

We have seen that there is reason for concern regarding the theoretical description of $p$-wave pions, which comprise the entire strong contribution to the LO asymmetry.  Because the total cross section is dominated by the RS diagram, small changes to the $p$-wave amplitudes are able to significantly modify the asymmetry while only slightly changing the total cross section.  As a temporary solution, we implemented a cutoff in the Delta diagram and thereby achieved acceptable agreement with the $p$-wave data.  Another solution to this problem is to use a coupled-channel $N\Delta$ potential for the initial-state.  This approach was taken by Ref. \citep{Filin:2009yh} who were able to achieve good fits to these data without a cutoff, since the OPE of the Delta diagram is then part of the wave function.

Another difference between our calculation and that of Ref. \cite{Filin:2009yh} is that in order to improve on the theoretical uncertainty ($\sim30\%$) of the Legendre coefficient $\alpha_0$, they used experimental data (from pionic deuterium) to obtain $\alpha_0=1.93\ \mu\text{b}$.  This is significantly larger than the theoretical NLO value we use, $\alpha_0=1.49\ \mu\text{b}$ ($\alpha_0=1.28\ \mu\text{b}$ for $\Lambda=417\ \text{MeV}$), and leads to a smaller value for the for the asymmetry.  Note that experiments for neutral \citep{Hutcheon:1989bt} and charged \citep{Heimberg:1996be} pion production found $\alpha_0=1.39\ \mu\text{b}$ and $\alpha_0=1.64\ \mu\text{b}$ (Coulomb corrected) respectively.  The final difference between our calculation and that of Ref. \cite{Filin:2009yh} is that they do not include any of the NLO asymmetry terms.

For the sake of comparison, we used our code to calculate the LO-only asymmetry with $\Lambda=417\ \text{MeV}$, using $\alpha_0=1.93$, and using Ref. \cite{Filin:2009yh}'s quoted values for $g_A$ and $f_\pi$.  These choices should reduce any differences between our calculations to those resulting from different wave functions.  For these choices we obtain $A_{fb}=14.0\frac{\delta m_N}{\text{MeV}}\times10^{-4}$, which is to be compared with their result of $A_{fb}=11.5\frac{\delta m_N}{\text{MeV}}\times10^{-4}$.  Although they did not present it this way, one can use the Cottingham sum rule along with their result to obtain $A_{fb}=23\times10^{-4}$, which only overestimates the data by less than one experimental uncertainty.

\subsection{Outlook}
Several issues remain to be understood theoretically.  Firstly, it appears that a contact term will be required to suppress the $^1S_0$ channel in the strong amplitude if one uses a purely $NN$ initial-state.  The interesting physics observation here is that the Delta part of the $NN$ wave function seems to be much more active than it is $NN$ scattering.  This contact term comes in at NNLO in the $p$-wave calculation, so it is more important at this time to come to a definitive conclusion on the IA reducibility issue.  The wave function corrections introduced in \cref{sec:diagrams} will be investigated in a more complete manner in the next chapter.

Secondly, the difference between the more recent experimental determination of $\alpha_0$ (from pionic deuterium) and the older $np\rightarrow d\pi^0$ data (which agrees with NLO theory) plays a large role in our over-prediction of the asymmetry.  This situation becomes even worse when the cutoff is used to decrease the $p$-wave amplitudes.  For these reasons, we conclude that further calculations are necessary.  In particular, one should extend the calculation to next order while examining both the power counting of recoil terms and the reducibility of loops.

More generally, the existence of multiple mass scales greatly complicates the power counting for this reaction, and it is clear that a converging expansion cannot yet be definitively claimed.  Another interesting aspect of this calculation is the use of a hybrid formalism.  One can argue that using phenomenological potentials to determine the $NN$ wave functions is equivalent to working to all orders in the EFT.  Thus there is a mismatch when the calculation of the operator is truncated at some order.  One way to remedy this situation (introduced by \citep{Park:2002yp}) is to use a cutoff when calculating the Fourier transforms of the operators.  This approach is taken in the next chapters.

\chapter[Impulse Reducibility: Wave Function Corrections]{Impulse approximation in the $\lowercase{np}\rightarrow\lowercase{d}\pi^0$ reaction reexamined}
\label{chap:wfncor}

The impulse approximation (one-body operator) in the $np\rightarrow d\pi^0$ reaction is reexamined with emphasis on the issues of reducibility and recoil corrections.  An inconsistency when one pion exchange is included in the production operator is demonstrated and then resolved via the introduction of ``wave function corrections" which nearly vanish for static nucleon propagators.  Inclusion of the recoil corrections to the nucleon propagators is found to change the magnitude and sign of the impulse production amplitude, worsening agreement with the experimental cross section by $\sim30\%$.  A cutoff is used to account for the phenomenological nature of the external wave functions, and is found to have a significant impact for $\Lambda\lesssim 2.5$ GeV.  This chapter is a modified version of our published paper \citep{Bolton:2010qu}; we have reorganized and rewritten parts in effort to make it flow better with the rest of this thesis.

\section{Introduction\label{sec:wfncorintro}}

As we have seen, there are two significant difficulties involved in pion production calculations.  Firstly, the large threshold momentum $\tilde{p}\sim\sqrt{m_\pi m_N}$ necessitates a reordering, sometimes called the MCS, of the $\chi$PT expansion.  Secondly, the identification of diagrams as either reducible or irreducible cannot be determined strictly by topology.  In \cref{sec:power counting} we reviewed the work of Ref. \citep{Lensky:2005jc}, which addressed these two difficulties in the NLO $pp\to d\pi^+$ calculation.  Pulling a pion exchange from the external wave functions into the term of the kernel containing the recoil $\pi\pi NN$ vertex resulted in an irreducible diagram.  This diagram cancelled other NLO loops that had an unphysical sensitivity to the wave functions in the convolution integral; the remainder of the new diagram was shown (in Ref. \citep{Lensky:2005jc}) to effectively enhance the LO $\pi\pi NN$ vertex by a factor of $4/3$.  In this paper we perform a similar investigation of reducibility in the impulse approximation.

An attempt to address this issue was put forth in our study (\cref{chap:csb}, published version \citep{Bolton:2009rq}) of charge symmetry breaking in $np\rightarrow d\pi^0$ where we introduced ``wave function corrections".  These corrections were calculated for the final state, and found to be a small fraction of the impulse diagram that they are correcting.  However, this calculation suffers from a particular approximation which we will describe.  Fixing this approximation has a very significant effect.  Furthermore, we show that the wave function corrections in the initial state are larger than one would expect in the MCS scheme.\footnote{The topic of this chapter is (once again) taken up in \cref{chap:nrred}, where we present a final definitive conclusion.  For completeness, we are leaving the arguments made in this paper (nearly) as in their published form.}

In \cref{sec:reaction} we review the $np\rightarrow d\pi^0$ reaction and the impulse approximation's role.  Then, \cref{sec:ope} examines the inconsistency that is found when one includes static OPE with the impulse approximation.  Also in this section, wave function corrections are presented as a solution to the problem.  \Cref{sec:recoil} discusses the correct implementation of the recoil corrections to the nucleon propagators.  The effects of including a cutoff are shown in \cref{sec:cutoff}.  Finally, we discuss the total cross section in \cref{sec:wfncorresults} and conclude in \cref{sec:summary}.

\section{Pion Production Review\label{sec:reaction}}

\begin{figure}
\centering
\includegraphics[height=2in]{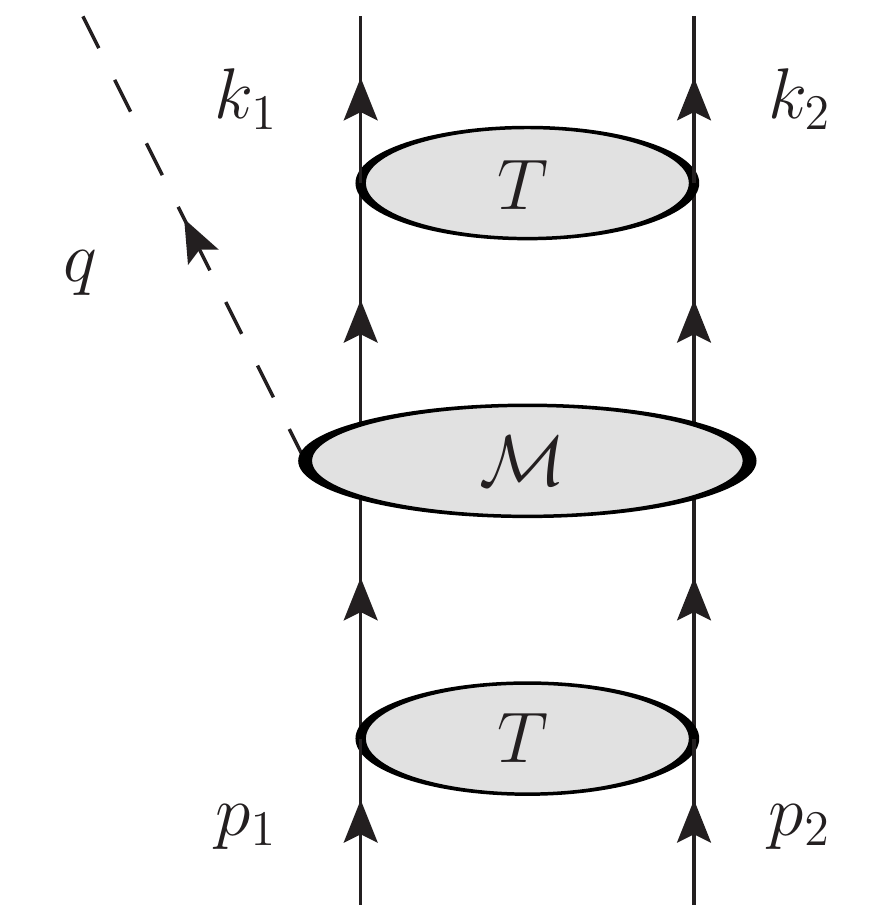}
\caption[Pion production operator]{Pion production operator.  Solid lines represent nucleons, dashed lines represent pions, and ovals represent interactions.\label{fig:momenta}}
\end{figure}

The pion production operator in momentum space depicted in Fig. \ref{fig:momenta} is a function of the pion momentum $\bfq$ and $\bfl=\bfk-\bfp$, where $\bfk$ ($\bfp$) refers to the final (initial) relative momentum of the nucleons.
The momentum transfer between the two nucleons in $\mathcal{M}$ is defined as $q'\equiv p_2-k_2$.  We use a phenomenological, non-relativistic potential ($V$) that is static: the energy of each individual nucleon is conserved by $V$.  Given this choice, working in the center of mass frame requires $q'\,^0=\omega_q/2$.  It should be noted that this is an approximation (the ``fixed kinematics approximation") that one needs to adopt in order to work in position space.  If one works in momentum space and fixes $q'\,^0$ via energy conservation at the $NN\pi$ vertex, this is called the ``equation of motion approximation."  For the RS diagram, Ref. \citep{Hanhart:2000wf} found that both of these approximations have problems, particularly when considering initial state interactions.  Nevertheless, to work within the hybrid formalism one of the approximations is required, and it appears that the former is preferable to the latter.

\begin{figure}
\centering
\includegraphics[width=\textwidth]{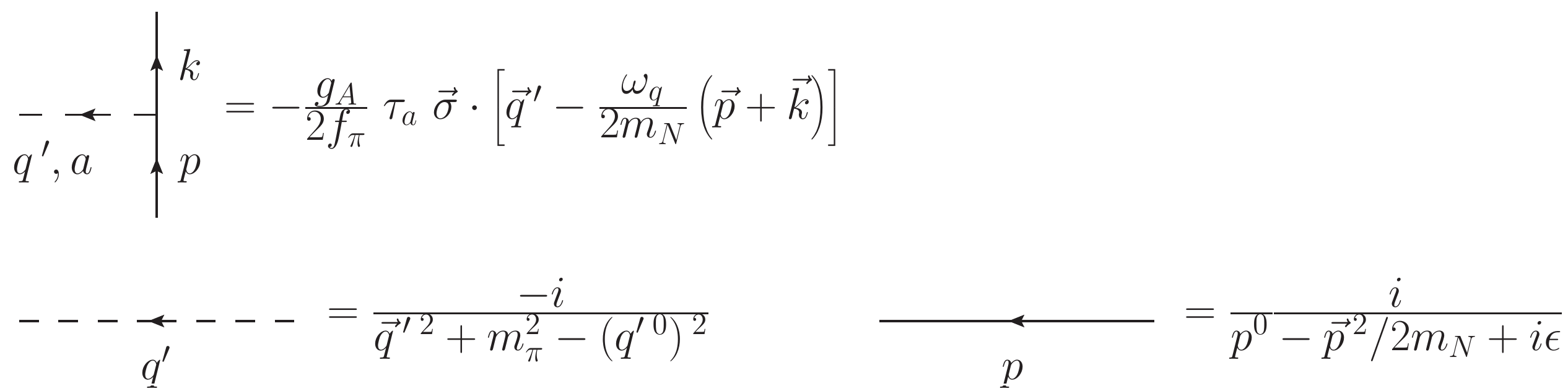}
\caption{Feynman rules\label{fig:rules}}
\end{figure}

In this work we employ threshold kinematics, where $q=(m_\pi,0)$, $q'=(m_\pi/2,\bfl)$, $p_{1,2}=(m_\pi/2,\pm\bfp)$, and $k_{1,2}=(0,\pm\bfk)$ with $|\bfp|=359\text{ MeV}$.  Also, at threshold only $s$-wave pions ($l_\pi=0$ with respect to the deuteron) are produced and the initial state is purely $^3P_1$ (see \cref{sec:selrules} for the details of this partial wave decomposition).  We use the hybrid methodology where ``operators" (two-particle irreducible diagrams) are calculated perturbatively (as in \cref{sec:method}) and then convolved (as in \cref{sec:hybrid}) with $NN$ wave functions which are obtained using phenomenological potentials (as in \cref{sec:nnwfns}).  \Cref{sec:lagrangian} discusses the chiral Lagrangian that defines the theory.  Using Eqs. (\ref{eq:l0}) and (\ref{eq:l1}), we obtain the Feynman rules shown in Fig. \ref{fig:rules}.

\begin{figure}
\centering
\includegraphics[height=2in]{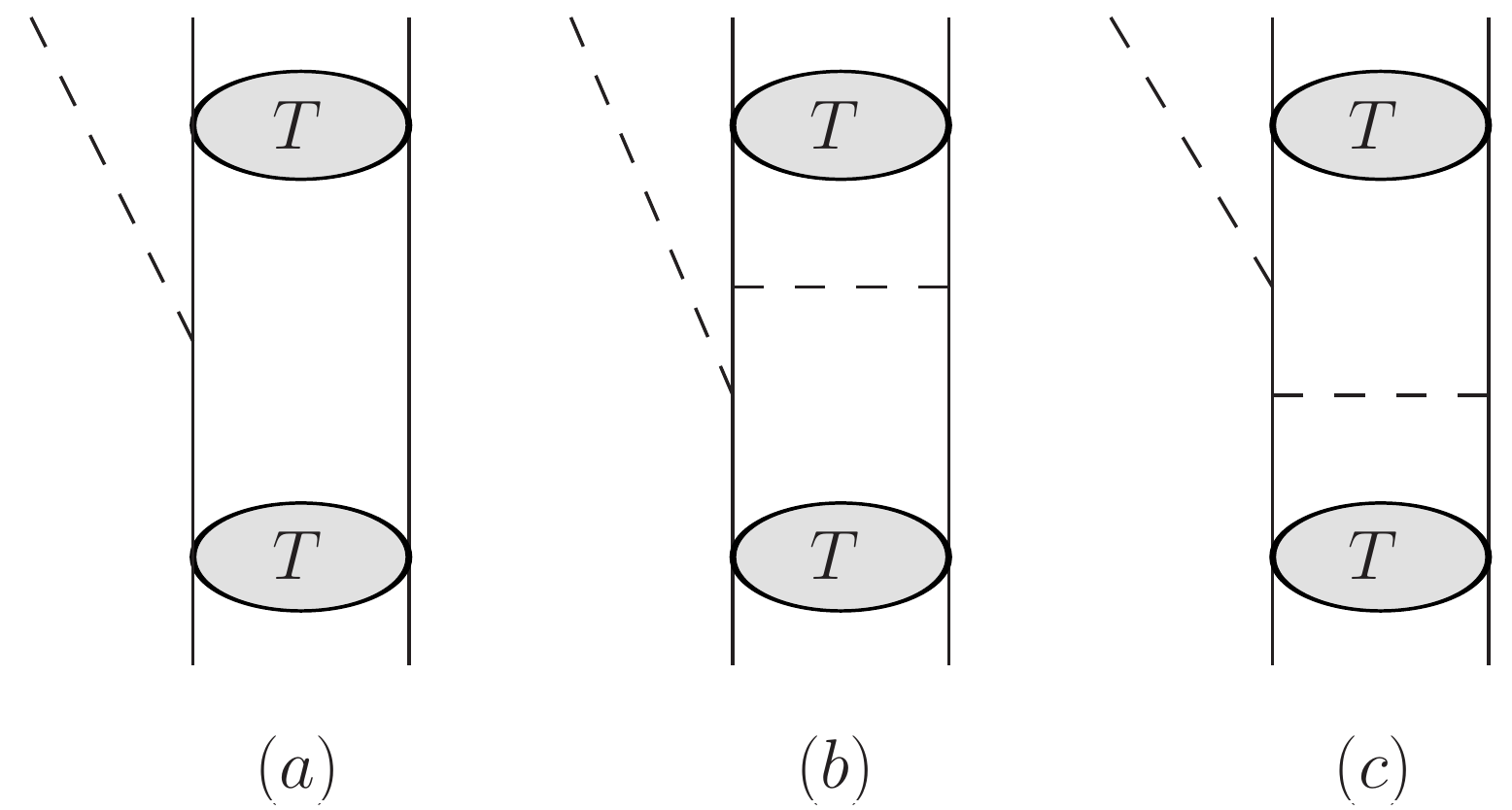}
\caption[Impulse approximation operator]{Impulse approximation operator alone (a), with OPE in the final state (b), and with OPE in the initial state (c).  Solid lines represent nucleons, dashed lines represent pions, and ovals represent the full $NN$ T-matrix.\label{fig:ia}}
\end{figure}

At leading order, $\mathcal{O}(\chi^1)$,\footnote{Recall that the expansion parameter is defined, $\chi\equiv\frac{\tilde{p}}{m_N}=\frac{m_\pi}{\tilde{p}}=\sqrt{m_\pi m_N}$.} the $s$-wave amplitude is dominated by the rescattering diagram, where a single pion is emitted from one nucleon and inelastically scattered by the other nucleon into an on-shell-produced pion.  The only other leading order $s$-wave diagram (recall that the Delta diagram is $p$-wave at threshold) is the recoil impulse approximation, shown in Fig. \ref{fig:ia}(a).
In \cref{sec:impdetails} we present the details of how the IA diagram is traditionally calculated; that is, ignoring the reducibility issues.  For the initial and final states, we make use of three different potentials: Argonne v18 \citep{Wiringa:1994wb}, Nijmegen II \citep{Stoks:1994wp}, and Reid '93 \citep{Stoks:1994wp}.  The results for the $s$-wave reduced matrix element $A_0$ (see \cref{sec:reduced}) are shown in Table \ref{tab:iaresults} along with the LO rescattering results.  These numbers have been calculated many times before; nothing new has been considered yet.
\begin{table}
\caption[Reduced matrix elements of the rescattering and impulse operators]{\label{tab:iaresults}Reduced matrix elements of the rescattering and impulse production operators for three different potentials.}
\begin{center}
\begin{tabular}{lccc}
\hline\hline
Diagram$\quad$ & Av18 & Nijm II & Reid '93\\ \hline
$A_0^{\text{res}}$ & 76.9 & 83.4 & 80.3\\
$A_0^{\text{imp}}$ & 4.9 & 1.3 & 3.5\\\hline\hline
\end{tabular}
\end{center}
\end{table}
Note that the rescattering results in \cref{tab:iaresults} include the factor of $4/3$ (which truly belongs at LO in the expansion) that we discussed in \cref{sec:wfncorintro}.  Finally, we point out that the experimental data (\cref{sec:wfncorresults}) imply a reduced matrix element of $80\leq A_0\leq94$.

Recall our previous discussions of the fact that the impulse approximation operator of Fig. \ref{fig:ia}(a) cannot formally be convolved with the initial and final states as described above because the nucleon emitting the pion cannot remain on-shell.  To put it another way, $q'\,^0$ vanishes in this diagram because there is no way for energy to be transferred.  The common approximation made in pion production calculations is to ignore this formal difficulty.  Also recall the prescription of Ref. \cite{Cohen:1995cc}, which pulled an OPE from the final state wave function (Fig. \ref{fig:ia}(b)) in order to argue that the impulse approximation was leading order.  This can be seen as an application of the Lippmann-Schwinger equation for the final state deuteron,
\begin{align}
\mid\psi_d\ra=GV\mid\psi_d\ra,\label{eq:LS}
\end{align}
where $G$ represents the two-nucleon propagator and $V$ represents the full potential.  In Figs. \ref{fig:ia}(b) and \ref{fig:ia}(c), we have replaced $V$ with one pion exchange.  This replacement is known to be a good approximation for the deuteron \citep{Ericson:1988gk}, but is not as valid for the initial state.  Including contributions to $V$ other one pion exchange is beyond the scope of this work.  \Cref{eq:LS} begs the question: are the diagrams in Figs. \ref{fig:ia}(a) and \ref{fig:ia}(b) the same size?  If so, we will be able to conclude that neglecting the required energy transfer in Fig. \ref{fig:ia}(a) does not significantly alter the true magnitude of the IA amplitude.

\section{Including One Pion Exchange\label{sec:ope}}

Calculation of Fig. \ref{fig:ia}(b) using Eq. (\ref{eq:LS}) is detailed in \cref{sec:opedetails}.  For lack of a better name, we will call this the ``OPE reducible" diagram.  In this calculation, we take $G=(E-H_0)^{-1}=(-E_d-\bfp^2/m_N)^{-1}$.  The energy of the exchanged pion in this case is taken to be $q'\,^0=0$.  This choice is consistent with the fact that the OPE is the first term in the $V$ of Eq. (\ref{eq:LS}), which should be the same $V$ that is used to generate the initial and final wave functions.  The results are shown in the first row of Table \ref{tab:opef}.
\begin{table}
\caption[Reduced matrix elements with final state OPE]{\label{tab:opef}Reduced matrix elements of the impulse approximation with OPE in the \textit{final} state [see Fig. \ref{fig:ia}(b)].}
\begin{center}
\begin{tabular}{lccc}
\hline\hline
Diagram & Av18 & Nijm II & Reid '93\\ \hline
$A_0^{\text{OPE,red,f}}$ & 75.2 & 64.6 & 79.3\\
$A_0^{\text{OPE,irr,f}}$ & 75.6 & 64.7 & 79.8\\
$A_0^{\text{OPE,irr,f}}-A_0^{\text{OPE,red,f}}\quad$ & 0.5 & 0.1 & 0.5\\ \hline\hline
\end{tabular}
\end{center}
\end{table}

We find an inconsistency between the impulse approximation (Fig. \ref{fig:ia}(a)) and OPE reducible (Fig. \ref{fig:ia}(b)) diagrams: although they are equivalent according to the Lippmann-Schwinger equation, they are of very different size numerically.  Using Av18, they are 4.9 and 75.2, respectively.  Of course this inconsistency is not surprising when one notes that three-momentum transfer is provided for in the latter diagram but not the former.

To resolve this problem, we reconsider the diagram in Fig. \ref{fig:ia}(b) as a fully irreducible operator, that is a member of the kernel prior to convolution with external wave functions.  This is justifiable in that the left intermediate nucleon is off-shell by $m_\pi/2$, more than the $m_\pi^2/m_N$ typical of reducible diagrams.  If we view it in this way, we are free to chose the energy of the exchanged pion to be $q'\,^0=m_\pi/2$ as mentioned in \cref{sec:reaction}.  The single-nucleon propagator for the left intermediate nucleon is taken from the rules shown in Fig. \ref{fig:rules}.  The calculation of the reduced matrix element for this ``OPE irreducible" operator is detailed in \cref{sec:opedetails} and the results are shown in the second row of Table \ref{tab:opef}.  We find that this diagram, which correctly accounts for energy transfer, is approximately equal to the OPE {\it re}ducible diagram.

The question remains: should Fig. \ref{fig:ia}(b) be included, and if so, how?  Until a clear procedure is defined for going from the full four-dimensional $\pi NN$ coupled-channels formalism to the more common three-dimensional uncoupled formalism, this question is open to interpretation.  We continue to take the view proposed in \cref{chap:csb}, which is that the OPE irreducible diagram should be included with the OPE reducible diagram subtracted off to prevent double counting.  This difference is shown in the third row of Table \ref{tab:opef}, and is referred to as the (nearly vanishing) ``wave function correction".  Note that the numerical cancellation is not as trivial as it appears.  Schematically, OPE is two derivatives acting on a Yukawa $e^{-mr}/mr$ that has different ranges for the reducible and irreducible cases.  In going from reducible to irreducible, the radial integral gets bigger because the range increases.  However, the derivatives bring down inverse powers of the range such that the overall amplitudes are similar in size.

Let us now discuss a new issue.  For $s$-wave pions at threshold we have a $\bsig\cdot(\bfp_i+\bfp_f)$ at the vertex where the pion is produced.  For this reason \cref{chap:csb} only considered OPE in the final state, assuming initial state OPE (Fig. \ref{fig:ia}(c)) to be suppressed by the small final state momentum $\bfk$.  However, one needs to be careful when applying power counting to calculations that involve external $NN$ wave functions.  At small distances the momenta of the nucleons (derivatives in position space) are distorted away from their constant values at asymptotically large distances.  As an example of this difficulty, it can be shown that at short enough distances the $\bfk^2/2m_N$ operator becomes larger than the $\bfp^2/2m_N$ operator in the context of the RS diagram.  For this reason, we also calculate Fig. \ref{fig:ia}(c) (for the details, see \cref{sec:opedetails}).  The results of this calculation are shown in Table \ref{tab:opei} where, again, the full wave function correction is nearly zero.
\begin{table}
\caption[Reduced matrix elements with initial state OPE]{\label{tab:opei}Reduced matrix elements of the impulse approximation with OPE in the \textit{initial} state (see Fig. \ref{fig:ia}(c)).}
\begin{center}
\begin{tabular}{lccc}
\hline\hline
Diagram & Av18 & Nijm II & Reid '93\\ \hline
$A_0^{\text{OPE,red,i}}$ & -11.2 & -23.7 & -15.0\\
$A_0^{\text{OPE,irr,i}}$ & -11.2 & -23.7 & -15.0\\
$A_0^{\text{OPE,irr,i}}-A_0^{\text{OPE,red,i}}\quad$ & $\sim$0 & $\sim$0 & $\sim$0\\ \hline\hline
\end{tabular}
\end{center}
\end{table}

\begin{figure}
\centering
\includegraphics[height=1.5in]{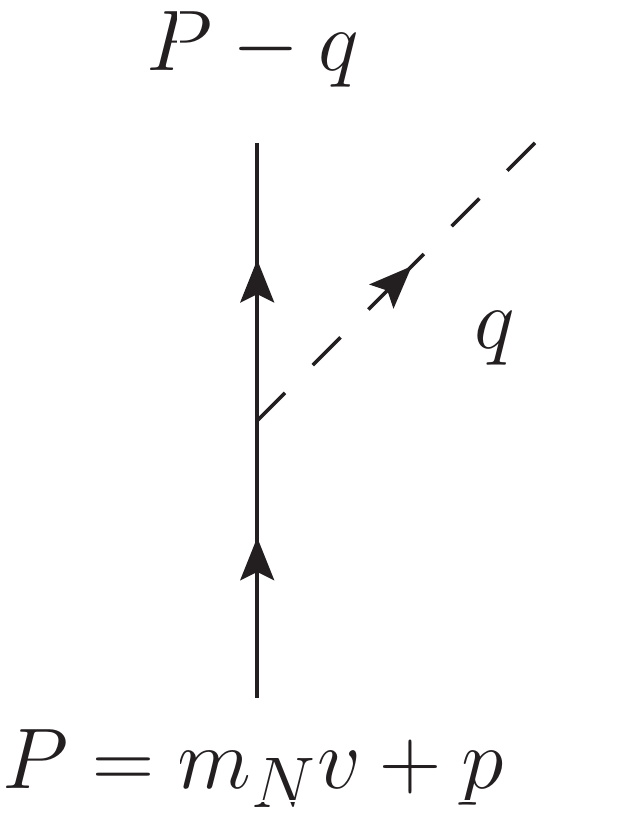}
\caption[Momenta of the nucleon propagator]{Momenta of the nucleon propagator where $v=(1,0,0,0)$ is used\label{fig:nprop}}
\end{figure}

There is another formal point to discuss with regard to the above calculations that was left out of \cref{chap:csb}.  Expressions for nucleon propagators in irreducible diagrams differ based on the power counting scheme used.  Consider the situation shown in Fig. \ref{fig:nprop}.
Starting from the full, relativistic propagator, the authors of Ref. \citep{Hanhart:2007mu} showed that, if $p^0\sim m_\pi$ and $\bfp\sim\sqrt{m_\pi m_N}$, the correct propagator after emitting the pion is
\begin{align}
\frac{i}{-q^0+(2\bfp\cdot\bfq-\bfq^2)/2m_N+i\epsilon},\label{eq:mcsprop}
\end{align}
where the second term in the denominator is the recoil correction, suppressed by one power of $\chi$ if $q^0\sim\sqrt{m_\pi m_N}$.  Note that we have ignored the resulting vertex corrections and anti-nucleon effects, both of which are suppressed for any choice of $q^0$.  Equation (\ref{eq:mcsprop}), which we will refer to as the ``new" method, comes in opposition to the ``old" method which derives the propagator from the non-relativistic chiral Lagrangian,
\begin{align}
\frac{i}{p^0-q^0-(\bfp-\bfq)^2/2m_N+i\epsilon},\label{eq:wprop}
\end{align}
where the third term comes from the NLO Lagrangian and is a candidate for promotion in MCS counting.  Although we will not present the argument of Ref. \citep{Hanhart:2007mu} in full, $\bfp^2/2m_N\approx m_\pi/2$ is clearly the connection between \cref{eq:mcsprop} and \cref{eq:wprop}.

Let us now consider the external pion vertex of Fig. \ref{fig:ia}(b), where in terms of the momenta shown in Fig. \ref{fig:nprop} we have $p=(m_\pi/2,\bfp)$ and $q=(m_\pi,0)$.  Since $p^0-q^0\sim m_\pi$, we promote the recoil corrections in the old propagator and find
\begin{align}\bs
iG_\text{new}^\text{irr,f}&=\frac{i}{-m_\pi}
\\
iG_\text{old}^\text{irr,f}&=\frac{i}{m_\pi/2-m_\pi-\bfp^2/2m_N}.
\es\end{align}
Next, consider the left-side OPE vertex of Fig. \ref{fig:ia}(c) where we have $p=(m_\pi/2,\bfp)$ and $q=(-\omega,\bfp-\bfk)$, and thus
\begin{align}\bs
iG_\text{new}^\text{irr,i}&=\frac{i}{\omega+(\bfp^2-\bfk^2)/2m_N+i\epsilon}
\\
iG_\text{old}^\text{irr,i}&=\frac{i}{m_\pi/2+\omega-\bfk^2/2m_N+i\epsilon}.\label{eq:newoldi}
\es\end{align}
Note that in the absence of distortions ($|\bfp|\approx\sqrt{m_\pi m_N}$, $|\bfk|\sim0$), $iG_\text{new}=iG_\text{old}$ for both the initial and final state propagators.  For the sake of clarity we will define as the ``free recoil approximation" (FRA) the use of these free particle values for the nucleon momenta.

\section{Nucleon Propagator Recoil\label{sec:recoil}}

In \cref{sec:ope}, the FRA was used for the recoil corrections to the nucleon propagators.  In this section we calculate the diagrams again, treating the momenta properly as operators instead of numbers.  In this section we use the old nucleon propagators for the irreducible diagrams.  In doing so, we avoid the $G^i_\text{new}$ of Eq. (\ref{eq:newoldi}), which would be difficult to evaluate exactly in position space.  It was pointed out in Ref. \citep{Bernard:1998sz} that the old nucleon propagators have formal convergence problems owing to the large external momenta.  Nevertheless, we expect to gain insight into the validity of the FRA using these propagators,
\begin{align}\bs
iG^\text{irr,f}&=\frac{i}{-m_\pi/2-\bfp^2/2m_N},
\\
iG^\text{irr,i}&=\frac{i}{m_\pi-\bfk^2/2m_N+i\epsilon}.\label{eq:npropnew}
\es\end{align}
Of course, according to MCS, $\bfk$ should not be counted as $\sim\sqrt{m_\pi m_N}$ and the $\bfk^2/2m_N$ term should therefore not appear as in Eq. (\ref{eq:npropnew}) until higher order.  We choose to retain it here as an investigation into the effects of the distortions.  For the reducible diagrams we continue to use $G^\text{red}=(E-H_0)^{-1}$.

\begin{figure}
\centering
\includegraphics[height=1.5in]{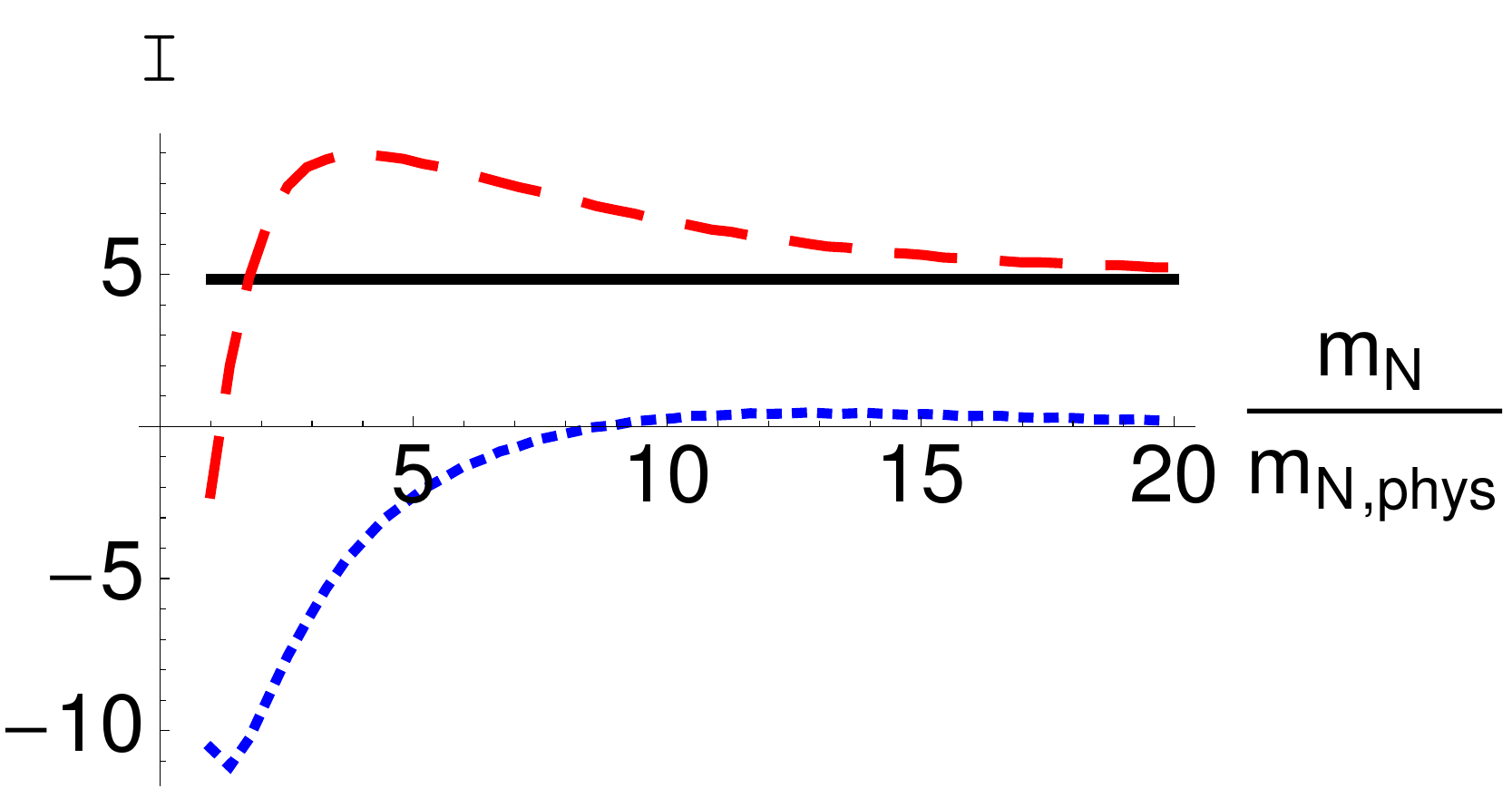}
\caption[Irreducible initial state OPE integral]{\label{fig:largeM}Irreducible initial state OPE integral as a function of $m_N$ using Av18.  The solid line displays the FRA result and the real (imaginary) part of the exact propagator result is shown as a dashed (dotted) curve.}
\end{figure}

The matrix elements can be calculated exactly in position space with Green function methods (see \cref{sec:recoildetails}).  The results are shown in Table \ref{tab:recoilprop}.
\begin{table}
\caption[Reduced matrix elements for the wave function corrections]{\label{tab:recoilprop}Reduced matrix elements for the wave function corrections with proper treatment of the momenta using the old expressions for the nucleon propagators.}
\begin{center}
\begin{tabular}{lccc}
\hline\hline
& Av18 & Nijm II & Reid '93\\ \hline
$A_0^{\text{OPE,irr,f}}$ & 80.8 & 70.8 & 89.1\\
$A_0^{\text{OPE,red,f}}$ & 92.4 & 81.2 & 103.3\\
$A_0^{\text{OPE,irr,f}}-A_0^{\text{OPE,red,f}}\quad$ & -11.7 & -10.4 & -14.2\\[.1in]
$A_0^{\text{OPE,irr,i}}$ & 5.1+24.5$i$ & 15.1+34.7$i$ & 7.8+27.8$i$ \\
$A_0^{\text{OPE,red,i}}$ & 16.2+8.2$i$ & 22.9+9.9$i$ & 18.2+8.7$i$ \\
$A_0^{\text{OPE,irr,i}}-A_0^{\text{OPE,red,i}}$ & -11.1+16.3$i$ & -7.7+24.8$i$ & -10.4+19.1$i$ \\ \hline\hline
\end{tabular}
\end{center}
\end{table}
We find that the final state wave function correction evaluated without the FRA gives a sizeable negative contribution of approximately $-10$.  Additionally, we find that the initial state corrections become as important as the \textit{lower-order} final state corrections.  To verify the surprising results of this calculation, we examine as an example the $m_N\rightarrow\infty$ limit of the radial integral for the irreducible initial state OPE in comparison with its analog from the previous section (which is independent of $m_N$ since $\bfk=0$ is used),
\begin{align}
\mathcal{I}\equiv\int dr\,r^2\left(\sqrt{2}\ddr\frac{u(r)}{r}+\left(\ddr+\frac{3}{r}\right)\frac{w(r)}{r}\right)G^\text{OPE,irr,i}\left(2f(m_\pi/2,r)+g(m_\pi/2,r)\right)R_i(r),
\end{align}
where the functions $f$ and $g$ are defined in \cref{sec:opedetails}.  As shown in Fig. \ref{fig:largeM}, in the large-$m_N$ limit the recoil term in the propagator vanishes and we recover the leading order result.

\section{\label{sec:cutoff}Cutoff Dependence}

It should not come as a surprise that odd things are happening in the short-distance part of the wave function, especially when we take derivatives.  In the hybrid formalism we are using, this domain of the wave function is calculated from a phenomenological potential: Woods-Saxon for Av18, one boson exchange for Nijm II, and Yukawa for Reid '93 (the very short range is exponential, exponential, and dipole, respectively).  Because these potentials are fitted to experimental phase shifts, the wave functions derived from them can be considered as infinitely high order in the EFT.  Thus one should consider using a cutoff to account for this mismatch between the operator and the wave functions.  Use of such a cutoff is referred to as EFT*, and was introduced in Ref. \citep{Park:2000ct}.

\begin{figure}
\centering
\includegraphics[height=1.5in]{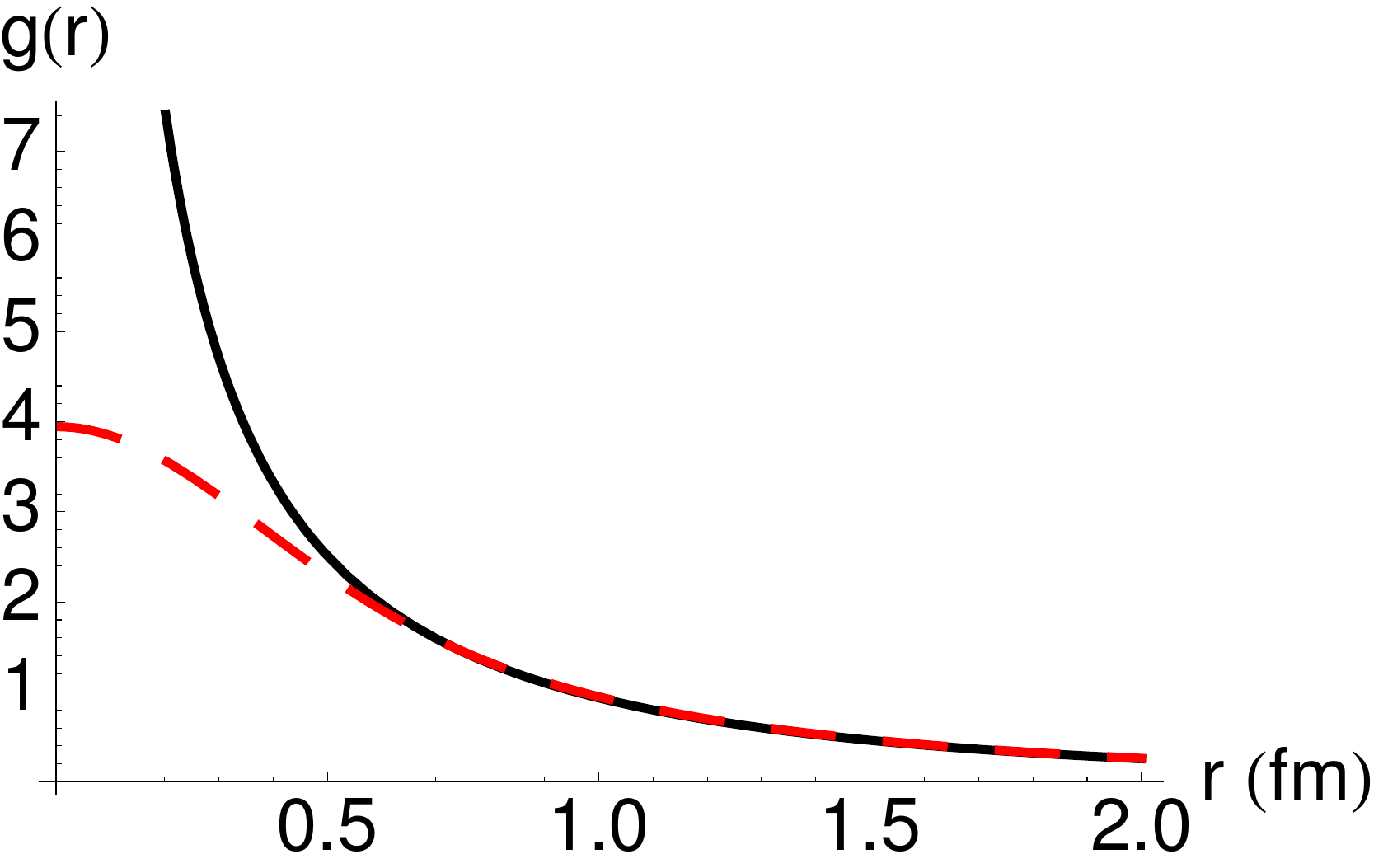}
\caption[Comparison of $g$ and $g_\Lambda$]{\label{fig:cutoffyukawa}Comparison of $g$ (solid curve) and $g_\Lambda$ (dashed curve) with $\omega=m_\pi/2$ and $\Lambda=1\text{ GeV}$.}
\end{figure}

\begin{figure}
\begin{center}
$\begin{array}{cc}
\includegraphics[width=.45\linewidth]{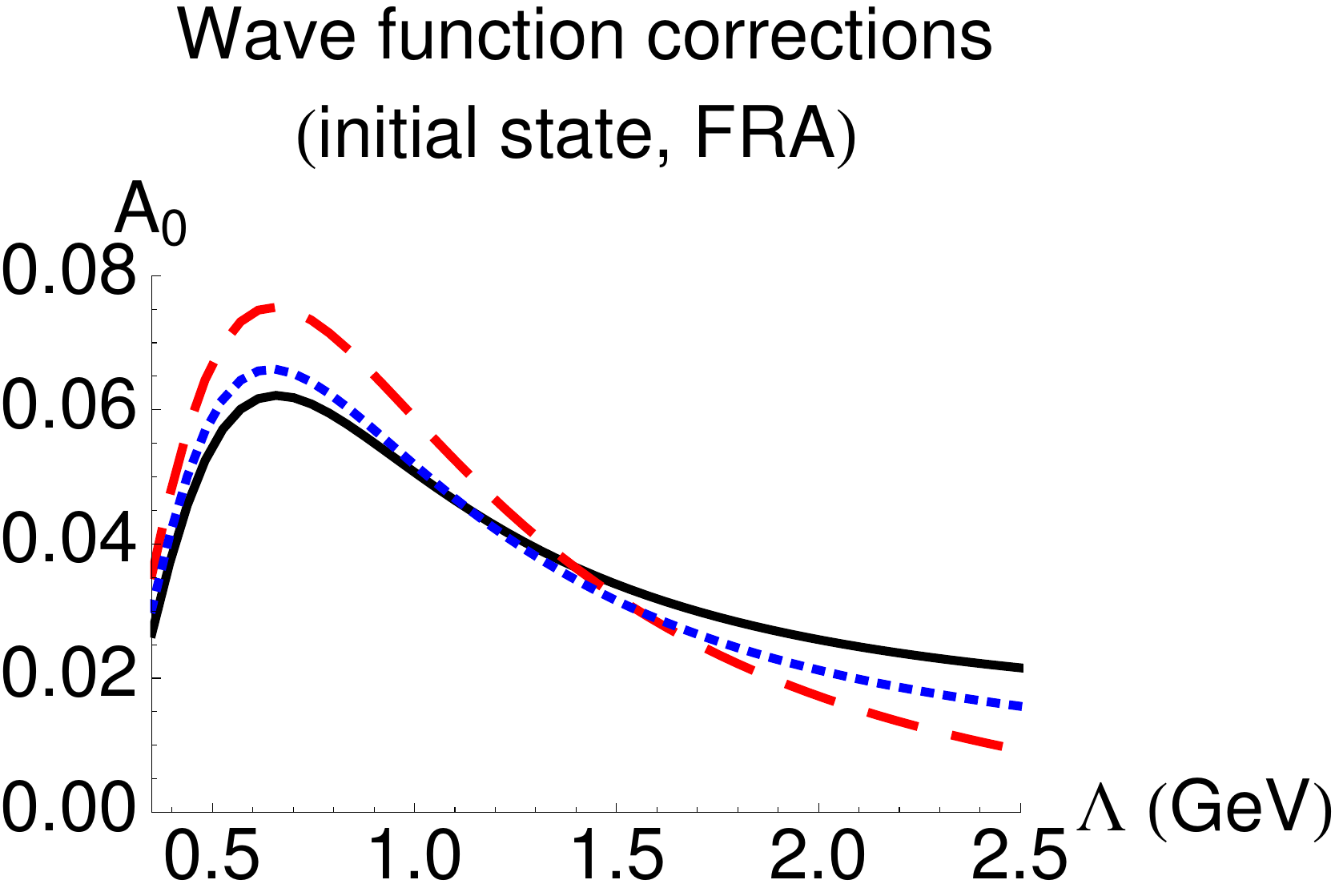} &
\includegraphics[width=.45\linewidth]{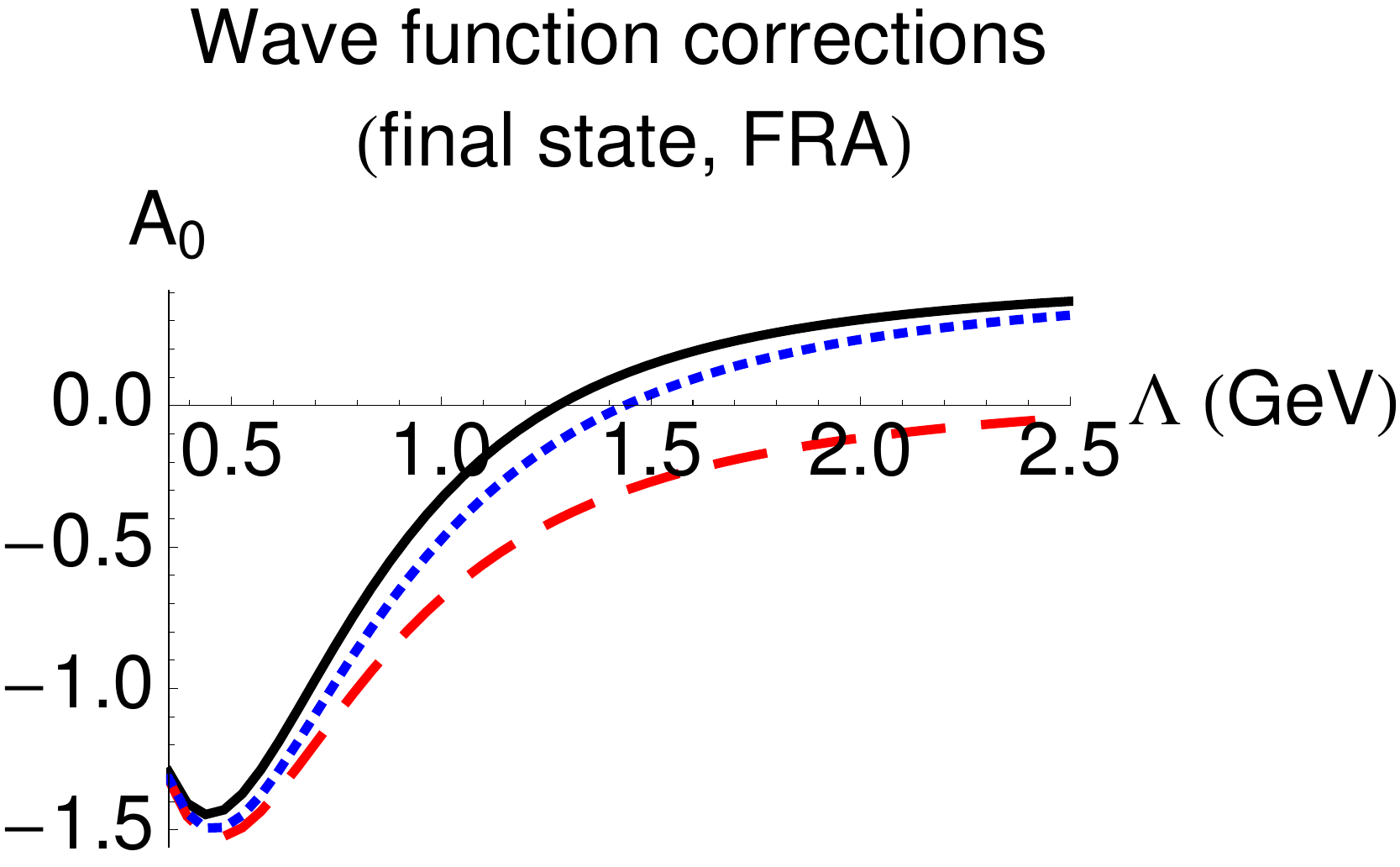} \\[.1in]
\includegraphics[width=.45\linewidth]{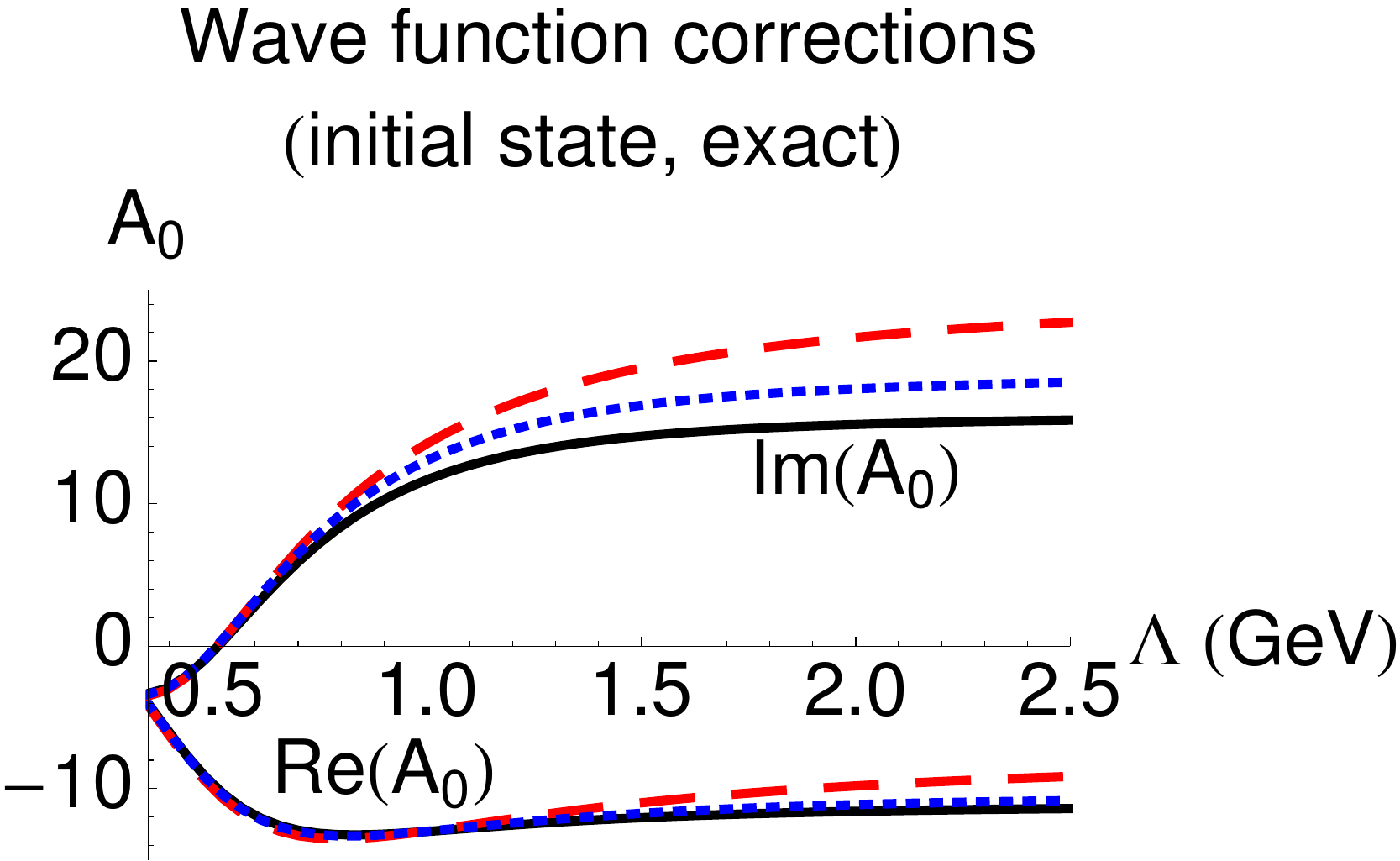} &
\includegraphics[width=.45\linewidth]{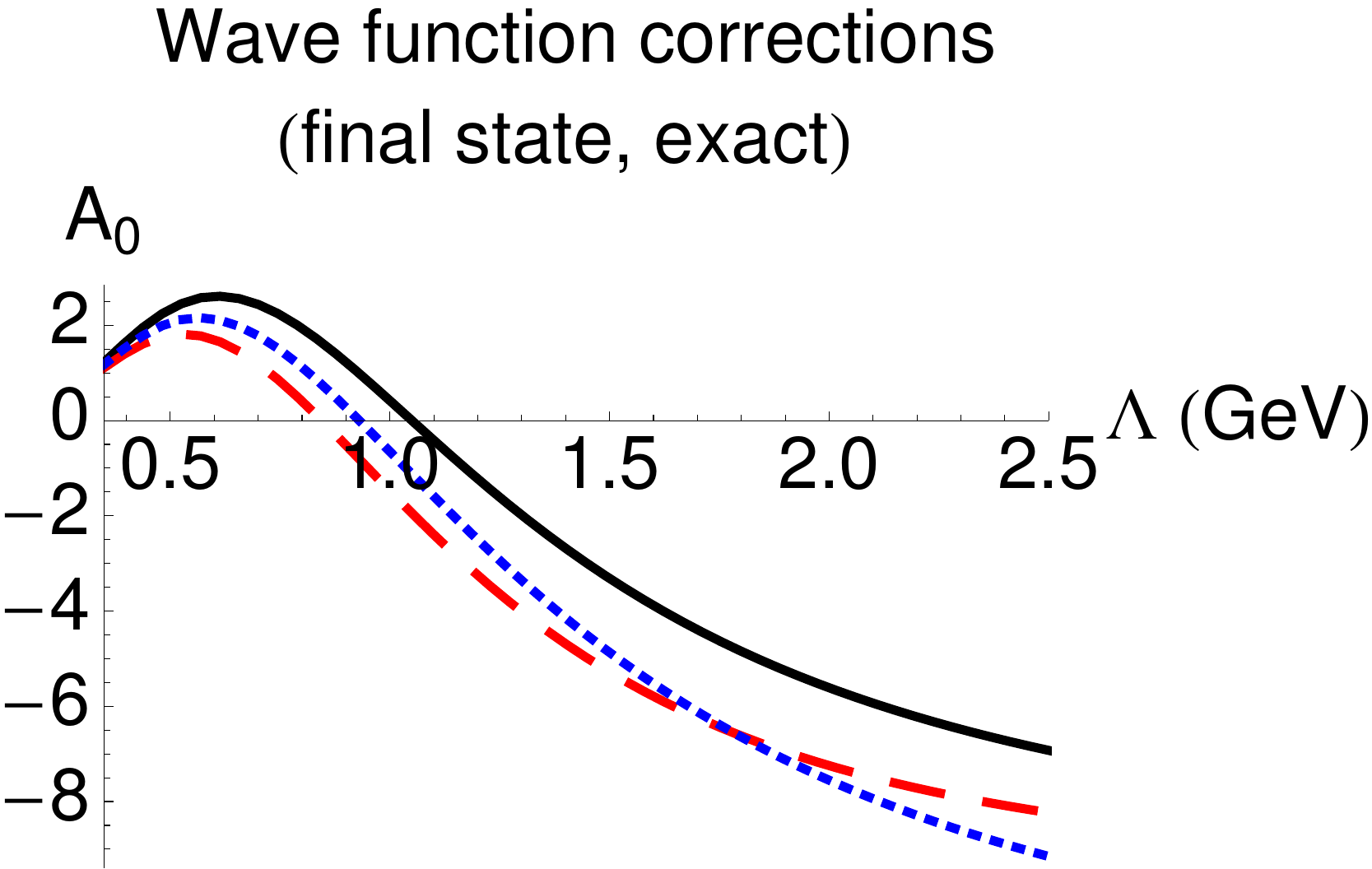}
\end{array}$
\caption[Cutoff dependence of various reduced matrix elements]{\label{fig:cutoffmxels}Cutoff dependence of various reduced matrix elements for Av18 (solid curve), NijmII (dashed curve), and Reid '93 (dotted curve).}
\end{center}
\end{figure}

In this section we investigate the effects of cutting off the convolution integrals that account for the the presence of initial and final states as discussed in \cref{sec:recoil}.  We use the procedure of Ref. \citep{Park:2002yp}, which modifies the Fourier transforms with a Gaussian cutoff,
\begin{align}\bs
\mathcal{M}(\bfr)&=\int\frac{d^3l}{(2\pi)^3}e^{i\bfl\cdot\bfr}S_\Lambda^2\left(\bfl^2\right)\mathcal{M}(\bfl)
\\
S_\Lambda\left(\bfl^2\right)&=\exp\left(-\frac{\bfl^2}{2\Lambda^2}\right).
\es\end{align}
Note that the traditional one-body IA is not affected by such a cutoff scheme.  For the OPE operators we define $g_\Lambda(\omega,r)$:
\begin{align}
\frac{\mo g_\Lambda(\omega,r)}{4\pi}\equiv\int\frac{d^3l}{(2\pi)^3}e^{i\bfl\cdot\bfr-\bfl^2/\Lambda^2}\frac{1}{\bfl^2+\mo^2}.\label{eq:gLdef}
\end{align}
The exact evaluation of this integral and of the derivatives required to compute the diagrams of this work are shown in \cref{sec:cutoffdetails}.  As desired, the cutoff regulates the behavior of $g(r)$ at the origin, as shown in Fig. \ref{fig:cutoffyukawa}.
The cutoff dependence of various reduced matrix elements is shown in Fig. \ref{fig:cutoffmxels}.

The fact that we observe significant cutoff dependence of the wave function corrections above the typical scale of $\sim$1 GeV is surprising.  Indeed, this sensitivity indicates the need for a counterterm because observables must be cutoff independent.  As pointed out in Ref. \citep{Gardestig:2005sn}, if one considers the difference of terms that comprise the wave function correction, 
\begin{align}
\frac{1}{\bfq'^2+m_\pi^2}-\frac{1}{\bfq'^2+3m_\pi^2/4}=-\frac{m_\pi^2/4}{\bfq'^2+m_\pi^2}\cdot\frac{1}{\bfq'^2+3m_\pi^2/4},
\end{align}
it can be argued that wave function correction is N$^2$LO in the MCS scheme.  However, if this view is to be accepted, the fact that the wave function corrections are much larger in magnitude than the LO impulse approximation should be considered surprising.

\section{\label{sec:wfncorresults}Cross Section Results}

Shown in Table \ref{tab:wfncorresults} is a summary of the findings discussed in this paper at $\Lambda=\infty$ and $\Lambda=1$ GeV along with the rescattering diagram.
\begin{table}
\caption{\label{tab:wfncorresults}Reduced matrix elements for three different potentials.}
\begin{center}
\resizebox{\textwidth}{!}{
\begin{tabular}{lcccccc}
\hline\hline
Diagram & \multicolumn{2}{c}{Av18} & \multicolumn{2}{c}{Nijm II} & \multicolumn{2}{c}{Reid '93}\\ \cline{2-3}\cline{4-5}\cline{6-7}
& $\Lambda=\infty$ & $\Lambda=1\text{ GeV}$ & $\Lambda=\infty$ & $\Lambda=1\text{ GeV}$& $\Lambda=\infty$ & $\Lambda=1\text{ GeV}$\\ \hline
Rescattering (NLO) & 76.9 & 76.2 & 83.4 & 81.5 & 80.3 & 79.1\\
Impulse & 4.9 & 4.9 & 1.3 & 1.3 & 3.5 & 3.5\\
Final wfn cor (FRA) & 0.5 & -0.3 & 0.1 & -0.7 & 0.5 & -0.5\\
Final wfn cor (exact) & -11.7 & 0.4 & -10.4 & -1.5 & -14.2 & -0.7\\
Initial wfn cor (FRA) & $\approx0$ & 0.1 & $\approx0$ & 0.1 & $\approx0$ & 0.1\\
Initial wfn cor (exact) & -11.1+16.3$i$ & -13.0+11.7$i$ & -7.7+24.8$i$ & -13.0+14.2$i$ & -10.4+19.1$i$ & -13.0+13.0$i$\\[.1in]
Total (FRA) & 82.2 & 80.8 & 84.8 & 82.2 & 84.2 & 82.2\\
Total (exact) & 59.0+16.3$i$ & 68.5+11.7$i$ & 66.7+24.8$i$ & 68.4+14.2$i$ & 59.2+19.1$i$ & 69.0+13.0$i$\\ \hline\hline
\end{tabular}
}
\end{center}
\end{table}

We also discuss the total cross section; as we recall from \cref{sec:kinoverview}, near threshold the total cross section is parametrized,
\begin{align}
\sigma=\frac{1}{2}\left(\alpha\eta+\beta\eta^3\right),\label{eq:sigtotch5}
\end{align}
where $q=m_\pi\eta$.  At threshold, one can only calculate $\alpha$,
\begin{align}
\alpha=\frac{m_\pi}{128\pi^2sp}|A_0|^2,\label{eq:alpha}
\end{align}
where $s=(m_\pi+m_d)^2$.  Also recall that charged pion production is related to neutral pion production by isospin symmetry.  This symmetry is the reason for the $1/2$ present in the definition of $\sigma$.  The most recent experimental data were shown in \cref{tab:exptch3}.

\begin{figure}
\begin{center}
\subfigure[\ FRA nucleon propagators]{\label{fig:totL-a}\includegraphics[width=.4\linewidth]{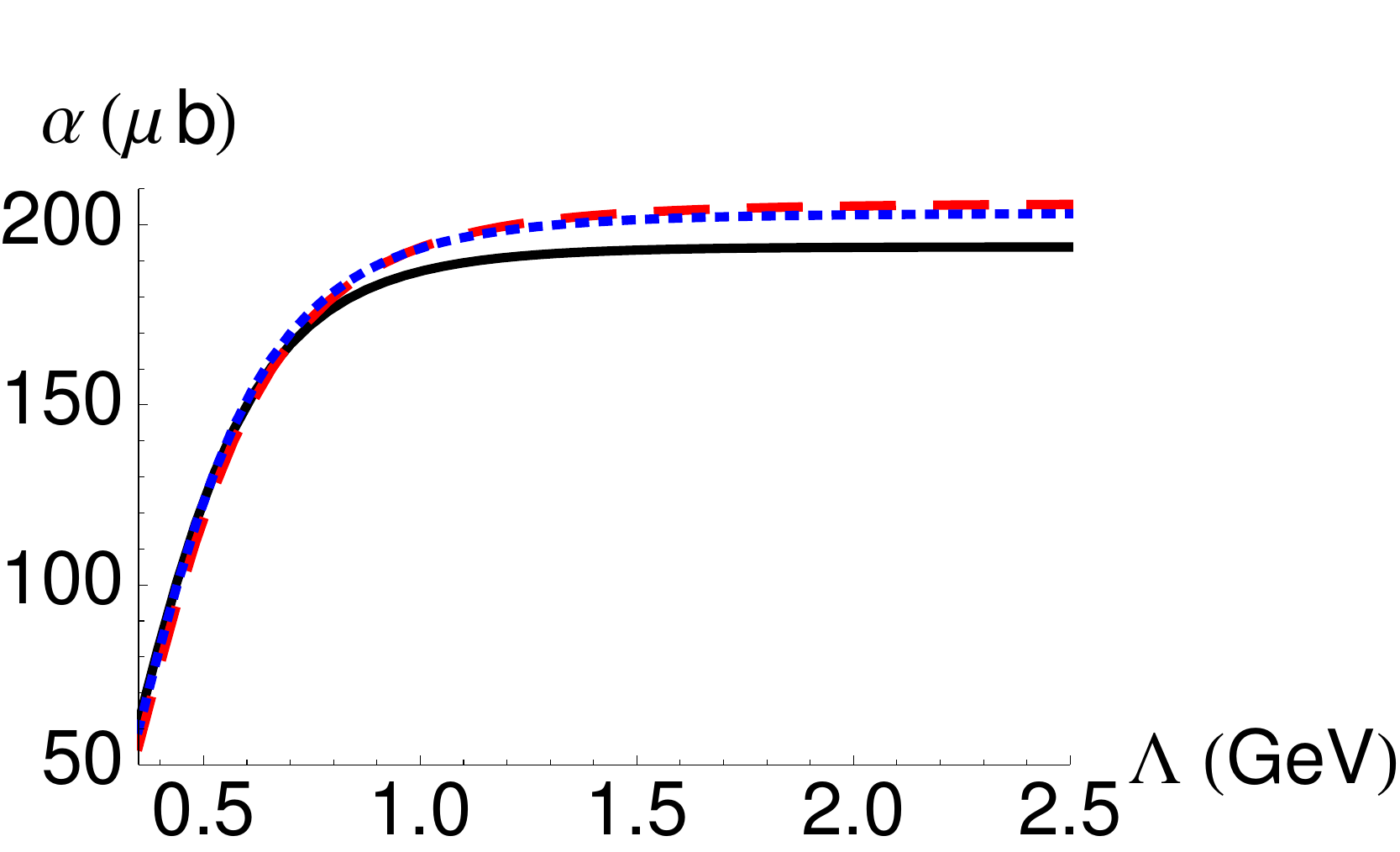}}
\hspace{.1\linewidth}
\subfigure[\ Exact nucleon propagators]{\label{fig:totL-b}\includegraphics[width=.4\linewidth]{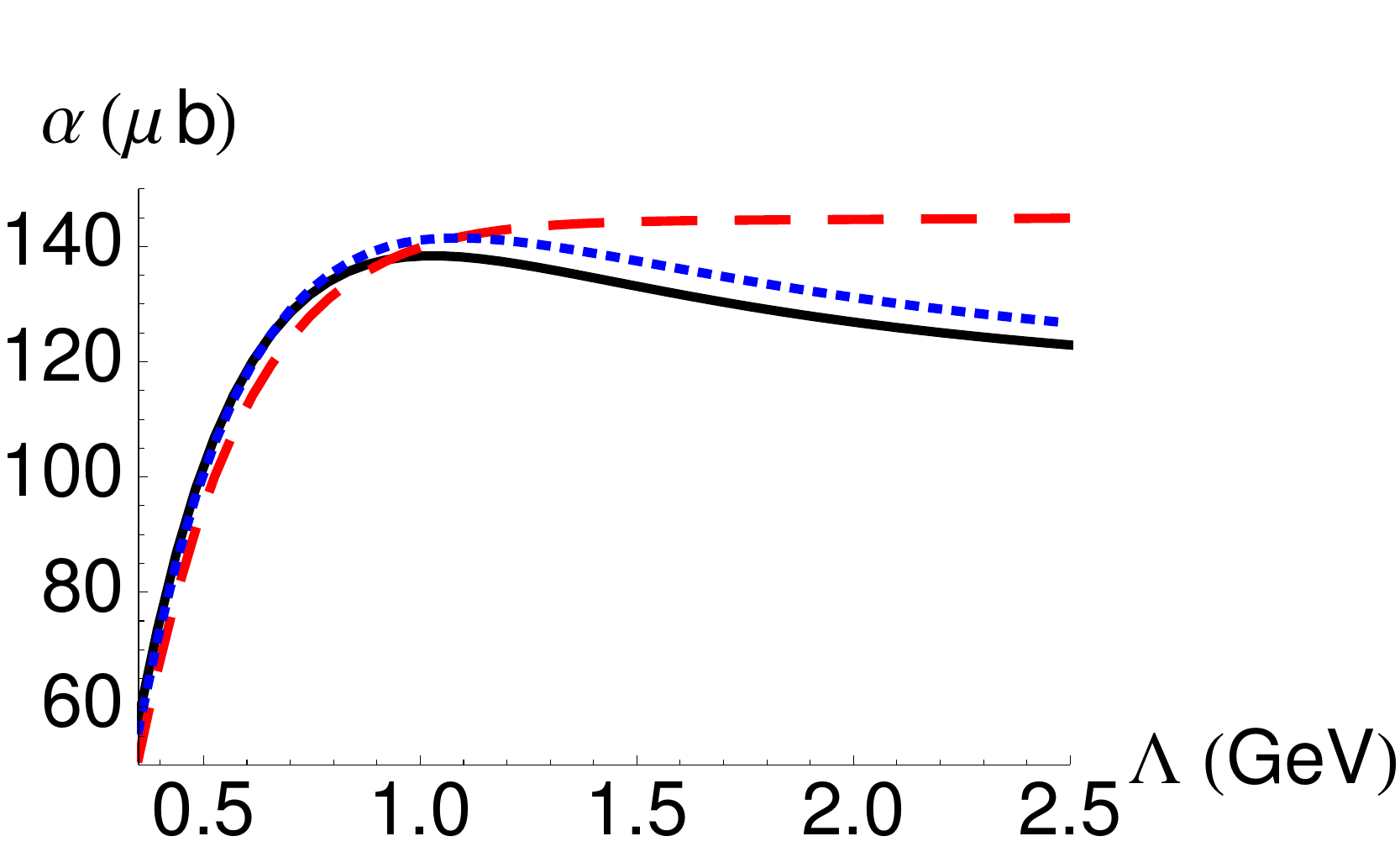}}
\end{center}
\caption[Cutoff dependence of the total cross section]{\label{fig:totL}Cutoff dependence of the total cross section.  Av18 (solid curve), NijmII (dashed curve), and Reid '93 (dotted curve).}
\label{fig:edge}
\end{figure}

The theoretical total cross section as a function of the cutoff is shown in Fig. \ref{fig:totL}.
The theoretical results include all diagrams up to $\mathcal{O}(\chi^2)$.  Thus theory can assign a rough uncertainty to the threshold cross section of $2\times\chi^2\approx30\%$.

\section{\label{sec:summary}Discussion}

Before the findings of this work, the total theoretical cross section at $\Lambda\approx1$ GeV [Fig. \ref{fig:totL-a}] was in agreement with the most recent experiment (fifth row of Table \ref{tab:exptch3}) at approximately the $1\sigma$ level.  Fixing the FRA approximation decreases the cross section, and if we stop here (Fig. \ref{fig:totL-b}), the agreement between theory and experiment becomes more tenuous (approximately the $2.5\sigma$ level).  A second conclusion of this work regards the MCS power counting, which dictates that $|\bfp|\sim\sqrt{m_\pi m_N}$ while $|\bfk|\sim m_\pi$.  We find that the wave function corrections of the initial state are of similar size to those of the final state once the FRA is removed, contradicting the previous sentence.  There exists a contact term (an $NNNN\pi$ vertex) at N$^2$LO along with tree-level diagrams proportional to the $c_i$ LECs and two-pion exchange loops.  Since all the LECs except the contact term are fixed by other data, it will be interesting to see if that contact term is of natural size.

The next chapter investigates further into the concept of reducibility.  Specifically, we will define a clear procedure for deciding what to include in the impulse diagram.  There has been a lack of consensus in the literature as to the inclusion of OPE, and if it is included, how that should be done.  Understanding this issue is important, not only for calculation of the total cross section but also for $p$-wave pion production, since the leading contribution to $p$-wave pion production comes from the impulse diagram.

\chapter[Impulse Reducibility: Non-Relativistic Reduction]{Impulse approximation in nuclear pion production reactions: absence of a one-body operator}
\label{chap:nrred}

The impulse approximation of pion production reactions is studied a final time by developing a relativistic formalism, consistent with that used to define the nucleon-nucleon potential.  For plane wave initial states we find that the usual one-body (1B) expression ${\cal O}_\text{1B}$ is replaced by ${\cal O}_\text{2B}=-iK(m_\pi/2){\cal O}_\text{1B}/m_\pi$, where $K(m_\pi/2)$ is the sum of all irreducible contributions to nucleon-nucleon scattering with energy transfer of $m_\pi/2$.  We show that ${\cal O}_\text{2B}\approx{\cal O}_\text{1B}$ for plane wave initial states.  For distorted waves, we find that the usual operator is replaced with a sum of two-body operators that is well approximated by the operator ${\cal O}_\text{2B}$.  Our new formalism solves the (traditionally ignored) problem of energy transfer forbidding a one-body impulse operator.  Using a purely one pion exchange deuteron, the net result is that the impulse amplitude for $np\to d\pi^0$ at threshold is enhanced by a factor of approximately two.  This amplitude is added to the larger ``rescattering" amplitude and, although experimental data remain in disagreement, the theoretical prediction of the threshold cross section is brought closer to (and in agreement with) the most recent data.  This chapter is a modified version of our paper \citep{Bolton:2010uj}; we have reorganized and rewritten parts in effort to make it flow better with the rest of this thesis.

\section{\label{sec:intro}Introduction}

In this chapter we take up once again the calculation of $NN\to d\pi$ with the pion in an $s$-wave.  We continue to focus on the contribution of the impulse approximation in which the produced pion does not interact at all with the spectator nucleon.  Recall that pion rescattering, not the IA, is known to make the largest contribution to the total cross section \citep{Koltun:1965yk}.  The $\Delta(1232)$ resonance is also known to contribute significantly to this observable, but in the $p$-wave channels.  Our motivation for the present study is to obtain increased precision in the total cross section calculation and to prepare for future application to other observables to which the IA contributes, such as $p$-wave pion production.

As we have seen, a challenge in the calculation of pion production is the presence of strongly interacting initial/final states.  When considering the IA diagram it is at first unclear whether OPE should be included in the kernel or not; historically, calculations have not done this, claiming instead that the necessary energy exchange is happening in the external wave functions.  This method was brought under question in \cref{chap:wfncor} (see also Ref. \citep{Gardestig:2005sn}).  Ideally, one would like to derive the correct method from a relativistic formalism that cleanly separates effects in wave functions from those appearing in the kernel.

\begin{figure}
\centering
\includegraphics[height=1.5in]{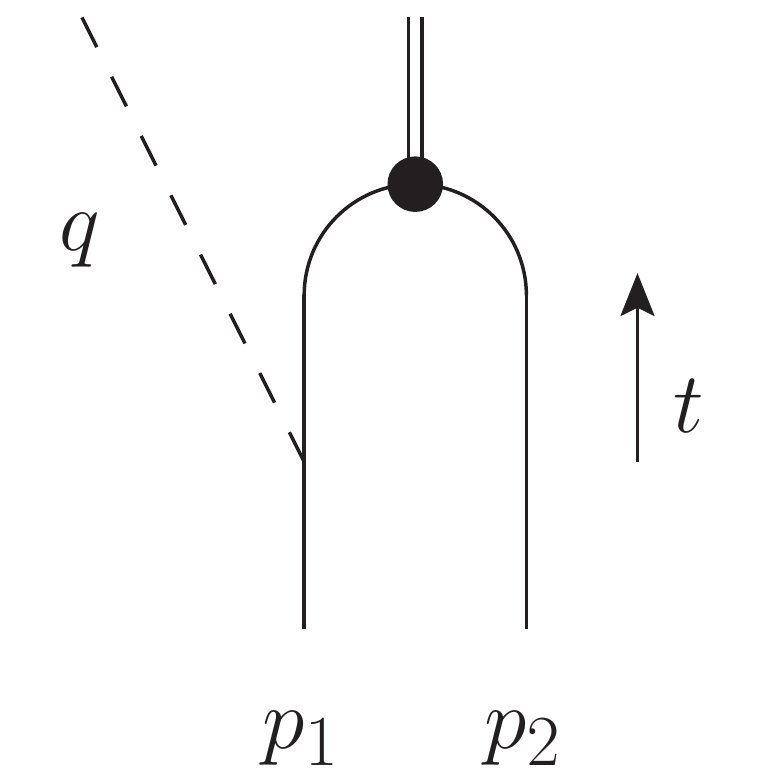}
\caption[Impulse approximation without initial state interactions]{\label{fig:iapw}Impulse approximation without initial state interactions.  Solid lines represent nucleons, dashed lines represent pions, and double solid lines represent a deuteron.}
\end{figure}

Consider the IA contribution to $NN\rightarrow d\pi$ in the plane wave (PW) approximation where initial state interactions are neglected (see Fig. \ref{fig:iapw}).  The amplitude for such a process has been estimated to go like ${\cal M}^{IA}\sim\frac{m_\pi}{m_N}\bsig\cdot\mathbf{p}_1\phi(p)\sim  \frac{m_\pi}{m_N}\sqrt{m_Nm_\pi}\phi(p) $, where $\phi(p)$ is the bound state wave function, evaluated in momentum space.
The suppression by $m_\pi^{3/2}$ was noted in Ref. \citep{Cohen:1995cc}, which also included an analysis that a more detailed treatment of the power counting based on including initial and final state interactions introduces a power of $1/m_\pi$ via an energy denominator such that the amplitude varies as $\sqrt{m_\pi}$.  Nevertheless, we see directly an explicit $m_\pi^{3/2}$ times $\phi(\sqrt{m\;m_\pi})$.

In the physical region where $m_\pi=140$ MeV, the wave function falls as a power of momentum greater than unity.  For small values of relative momentum, the deuteron wave function also falls more rapidly than an inverse power of its argument.  If one takes $m_\pi$ to be small, the deuteron remains weakly bound \citep{Beane:2002vs, Bulgac:1997ji} and therefore its momentum wave function will also fall rapidly in the chiral limit.  Thus the power counting can only be considered a very rough estimate.  If we follow \citep{Cohen:1995cc}, the impulse term is a leading order term, but the deuteron wave function is quite small for physical values of $p$ and there is also a substantial cancellation between the deuteron $s$- and $d$-states.  Thus this term's contribution to the cross section \citep{Koltun:1965yk} is small and there is a contradiction between power counting expectations and realistic calculations.

This contradiction was also discussed at length in \cref{chap:wfncor} where we introduced ``wave function corrections" as a possible solution.  This proposal included OPE with an energy transfer of $m_\pi/2$ in the impulse kernel, but then subtracted off a similar diagram with static OPE in order to prevent double counting.  The result depended strongly on the treatment of the intermediate off-shell nucleon propagator and no definitive conclusion was reached.  This chapter is intended to settle the debate regarding the inclusion of OPE in the impulse approximation.  We demonstrate, by starting from a consistent relativistic formalism, that non-static OPE is to be included with no subtraction necessary; the impulse amplitude that should be used is given in Eq. (\ref{eq:mdwfinal}).  Furthermore, we show that the traditional approach of using a one-body kernel is correct only in the absence of initial state interactions.

The key issue in understanding the energy transfer mechanism of the IA is the reduction of the hybrid formalism from {\it four} to {\it three} dimensions.  In \cref{sec:bs} we review aspects of the four-dimensional Bethe-Salpeter (BS) formalism for the two-nucleon problem.  \Cref{sec:nnpiamp} presents the $N\to N\pi$ operator and \cref{sec:pw} shows that for plane wave $NN\to d\pi$, the traditional IA is approximately valid.  Next, \cref{sec:distortions} considers the full distorted-wave amplitude by calculating the corresponding loop diagram, including the effects of the non-zero time components of the momenta of the exchanged mesons.  We are then able to interpret the resulting three-dimensional distorted-wave amplitude as a sum of two-body operators.  We demonstrate the new formalism by explicitly evaluating $s$-wave $NN\rightarrow d\pi$ amplitudes at threshold.  To aid the flow of the arguments, approximations made in this section are verified to be sub-leading in \cref{sec:isi,sec:fsi,sec:sigma}.  A comparison with experimental cross section data is made in \cref{sec:discussionch6}, where we also discuss implications and future directions.

\section{\label{sec:bs}Bethe-Salpeter Basics}

Recall the definition of the nucleon-nucleon potential from the Bethe-Salpeter formalism. We follow the approach of Partovi and Lomon \citep{Partovi:1969wd} and also consider the relationship between the Bethe-Salpeter wave function and the usual equal time wave function as recently discussed in Ref. \citep{Miller:2009fc}.

Partovi and Lomon write the Bethe-Salpeter equation for the nucleon-nucleon scattering amplitude $\cal M$ as 
\begin{align}
{\cal M}=K +K G {\cal M},\label{eq:bs}
\end{align}
where $K$ is the sum of all irreducible diagrams.  The quantities ${\cal M}$ and $K$ depend on the total four-momentum $P_\text{tot}$ and the relative four-momentum  $k$.  The two individual momenta are $p_{1,2}=P_\text{tot}/2\pm k$ and $G$ is the product of two Feynman propagators:
\begin{align}
G =\left(\frac{i}{\slashed{p}_1 -m_N +i\epsilon}\right)_1 \; \left(\frac{i}{\slashed{p}_2 -m_N +i\epsilon}\right)_2=G_1G_2.
\end{align}
The quantities ${\cal M}$ and $K$ differ from those of \citep{Partovi:1969wd} by a factor of $-i/(2\pi)$.  Partovi and Lomon replace the relativistic $G$ by the Lippmann-Schwinger propagator $g$ for two particles.  For scalar particles, $g$ is obtained from $G$ by integrating over the zero'th (energy)  component of one of the two particles \citep{Miller:2009fc}. For fermions, one must also project onto the positive energy sub-space of both particles.  This is accomplished in the center of mass frame by taking \citep{Partovi:1969wd}
\begin{align}
g(k|P_\text{tot})=2\pi i\frac{[\gamma^0 E(\bfk)-\bgamma\cdot\bfk +m_N]_1[\gamma^0 E(\bfk)+\bgamma\cdot\bfk +m_N]_2}{E(\bfk)(P_\text{tot}^2-4m_N^2-4\bfk^2+i\epsilon)}\delta(k^0) ,\label{eq:g}
\end{align}
where $E(\bfk )\equiv \sqrt{\bfk^2+m_N^2}$.  Note that $g$ contains the important two-nucleon unitary cut.  The non-relativistic potential $U$ is defined so as to reproduce the correct on-shell $NN$ scattering amplitude $\cal M$ using the Lippmann-Schwinger (LS) equation
\begin{align}
{\cal M}=U+Ug {\cal M}.\label{eq:ls}
\end{align}
The quantity $U$ is obtained by equating the ${\cal M}$ of Eq. (\ref{eq:bs}) with that of Eq. (\ref{eq:ls}) to find \citep{Partovi:1969wd}
\begin{align}
U=K+K(G-g)U.\label{eq:udef}
\end{align}
In solving Eq. (\ref{eq:ls}) for the on-energy shell scattering amplitude, $U$ never changes the value of the relative energy $k^0$ away from 0.  Equations (\ref{eq:ls}) and (\ref{eq:udef}) are consistent with  Weinberg power counting in which one calculates the potential using chiral perturbation theory and then solves the LS equation to all orders.  The term $G-g$ may be thought of a purely relativistic effect arising from off-shell (short-lived) intermediate nucleons, and in the present context a perturbative effect.

\begin{figure}
\centering
\includegraphics[height=1.25in]{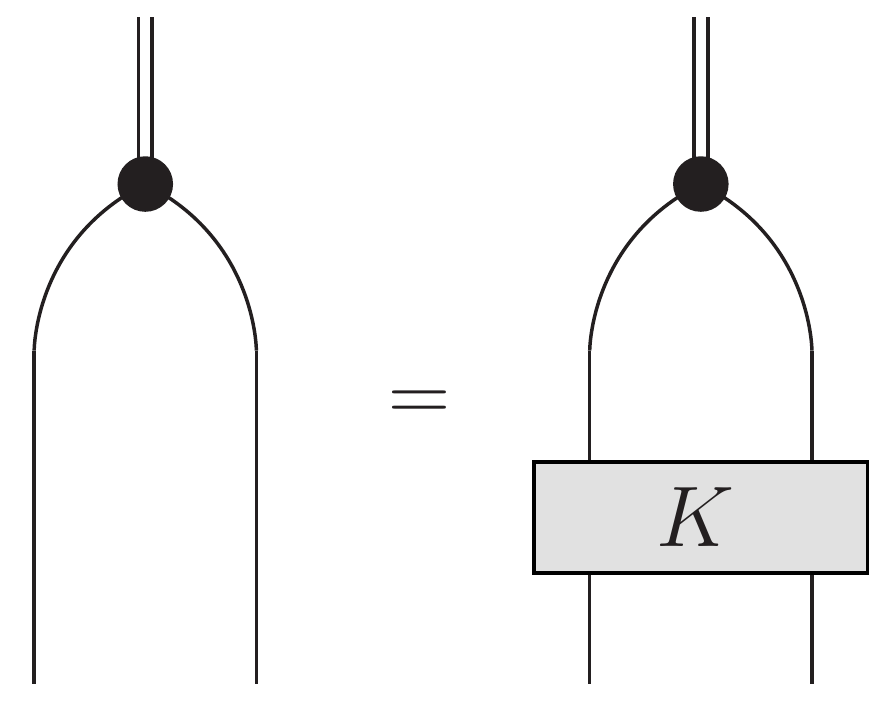}
\caption{\label{fig:dBS}Bethe-Salpeter equation near the deuteron pole.}
\end{figure}

\begin{figure}
\begin{center}
\subfigure[\ $NN$ scattering]{\label{fig:scattBS}\includegraphics[height=1in]{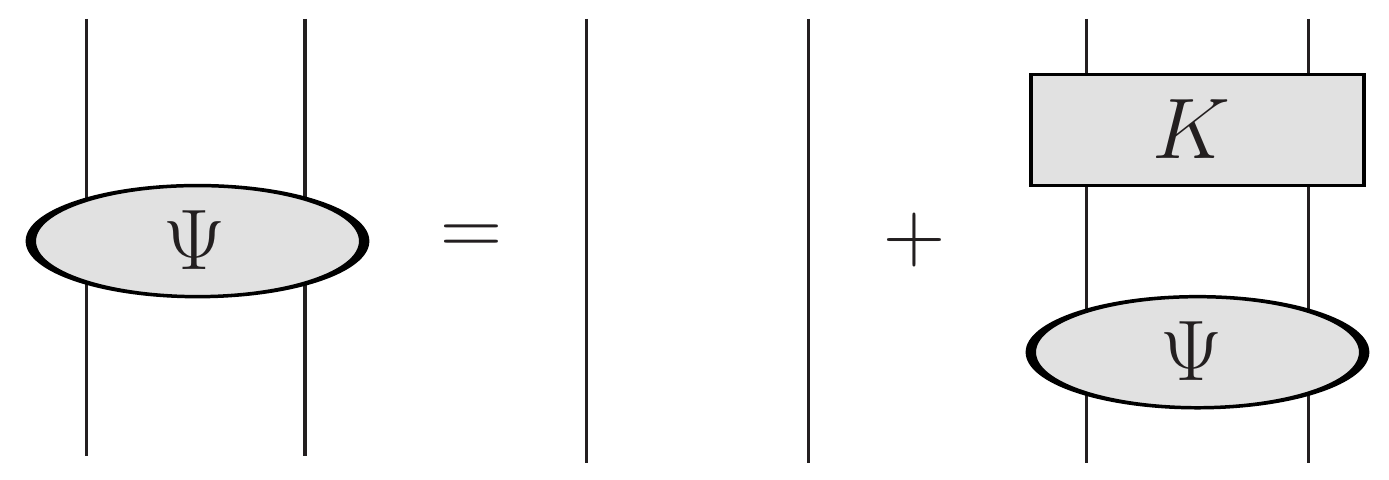}}
\hspace{.15\linewidth}
\subfigure[\ Deuteron]{\label{fig:dwfn}\includegraphics[height=1in]{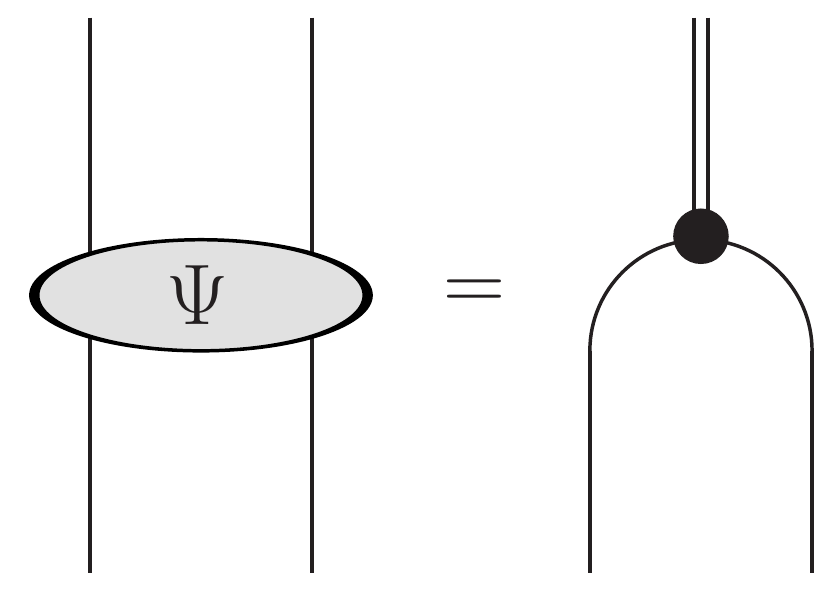}}
\end{center}
\caption{\label{fig:BSwfns}Bethe-Salpeter wave functions.}
\end{figure}

Consider the deuteron wave function in the final state of a pion production reaction. For $P^2$ near the pole position, the second term of Eq. (\ref{eq:bs}) dominates and we replace the scattering amplitude with the vertex function $\Gamma$:  ${\cal M}\rightarrow \Gamma$, and
\begin{align}
\Gamma=KG\Gamma.
\end{align}
This equation is shown pictorially in Fig. \ref{fig:dBS}.
The Bethe-Salpeter wave function $\Psi$ is defined as $G\Gamma$ so that 
\begin{align}
\Psi=G\Gamma=GK\Psi.\label{eq:bsb}
\end{align}
The wave functions of the scattering state and the deuteron are shown in Fig. \ref{fig:BSwfns}.
If one uses Eq. (\ref{eq:ls}), the bound-state wave function $\phi$ is obtained by solving the equation
\begin{align}
\phi=gU\phi=gUgU\phi.
\end{align}
The second equation shows that $U$ also is evaluated at vanishing values of time components of the relative momenta. We will treat the amplitudes $\Psi,\phi$ and the state vectors $|\Psi\rangle,|\phi\rangle$ (either bras or kets) as interchangeable. 

The next step is to relate $\Psi$ with $\phi$, which can be thought of as the usual bound-state wave function.
 This is most easily accomplished by using the projection operator $P$ on the product space of two positive-energy on-mass-shell nucleons.  We then have
\begin{align}
PG=GP\equiv G_P=g,
\end{align}
with the last step resulting from the explicit appearance of two positive-energy projection operators for on-mass-shell nucleons in Eq. (\ref{eq:g}).  We define $Q=I-P$ and use the notation $\Psi_{P}\equiv {P}\Psi,\Psi_{Q}\equiv {Q}\Psi$ and $PKP\equiv K_{PP},PKQ\equiv K_{PQ}\;,{etc}.$  The $Q$-space includes all terms with one or both nucleons off the mass-shell.  The amplitude $\Psi_P$ contains the ordinary nucleonic degrees of freedom so  one expects that it corresponds to $\phi$. This is now shown explicitly.  Use $I=P+Q$ in Eq. (\ref{eq:bsb}) and multiply by $P$ and then also by $Q$ to obtain the coupled-channel version of the relativistic bound state equation:
\begin{align}
\Psi_P&=G_PK_{PP}\Psi_P+G_PK_{PQ}\Psi_Q\label{eq:psip}
\\
\Psi_Q&=G_QK_{QP}\Psi_P+G_QK_{QQ}\Psi_Q.\label{eq:psiq}
\end{align}
Solving Eq. (\ref{eq:psiq}) for $\Psi_Q$ and using the result in Eq. (\ref{eq:psip}) gives
\begin{align}
\Psi_Q&=[1-G_QK_{QQ}]^{-1}G_QK_{QP}\Psi_P
\\
\Psi_P&=G_P\left(K_{PP} +K_{PQ}[G_Q^{-1}-K_{QQ}]^{-1}K_{QP}\right)\Psi_P,\label{eq:peq}
\end{align}
but one can multiply Eq. (\ref{eq:udef}) by $P\cdots P$ \textit{etc}. to obtain the result
\begin{align}
U_{PP} =K_{PP} +K_{PQ}[G_Q^{-1}-K_{QQ}]^{-1}K_{QP},
\end{align}
thus Eq. (\ref{eq:peq}) can be re-expressed as
\begin{align}
\Psi_P=G_PU_{PP}\Psi_P=gU\Psi_P.
\end{align}
This last equation is identical to Eq. (\ref{eq:bsb}).  Thus we have the result that 
\begin{align}
\Psi_P=\phi.
\end{align}
$\Psi_P$ is not the complete wave function, but we expect that $\Psi_Q$ is a perturbative correction because the deuteron is basically a non-relativistic system.

\section{\label{sec:nnpiamp}The \texorpdfstring{$N\rightarrow N\pi$}{N --> N pi} Amplitude}

We now turn to the application of the Bethe-Salpeter formalism to the problem of threshold pion production.  First, we remind the reader of the one-body pion production operator in HB$\chi$PT, which was derived in \cref{sec:HBChPT,sec:lagrangian}.  Here we repeat the derivation without making (yet) the non-relativistic reduction.  The nucleon field is split into its heavy ($H_v$) and light ($N_v$) components,
\begin{align}\bs
\Psi(x)&=e^{-im_Nv\cdot x}\left(N_v(x)+H_v(x)\right)
\\
N_v(x)&=e^{im_Nv\cdot x}P_+\Psi(x)
\\
H_v(x)&=e^{im_Nv\cdot x}P_-\Psi(x)
\es\end{align}
where $P_\pm=(1\pm\slashed{v})/2$ and $v$ is the velocity vector satisfying $v^2=1$ and chosen in this work to be $v=(1,{\bf 0})$.  The heavy component is integrated out of the path integral and the resulting free equation of motion for the light component has a solution,
\begin{align}
N(x)=\sqrt{E+m_N}\begin{pmatrix}\chi\\0\end{pmatrix}e^{-i(E-m_N)t+i\bfp\cdot\bfx},
\end{align}
where $E=\sqrt{\bfp\,^2+m_N^2}$ and $\chi$ is a two-component Pauli spinor.  In \cref{sec:nnpi} we show that the LO Feynman rule for the $s$-wave $N\rightarrow N\pi$ amplitude vanishes at threshold and that the NLO rule is
\begin{align}
\mathcal{O}_\pi=-i\frac{m_\pi}{2m_N}\frac{g_A}{2f_\pi}\gamma^5\gamma_i\gamma^0(\overrightarrow{\bnabla}-\overleftarrow{\bnabla})_i\tau_a,\label{eq:vertex}
\end{align}
where the derivatives act on the nucleon wave functions.

\section{\label{sec:pw}The \texorpdfstring{$NN\to d\pi$}{NN --> d pi} Reaction: Plane Wave Initial States}

Traditionally, the impulse approximation to pion production is calculated by using the operator of Eq. (\ref{eq:vertex}) as the irreducible kernel to be evaluated between non-relativistic nucleon-nucleon wave functions for the initial and final states.  Between two-component nucleon spinors, $\gamma^5\gamma_i\gamma^0\rightarrow\sigma_i$, so
\begin{align}
\mathcal{M}^\text{PW}_\text{1B}=\la\phi|\left[-i\frac{m_\pi}{2m_N}\frac{g_A}{2f_\pi}\bsig\cdot\left(\overrightarrow{\bnabla}-\overleftarrow{\bnabla}\right)\tau_{a}\right]|p_1,p_2\ra,\label{eq:Mapprox}
\end{align}
where the superscript on ${\cal M}$ indicates that we have neglected initial state interactions.  Next, we show that Eq. (\ref{eq:Mapprox}) is only an approximation to the full impulse amplitude derived from the relativistic Bethe-Salpeter formalism.  We will see that this approximation is only valid in the absence of initial state interactions.

For the case of plane waves in both the initial and final states, a one-body operator is forbidden by energy-momentum conservation,
\begin{align}
\la p_3,p_4|{\cal O}_\pi|p_1,p_2\ra=0,
\end{align}
with all the $p_i$ on mass shell.  The correct formalism must be able to explain the required energy transfer.  Our primary thesis is that the diagram of Fig. \ref{fig:iapw} must be obtained from the Feynman rules as
\begin{align}
{\cal M}^\text{PW}_\text{2B}=\la\Gamma|G_1{\cal O}_\pi|p_1,p_2\ra=\left\la\Psi\left|K(m_\pi/2)\,G_1{\cal O}_\pi\right|p_1,p_2\right\ra,\label{eq:mpw}
\end{align}
where $G_1$ is the Feynman propagator of the intermediate off-shell nucleon and $K(m_\pi/2)$ is the sum of all irreducible diagrams with energy transfer of $m_\pi/2$.  The second equality of Eq. (\ref{eq:mpw}) results from the relation between $\Gamma$ and $\Psi$ in Eq. (\ref{eq:bsb}).  This manipulation is necessary because $\la\phi|$ will be used for evaluation instead of $\la\Psi|$, meaning that the relative energy must remain zero in the final state.  Thus the full kernel for pion production via the impulse approximation is $KG_1{\cal O}_\pi$ rather than just ${\cal O}_\pi$.  Because $KG_1{\cal O}_\pi$ is a two-body operator, the momentum mismatch which suppresses the IA in the traditional treatment is removed.  

There are two points to emphasize here.  Firstly, this treatment is not equivalent to the heavy meson exchange operators of Refs. \citep{Horowitz:1993sh, Lee:1993xh} which are intended to account for the relativistic initial and final state interactions not present in phenomenological potentials.  Secondly, although the assertion of Eq. (\ref{eq:mpw}) greatly changes the way impulse pion production is calculated, one should not perform the same manipulations for the similar impulse approximation to photo-disintegration.  The reason for this is simply that near threshold the nucleon remains essentially on-shell and the diagram is therefore clearly reducible.

Next, we use $\Psi=\Psi_P+\Psi_Q=\phi+\Psi_Q$ and focus on the $\phi$ term; the other term contains non-nucleonic physics and may be treated as a correction.  Thus the impulse approximation is given by
\begin{align}
{\cal M}^\text{PW}_\text{2B}\approx\la\phi|K(m_\pi/2)\,G_1{\cal O}_\pi|p_1p_2\ra.\label{eq:Mia}
\end{align}

Consider the spacetime structure of the product, $G_1{\cal O}_\pi$.  The relativistic propagator $G_1$ is decomposed into three terms: $1$, $\gamma^0$, and $\gamma^i$.  Between two-component nucleon spinors
\begin{align}\bs
\gamma^5\gamma_i\gamma^0&\rightarrow\sigma_i
\\
\gamma^0\gamma^5\gamma_i\gamma^0&\rightarrow\sigma_i
\\
\gamma^i\gamma^5\gamma_j\gamma^0&\rightarrow0,\label{eq:nrelred}
\es\end{align}
and so we can make the replacement
\begin{align}\bs
G_1{\cal O}_\pi=i\frac{\slashed{p}_1-\slashed{q}+m_N}{(p_1-q)^2-m_N^2+i\epsilon}{\cal O}_\pi&\rightarrow i\frac{E(\bfp_1)-m_\pi+m_N}{-2E(\bfp_1)m_\pi+m_\pi^2+i\epsilon}{\cal O}_\pi
\\
&=\frac{i}{-m_\pi}\left(1-\frac{m_\pi}{4m_N}\right){\cal O}_\pi,
\es\end{align}
where in the second line we have used that $E(\bfp_1)=m_N+m_\pi/2$ at threshold.  Note that this propagator agrees with that obtained from the Feynman rules for B$\chi$PT at LO.

\begin{figure}
\centering
\includegraphics[height=1.5in]{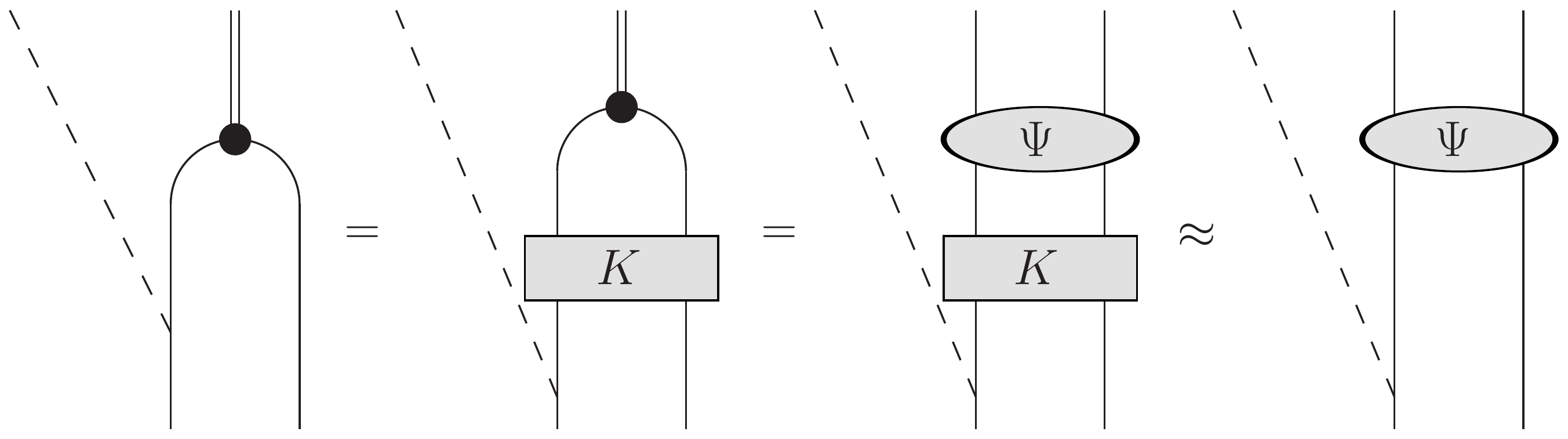}
\caption[Bethe-Salpeter formalism applied to pion production]{\label{fig:iapwBS}Bethe-Salpeter formalism applied to pion production for plane wave initial states.}
\end{figure}

In order to make connection with the traditional Eq. (\ref{eq:vertex}), we use the approximations $K\approx U$ [corrections are ${\cal O}(g-G)$] and $G_1\approx-i/m_\pi$ [corrections are ${\cal O}(m_\pi/m_N)$].  Putting these substitutions into Eq. (\ref{eq:Mia}), 
\begin{align}
{\cal M}^\text{PW}_\text{2B}\approx\la\phi|\left[-\frac{iU\left(\frac{m_\pi}{2}\right)}{m_\pi}{\cal O}_\pi\right]|p_1p_2\ra.\label{eq:mpwfinal}
\end{align}
The quantity $U$ is related to the potential energy by $U=-iV$.  Ignoring the fact that $U$ should be evaluated for non-zero energy transfer, we use the equal-time Schr\"{o}dinger equation to replace $V\rightarrow-E_d-p^2/m_N$ and then neglect the binding energy to find ${\cal M}^\text{PW}_\text{2B}\approx{\cal M}^\text{PW}_\text{1B}$.  This means that for a PW initial state, the traditional impulse approximation should be roughly adequate.  This is borne out in the actual calculation of the reduced matrix elements for Eqs. (\ref{eq:Mapprox}) and (\ref{eq:mpwfinal}),
\begin{align}\bs
A_\text{1B}^\text{PW}&=-24.0
\\
A_\text{2B}^\text{PW}&=-25.6,
\es\end{align}
where we have used \cref{sec:reduced}'s definition of the reduced matrix element (we suppress the subscript on \cref{sec:reduced}'s $A_0$ for clarity) and used the same static phenomenological potential for $V$ (here, Argonne v18 \citep{Wiringa:1994wb}) that is used to calculate the wave functions.  See Fig. \ref{fig:iapwBS} for a pictorial description of this section.

\begin{figure}
\centering
\includegraphics[height=1.5in]{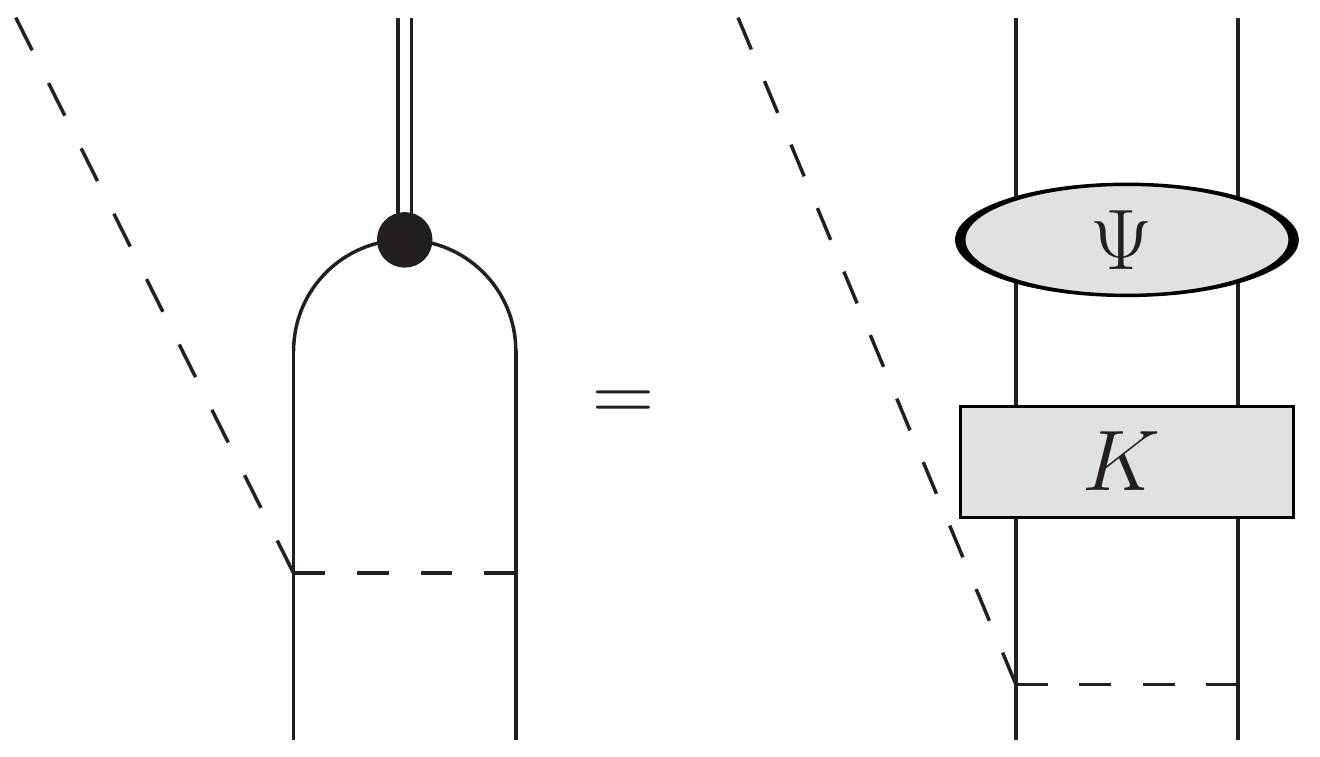}
\caption[Bethe-Salpeter equation in the rescattering amplitude]{\label{fig:rsBS}Use of the Bethe-Salpeter equation in the rescattering amplitude}
\end{figure}

It is important to note that the Bethe-Salpeter equation can also be used for the RS diagram as shown in Fig. \ref{fig:rsBS}.
In fact, the diagram on the right in Fig. \ref{fig:rsBS} (with $K$ approximated by OPE) is one of the two topologically reducible diagrams used by Ref. \citep{Lensky:2005jc} to resolve the problem arising from calculation of NLO loops as we have discussed in previous chapters.

In the next section, we will show that for distorted wave initial states, Eq. (\ref{eq:mpwfinal}) is replaced by 
\begin{align}
{\cal M}^\text{DW}_\text{2B}\approx{}_f\la\phi|\left[-\frac{iU\left(\frac{m_\pi}{2}\right)}{m_\pi}{\cal O}_\pi+{\cal O}_\pi\frac{iU\left(\frac{m_\pi}{2}\right)}{m_\pi}\right]|\phi\ra_i,\label{eq:mdwfinal}
\end{align}
where the first term contributes at leading order in the theory and the second term at next-to-leading order.

\section{\label{sec:distortions}The \texorpdfstring{$NN\to d\pi$}{NN --> d pi} Reaction: Distorted Wave Initial States}
\subsection{\label{sec:dwdef}Definition of distorted wave operator}

There is no reason to expect the result ${\cal M}^\text{PW}_\text{2B}\approx{\cal M}^\text{PW}_\text{1B}$ to carry over for a distorted wave (DW) initial state where $\bfp^2=m_\pi m_N$ no longer holds.
Indeed, we will show that the traditional expression for the impulse approximation does not hold for DW amplitudes.

The fully-relativistic initial-state wave function is denoted $|\Psi\ra_i$,
\begin{align}
|\Psi\ra_i=|p_1,p_2\ra+GK|\Psi\ra_i,
\end{align}
where the first term is exactly the initial state used in the definition of ${\cal M}^\text{PW}$ of Eqs. (\ref{eq:Mapprox}) and (\ref{eq:mpw}).  The complete DW impulse operator is defined as,
\begin{align}
{\cal M}^\text{DW}={\cal M}^\text{PW}+{\cal M}^\text{ISI}.\label{eq:mdw}
\end{align}
The second term includes the production operator $KG_1{\cal O}_\pi$ from Eq. (\ref{eq:mpw}) along with initial state interactions,
\begin{align}\bs
{\cal M}^\text{ISI}_\text{2B}&={}_f\la\Psi|KG_1{\cal O}_\pi GK|\Psi\ra_i
\\
&\approx{}_f\la\phi|KG_1{\cal O}_\pi GK|\phi\ra_i,\label{eq:misi}
\es\end{align}
where in the second line we have once again used $\Psi=\phi+\Psi_Q$ and neglected the $Q$-space.

\begin{figure}
\centering
\includegraphics[height=3in]{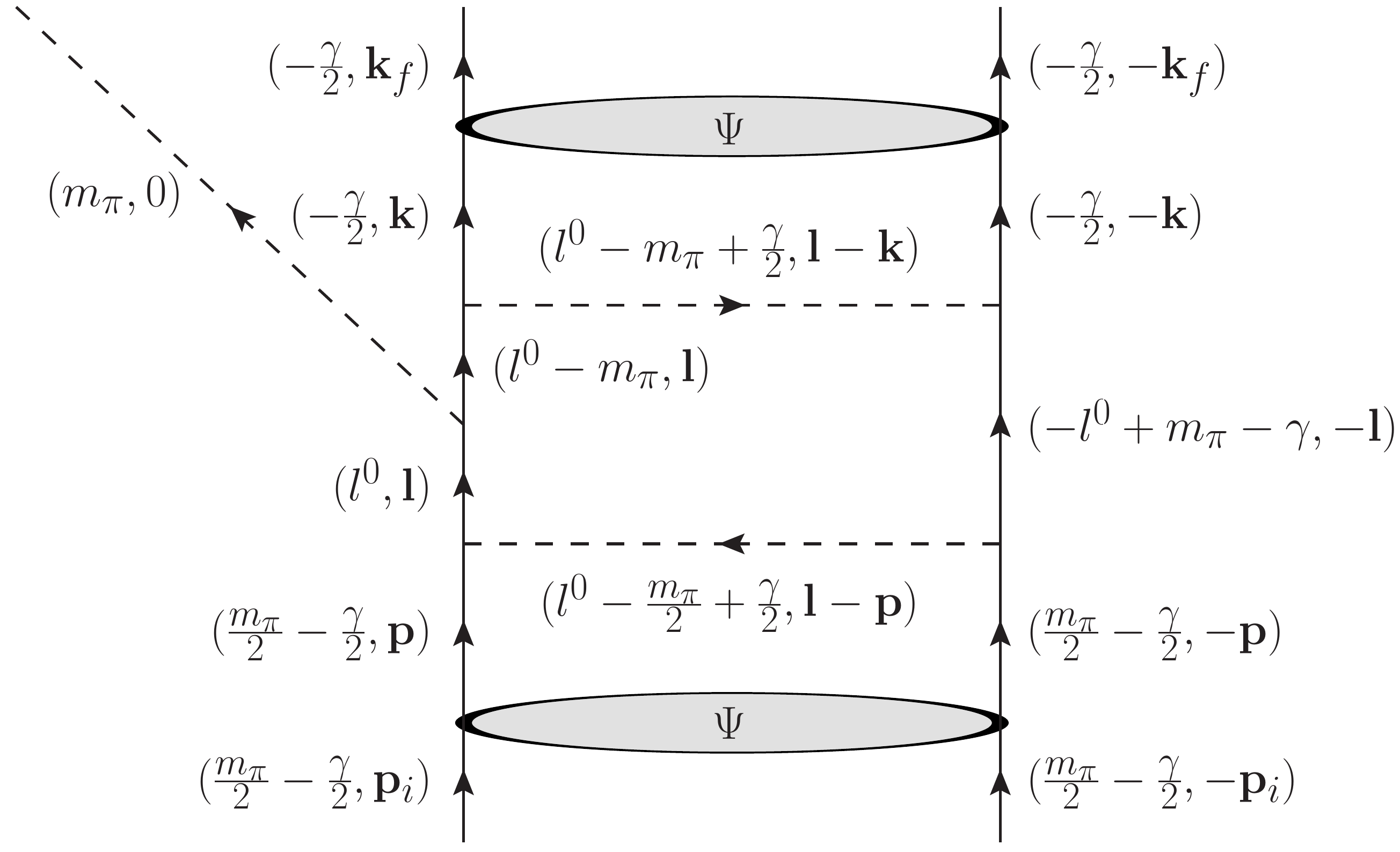}
\caption[Impulse approximation using distorted waves]{Impulse approximation using distorted waves.  Solid lines represent nucleons, dashed lines represent pions, and ovals represent wave functions.}
\label{fig:loop}
\end{figure}

As noted by Ref. \citep{Hanhart}, the kernel of Eq. (\ref{eq:misi}) is a loop integral which is shown in Fig. \ref{fig:loop} with $K$ being approximated by OPE.
Note that four-momenta are conserved at every vertex.  One pion exchange is the first contribution to $K$ in $\chi$PT besides a short range operator which is irrelevant for the $s$-wave $NN\to d\pi$ amplitude (see \cref{sec:lagrangian}).  Nevertheless, one must exercise caution due to the large expansion parameter of pion production.  To this end, we employ the deuteron of Ref. \citep{Friar:1984wi} which is calculated from a purely-OPE potential with suitable form factors.  As discussed in \cref{sec:deuteron}, this deuteron wave function is quite accurate and increases the rescattering amplitude by only 3\% over a phenomenological deuteron.  Having then employed this deuteron wave function in the calculation of the traditional DW impulse approximation, we will be able to avoid any complications from higher order parts of the potential in our subsequent investigation of the two-body operator of Eq. (\ref{eq:misi}).  In other words, although the full potential must be present in an exact calculation, we expect to gain insight into the correct formalism by using an OPE-only deuteron.  We continue to use the phenomenological potentials for the initial state.  To verify that the use of $K=\text{OPE}$ in the initial state does not spoil our results too much, \cref{sec:sigma} examines heavy meson exchange in the initial state.  As will be discussed, this effect is parametrically suppressed.

Note that the relative momenta of the nucleons before and after the loop ($\bfp$ and $\bfk$) are external momenta to the loop integral over $l=(l^0,\bfl)$, but are eventually integrated over in the hybrid formalism.  Let us focus solely on the energy part of the loop integral and ignore the vertex factors and overall constants.  We define the integral $I$,
\begin{align}\bs
I={}&i^5\int\frac{dl^0}{2\pi}\frac{1}{l^0-E+i\epsilon}\frac{1}{l^0-m_\pi-E+i\epsilon}\frac{1}{-l^0+m_\pi-\gamma-E+i\epsilon}\frac{1}{l^0-m_\pi/2-\gamma/2+\omega_i-i\epsilon}
\\
{}&\times\frac{1}{l^0-m_\pi/2-\gamma/2-\omega_i+i\epsilon}\frac{1}{l^0-m_\pi+\gamma/2+\omega_f-i\epsilon}\frac{1}{l^0-m_\pi+\gamma/2-\omega_f+i\epsilon},\label{eq:loop}
\es\end{align}
where $\omega_i^2=(\bfl-\bfp)^2+m_\pi^2$ is the on-shell energy of the initial-state pion, $\omega_f^2=(\bfl-\bfk)^2+m_\pi^2$ is the on-shell energy of the final state pion, and $E=\bfl^2/2m_N$ is the kinetic energy of a single intermediate nucleon.  Note that $\bfp_i^2\approx m_\pi m_N-\gamma m_N$ and $\bfk_f^2\approx-\gamma m_N$.

It is straightforward to show that if the energy components of the exchanged pions in the above loop are set to zero (violating conservation of four-momentum), one obtains the traditional impulse approximation.  In this case, the pion energy denominators are pulled out of the integral which is then evaluated by closing the contour in the lower half plane,
\begin{align}
I_\text{1B}=\left[\frac{1}{-\omega_f^2}\frac{1}{-\gamma-\bfl^2/m_N}\right]\left[\frac{1}{m_\pi-\gamma-\bfl^2/m_N}\frac{1}{-\omega_i^2}\right].
\end{align}
The quantity in the first set of brackets can be recognized as the product of OPE with the final state wave function while the second set is the product of the initial state wave function with OPE.  This is precisely the operator that the traditional evaluation includes.

\subsection{\label{sec:topt}Reduction to time ordered perturbation theory}
\begin{figure}
\centering
\includegraphics[height=1.5in]{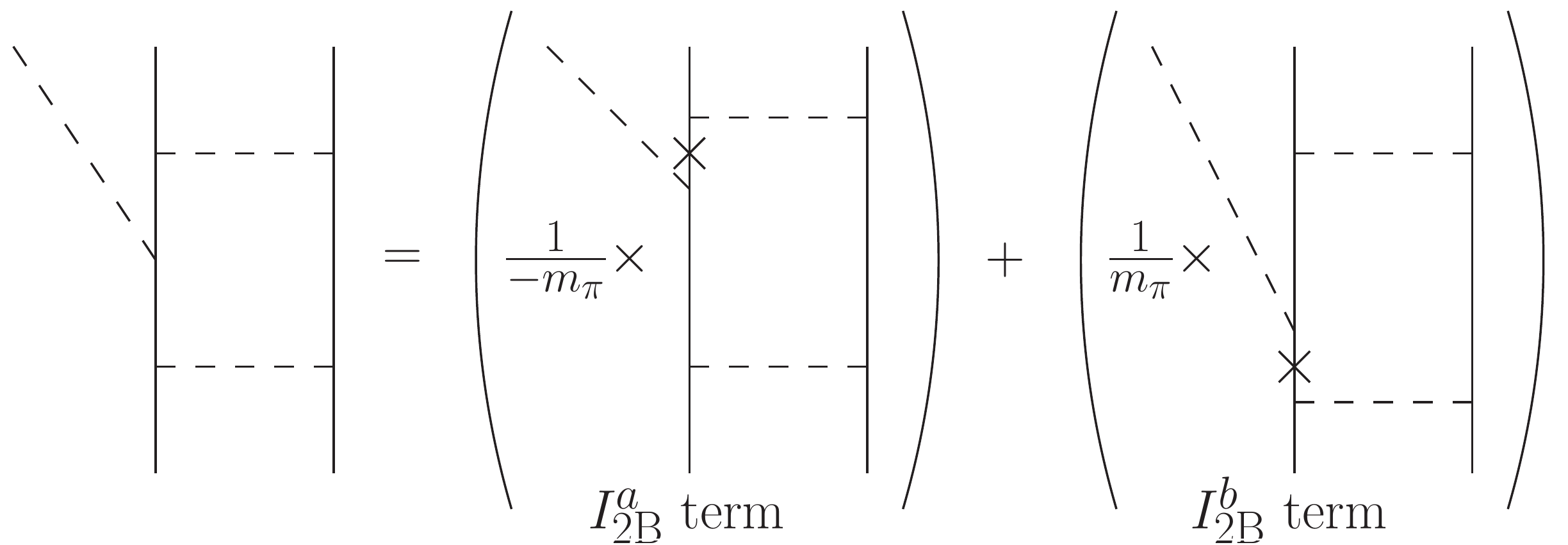}
\caption[Definition of the two terms in Eq. (\ref{eq:split2})]{Definition of the two terms in Eq. (\ref{eq:split2}).  Crosses represent the propagators which are absent due to the partial fractions decomposition.}
\label{fig:split}
\end{figure}

Our goal is to evaluate the integral in Eq. (\ref{eq:loop}), showing that it is a sum of diagrams from time ordered perturbation theory (TOPT) terms which can be combined to obtain Eq. (\ref{eq:mdwfinal}).  To begin, we rewrite the first two factors as a sum,
\begin{align}
\frac{1}{l^0-E+i\epsilon}\frac{1}{l^0-m_\pi-E+i\epsilon}=\frac{1}{-m_\pi}\left(\frac{1}{l^0-E+i\epsilon}-\frac{1}{l^0-m_\pi-E+i\epsilon}\right).\label{eq:split}
\end{align}
This is the key to our method because after making this split, we see two terms which each have the propagator structure of a rescattering box loop.  Consider the first term in Eq. (\ref{eq:split}); this loop integral looks like a two-body operator multiplied by $\frac{1}{-m_\pi}$ and augmented with an initial-state interaction.  The second term looks like the same with final-state interaction.  We define these two integrals to be $I_\text{2B}^a$ and $I_\text{2B}^b$, respectively,
\begin{align}
I_\text{2B}=I_\text{2B}^a+I_\text{2B}^b.\label{eq:split2}
\end{align}
Figure \ref{fig:split} illustrates the splitting described in Eq. (\ref{eq:split2}).

\begin{figure}
\centering
\includegraphics[width=\linewidth]{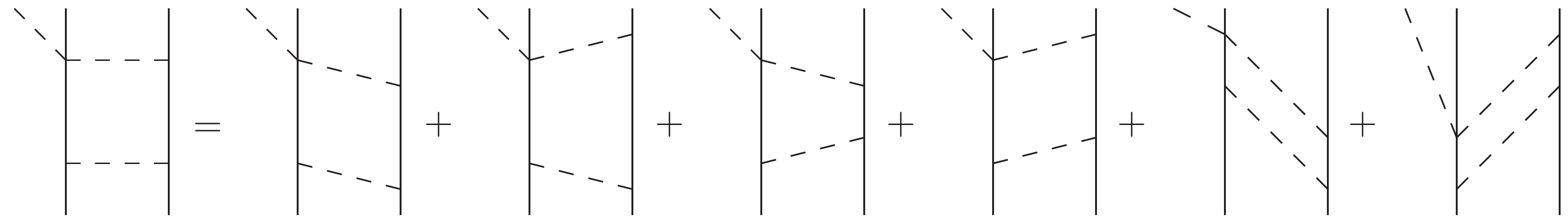}
\caption{TOPT terms resulting from the $I_\text{2B}^a$ integral.}
\label{fig:topt}
\end{figure}

Next, we perform partial fraction decomposition on each of the pion propagators, splitting each of the two terms into four terms.  Then, we continue the decomposition process until each term can be expressed as a single residue.  For $I_\text{2B}^a$ we will isolate the poles containing $\omega_f$ and then close the contour around them (for $I_\text{2B}^b$, the $\omega_i$ poles are isolated).  By isolating the poles in this way, the resulting expression is easily recognized as the sum of six TOPT terms.  For clarity, we show these terms pictorially for $I_\text{2B}^a$ in Fig. \ref{fig:topt} where we have left the overall $\frac{1}{-m_\pi}$ implicit.
We assume for now that the stretched box diagrams are small, as they were in the rescattering toy model investigation \citep{Hanhart:2000wf} and denote the sum of the four remaining terms with a $\mathring{I}$.

Finally, motivated by the interpretation which is presented in the next section, we algebraically re-combine these four terms to find
\begin{align}
\mathring{I}_\text{2B}^a&=\frac{1}{(m_\pi/2)^2-(\omega_f+\delta_a(\bfl))^2}\left[1+\frac{\delta_a(\bfl)}{\omega_f}\right]\frac{1}{-m_\pi}\left[1-\frac{\delta_a(\bfl)}{\omega_i+\delta_a(\bfl)}\right]\frac{1}{m_\pi-2E-\gamma}\frac{1}{-\omega_i^2}\label{eq:isifinal}
\\
\mathring{I}_\text{2B}^b&=\frac{1}{-\omega_f^2}\frac{1}{-2E-\gamma}\left[1-\frac{\delta_b(\bfl)}{\omega_f+\delta_b(\bfl)}\right]\frac{1}{m_\pi}\left[1+\frac{\delta_b(\bfl)}{\omega_i}\right]\frac{1}{(m_\pi/2)^2-(\omega_i+\delta_b(\bfl))^2},\label{eq:fsifinal}
\end{align}
where we have separated out terms involving $\delta_a$ and $\delta_b$,
\begin{align}\bs
\delta_a(\bfl)&=\frac{\bfl^2}{2m_N}-\frac{m_\pi}{2}+\frac{\gamma}{2}
\\
\delta_b(\bfl)&=\frac{\bfl^2}{2m_N}+\frac{\gamma}{2},
\es\end{align}
because (as will be shown in the next section) they are sub-leading and we will neglect them in the main body of this work.  The only approximation made in the evaluation of the loop integral to obtain Eqs. (\ref{eq:isifinal}) and (\ref{eq:fsifinal}) is to neglect the stretched boxes.  Let us pause to summarize what we have done so far: (1) the DW amplitude was written down as a loop integral, (2) partial fractions was used to split the product of the two nucleon propagators into a sum $I_\text{2B}^a+I_\text{2B}^b$, (3) the loop integrals were evaluated and the result expressed in terms of TOPT diagrams, and (4) the TOPT diagrams were algebraically combined into a form useful for the following interpretation.

\subsection{\label{sec:interp}Interpretation}

\begin{figure}
\begin{center}
\subfigure[\ ${\cal O}_\text{2B}^a$ of Eq. (\ref{eq:prodisi2})]{\label{fig:iaf}\includegraphics[height=1.5in]{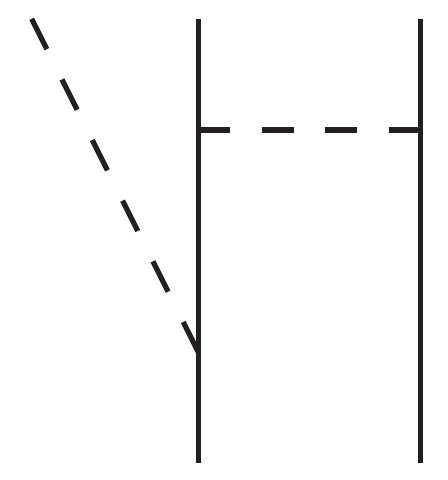}}
\hspace{.15\linewidth}
\subfigure[\ ${\cal O}_\text{2B}^b$ of Eq. (\ref{eq:prodfsi2})]{\label{fig:iai}\includegraphics[height=1.5in]{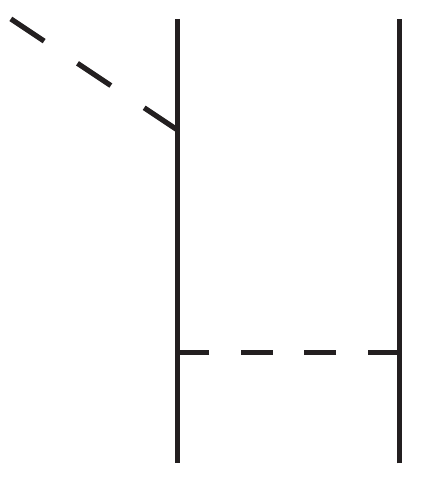}}
\end{center}
\caption[Two-body impulse production operators]{\label{fig:prodop}Two-body impulse production operators (pion exchange is non-static).}
\end{figure}

Although not obvious at first sight, convolution of the operator corresponding to Eq. (\ref{eq:isifinal}) with wave functions as defined in Eqs. (\ref{eq:mdw}) and (\ref{eq:misi}) results in an amplitude approximately equivalent to that which one obtains by using the operator shown in Fig. \ref{fig:iaf}.  The same is true of Eq. (\ref{eq:fsifinal}) with Fig. \ref{fig:iai}, and together they replace the traditional (one-body) impulse approximation with Eq. (\ref{eq:mdwfinal}).  Furthermore, the operator that results from Eq. (\ref{eq:fsifinal}) is expected to be small by power counting arguments.  The task of this subsection is to verify these statements in detail.

In Eq. (\ref{eq:isifinal}) the factor $(m_\pi-2E-\gamma)^{-1}(-\omega_i^2)^{-1}$ is interpreted as the product of the two-nucleon initial-state wave function with static OPE.  This is the statement that
\begin{align}
\frac{1}{m_\pi-2E-\gamma}\frac{1}{-\omega_i^2}=gU^\text{OPE}.
\end{align}
This factor can be absorbed (after adding in the PW term) using the zero-relative-energy Lippmann-Schwinger equation that is employed by the phenomenological potentials we are using.  We will continue to refer to the initial wave function as a function of $\bfp$ and $\bfp_i$, so absorbing this factor means that we set $\bfl=\bfp$.

Likewise, in Eq. (\ref{eq:fsifinal}) the factor $(-\omega_f^2)^{-1}(-2E-\gamma)^{-1}$ is interpreted as the product of static OPE with the two-nucleon final-state wave function: $U^\text{OPE}g$.  Absorbing this factor into the wave function, we set $\bfl=\bfk$.  The remaining factors of $\mathring{I}_\text{2B}^a$ and $\mathring{I}_\text{2B}^b$ become the two-body impulse production operators,
\begin{align}
{\cal O}_\text{2B}^a&=\frac{\bsig_1\cdot(\bfp-\bfk)\bsig_2\cdot(\bfk-\bfp)}{(m_\pi/2)^2-(\omega_f+\delta_a(\bfp))^2}\left[1+\frac{\delta_a(\bfp)}{\omega_f}\right]\frac{1}{-m_\pi}\left[1-\frac{\delta_a(\bfp)}{\omega_i+\delta_a(\bfp)}\right]\bfS\cdot\bfp\label{eq:prodisi}
\\
{\cal O}_\text{2B}^b&=\bfS\cdot\bfk\left[1-\frac{\delta_b(\bfk)}{\omega_f+\delta_b(\bfk)}\right]\frac{1}{m_\pi}\left[1+\frac{\delta_b(\bfk)}{\omega_i}\right]\frac{\bsig_1\cdot(\bfp-\bfk)\bsig_2\cdot(\bfk-\bfp)}{(m_\pi/2)^2-(\omega_i+\delta_b(\bfk))^2},\label{eq:prodfsi}
\end{align}
where we have now made explicit the momentum dependences of the vertices and used ${\bf S}=(\bsig_1+\bsig_2)/2$.  It is also important to include form factors in the OPE which match those of the wave functions.  These form factors are present in our calculation even though we leave them out of this expression for the sake of generality.

Next, note that in the evaluation of the matrix element using Eq. (\ref{eq:prodisi}), the initial state wave function is peaked about its plane wave value $\bfp\approx\bfp_i$, and thus $E\approx m_\pi/2-\gamma/2$ and $\delta_a(\bfp)\approx0$.  On the other hand, in Eq. (\ref{eq:prodfsi}), we have $\bfk\approx\bfk_i$ and $E\approx-\gamma/2$ and $\delta_b(\bfk)\approx0$.  If we were to neglect all the $\delta$'s, we would have
\begin{align}
{\cal O}_\text{2B}^a&\approx\frac{\bsig_1\cdot(\bfp-\bfk)\bsig_2\cdot(\bfk-\bfp)}{(m_\pi/2)^2-\left((\bfp-\bfk)^2+m_\pi^2\right)}\frac{1}{-m_\pi}\bfS\cdot\bfp\label{eq:prodisi2}
\\
{\cal O}_\text{2B}^b&\approx\bfS\cdot\bfk\frac{1}{m_\pi}\frac{\bsig_1\cdot(\bfp-\bfk)\bsig_2\cdot(\bfk-\bfp)}{(m_\pi/2)^2-\left((\bfp-\bfk)^2+m_\pi^2\right)},\label{eq:prodfsi2}\end{align}
which suggests that these operators can be approximately interpreted as the diagrams in Fig. \ref{fig:prodop}.
Thus we have finally obtained our central result, [Eq. (\ref{eq:mdwfinal})], which states that the correct impulse approximation is a two-body operator.  The contribution to pion production given in Eq. (\ref{eq:mdwfinal}) is not replacing the rescattering diagram (which is also two-body), but rather replacing the traditional contribution which has been referred to as the impulse approximation (or direct production).  Note that if we assign standard pion production power counting to these diagrams, Fig. \ref{fig:iaf} is $\mathcal{O}\left(\sqrt{\frac{m_\pi}{m_N}}\right)$ while \cref{fig:iai} is $\mathcal{O}\left(\frac{m_\pi}{m_N}\right)$.  In the next section the approximate expressions given in Eqs. (\ref{eq:prodisi2}) and (\ref{eq:prodfsi2}) are numerically evaluated.  Nevertheless, we acknowledge the importance of verifying that the $\delta$ terms are indeed small and relegate that discussion to \cref{sec:isi,sec:fsi}.

\subsection{\label{sec:evaluation}Evaluation of two-body operators}

Next, we calculate the threshold $s$-wave $np\to d\pi^0$ amplitudes corresponding to Eqs. (\ref{eq:prodisi2}) and (\ref{eq:prodfsi2}).  We do not present the details here as most are given in \cref{sec:opedetails}.  Again, we remind the reader that for the sake of consistency we use a deuteron wave function calculated from a purely-OPE potential (with form factors as described in \cref{sec:deuteron}).  For the initial-state distorted waves, we use three different phenomenological potentials (Av18 \citep{Wiringa:1994wb}, Nijmegen II \citep{Stoks:1994wp}, and Reid `93 \citep{Stoks:1994wp}).  In Table \ref{tab:results}, we display the results in terms of the reduced matrix elements of \cref{sec:reduced}.
\begin{table}
\caption[Threshold reduced matrix elements]{\label{tab:results}Threshold reduced matrix elements calculated with an OPE deuteron and various phenomenological initial states.  The first row shows the traditional impulse approximation (one-body) while the second and third show our replacement (two-body).}
\begin{center}
\renewcommand{\tabcolsep}{3mm}
\begin{tabular}{cccc}
\hline\hline
& Av18 & Reid '93 & Nijm II\\ \hline
$A^\text{DW}_\text{1B}$ & 8.3 & 7.1 & 5.4\\
$A_\text{2B}^\text{DW,a}$ & 17.4 & 13.5 & 7.8\\
$A_\text{2B}^\text{DW,b}$ & $-1.5$ & $-2.2$ & $-6.9$\\ \hline\hline
\end{tabular}
\end{center}
\end{table}

The first row of Table \ref{tab:results} gives the traditional (one-body) impulse approximation, which is slightly bigger than in \cref{chap:wfncor} due to the use of the OPE deuteron.  The next row shows that the new two-body operator (at leading order) is roughly twice as large as the traditional calculation it is replacing.  We mention here that the significant cancellation between deuteron $s$- and $d$-states remains, keeping the impulse amplitude smaller than rescattering; however, the cancellation is less complete when using our new two-body operator.  The final row verifies that the ${\cal O}_\text{2B}^b$ diagram is smaller than the ${\cal O}_\text{2B}^a$ diagram, as dictated by the power counting.  The Nijmegen II potential provides a bit of deviation from these results, and it will be interesting to investigate other potentials to determine the true model dependence of this calculation.  In finding these results, it is important that the pion propagators of Eqs. (\ref{eq:prodisi2}) and (\ref{eq:prodfsi2}) be implemented in a manner consistent with the potential used for the wave function of Fig. \ref{fig:loop}.  Namely, the cutoff procedure of the convolution integral with form factors needs to match that by which the potential was constructed.  \Cref{sec:deuteron} contains the details of this procedure.

Our conclusion is that the traditional impulse approximation is an underestimate.  While it is true that several approximations were made in order to permit final expressions as simple as Eqs. (\ref{eq:prodisi2}) and (\ref{eq:prodfsi2}), we believe this conclusion to be sound.  The $\delta$ terms do not defy their classification as sub-leading (see \cref{sec:isi,sec:fsi}), and \cref{sec:sigma} shows that using $K=\text{OPE}$ in the initial state is at least reasonable.  In summary, we simply claim that Eq. (\ref{eq:prodisi2}) is the new impulse approximation at leading order in the effective field theory.  The corrections in the aforementioned appendices, in addition to Eq. (\ref{eq:prodfsi2}) contribute to the next-to-leading order calculation, which needs to be systematically considered in a later work.

Finally, it is important to note that although the OPE deuteron reproduces the phenomenological results for the rescattering diagram quite well, the numbers in this section are greatly changed if a phenomenological deuteron is used.  Using Av18 we find  $A^\text{DW}_\text{1B}=4.9$, and by using the cutoff procedure of Av18 for the two-body operators, we find $A_\text{2B}^\text{DW,a}=33.5$, $A_\text{2B}^\text{DW,b}=-2.8$.  Thus, the ratio of our new two-body operator to the traditional impulse operator is $\sim7$ instead of the $\sim2$ presented above.  At this time one is faced with a choice of either: (1) using a ``correct" phenomenological deuteron and leaving out parts of the potential when calculating the two-body kernel or (2) using an inexact OPE deuteron with a completely self-consistent kernel.  For the time being, we believe the latter to be more trustworthy, if not ideal.

\section{\label{sec:discussionch6}Discussion}

Recall from \cref{tab:exptch3} the results for the total cross section obtained by the five most recent experiments in terms of the parameters $\alpha$ and $\beta$ of \cref{eq:alphadefch3}.  Since the present calculation is performed at threshold ($\eta=0$), we compute only the value of $\alpha$,
\begin{align}
\alpha=\frac{m_\pi}{128\pi^2sp}\left|A\right|^2.\label{eq:alphach6}
\end{align}
For ease of comparison, we invert Eq. (\ref{eq:alphach6}), plug in the results of the mentioned experiments, and propagate the errors to find Table \ref{tab:expt}.
\begin{table}
\caption[Experimental threshold reduced matrix elements]{\label{tab:expt}Threshold reduced matrix elements extracted from experiment}
\begin{center}
\begin{tabular}{lc}
\hline\hline
Experiment & $A^\text{expt}$\\ \hline
$np\rightarrow d\pi^0$ (TRIUMF, 1989) \citep{Hutcheon:1989bt} & $80.1\pm1.1$\\
$\pi^+d\to pp$ (1991) \citep{Ritchie:1991rc} & $77.9\pm0.7$\\
$\vec{p}p\rightarrow d\pi^+$ (IUCF, 1996) \citep{Heimberg:1996be} & $85.2\pm1.0$\\
$pp\rightarrow d\pi^+$ (COSY, 1998) \citep{Drochner:1998ja} & $84.6\pm1.9$\\
Pionic deuterium (PSI, 2010) \citep{Strauch:2010rm}$\quad$ & $93.8^{+0.9}_{-2.0}$\\ \hline\hline
\end{tabular}
\end{center}
\end{table}

The full theoretical amplitude includes not only the impulse diagram but also the rescattering diagram, which is given in Table \ref{tab:tot} along with the total amplitude using either the traditional one-body or the leading-order two-body impulse diagram.
\begin{table}
\caption[Rescattering and total reduced matrix elements]{\label{tab:tot}Rescattering and total reduced matrix elements for a variety of potentials.  The second line shows the traditional calculation (with a one-body IA) while the third shows our replacement (with a two-body IA).}
\begin{center}
\begin{tabular}{lccc}
\hline\hline
& Av18 & Reid '93 & Nijm II\\ \hline
RS & 69.8 & 72.1 & 74.0\\
RS + IA (1B) & 78.1 & 79.2 & 79.4\\
RS + IA (2B)$\quad$ & 87.2 & 85.6 & 81.8\\ \hline\hline
\end{tabular}
\end{center}
\end{table}
The uncertainty in an effective field theory calculation is estimated by the power counting scheme.  In this work, we have included both the rescattering and the impulse diagrams up to ${\cal O}\left(\sqrt{m_\pi/m_N}\right)$.  Therefore one might assign an uncertainty of $m_\pi/m_N=14\%$ to the calculation but stress that such an estimate based solely on power counting is rough at best.  Taking this uncertainty, we see that the theory update presented here changes the situation from under-prediction of the most recent pionic deuterium experiment by $\sim1.3\,\sigma$, to under-prediction by $\sim0.7\,\sigma$.

In summary, we have developed a consistent formalism that allows one to separate effects of the kernel from those of the wave functions, finding a new impulse approximation kernel.  This two-body operator, given in Eq. (\ref{eq:mdwfinal}), replaces the traditional one-body impulse approximation and is the central result of the present work.  We numerically investigated the simplest example ($s$-wave $NN\rightarrow d\pi$) and found the impulse amplitude to be increased by a factor of roughly two over the traditional amplitude.  This calculation was performed with a regulated OPE deuteron which has advantages and disadvantages as described in the body of this work.  Rescattering remains the dominant contribution to the cross section.  We find that the updated total cross section is $\sim10\%$ larger than before and is in agreement with experiment at leading order.  We verified that corrections to the new impulse approximation (which together with other loops and counterterms will contribute at next-to-leading order) do not destroy these results.

These findings suggest several directions for future research.  Firstly, one needs to develop a power counting scheme for the ``Q space" discussed in \cref{sec:bs}.  Secondly, the significant model dependence of the new formulation of the impulse approximation needs to be investigated in a renormalization group invariant way.  Thirdly, it will be very interesting to see the impact of this increased impulse amplitude on the $pp\rightarrow pp\pi^0$ cross section which is suppressed due to the absence of rescattering.  Finally, one could look at the energy dependence ($p$-wave pions) of $NN\rightarrow NN\pi$, for which there is an abundance of experimental data.
\chapter{Discussion}
\label{chap:discussion}

In this thesis, we have reported on two studies: \cref{chap:csb} described the calculation of the asymmetry of $np\to d\pi^0$ resulting from CSB (most importantly the $\delta m_N$ interaction), \cref{chap:wfncor,chap:nrred} eventually led to the discovery that the hybrid formalism forbids the traditional one-body impulse approximation, replacing it with a two-body operator.  Let us now summarize these results and discuss the outlook for these topics.

We performed the asymmetry calculation up to partial-NLO and found $A^\text{thy}_\text{fb}=40.2\times10^{-4}$ which is to be compared with $A^\text{expt}_\text{fb}=[17.2\pm9.7]\times10^{-4}$.  To obtain our result for the asymmetry, we used the Cottingham sum rule to fix $\delta m_N$; the other CSB coupling, $\beta_1$, was fixed by assuming it to originate from an intermediate $\eta$ which fluctuates into a $\pi^0$.  Both of these couplings are not precisely known, and it is possible that revised values could improve our calculation's agreement with experiment.  Nevertheless, the disagreement in $A_\text{fb}$ we see when using the best available CSB parameters is significant.  It is important to understand to what extent this disagreement is due to the CSB couplings themselves, and to what extent it is due to incompletely-understood theoretical issues.

In \cref{chap:csb}, an unexpectedly large amplitude in the ${}^1S_0$ channel led to a disagreement with $p$-wave data and was fixed with an ad-hoc cutoff.  The fact that Ref. \citep{Filin:2009yh}, with its $N\Delta$ coupled-channel wave functions, did not find a similarly large amplitude suggests that, with our $NN$ wave functions, we need to perform the $p$-wave calculation to higher order in the EFT.  As we discussed, the first set of loops vanish.  The NNLO calculation includes various tree-level diagrams (involving the known $c_i$ LECs) along with a $\pi N^4$ contact interaction (involving the unknown $d$ LEC).  Recently, Ref. \cite{Baru:2009fm} performed this NNLO calculation and fit $d$ with the relevant experimental data.  The determination of $d$ requires an unambiguous knowledge of all other $p$-wave diagrams up to the order where it appears.  In this thesis we have uncovered a problem with the way the IA diagram has been calculated.  Therefore, a new determination of $d$ should now be performed.

A second difficulty encountered in the CSB study regards the total cross section of $np\to d\pi^0$.  In this thesis, we applied recent power counting and reducibility developments to calculate a total cross section roughly in agreement with collider data.  At the same time, pionic deuterium experiments have been shown to imply a cross section well above this prediction.  The study of Ref. \citep{Filin:2009yh} used the cross section obtained from the atomic experiment as an ingredient in their asymmetry calculation.  This fact accounts for most of the discrepancy between our two calculations.  Why do the atomic experiments disagree with the collider ones by such a large amount?

As part of the CSB study of \cref{chap:csb}, we became interested in the energy transfer mechanism of the IA diagram.  A na\"ive attempt at a solution (wave function corrections) was made in that chapter.  This solution was investigated in greater detail in \cref{chap:wfncor} and the results were not satisfactory.  Significant model dependence was observed resulting from the use of different phenomenological potentials.  With a correctly implemented hybrid approach (including correct cutoffs/form factors), this should not be the case.  Additionally, a concerning departure from the MCS was seen when the intermediate nucleon propagators were treated properly.  Rather than attempt to fix these issues from within the wave function correction paradigm, we took a step back and thought more deeply about the hybrid formalism.

In \cref{chap:nrred}, we investigated the algebraic reduction of the full 4D loop-integral which occurs in a distorted-wave calculation.  We were able to clearly (even numerically) identify the traditional IA as an approximation that is only valid in the absence of initial-state interactions.  By making well-controlled approximations, we were then able to identify the true distorted-wave IA: a two-body operator which replaces the old one-body operator.  We are confident in our identification of a new LO expression, but the reduction from the relativistic formalism to the standard hybrid formalism involved a large number of approximations which remain to be classified in the EFT expansion.  There is a significant task ahead for a complete calculation.

Despite these difficulties in the IA study, the result remains that the $s$-wave IA amplitude is enhanced by a factor of approximately two.  This promises a significant numerical change to the calculation of $p$-wave pion production (recall that for $p$-wave pions, the IA is {\it not} washed out by a large RS amplitude).  A recalculation of both the $l=2$ Legendre coefficient and the analyzing power, discussed in \cref{sec:pwave}, is the natural place to begin.  We will then be prepared for another recalculation of the asymmetry.

A different avenue to pursue when applying the new formalism is the theory of the $pp\to pp\pi^0$ reaction, which, as discussed in \cref{sec:history} is quite sensitive to the IA.  Here one needs to re-evaluate the two-body $s$-wave amplitude of \cref{chap:nrred} between the initial ${}^3P_0$ and the final ${}^1S_0$ $pp$ scattering wave functions, while correctly accounting for the three-body phase space.  Because it has been known for a long time that the traditional one-body IA is too small to account for the total cross section by itself, it will be interesting to see how much bigger the new two-body operator is.

\printendnotes

%
\bibliography{References}

%

\appendix
\raggedbottom\sloppy
 
 \chapter{General Appendices}
 \label{chap:general}

\section{Fundamental Constants and Parameters}

Some of the constants used in this thesis are given in \cref{eq:parameters}.
\begin{equation}
\begin{aligned}
m_{\pi^0}&=134.98\text{ MeV} &\qquad g_A&=1.32\\
m_N&=938.92\text{ MeV} & f_\pi&=92.4\text{ MeV}\\
m_\Delta&=1232\text{ MeV} & h_A&=2.1g_A\\
E_b&=2.224\text{ MeV} & \hbar c&=197.327\text{ MeV}\cdot\text{fm}\label{eq:parameters}
\end{aligned}
\end{equation}
Note that different values for $g_A$ and $f_\pi$ were used in \cref{chap:csb}.  The LO values in \cref{eq:parameters} are more apt for \cref{chap:wfncor,chap:nrred} because they are LO calculations.

\section{Lagrange Densities\label{sec:lagrangian}}

This appendix summarizes the Lagrange densities developed and used in this thesis.  Most of these expressions were developed in Ch. \ref{chap:chpt} in terms of the pion matrix $U$.  For the present purposes it is sufficient to expand up to two-pion interactions.  Terms with more than two pions, as well as other terms not used in this thesis are omitted, and for this reason we write a $\ldots$ at the end of each expression.  In this section the superscript in ${\cal L}^{(n)}$, rather than indicating the number of derivatives or powers of $\sqrt{m_q}$, indicates the index of a Lagrange density,
\begin{align}
\nu=d+\frac{f}{2}-2\label{eq:index},
\end{align}
where $d$ is the sum of the number of derivatives and powers of $m_\pi$, and $f$ is the number of fermion fields.  This represents the standard power counting for nuclear physics.

The starting point is the general Lagrange densities in \cref{eq:chptl2,eq:hbchptl1,eq:hbchptl2,eq:deltakin,eq:lpiND1,eq:lpiND2}, to which we apply the expansions,
\begin{align}\bs
U&=1+\frac{i}{f_\pi}\tau_a\pi_a-\frac{1}{2f_\pi^2}(\delta_{ab}+i\epsilon_{abc}\tau_c)\pi_b\pi_c+\ldots
\\
\frac{i}{2}(u^\dagger\partial_\mu u+u\partial_\mu u^\dagger)&=-\frac{1}{4f_\pi^2}\epsilon_{abc}\tau_a\pi_b\partial_\mu\pi_c+\ldots
\\
i(u^\dagger\partial_\mu u-u\partial_\mu u^\dagger)&=-\frac{1}{f_\pi}\tau_a\partial_\mu\pi_a+\ldots.
\es\end{align}
We also rewrite the quark mass matrix $M=(m_u+m_d)/2+(m_u-m_d)/2\tau_3$, use $B(m_u+m_d)=m_\pi^2$, and employ $v=(1,{\bf 0})$ in the spin-3/2 projection operators,
\begin{align}\bs
P^{3/2}_{00}&\overset{v=(1,{\bf 0})}{=}0=P^{3/2}_{0i}=P^{3/2}_{i0}
\\
P^{3/2}_{ij}&\overset{v=(1,{\bf 0})}{=}-\delta_{ij}-\frac{1}{3}\gamma_i\gamma_j.
\es\end{align}
Furthermore, employing the heavy-baryon formalism, one is free to replace $\gamma_i\gamma_j\rightarrow -\sigma_i\sigma_j$.  Finally, it is common to express the Lagrange density in terms of an un-projected field $\Delta^i_a$ and leave the (iso)spin {\it transition} matrices $(T_{ab})S_{ij}$ explicit,
\begin{align}\bs
T_{ab}&=\delta_{ab}-\frac{1}{3}\tau_a\tau_b
\\
S_{ij}&=\delta_{ij}-\frac{1}{3}\sigma_i\sigma_j,
\es\end{align}
which satisfy $(TT^\dagger)_{ab}=T_{ab}$ and $(SS^\dagger)_{ij}=S_{ij}$.

The $\nu=0$ Lagrangian of HB$\chi$PT (with isovectors in $\mathbf{bold}$ font) with the Delta included as an explicit degree of freedom is
\begin{align}\bs
\mathcal{L}^{\left(0\right)}={}&\frac{1}{2}\left(\partial\bpi\right)^2-\frac{1}{2}m_\pi^2\bpi^2+N^\dagger i\partial_0N
\\
{}&-\frac{1}{4f_\pi^2}N^\dagger\left(\btau\cdot\left(\bpi\times\dot{\bpi}\right)\right)N+\frac{g_A}{2f_\pi}N^\dagger\left(\btau\cdot\vec{\sigma}\cdot\vec{\nabla}\bpi\right)N
\\
{}&+\Delta^\dagger\left(i\partial_0-\delta\right)\Delta+\frac{h_A}{2f_\pi}\left[N^\dagger\nabla_i\pi_aS_{ij}T_{ab}\Delta^j_b+H.c.\right]+...,
\label{eq:l0}
\es\end{align}
where $\btau$ and $\vec{\sigma}$ are the Pauli matrices acting on the isospin and spin of a single nucleon.

The $\nu=1$ Lagrangian includes propagator corrections, recoil terms, and the leading s-wave CSB operator
\begin{align}\bs
\mathcal{L}^{\left(1\right)}={}&\frac{1}{2m_N}N^\dagger\vec{\nabla}^2N+\frac{1}{2m_N}\left[\frac{1}{4f_\pi^2}iN^\dagger\btau\cdot\left(\bpi\times\vec{\nabla}\bpi\right)\cdot\vec{\nabla}N-\frac{g_A}{2f_\pi}iN^\dagger\btau\cdot\dot{\bpi}\vec{\sigma}\cdot\vec{\nabla}N+H.c.\right]
\\
{}&+\frac{1}{2m_N}\Delta^\dagger\vec{\nabla}^2\Delta-\frac{1}{2m_N}\frac{2h_A}{2f_\pi}\left[iN^\dagger\dot{\pi}_aS_{ij}T_{ab}\nabla^i\Delta^j_b+H.c.\right]
\\
{}&+\frac{\delta m_N}{2}N^\dagger\left[\tau_3-\frac{2}{4f_\pi^2}\pi_3\btau\cdot\bpi\right]N+\frac{\overline{\delta} m_N}{2}N^\dagger\left[\tau_3+\frac{2}{4f_\pi^2}\left(\pi_3\btau\cdot\bpi-\tau_3\bpi^2\right)\right]N+...\,.\label{eq:l1}
\es\end{align}
As previously mentioned, several terms have been omitted from this expression.  Two pion seagull terms with both derivatives on the pion fields do not get promoted in the MCS scheme.  This same property applies to terms with the $c_i$ low energy constants that appear at this order.  Terms with the $d_i$ low energy constants do not contribute to s-wave pion production.  Finally, the $NNNN$ contact terms $C_{S,T}$ do not contribute because we are using a potential with a repulsive core [$R_i(r)R_f(r)\rightarrow0$ as $r\rightarrow0$ for $l_i=1$, $l_f=0$].

Although there are a host of $\nu=2$ terms, we just list the CSB term relevant for this calculation
\begin{align}
\mathcal{L}^{\left(2\right)}=\frac{\beta_1}{2f_\pi}N^\dagger\vec{\sigma}\cdot\vec{\nabla}\pi_3N+...\,.
\label{eq:l2}
\end{align}

\section{\label{sec:nnwfns}Calculation of Nucleon-Nucleon Wave Functions}

\subsection{Problem definition}
In this section, we describe the numerical exercise of calculating scattering and bound state wave functions from phenomenological potentials.  The appendices of Ref. \citep{Machleidt:1987hj,Wong:1998ex} very helpful to this end.  The starting point is the time-independent Schr\"{o}dinger equation for a state with spin $S$, orbital angular momentum $L$, total angular momentum $J$, and magnetic quantum number $m_J$,
\begin{align}
\left(-\frac{1}{m_N}\bnabla^2+\hat{V}(\bsig_1,\bsig_2,\bfr)\right)\frac{u_{LJ}(r)}{p r}\mid{}^{2S+1}L_J,m_J(\hat{\bfr})\ra=\frac{p^2}{m_N}\frac{u_{LJ}(r)}{p r}\mid{}^{2S+1}L_J,m_J(\hat{\bfr})\ra,
\end{align}
where $m_N$ is the twice the reduced mass for the nucleon-nucleon system, $\bfr=(\bfr_1-\bfr_2)$ is the relative coordinate, and $p^2/m_N$ is the kinetic energy in terms of the relative momentum $\bfp=(\bfp_1-\bfp_2)/2$.  The spin-angle wave function is defined according to,
\begin{align}
\mid{}^{2S+1}L_J,m_J(\hat{\bfr})\ra=\sum_{m_S}\mid\la Sm_S,Lm_J-m_S\mid Jm_J\ra Y_{Lm_J-m_S}(\hat{\bfr})\mid Sm_s\ra,
\end{align}
where the $\la\cdot\cdot\mid\cdot\ra$ indicates a Clebsch-Gordan coefficient.  Phenomenological potentials are available in the form of data tables for each individual partial wave, i.e. $V(r)$ for $r=0.01,\ 0.02,\ ,\ldots,\ 15.0$ fm,
\begin{align}
V_{SLJ}(r)=\int d\hat{\bfr}\la{}^{2S+1}L_J,m_J(\hat{\bfr})\mid \hat{V}(\bsig_1,\bsig_2,\bfr)\mid{}^{2S+1}L_J,m_J(\hat{\bfr})\ra
\end{align}
The deuteron has the added complication of being a ${}^3S_1-{}^3D_1$ coupled channel $np$ state; the unbound version of this state is also relevant for CSB matrix elements.  In this case the makers of the potential typically provide the off-diagonal term,
\begin{align}\bs
V_{\epsilon}(r)&=\int d\hat{\bfr}\la{}^3D_1,m(\hat{\bfr})\mid\hat{V}(\bsig_1,\bsig_2,\bfr)\mid{}^3S_1,m(\hat{\bfr})\ra
\\
&=\int d\hat{\bfr}\la{}^3S_1,m(\hat{\bfr})\mid\hat{V}(\bsig_1,\bsig_2,\bfr)\mid{}^3D_1,m(\hat{\bfr})\ra.
\es\end{align}

\subsection{Solution for uncoupled channels}
First we will describe the method for finding the wave functions for the channels in which the potential is diagonal.  In this case, the Schr\"{o}dinger equation reduces to,
\begin{align}
\left[-\frac{1}{m_N}\left(\frac{1}{r}\frac{\partial^2}{\partial r^2}r-\frac{l(l+1)}{r^2}\right)+V(r)-\frac{p^2}{m_N}\right]\frac{u_l(r)}{pr}=0.\label{eq:fullode}
\end{align}
Since we will be integrating this equation out numerically, what we need are the solutions to this equation in both the $r\to r_\text{min}$ limit where the centrifugal barrier dominates,
\begin{align}
\left(\frac{\partial^2}{\partial r^2}-\frac{l(l+1)}{r^2}\right)\frac{u_l(r)}{p}=0\quad\Rightarrow\quad\frac{u_l(r)}{pr}=Ar^l+\frac{B}{r^{l+1}},\label{eq:smallbc}
\end{align}
and the $r\to r_\text{max}$ limit where the potential vanishes,
\begin{align}
\left(\frac{\partial^2}{\partial r^2}-\frac{l(l+1)}{r^2}+p^2\right)\frac{u_l(r)}{p}=0\quad\Rightarrow\quad\frac{u_l(r)}{pr}=Cj_l(pr)+Dn_l(pr),\label{eq:largebc}
\end{align}
where $j_l(n_l)$ is the spherical Bessel (Neumann) function.  Note that we can discard the irregular solutions (the $B$ and $D$ terms).  Also note that the situation is different for $l=0$ which has no centrifugal barrier; here, we find the solution near the origin to be $\propto\sinh(\xi r)$ with $\xi^2=m_NV(r_\text{min})-p^2$.  The effect of the potential in distorting these solutions (at least for elastic scattering) is to introduce a phase shift at large $r$,
\begin{align}
j_l(pr)\approx \frac{\sin(pr-\frac{l\pi}{2})}{pr}\quad\Rightarrow\quad\frac{u_l(r)}{pr}\overset{r\to\infty}{\to}C\frac{\sin(pr-\frac{l\pi}{2}+\delta_l)}{pr}.
\end{align}
A few lines of algebraic manipulation then yields,
\begin{align}
\frac{u_l(r)}{pr}\overset{r\to\infty}{\to}C(\cos(\delta_l)j_l(pr)-\sin(\delta_l)n_l(pr)).
\end{align}

At this point, we are able to perform the numerical integration via the following algorithm:
\begin{itemize}
\item{For $l\neq0$: Taking $u_l(r_\text{min})=u^*$ and $u'_l(r_\text{min})=\frac{(l+1)u^*}{r_\text{min}}$ as the boundary conditions, integrate the full differential equation of \cref{eq:fullode} from $r_\text{min}$ to $r_\text{max}$.}
\item{For $l=0$: Taking $u_0(r_\text{min})=u^*$ and $u'_0(r_\text{min})=u^*\xi\coth(\xi r_\text{min})$ as the boundary conditions, integrate the full differential equation of \cref{eq:fullode} from $r_\text{min}$ to $r_\text{max}$.}
\item{Solve for the overall normalization and phase shift by matching onto the asymptotic form given by \cref{eq:largebc} and its derivative.}
\end{itemize}

\subsection{Solution for coupled channel}
For the coupled channels we find,
\begin{align}
\left[-\frac{1}{m_N}\frac{1}{r}\frac{\partial^2}{\partial r^2}r+V_0(r)-\frac{p^2}{m_N}\right]\frac{u(r)}{pr}&=-V_\epsilon(r)\frac{w(r)}{pr}\label{eq:seqn}
\\
\left[-\frac{1}{m_N}\left(\frac{1}{r}\frac{\partial^2}{\partial r^2}r-\frac{6}{r^2}\right)+V_2(r)-\frac{p^2}{m_N}\right]\frac{w(r)}{pr}&=-V_\epsilon(r)\frac{u(r)}{pr},\label{eq:deqn}
\end{align}
where $u(w)$ is the $l=0(2)$ solution.  \Cref{eq:seqn,eq:deqn} admit two solutions $\{u_1,w_1\},\ \{u_2,w_2\}$.  One may obtain independent solutions by solving the equations once with the boundary condition $u(r_\text{min})=0$ and then again with $w(r_\text{min})=0$.  These two solutions are combined in two different linear combinations to find the two true physical solutions $\{u_\alpha,w_\alpha\},\ \{u_\beta,w_\beta\}$,
\begin{equation}
\begin{aligned}
u_\alpha&=Au_1+Bu_2 \\ w_\alpha&=Aw_1+Bw_2
\end{aligned} \qquad\qquad
\begin{aligned}
u_\beta&=Cu_1+Du_2 \\ w_\beta&=Cw_1+Dw_2.
\end{aligned}\label{eq:coupledsol}
\end{equation}

The free solution for the $\alpha$ channel can be expressed as a sum of incoming and outgoing spherical waves,
\begin{align}
\frac{u_\alpha(r)}{pr}&\to ah_0^{(1)}(pr)+bh_0^{(2)}(pr)
\\
\frac{w_\alpha(r)}{pr}&\to ch_2^{(1)}(pr)+dh_2^{(2)}(pr),
\end{align}
where, because the normalization will be fixed later, we can set $b=1$.  The incoming and outgoing solutions are related by the scattering matrix $S$ according to,
\begin{align}
\begin{pmatrix}a\\c\end{pmatrix}=S\begin{pmatrix}b\\d\end{pmatrix}.
\end{align}
In the ``Stapp" convention \citep{Stapp:1956mz} which yields the ``nuclear bar" phase shifts, the scattering matrix is written,
\begin{align}
S=\begin{pmatrix}
e^{2i\delta_\alpha}\cos(2\epsilon) & ie^{i(\delta_\alpha+\delta_\beta)}\sin(2\epsilon) \\
ie^{i(\delta_\alpha+\delta_\beta)}\sin(2\epsilon) & e^{2i\delta_\beta}\cos(2\epsilon)
\end{pmatrix},
\end{align}
where $\delta_\alpha(\delta_\beta)$ is the phase shift of the $l=0(2)$ solution and $\epsilon$ is the mixing angle.\footnote{The older ``eigen" phase shifts \citep{Blatt:1952zza} use a scattering matrix,
\begin{align*}
S=\begin{pmatrix}
e^{2i\delta_\alpha}\cos^2\epsilon+e^{2i\delta_\beta}\sin^2\epsilon & \left(e^{2i\delta_\alpha}-e^{2i\delta_\beta}\right)\sin\epsilon\cos\epsilon \\
\left(e^{2i\delta_\alpha}-e^{2i\delta_\beta}\right)\sin\epsilon\cos\epsilon & e^{2i\delta_\alpha}\sin^2\epsilon+e^{2i\delta_\beta}\cos^2\epsilon
\end{pmatrix}.
\end{align*}}

The final step is to require that (by definition) in the $\alpha$ channel, the $u$ solution must have an outgoing wave with amplitude $e^{2i\delta_\alpha}$; this means that we set $a=e^{2i\delta_\alpha}$ and solve for $c$ and $d$,
\begin{alignat}{3}
e^{2i\delta_\alpha}&=e^{2i\delta_\alpha}\cos(2\epsilon)+ie^{i(\delta_\alpha+\delta_\beta)}\sin(2\epsilon)d\quad &\Rightarrow\quad d&=-ie^{i(\delta_\alpha-\delta_\beta)}\tan\epsilon
\\
c&=ie^{i(\delta_\alpha+\delta_\beta)}\sin(2\epsilon)+e^{2i\delta_\beta}\cos(2\epsilon)d\quad &\Rightarrow\quad c&=ie^{i(\delta_\alpha+\delta_\beta)}\tan\epsilon.
\end{alignat}
A few more lines of algebra yield the final asymptotic forms for the $\alpha$ channel (up to an overall normalization),
\begin{align}\bs
\frac{u_\alpha(r)}{pr}&\to\frac{\sin(pr+\delta_\alpha)}{pr}
\\
\frac{w_\alpha(r)}{pr}&\to\tan\epsilon\frac{\cos(pr+\pi+\delta_\beta)}{pr}.\label{eq:coupledasy}
\es\end{align}
To obtain the $\beta$ channel solutions, we require that the outgoing wave part of $u_\beta$ have amplitude $e^{2i\delta_\beta}$.  The algebra then proceeds analogous to the $\alpha$ channel we just described.

Using \cref{eq:coupledsol,eq:coupledasy}, their derivatives, and the corresponding $\beta$ channel equations, one is able to solve for the phase shifts normalizations necessary to specify the coupled-channel wave functions.  We refrain from displaying the rest of the formulas here and mention one final channel, the deuteron.

The deuteron is the bound state of the scattering channel we just described.  Its wave functions' asymptotic forms do not involve any phase shifts and can be simply expressed,
\begin{align}\bs
u(r)&\to Ae^{-\gamma r}
\\
w(r)&\to\eta A\left(1+\frac{3}{\gamma r}+\frac{3}{\gamma^2r^2}\right)e^{-\gamma r},
\es\end{align}
where $\gamma=\sqrt{m_NE_b}$ is the binding momentum and $\eta$ is the asymptotic $d$ to $s$ ratio.

\section{\label{sec:redmxels}Spin-Angle Reduced Matrix Elements}

\subsection{\label{sec:WEthm}Wigner-Eckart Theorem}
When performing the calculations in this thesis, we have taken care to consistently apply partial wave expansions.  As part of such a calculation, on encounters matrix elements of the form,
\begin{align}
\la(S'L')J'm_J\mid T^k_q\mid(SL)Jm_J\ra,
\end{align}
where $T^k_q$ is a tensor operator of rank $k$ and projection $m$ which is created by coupling one of the spin vectors, $\bfS=(\bsig_1+\bsig_2)/2$ or $\bfE=(\bsig_1-\bsig_2)/2$, to one of the spherical harmonics $Y_{lm}\orh$,
\begin{align}
T^k_q=\sum_{m_s,m_l}\la 1m_s,lm_l\mid kq\ra{\cal S}_{m_s}Y_{lm_l}\orh,\label{eq:Tcoupled}
\end{align}
where ${\cal S}$ represents either $\bfS$ or $\bfE$ and the angular momentum addition rules must be satisfied: $k\geq|l-1|$, $k\leq l+1$, and $m_s+m_l=q$.  One must take care to use spherical tensors; a rank-one spherical tensor is related to its Cartesian components by
\begin{align}
V_{\pm 1}=\mp\frac{1}{\sqrt{2}}(V_x\pm iV_y),\qquad V_0=V_z.
\end{align}
This connection allows one to relate the position vector to the spherical harmonic $Y_1$,
\begin{align}
r_\mu=\sqrt{\frac{4\pi}{3}}rY_{1\mu}\orh.
\end{align}

The Wigner-Eckart Theorem states that a general matrix element $\la j'm'\mid T^k_q\mid jm\ra$ can be expressed as the product of a ``reduced matrix element" $\la j'\mid\mid T^k\mid\mid j\ra$, which does not depend on the projections, and a Clebsch-Gordan coefficient,
\begin{align}
\la j'm'\mid T^k_q\mid jm\ra=\frac{\la jm,kq\mid j'm'\ra}{\sqrt{2j'+1}}\la j'\mid\mid T^k\mid\mid j\ra.
\end{align}
In this way, one only needs to calculate the reduced matrix element once, simplifying calculations (such as the present one) that involve sums over the projections.  The reduced matrix elements can be obtained by evaluating one particular matrix element explicitly.  The reduced matrix elements we will be using are,
\begin{align}\bs
\la1\mid\mid\bfS\mid\mid1\ra&=\sqrt{6}\\
\la1\mid\mid\bfE\mid\mid0\ra&=\sqrt{3}\\
\la0\mid\mid\bfE\mid\mid1\ra&=-\sqrt{3}\\
\la l'\mid\mid Y_L\mid\mid l\ra&=\frac{\hat{L}\hat{l}}{\sqrt{4\pi}}\la l0,L0\mid l'0\ra,
\es\end{align}
where $\hat{J}=\sqrt{2J+1}$.

When the tensor is formed by coupling two individual tensors which act on different spaces as in \cref{eq:Tcoupled}, one uses a general formula which is worked out in many textbooks on angular momentum (for example, see Ref. \citep{Wong:1998ex}),
\begin{align}
\la S'L'J'\mid\mid T_k\mid\mid SLJ\ra=\hat{J'}\hat{k}\hat{J}\left\{\begin{matrix}S'&L'&J'\\S&L&J\\{\cal S}&l&k\end{matrix}\right\}\la S'\mid\mid{\cal S}\mid\mid S\ra\,\la L'\mid\mid Y_l\mid\mid L\ra,\label{eq:9j}
\end{align}
where the 3x3 object enclosed in the braces is a 9j-symbol (a sum of products of Clebsch-Gordan coefficients).  This expression becomes even simpler when the $T^k_q$ is a scalar product\footnote{In terms of spherical components a scalar product is given by ${\cal S}\cdot V=\sum_m(-1)^m{\cal S}_mV_{-m}$.} ${\cal S}\cdot\hat{\bfr}$,
\begin{align}
\la S'L'J'\mid\mid{\cal S}\cdot\hat{\bfr}\mid\mid SLJ\ra=\delta_{J'J}(-1)^{S+L'+J}\hat{J}\left\{\begin{matrix}S'&L'&J'\\L&S&1\end{matrix}\right\}\la S'\mid\mid{\cal S}\mid\mid S\ra\,\la L'\mid\mid\hat{\bfr}\mid\mid L\ra,\label{eq:6j}
\end{align}
where the 3x2 object enclosed in the braces is a 6j-symbol (another sum of products of Clebsch-Gordan coefficients).

\subsection{\label{sec:examplemxel}Evaluation of relevant reduced matrix elements}
To illustrate the methods described in \cref{sec:WEthm}, this section will tabulate the reduced matrix elements which are necessary to calculate the pion rescattering diagram amplitudes.  This diagram is treated as an example calculation in \cref{sec:method}.  First consider the operator $\bfS\cdot\hat{\bfr}$ which contributes to $s$-wave pion production (${}^3P_1\to{}^3S_1,\,{}^3D_1$),
\begin{align}\bs
\la(10)1\mid\mid\bfS\cdot\hat{\bfr}\mid\mid(11)1\ra&=\sqrt{2}\\
\la(12)1\mid\mid\bfS\cdot\hat{\bfr}\mid\mid(11)1\ra&=1.
\es\end{align}

Next consider the operator $\bfE\cdot\hat{\bfr}\,Y_{1}\orh$ which contributes to $p$-wave pion production (${}^1S_0,\,{}^1D_2\to{}^3S_1,\,{}^3D_1$).  One way to evaluate this matrix element is to insert a complete set of states to the right of the $\bfE\cdot\hat{\bfr}$.  This set will collapse to one term, ${}^1P_1$,
\begin{align}\bs
\la(1L')1\mid\mid\bfE\cdot\hat{\bfr}\,Y_{1}\orh\mid\mid(0J)J\ra&=
\sum_{SLJ}\la(1L')1\mid\mid\bfE\cdot\hat{\bfr}\mid\mid(SL)J\ra\,\la(SL)J\mid\mid Y_{1}\orh\mid\mid(0J)J\ra\\
&=\la(1L')1\mid\mid\bfE\cdot\hat{\bfr}\mid\mid(01)1\ra\,\la(01)1\mid\mid Y_{1}\orh\mid\mid(0J)J\ra.
\es\end{align}
Now one simply needs to evaluate the individual reduced matrix elements\footnote{The uncoupled tensor operator $Y_1$ between coupled states can still be evaluated using \cref{eq:9j} by recognizing that the spin-space operator is simply the identity operator.},
\begin{equation}
\begin{aligned}
\la(10)1\mid\mid\bfE\cdot\hat{\bfr}\mid\mid(01)1\ra&=1 &\quad \la(01)1\mid\mid Y_1\mid\mid(00)0\ra&=\sqrt{\frac{3}{4\pi}}
\\
\la(12)1\mid\mid\bfE\cdot\hat{\bfr}\mid\mid(01)1\ra&=-\sqrt{2} &\quad \la(01)1\mid\mid Y_1\mid\mid(02)2\ra&=-\sqrt{\frac{6}{4\pi}}.
\end{aligned}
\end{equation}
 \chapter[\texorpdfstring{\Cref{chap:csb}}{Ch. 4} Details]{Ch. \lowercase{\ref{chap:csb}} Details}
 \label{chap:csbdetails}

\section{Defining Reduced Matrix Elements\label{sec:reduced}}

One can show that for s-wave pions, the production operator is always either a scalar or a rank-two tensor while for p-wave pions, it is always a rank-one tensor.  This guides the following definition of the reduced matrix elements.  Note that the Clebsch-Gordan coefficients which will be summed over as well as the Spherical Harmonics describing the angular distribution of the differential cross section are ``pulled out."  First we define the strong reduced matrix elements,
\begin{align}\bs
\left\la f\mid\hat{\mathcal{M}}^{str}_{l_\pi=0}\mid i\right\ra={}&\left(\frac{1}{\sqrt{3}}A_0+\frac{\left\la1\,m_f,2\,0\mid1\,m_f\right\ra}{\sqrt{3}}A_2\right)e^{i\delta_1}\text{cg}_1
\\
{}&\times\left\la1\,m_s,1\,m_f-m_s\mid1\,m_f\right\ra Y^{1\,*}_{m_f-m_s}(\hat{p})\es\label{eq:redstrs}
\\
\left\la f\mid\hat{\mathcal{M}}^{str}_{l_\pi=1}\mid i\right\ra={}&\frac{1}{\sqrt{3}}Be^{i\delta_0}\text{cg}_0\delta_{m_f,0}Y^0_0(\hat{p})+\frac{\left\la2\,m_f,1\,0\mid1\,m_f\right\ra}{\sqrt{3}}Ce^{i\delta_2}\text{cg}_0Y^{2\,*}_{m_f}(\hat{p}),\label{eq:redstrp}
\end{align}
where $m_s=m_1+m_2$, $\text{cg}_0=\cgz$, and $\text{cg}_1=\left\la1/2\,m_1,1/2\,m_2\mid1\,m_s\right\ra$.  $A_0$ and $A_2$ are the s-wave reduced matrix elements, and $B$ and $C$ are the p-wave reduced matrix elements.  To clarify the notation consider $A_2$, for example.
\begin{align}
A_2=\int dr\,r^2\left(\frac{u_d(r)}{r}\la(10)1\mid\mid+\frac{w_d(r)}{r}\la(12)1\mid\mid\right)\hat{\mathcal{M}}^{str}_{l_\pi=0,J=2}\left(4\pi\,i\frac{u_{1,1}(r)}{pr}\mid\mid(11)1\ra\right),
\end{align}
where the subscript $J=2$ on the $\hat{\mathcal{M}}$ indicates that we are using the portion of the operator that is a rank-two tensor in the space of total angular momentum.  Note also that we have used the following general definition of a reduced matrix element,
\begin{align}
\la(S'L')J'm_J'\mid T^k_q\mid(SL)Jm_J\ra\equiv\frac{\la Jm_J,kq\mid J'm_J'\ra}{\sqrt{2J'+1}}\la(S'L')J'\mid\mid T^k\mid\mid (SL)J\ra.
\end{align}
Similarly for the CSB reduced matrix elements,
\begin{align}
\left\la f\mid\hat{\mathcal{M}}^{csb}_{l_\pi=0}\mid i\right\ra={}&\left(\frac{1}{\sqrt{3}}\overline{A}_0+\frac{\left\la1\,m_f,2\,0\mid1\,m_f\right\ra}{\sqrt{3}}\overline{A}_2\right)e^{i\overline{\delta}_1}\text{cg}_0Y^{1\,*}_{m_f}(\hat{p})\label{eq:redcsbs}
\\ \bs
\left\la f\mid\hat{\mathcal{M}}^{csb}_{l_\pi=1}\mid i\right\ra={}&\frac{\left\la1\,m_f,1\,0\mid1\,m_f\right\ra}{\sqrt{3}}\left(\overline{B}_\alpha e^{i\overline{\delta}_\alpha}+\overline{B}_\beta e^{i\overline{\delta}_\beta}\right)\text{cg}_1\delta_{m_f,m_s}Y^0_0(\hat{p})
\\
{}&+\frac{\left\la1\,m_f,1\,0\mid1\,m_f\right\ra}{\sqrt{3}}\left(\overline{C}_\alpha e^{i\overline{\delta}_\alpha}+\overline{C}_\beta e^{i\overline{\delta}_\beta}\right)\text{cg}_1
\\
{}&\times\left\la1\,m_s,2\,m_f-m_s\mid1\,m_f\right\ra Y^{2\,*}_{m_f-m_s}(\hat{p})
\\
{}&+\frac{\left\la2\,m_f,1\,0\mid1\,m_f\right\ra}{\sqrt{3}}\overline{D}e^{i\overline{\delta}_2}\text{cg}_1\left\la1\,m_s,2\,m_f-m_s\mid2\,m_f\right\ra Y^{2\,*}_{m_f-m_s}(\hat{p}),\label{eq:redcsbp}
\es\end{align}
where $\overline{A}_0$ and $\overline{A}_2$ are the s-wave reduced matrix elements and $\overline{B}$, $\overline{C}$, and $\overline{D}$ are the p-wave reduced matrix elements.  Also note that the strong phase shifts have been denoted $\delta_L$ for each of the three initial channels, the CSB phase shifts are denoted $\overline{\delta}_L$ for the $^1P_1$ and $^3D_2$ channels, and the coupled channel phase shifts are $\overline{\delta}_\alpha$ and $\overline{\delta}_\beta$.  Since the $^3S_1$ and $^3D_1$ channels are coupled, $\overline{B}$ and $\overline{C}$ are split into $\alpha$ and $\beta$ parts which have different phase shifts in the presence of initial state interactions.

Finally, for comparison purposes we include a translation between our reduced matrix elements and those of Ref. \citep{Baru:2009fm},
\begin{align}\bs
C_0&=-\frac{1}{\sqrt{4\pi}}A_0e^{i\delta_0}
\\
C_1&=-i\frac{1}{\sqrt{6\pi}}Be^{i\delta_1}
\\
C_2&=i\sqrt{\frac{3}{4\pi}}Ce^{i\delta_2},
\es\end{align}
where the ``C's" are $pp\to d\pi^+$ amplitudes and isospin symmetry has been used to determine the translations.

\section{Diagram Technique\label{sec:method}}

\begin{figure}
\centering
\includegraphics[height=2in]{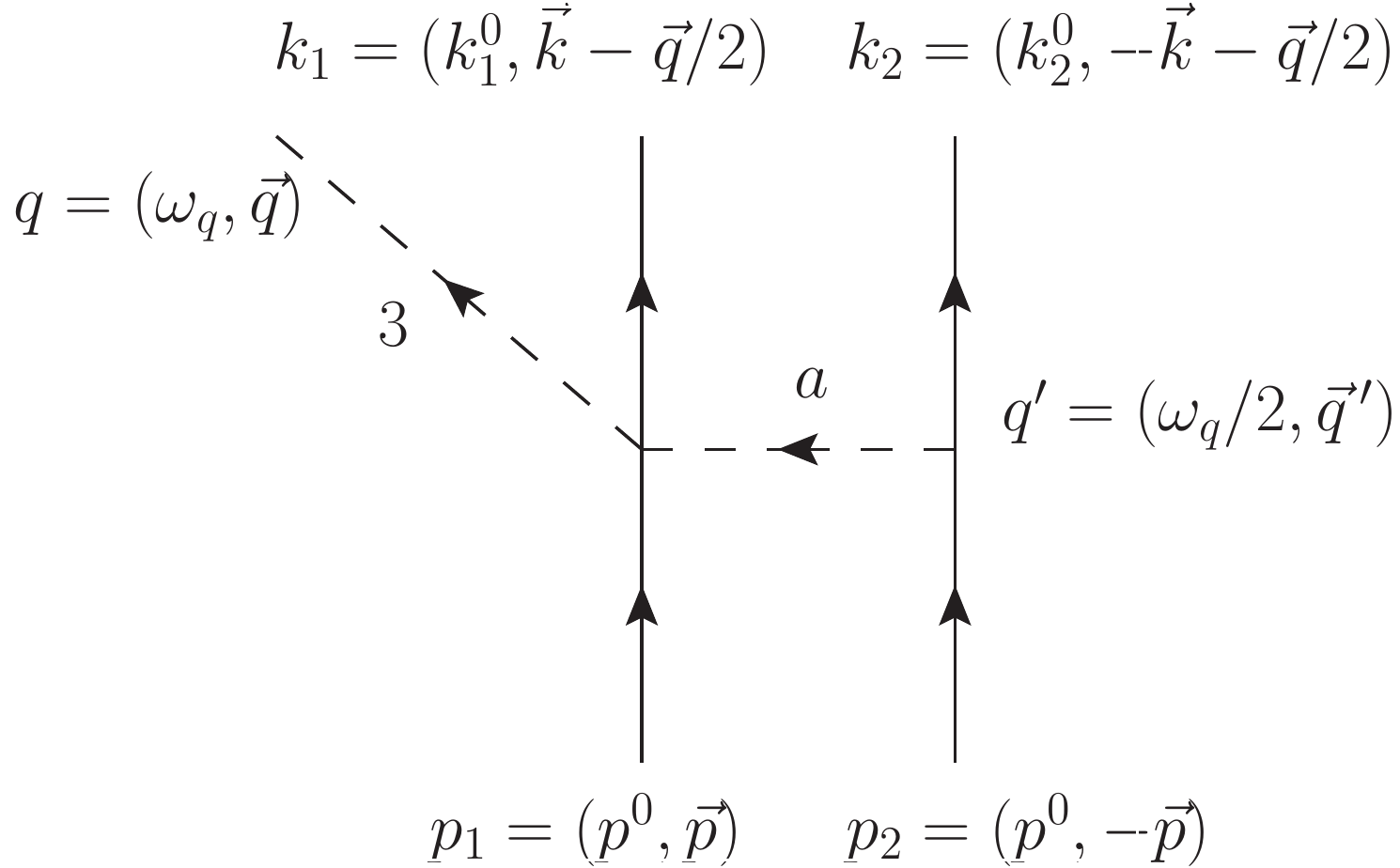}
\caption[Strong rescattering diagram]{Strong rescattering diagram.  Solid lines represent nucleons, dashed lines represent pions, and the pions' isospin z-components are 3 and a.\label{fig:strongrs}}
\end{figure}

To establish the diagram technique consider Fig. \ref{fig:strongrs} in light of Eq. (\ref{eq:mft}).
We write down the amplitude using momentum space Feynman rules, Fourier transform, and then convolve with the initial and final state wave functions.  The left line is taken to be ``nucleon 1" and we make the approximation that the exchanged pion carries half of the produced pion's energy, $q'=(\omega_q/2,\bfq')$.  Momentum conservation gives $\bfq'=\bfk-\bfp+\bfq/2$.  According to Eq. (\ref{eq:l0}) the WT vertex contributes $1/(4f_\pi^2)\,\tau_{1,b}\,\epsilon_{a3b}\,(\omega_q/2+\omega_q)$ while the $\pi NN$ vertex contributes $g_A/(2f_\pi)\,\tau_{2,a}\,\bsig_2\cdot(-\bfq')$.  The momentum space propagator is $-i/(\bfq'^2+\mu^2)$ where $\mu^2=m_\pi^2-\omega_q^2/4$ is the effective mass of the rescattered pion.  Next, as discussed in Sec. \ref{sec:hybrid}, we Fourier transform with respect to $\bfl=\bfk-\bfp$,
\begin{align}\bs
\int \frac{d^3l}{(2\pi)^3}\,e^{i\bfl\cdot\bfr}\frac{\bsig_2\cdot\bfq'}{\bfq'^2+\mu^2}&=e^{-i\bfq\cdot\bfr/2}\int\frac{d^3q'}{(2\pi)^3}\,e^{i\bfq'\cdot\bfr}\frac{\bsig_2\cdot\bfq'}{\bfq'^2+\mu^2}
\\
&=e^{-i\bfq\cdot\bfr/2}\,\bsig_2\cdot(-i\bnabla)\,\frac{e^{-\mu r}}{4\pi r}
\\
&=\frac{i\mu}{4\pi}e^{-i\bfq\cdot\bfr/2}\,h(r)\,\bsig_2\cdot\hat{r},\label{eq:rsft}
\es\end{align}
where $h(r)\equiv (1+1/\mu r)e^{-\mu r}/r$.

The deuteron has isospin 0 and the $np$ wave fuction includes a $T=1$, $T_z=0$ isospinor $\left|1,0\right\ra$, and thus
\begin{align}
\la0,0\mid i\epsilon_{a3b}\tau_{1,b}\tau_{2,a}\mid1,0\ra=-2.
\end{align}
At this point, we have (defining $\hat{\mathcal{M}}'=\hat{\mathcal{M}}/\sqrt{2E_1\,2E_2\,2E_d}\,$)
\begin{align}
\left\la0,0\mid\hat{\mathcal{M}'_L}\mid1,0\right\ra=-i\frac{g_A}{2f_\pi}\,\frac{3\omega_q/2}{8\pi f_\pi^2}\,\mu\,h(r)\,e^{-i\bfq\cdot\bfr/2}\,\bsig_2\cdot\hat{r}.
\label{eq:mstrrsl}
\end{align}
To calculate the diagram with the WT vertex on nucleon 2 we consider how each part of the left side of Eq. (\ref{eq:mstrrsl}) transforms under $1\leftrightarrow2$.  Since the strong part of the Lagrangian is invariant under isospin, $\hat{\mathcal{M}}$ is invariant.  The initial isospin ket, $\left|1,0\right\ra$ is invariant as well, but $\left|0,0\right\ra\rightarrow-\left|0,0\right\ra$.  Also note that $\bfr\rightarrow-\bfr$.  Thus,
\begin{align}
\left\la0,0\mid\hat{\mathcal{M}'_R}\mid1,0\right\ra=-i\frac{g_A}{2f_\pi}\,\frac{3\omega_q/2}{8\pi f_\pi^2}\,\mu\,h(r)\,e^{i\bfq\cdot\bfr/2}\,\bsig_1\cdot\hat{r}.
\end{align}

Defining $\bfS\equiv(\bsig_1+\bsig_2)/2$, $\bfE\equiv(\bsig_1-\bsig_2)/2$, and
\begin{align}\bs
\mathcal{E}&\equiv\text{exp}(i\bfq\cdot\bfr/2)+\text{exp}(-i\bfq\cdot\bfr/2)
\\
\mathcal{O}&\equiv\text{exp}(i\bfq\cdot\bfr/2)-\text{exp}(-i\bfq\cdot\bfr/2),\label{eq:eando}
\es\end{align}
we have the complete rescattering operator,
\begin{align}
\left\la0,0\mid\hat{\mathcal{M}}'_{RS}\mid1,0\right\ra=-i\,\gamma_{RS}\,h(r)\left(\mathcal{E}\,\bfS\cdot\hat{r}+\mathcal{O}\,\bfE\cdot\hat{r}\right),\label{eq:mstrrs}
\end{align}
where
\begin{align}
\gamma_{RS}\equiv\frac{g_A}{2f_\pi}\frac{3\omega_q/2}{8\pi f_\pi^2}\,\mu.
\end{align}

To proceed, we preform a partial wave expansion on $\mathcal{E}$ and $\mathcal{O}$ and just keep the leading term.  Note that we use the coordinate system defined by $\hat{q}=\hat{z}$.  Then we calculate the spin-angle matrix elements of the rank zero $\mathcal{E}\,\bfS\cdot\hat{r}$ and the rank one $\mathcal{O}\,\bfE\cdot\hat{r}$ operators (see \cref{sec:examplemxel}).  The final expression for $\langle f\mid\hat{\mathcal{M}}\mid i\rangle$ simplifies to a radial integral which is computed numerically.  Note that the first term in Eq. (\ref{eq:mstrrs}) corresponds to s-wave pions because $\mathcal{E}$ carries $L=0$,
\begin{align}\bs
{\cal E}&=2\times4\pi\,j_0\left(\frac{qr}{2}\right)Y_{00}(\hat{\bfr})Y^*_{00}(\hat{\bfq})+\ldots\\
&=2\,j_0\left(\frac{qr}{2}\right),
\es\end{align}
while $\hat{r}$ carries $L=1$,
\begin{align}\bs
{\cal O}={}&2\times4\pi\,i\,j_1\left(\frac{qr}{2}\right)\sum_{m}Y_{1m}(\hat{\bfr})Y^*_{1m}(\hat{\bfq})+\ldots\\
\overset{\hat{\bfq}=\hat{z}}{=}{}&2\sqrt{12\pi}\,i\,j_1\left(\frac{qr}{2}\right)Y_{10}(\hat{\bfr}),
\es\end{align}
and thus the operator will change the parity.  Likewise, the second term corresponds to p-wave pions.  In terms of the reduced matrix elements of Eqs. (\ref{eq:redstrs}) and (\ref{eq:redstrp}), we have
\begin{align}\bs
A_0^{RS}&=\sqrt{2E_1\,2E_2\,2E_d}\,8\pi\,\gamma_{RS}\,\sqrt{2}\,K_1
\\
A_2^{RS}&=0
\\
B^{RS}&=\sqrt{2E_1\,2E_2\,2E_d}\,8\pi\,\gamma_{RS}\,\sqrt{3}\,K_0
\\
C^{RS}&=\sqrt{2E_1\,2E_2\,2E_d}\,8\pi\,\gamma_{RS}\,\sqrt{6}\,K_2,\label{eq:rsresults}
\es\end{align}
where the integrals are defined
\begin{align}\bs
K_1&\equiv\int dr\,r^2\left(\frac{u_d(r)}{r}+\frac{w_d(r)}{\sqrt{2}r}\right)j_0\left(\frac{qr}{2}\right)\,h(r)\frac{u_{1,1}(r)}{pr}
\\
K_0&\equiv\int dr\,r^2\left(\frac{u_d(r)}{r}-2\frac{w_d(r)}{\sqrt{2}r}\right)j_1\left(\frac{qr}{2}\right)\,h(r)\frac{u_{0,0}(r)}{pr}
\\
K_2&\equiv\int dr\,r^2\left(\frac{u_d(r)}{r}-2\frac{w_d(r)}{\sqrt{2}r}\right)j_1\left(\frac{qr}{2}\right)\,h(r)\frac{u_{2,2}(r)}{pr}.\label{eq:rsintegrals}
\es\end{align}

\section{\label{sec:observables}Observables}

One experimental observable is the analyzing power, $A_y$, defined in Eq. (\ref{eq:aydef}).  In the strong sector, we find (neglecting $A_2$ which is numerically small)
\begin{align}
A_y=\frac{\sqrt{3}\cos(\phi)\sin(\theta)A_0\left(\sqrt{2}B\sin(\delta_0-\delta_1)+C\sin(\delta_2-\delta_1)\right)}{3A_0^2+B^2+C^2+\left(C^2-2\sqrt{2}BC\cos(\delta_2-\delta_0)\right)P_2(\cos\theta)},\label{eq:ay}
\end{align}
where the angular dependence is that of the nucleon relative momentum, $\bfp$, with respect to the pion momentum, $\hat{q}=\hat{z}$.  To compare with experimental results, which use $\hat{p}=\hat{z}$ and $\phi_\pi=0$, we need to set $\phi_N=\pi$ and $\theta_N=\theta_\pi$.

To calculate the differential cross section as well as the asymmetry, we need to square the sum of all the matrix elements, sum over $m_f$ and average over $m_1$ and $m_2$.  First we define
\begin{align}
\frac{1}{4}\sum\left|\left\la f\mid\hat{\mathcal{M}}^{tot}\mid i\right\ra\right|^2=M_0+M_1P_1(\cos\theta)+M_2P_2(\cos\theta)+M_3P_3(\cos\theta),
\end{align}
so that
\begin{align}
\sigma&=\frac{|\bfq|}{64\,\pi^2\,s\,|\bfp|}4\pi M_0\label{eq:sigmadef}
\\
A_{fb}&=\frac{M_1-\frac{1}{4}M_3}{2M_0}.\label{eq:afbdef}
\end{align}
The results for the required quantities are
\begin{align}\bs
M_0={}&\frac{1}{48\,\pi}\left[3(A_0^2 + \overline{A}_0^2) + \frac{3}{5} (A_2^2 + \overline{A}_2^2) + \left(B^2 + \overline{B}_\alpha^2 + \overline{B}_\beta^2 + C^2 + \overline{C}_\alpha^2+\overline{C}_\beta^2 + \overline{D}^2\right)\right.
\\
{}&+\left.2\left(\overline{B}_\alpha\overline{B}_\beta + \overline{C}_\alpha\overline{C}_\beta\right)\cos\left(\overline{\delta}_\alpha-\overline{\delta}_\beta\right)\right]\label{eq:m0}
\es \\ \bs
M_1={}&\frac{\sqrt{3}}{24\,\pi}\left[B\left(\overline{A}_0-2\sqrt{\frac{1}{10}}\overline{A}_2\right)\cos(\overline{\delta}_1-\delta_0)-\sqrt{2}C\left(\overline{A}_0-\frac{1}{5\sqrt{10}}\overline{A}_2\right)\cos(\overline{\delta}_1-\delta_2)\right.
\\
{}&+\left(A_0+\frac{1}{\sqrt{10}}A_2\right)\left(\left(\overline{B}_\alpha+\frac{1}{\sqrt{2}}\overline{C}_\alpha\right)\cos(\overline{\delta}_\alpha-\delta_1)+(\alpha\to\beta)\right)
\\
{}&-\left.\sqrt{\frac{3}{2}}\left(A_0-\frac{1}{5\sqrt{10}}A_2\right)\overline{D}\cos(\overline{\delta}_2-\delta_1)\right]\label{eq:m1}
\es \\ \bs
M_2={}&\frac{1}{8\sqrt{10}\,\pi}\left[(A_0 A_2-2\overline{A}_0\overline{A}_2)-\frac{1}{2\sqrt{10}}(A_2^2-2\overline{A}_2^2)\right.
\\
{}&+\frac{5}{3\sqrt{10}}(C^2-\frac{1}{2}\overline{C}_\alpha^2-\frac{1}{2}\overline{C}_\beta^2+\frac{1}{2}\overline{D}^2-\overline{C}_\alpha\overline{C}_\beta\cos(\overline{\delta}_\alpha-\overline{\delta}_\beta))
\\
{}&-\frac{\sqrt{5}}{3}\left(2BC\cos(\delta_2-\delta_0)-\overline{B}_\alpha\overline{C}_\alpha-\overline{B}_\beta\overline{C}_\beta-(\overline{B}_\alpha\overline{C}_\beta+\overline{B}_\beta\overline{C}_\alpha)\cos(\overline{\delta}_\alpha-\overline{\delta}_\beta)\right)
\\
{}&\left.-\sqrt{\frac{5}{3}}\left(\overline{B}_\alpha+\frac{1}{\sqrt{2}}\overline{C}_\alpha\right)\overline{D}\cos(\overline{\delta}_2-\overline{\delta}_\alpha)-(\alpha\to\beta)\right]\label{eq:m2}
\es \\ \bs
M_3={}&\frac{3}{40\,\pi}\sqrt{\frac{3}{5}}\left[C\overline{A}_2\cos(\overline{\delta}_1-\delta_2)-\frac{1}{\sqrt{3}}A_2\overline{D}\cos(\overline{\delta}_2-\delta_1)\right].\label{eq:m3}
\es\end{align}

Disregarding the small $M_3$ term, the asymmetry is proportional to Eq. (\ref{eq:m1}).  The physical content of Eq. (\ref{eq:m1})'s first line is the interference of strong $p$-wave pions with CSB $s$-wave pions and the second and third lines are strong $s$-wave and CSB $p$-wave.

Table \ref{tab:strmxels} shows the strong reduced matrix elements and Table \ref{tab:csbmxels} the CSB reduced matrix elements.  The CSB rescattering numbers were calculated including the new contributions discovered by Ref. \citep{Filin:2009yh}.  The $np$ phase shifts (in radians) which appear in the cross section according to Eqs. (\ref{eq:m0}-\ref{eq:m3}) are given in Tables \ref{tab:strongdeltas} and \ref{tab:csbdeltas}.

\begin{table}
\caption{\label{tab:strmxels}Strong reduced matrix elements}
\begin{center}
\begin{tabular}{lcccc}
\hline\hline
Diagram & $A_0$ & $A_2$ & $B$ & $C$ \\ \hline
Impulse (w/ wfn corr) & 0 & 0 & -6.59 & 32.85 \\
Impulse Recoil & 5.62 & 0.21 & 1.17 & -8.56 \\
RS (w/ Recoil) & 81.12 & 0 & 0.54 & 1.66 \\
Delta (no cutoff) & 0 & 0 & -33.94 & 37.96 \\
Delta ($\Lambda=417\ \text{MeV}$)$\quad$ & 0 & 0 & -10.48 & 21.60 \\ \hline\hline
\end{tabular}
\end{center}
\end{table}

\begin{table}
\caption{\label{tab:csbmxels}CSB reduced matrix elements}\begin{center}
\begin{tabular}{lccccccc}
\hline\hline
Diagram & $\overline{A}_0$ & $\overline{A}_2$ & $\overline{B}_\alpha$ & $\overline{B}_\beta$ & $\overline{C}_\alpha$ & $\overline{C}_\beta$ & $\overline{D}$ \\ \hline
Impulse ($\times\frac{1}{\beta_1}$)$\quad$ & 0 & 0 & 12.23 & 29.72 & -7.80 & -15.05 & -28.30 \\
RS ($\times\frac{100\,\text{MeV}}{\delta m_N^{str}}$) & -28.83 & 0 & -1.37 & 1.79 & 1.48 & 1.89 & -4.92 \\
Delta ($\times\frac{1}{\beta_1}$) & 0 & 0 & 12.60 & -7.71 & -8.47 & -2.86 & 22.37 \\ \hline\hline
\end{tabular}
\end{center}
\end{table}

\begin{table}
\begin{center}
\begin{minipage}{0.4\linewidth}
\caption{\label{tab:strongdeltas}Strong phase shifts}
\centering
\begin{tabular}{lc} \hline\hline
$\delta_1\quad$ & -0.47\\
$\delta_0$ & -0.044\\
$\delta_2$ & 0.16\\ \hline\hline
\end{tabular}
\end{minipage}
\begin{minipage}{0.4\linewidth}
\centering
\caption{\label{tab:csbdeltas}CSB phase shifts}
\begin{tabular}{lc} \hline\hline
$\overline{\delta}_1$ & -0.44\\
$\overline{\delta}_\alpha$ & 0.19\\
$\overline{\delta}_\beta\quad$ & -0.43\\
$\overline{\delta}_2$ & 0.44\\ \hline\hline
\end{tabular}
\end{minipage}
\end{center}
\end{table}
\chapter[\texorpdfstring{\Cref{chap:wfncor}}{Ch. 5} Details]{Ch. \lowercase{\ref{chap:wfncor}} Details}

\section{Impulse Approximation Details\label{sec:impdetails}}

This Section begins with a presentation of the details of the calculation of the traditional IA pion production amplitude.  Evaluating the isospin matrix element
\begin{align}
\la00\mid\tau_{1,3}\mid10\ra=1\label{eq:isospin}
\end{align}
and using the vertex rule shown in Fig. \ref{fig:rules}, we obtain for Fig. \ref{fig:ia}(a) at threshold
\begin{align}
\left\la00\mid\hat{\mathcal{M}}'_L(\bfp,\bfk)\mid10\right\ra=\frac{g_A}{2f_\pi}\frac{m_\pi}{2m_N}\bsig_1\cdot(\bfp+\bfk)(2\pi)^3\delta^3(\bfp-\bfk),\label{eq:mmom}
\end{align}
where $\hat{\mathcal{M}}'=\hat{\mathcal{M}}/\sqrt{2m_N\,2m_P\,2m_d}\equiv\hat{\mathcal{M}}/N$.  Since we are using position space $np$ wave functions, we Fourier transform the matrix element with respect to $\bfl=\bfk-\bfp$, which is identical to $\bfq'$ at threshold,
\begin{align}
\int\frac{d^3l}{(2\pi)^3}\,e^{i\bfl\cdot\bfr}(2\pi)^3\delta^3(\bfp-\bfk)=1\label{eq:ft1}
\end{align}
Note that we group the $\bfp$ and $\bfk$ with their respective wave functions prior to performing the Fourier transform
\begin{align}
\bsig_1\cdot(\bfp+\bfk)\rightarrow\bsig_1\cdot(-i\overrightarrow{\nabla}_{np}+i\overleftarrow{\nabla}_d).\label{eq:ft2}
\end{align}
Thus the full position space operator is
\begin{align}
\left\la00\mid\hat{\mathcal{M}}'_{L}(\bfr)\mid10\right\ra=-i\frac{g_A}{2f_\pi}\frac{m_\pi}{2m_N}\bsig_1\cdot(\overrightarrow{\nabla}_{np}-\overleftarrow{\nabla}_d).\label{eq:mpos}
\end{align}
To calculate the diagram with rescattering on the other nucleon, we consider how each part of the left side of Eq. (\ref{eq:mpos}) transforms under $1\leftrightarrow2$.  Since the strong part of the Lagrangian is invariant under isospin, $\hat{\mathcal{M}}$ is invariant.  The initial isospin ket $\left|1,0\right\ra$ is invariant as well, but $\left|0,0\right\ra\rightarrow-\left|0,0\right\ra$ and $\bnabla\rightarrow-\bnabla$.  Thus,
\begin{align}
\left\la00\mid\hat{\mathcal{M}}'_{L+R}(\bfr)\mid10\right\ra=-2i\frac{g_A}{2f_\pi}\frac{m_\pi}{2m_N}\bfS\cdot(\overrightarrow{\nabla}_{np}-\overleftarrow{\nabla}_d).\label{eq:mpos2}
\end{align}

The final spin-angle wave function is that of the deuteron, while the initial state for $s$-wave pion production is solely $^3P_1$,
\begin{align}\bs
\mid f(\bfr)\ra&\equiv\frac{u(r)}{r}\mid^3S_1\ra+\frac{w(r)}{r}\mid^3D_1\ra
\\
\mid i(\bfr)\ra&\equiv4\pi i\,\frac{u_{1,1}(r)}{pr}\mid^3P_1\ra,\label{eq:states}
\es\end{align}
where we have absorbed the unobservable (since there is only one initial channel available) phase into the definition of the matrix element.  The spin-angle matrix elements are calculated,
\begin{align}
\la f(\bfr)\mid\mid\bfS\cdot\left(\overrightarrow{\nabla}-\overleftarrow{\nabla}\right)\mid\mid i(\bfr)\ra=4\pi i\left[R_f(r)\frac{\partial R_i(r)}{\partial r}+R_{f,2}(r)\frac{2}{r}R_i(r)-\frac{\partial R_f(r)}{\partial r}R_i(r)\right],\label{eq:spinanglegrad}
\end{align}
where $R_f(r)\equiv\sqrt{2}u(r)/r+w(r)/r$, $R_{f,2}(r)\equiv\sqrt{2}u(r)/r-2w(r)/r$, and $R_i(r)\equiv u_{1,1}(r)/pr$.

Using Eqs. (\ref{eq:mpos2}) and (\ref{eq:spinanglegrad}), we have the final result for the reduced matrix element,
\begin{align}\bs
A_0^\text{imp}&\equiv\int dr\,r^2\left(\la00\mid\otimes\la f(\bfr)\mid\mid\right)\hat{\mathcal{M}}(\bfr)\left(\mid\mid i(\bfr)\ra\otimes\mid10\ra\right)
\\
&=N8\pi\frac{g_A}{2f_\pi}\frac{m_\pi}{2m_N}K,
\es \\
K&\equiv\int dr\,r^2\left[R_f(r)\frac{\partial R_i(r)}{\partial r}+R_{f,2}(r)\frac{2}{r}R_i(r)-\frac{\partial R_f(r)}{\partial r}R_i(r)\right].
\end{align}

\section{Including OPE Details\label{sec:opedetails}}
\subsection{Reducible OPE\label{sec:opedetails1}}

Taking just the $\bfq'$ terms at the OPE vertices, Fig. \ref{fig:ia}(b) is given by
\begin{align}\bs
\hat{\mathcal{M}}'(\bfp,\bfk)={}&\left(-\frac{g_A}{2f_\pi}\right)^3\btau_1\cdot\btau_2\,\bsig_1\cdot(-\bfq')\frac{-i}{\bfq'\,^2+\mu(0)^2}\bsig_2\cdot\bfq'
\\
{}&\times\tau_{1,3}\,\frac{i}{-E_d-\bfp^2/m_N}\bsig_1\cdot\left(-\frac{m_\pi}{2m_N}2\bfp\right),\label{eq:mwfn}
\es\end{align}
where $\mo^2\equiv m_\pi^2-\omega^2$.  Adding to this expression emission from the right nucleon and approximating $\bfp^2=m_\pi m_N$ as discussed at the end of Sec. \ref{sec:ope}, we find
\begin{align}\bs
\la00\mid\hat{\mathcal{M}}'(\bfp,\bfk)\mid10\ra&=\frac{12g_A^3}{8f_\pi^3}\frac{m_\pi}{2m_N}\bsig_1\cdot\bfq'\bsig_2\cdot\bfq'\frac{1}{\bfq'\,^2+\mu(0)^2}\frac{1}{-E_d-m_\pi}\bfS\cdot\bfp,
\\
\la00\mid\hat{\mathcal{M}}'(\bfr)\mid10\ra&=\frac{ig_A^3}{8\pi f_\pi^3}\frac{m_\pi}{2m_N}\mu(0)^3\left(S_{12}f(\omega,r)+\bsig_1\cdot\bsig_2\,g(\omega,r)\right)\frac{1}{-E_d-m_\pi}\bfS\cdot\bnabla,\label{eq:mwfntot}
\es\end{align}
where the $\bnabla$ acts on the initial $np$ wave function, $S_{12}=3\bsig_1\cdot\hat{r}\bsig_2\cdot\hat{r}-\bsig_1\cdot\bsig_2$ is the normal tensor operator, and
\begin{align}\bs
g(\omega,r)&=\frac{e^{-\mo r}}{\mo r},
\\
f(\omega,r)&=\left(1+\frac{3}{\mo r}+\frac{3}{(\mo r)^2}\right)\frac{e^{-\mo r}}{\mo r}
\es\end{align}
come from the Fourier transform [see Eq. (\ref{eq:ft1})] of the pion propagator.  Next, we evaluate
\begin{align}
\left(\frac{u(r)}{r}\la^3S_1\mid+\frac{w(r)}{r}\la^3D_1\mid\right)\left(S_{12}f(\omega,r)+\bsig_1\cdot\bsig_2\,g(\omega,r)\right)\equiv\frac{\tilde{u}(\omega,r)}{r}\la^3S_1\mid+\frac{\tilde{w}(\omega,r)}{r}\la^3D_1\mid,\label{eq:ope}
\end{align}
where
\begin{align}\bs
\frac{\tilde{u}(\omega,r)}{r}&=\frac{u(r)}{r}g(\omega,r)+2\sqrt{2}\frac{w(r)}{r}f(\omega,r),
\\
\frac{\tilde{w}(\omega,r)}{r}&=\frac{w(r)}{r}\left(g(\omega,r)-2f(\omega,r)\right)+2\sqrt{2}\frac{u(r)}{r}f(\omega,r).
\es\end{align}
Thus,
\begin{align}\bs
\la f(\bfr)\mid\mid{}&\left(S_{12}f(\omega,r)+\bsig_1\cdot\bsig_2\,g(\omega,r)\right)\bfS\cdot\bnabla\mid\mid i(\bfr)\rangle
\\
{}&=4\pi i\left(\frac{\tilde{u}(\omega,r)}{r}\sqrt{2}\left(\ddr+\frac{2}{r}\right)+\frac{\tilde{w}(\omega,r)}{r}\left(\ddr-\frac{1}{r}\right)\right)R_i(r),\label{eq:redwfn}
\es\end{align}
and we finally arrive at the full reduced matrix element,
\begin{align}\bs
A_0^\text{OPE,red,f}&=-N\frac{g_A^3}{2f_\pi^3}\frac{m_\pi}{2m_N}\frac{\mu(0)^3}{-E_d-m_\pi}L^f(0),
\\
L^f(\omega)&=\int dr\,r^2\left[\frac{\tilde{u}(\omega,r)}{r}\sqrt{2}\left(\ddr+\frac{2}{r}\right)+\frac{\tilde{w}(\omega,r)}{r}\left(\ddr-\frac{1}{r}\right)\right]R_i(r).
\es\end{align}

\subsection{Irreducible OPE\label{sec:opedetails2}}

Finally, as described in Sec. \ref{sec:ope}, for the irreducible diagram we use $(-m_\pi/2-\bfp^2/2m_N)^{-1}\approx(-m_\pi)^{-1}$ for the intermediate nucleon propagator and take $\omega=m_\pi/2$.
\begin{align}
A_0^\text{OPE,irr,f}=-N\frac{g_A^3}{2f_\pi^3}\frac{m_\pi}{2m_N}\frac{\mu(m_\pi/2)^3}{-m_\pi}L^f(m_\pi/2).
\end{align}

\subsection{Initial state OPE\label{sec:opedetails3}}

For OPE in the initial state, the isospin matrix element is $\la 00\mid\tau_{1,3}\btau_1\cdot\btau_2\mid10\ra=1$, and because the initial state consists of just one channel, $^3P_1$,
\begin{align}
\left(S_{12}f(\omega,r)+\bsig_1\cdot\bsig_2\,g(\omega,r)\right)\mid^3P_1\ra=\left(2f(\omega,r)+g(\omega,r)\right)\mid^3P_1\ra.\nonumber
\end{align}
Evaluating the $\bfS\cdot\overleftarrow{\nabla}$ reduced matrix elements, we find
\begin{align}\bs
A_0^\text{OPE,red,i}&=-N\frac{g_A^3}{2f_\pi^3}\frac{m_\pi}{2m_N}\frac{\mu(0)^3/3}{m_\pi}L^i(0),
\\
A_0^\text{OPE,irr,i}&=-N\frac{g_A^3}{2f_\pi^3}\frac{m_\pi}{2m_N}\frac{\mu(m_\pi/2)^3/3}{m_\pi}L^i(m_\pi/2),
\\
L^i(\omega)&=\int dr\,r^2\left(\sqrt{2}\ddr\frac{u(r)}{r}+\left(\ddr+\frac{3}{r}\right)\frac{w(r)}{r}\right)\left(2f(\omega,r)+g(\omega,r)\right)R_i(r).\label{eq:a0wfni}
\es\end{align}

\section{Exact Wave Function Corrections Details\label{sec:recoildetails}}

Consider the nucleon propagator for reducible OPE in the initial state.  Pulling out a $-m_N$ and expanding this function in spherical coordinates, we have
\begin{align}\bs
iG_0(\bfr,\bfr')&=-m_N\int\frac{d^3k}{(2\pi)^3}e^{-i\bfk\cdot(\bfr-bfr')}\frac{i}{\bfk^2-\xi^2-i\epsilon}
\\
&=-2im_N\frac{e^{i\xi |\bfr-\bfr'|}}{4\pi|\bfr-\bfr'|}
\\
&=2m_N\xi\sum_{l,m}j_l(\xi r_<)h_l^{(1)}(\xi r_>)Y^{l\,*}_m(\hat{r}')Y^l_m(\hat{r}),\label{eq:ghelm}
\es\end{align}
where $\xi=\sqrt{m_\pi m_N}$ and $r_<(r_>)$ is the lesser (greater) of $|\bfr|,|\bfr'|$.  This spherical partial wave expansion was derived from the differential equation
\begin{align}
\left(-\frac{1}{r}\frac{\partial^2}{\partial r^2}r+\frac{l(l+1)}{r^2}-\xi^2\right)\mathcal{G}(r,r')=\frac{\delta(r-r')}{rr'},
\end{align}
where $iG_0=-im_N\mathcal{G}$.  First, one solves the homogeneous equation and requires both finiteness at the origin and outgoing wave behavior for large $r$.  Thus, $\mathcal{G}(r,r')=Aj_l(\xi r_<)h_l^{(1)}(\xi r_>)$.  Next, the boundary condition at $r=r'$ is obtained by integrating the differential equation across the boundary.  In terms of $g(r,r')=rr'\mathcal{G}(r,r')$,
\begin{align}
\frac{\partial}{\partial r}g_>(r,r')|_{r=r'+\epsilon}-\frac{\partial}{\partial r}g_<(r,r')|_{r=r'-\epsilon}=-1,
\end{align}
which yields $A=i\xi$.  At this point in the diagram, the two-nucleon state is still $^3P_1$, so we preform one of the angular integrals and obtain
\begin{align}
iG_0(\bfr,\bfr')\rightarrow m_N\xi j_1(\xi r_<)h_1^{(1)}(\xi r_>).\label{eq:gfn}
\end{align}
Thus,
\begin{align}\bs
A_0^\text{OPE,red,i}(m_\pi/2)={}&-N\frac{g_A^3}{2f_\pi^3}\frac{m_\pi}{2m_N}\frac{\mu(0)^3}{3}\left(-im_N\xi L^i(0)\right)
\\
L^i(m_\pi/2)={}&\int dr\,dr'\,r^2r'^{2}\left(\sqrt{2}\ddr\frac{u(r)}{r}+\left(\ddr+\frac{3}{r}\right)\frac{w(r)}{r}\right)
\\
{}&\times j_1(\xi r_<)h_1^{(1)}(\xi r_>)\left(2f(m_\pi/2,r')+g(m_\pi/2,r')\right)R_i(r').
\es\end{align}

For the irreducible initial state OPE, the only difference is that a $-2m_N$ gets pulled out and the momentum becomes $\xi'=\sqrt{2m_\pi m_N}$,
\begin{align}
A_0^\text{OPE,irr,i}(m_\pi/2)=-N\frac{g_A^3}{2f_\pi^3}\frac{m_\pi}{2m_N}\frac{\mu(m_\pi/2)^3}{3}\left(-2im_N\xi' L^i(m_\pi/2)\right)
\end{align}
For the final state OPE, we can obtain the correct Green function from Eq. (\ref{eq:gfn}) by letting $\xi\rightarrow i\xi$ and using the correct $l$ for the term under consideration.

\section{Cutoff Details\label{sec:cutoffdetails}}

In this section we display the exact expressions needed to implement the Gaussian cutoff of Sec \ref{sec:cutoff}.  For the OPE diagrams, the integral of Eq. (\ref{eq:gLdef}) is evaluated,
\begin{align}
g_\Lambda(\omega,r)=\frac{1}{2}e^{\mo^2/\Lambda^2}\left[\frac{e^{-\mo r}}{\mo r}\text{erfc}\left(-\frac{\Lambda r}{2}+\frac{\mo}{\Lambda}\right)-\frac{e^{\mo r}}{\mo r}\text{erfc}\left(\frac{\Lambda r}{2}+\frac{\mo}{\Lambda}\right)\right]\label{eq:gcutoff}.
\end{align}
One also needs derivatives of Eq. (\ref{eq:gcutoff}),
\begin{align}
\bsig_1\cdot\bnabla\bsig_2\cdot{}\bnabla g_\Lambda(\omega,r)=\frac{\mo^3}{3}\left(S_{12}f_\Lambda(\omega,r)+\bsig_1\cdot\bsig_2l_\Lambda(\omega,r)\right),
\end{align}
where,
\begin{align}\bs
f_\Lambda(\omega,r)={}&\frac{1}{2}e^{\mo^2/\Lambda^2}\left[\left(1+\frac{3}{\mo r}+\frac{3}{(\mo r)^2}\right)\text{erfc}\left(-\frac{\Lambda r}{2}+\frac{\mo}{\Lambda}\right)\right.
\\
{}&\left.-\frac{\Lambda}{\sqrt{\pi}\mo}\left(\frac{\Lambda^2}{2\mo^2}\mo r+1+\frac{3}{\mo r}\right)e^{-\left(-\frac{\Lambda r}{2}+\frac{\mo}{\Lambda}\right)^2}\right]\frac{e^{-\mo r}}{\mo r}
\\
{}&+\left(\mu\rightarrow\mu\text{ and }\Lambda\rightarrow-\Lambda\right)
\es \\ \bs
l_\Lambda(\omega,r)={}&\frac{1}{2}e^{\mo^2/\Lambda^2}\left[\text{erfc}\left(-\frac{\Lambda r}{2}+\frac{\mo}{\Lambda}\right)\right.
\\
{}&\left.-\frac{\Lambda}{\sqrt{\pi\mo}}\left(\frac{\Lambda^2}{2\mo^2}\mo r+1\right)e^{-\left(-\frac{\Lambda r}{2}+\frac{\mo}{\Lambda}\right)^2}\right]\frac{e^{-\mo r}}{\mo r}
\\
{}&+\left(\mu\rightarrow\mu\text{ and }\Lambda\rightarrow-\Lambda\right).
\es\end{align}
\chapter[\texorpdfstring{\Cref{chap:nrred}}{Ch. 6} Details]{Ch. \lowercase{\ref{chap:nrred}} Details}

\section{\texorpdfstring{$N\rightarrow N\pi$}{N --> N pi} from B$\chi$PT\label{sec:nnpi}}

Recall that the LO $NN\pi$ interaction reads,
\begin{align}
\mathcal{L}^{(0)}\subset\overline{N}\frac{g_A}{2}\slashed{u}_\perp\gamma^5N,
\end{align}
where $u_{\perp,\mu}=u_\mu-v\cdot u\,v_\mu$, $u_\mu=i(u^\dagger\partial_\mu u-u\partial_\mu u^\dagger)$, and $u^2=e^{i\tau_a\pi_a/f_\pi}$.  Expanding, we find
\begin{align}\bs
u_\mu&=-\frac{\tau_a}{f_\pi}\partial_\mu\pi_a
\\
\slashed{u}_\perp&=\gamma_0(u_0-u_0\cdot1)-\gamma_i\left(-\frac{\tau_a}{f_\pi}\partial_i\pi_a\right)\qquad\qquad i=1,2,3
\\
&=\frac{\tau_a}{f_\pi}\gamma_i\partial_i\pi_a,
\es\end{align}
and thus the Feynman rule for an outgoing pion with momentum $q$ and isospin $a$ is
\begin{align}
{\cal O}_\pi^{(0)}=-i\left(\frac{g_A}{2f_\pi}\gamma_i\gamma^5\right)\left(iq_i\right)\tau_a.
\end{align}

At threshold, the pion four-momentum is $q=(m_\pi,0,0,0)$ making ${\cal O}_\pi^{(0)}=0$.  This reflects the fact (well-known from current algebra) that threshold pion production proceeds via the off-diagonal, and therefore $1/m_N$ suppressed, interaction $g_\pi\gamma^5\gamma^0q^0\tau$.  In the effective theory, this recoil correction shows up in the NLO Lagrangian
\begin{align}
\mathcal{L}^{(1)}\subset-i\frac{g_A}{2m_N}\overline{N}\left\{v^\mu u_\mu,S^\mu\partial_\mu\right\}N,
\end{align}
where the spin vector is $S^\mu=-\frac{1}{2}\gamma^5(\gamma^\mu\slashed{v}-v^\mu)$.  Thus the Feynman rule is
\begin{align}\bs
\mathcal{O}_\pi^{(1)}&=-i\left(-i\frac{g_A}{2m_N}\right)\left[-\frac{\tau_a}{f_\pi}(-im_\pi)\right]\left[\frac{1}{2}\gamma^5\gamma_i\gamma^0(\overrightarrow{\bnabla}-\overleftarrow{\bnabla})_i\right]
\\
&=-i\frac{m_\pi}{2m_N}\frac{g_A}{2f_\pi}\gamma^5\gamma_i\gamma^0(\overrightarrow{\bnabla}-\overleftarrow{\bnabla})_i\tau_a,\label{eq:feynrule}
\es\end{align}
where the derivatives act on the nucleon wave functions.

\section{\label{sec:deuteron}One Pion Exchange Deuteron}

\begin{figure}
\centering
\includegraphics[height=1.5in]{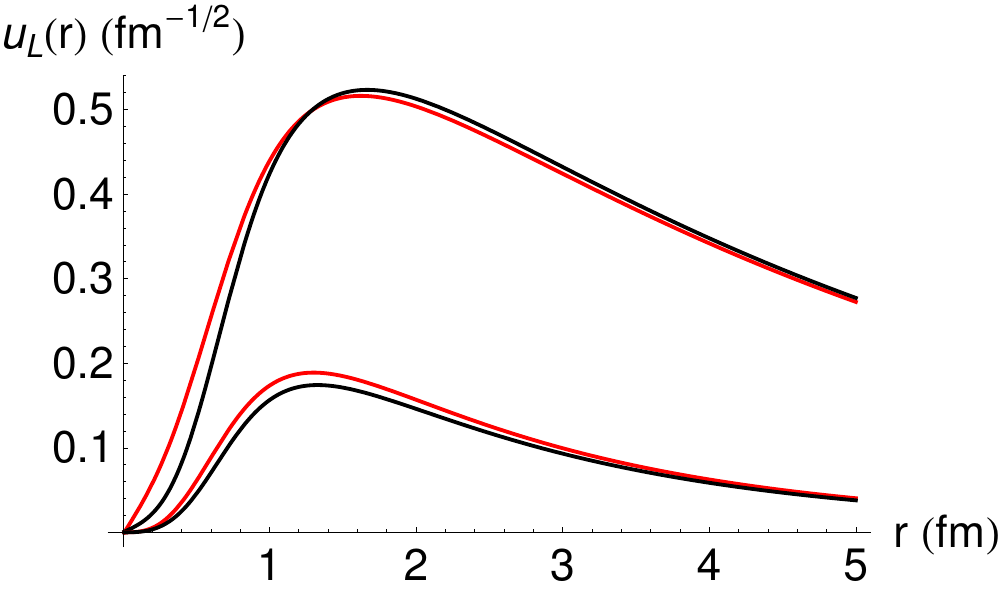}
\caption[Deuteron $s$- and $d$-state wave functions]{\label{fig:opedeuteron}Deuteron $s$- and $d$-state wave functions (the $s$-state is larger).  The potentials used to calculate the wave functions are Av18 (black) and the cutoff OPE described in this section (red).}
\end{figure}

In this Appendix, we present the method by which the deuteron wave function is calculated for use in Sec. \ref{sec:distortions}.  This method is taken directly from the work of Friar, Gibson, and Payne \citep{Friar:1984wi}.  The OPE potential is defined to have central ($Y$) and tensor ($T$) parts,
\begin{align}
V_\pi(\bfr)=f^2m_\pi\frac{\tau_{1,a}\tau_{2,a}}{3}\left[\bsig_1\cdot\bsig_2Y(r)+S_{12}T(r)\right],\label{eq:opedef}
\end{align}
where $f^2=0.079$ (to be distinguished from $f_\pi$) measures the strength of the pion-nucleon coupling and $S_{12}$ is the standard tensor operator.  The deuteron has isospin zero and spin one, so we have
\begin{align}
V_\pi(\bfr)=-f^2m_\pi\left[Y(r)+S_{12}T(r)\right].
\end{align}
The $Y$ and $T$ functions are expressed as derivatives of the Fourier transform of the pion propagator,
\begin{align}\bs
Y(r)&=h_0''(x)-h_0'(x)/x
\\
T(r)&=h_0''(x)+2h_0'(x)/x
\\
h_0(x)&=\frac{4\pi}{(2\pi)^3m_\pi}\int d^3q\frac{e^{-i\bfq\cdot\bfr}}{\bfq^2+m_\pi^2}F^2_{\pi NN}(\bfq^2),\label{eq:h0}
\es\end{align}
where $x=m_\pi r$ and $F_{\pi NN}$ is the form factor for which we use,
\begin{align}
F_{\pi NN}(\bfq^2)=\left(\frac{\Lambda^2-m_\pi^2}{\bfq^2+\Lambda^2}\right)^n.
\end{align}
In Ref. \citep{Friar:1984wi}, it is shown that
\begin{align}\bs
Y(r)&=\frac{e^{-x}}{x}-\beta^3e^{-\beta x}\sum_{i=0}^{2n-1}\frac{\xi^i}{i!}\left(\delta_i(\beta x)-2i\delta_{i-1}(\beta x)\right)
\\
T(r)&=\frac{e^{-x}}{x}\left(1+\frac{3}{x}+\frac{3}{x^2}\right)-\beta^3e^{-\beta x}\sum_{i=0}^{2n-1}\frac{\xi^i}{i!}\left[\delta_i(\beta x)-(2i-3)\delta_{i-1}(\beta x)\right]
\es\end{align}
where $\beta=\Lambda/m_\pi$ and $\xi=(\beta^2-1)/2\beta^2$ and the $\delta_i$ are defined by
\begin{align}
\delta_{i+1}(\beta x)=(2i-1)\delta_i(\beta x)+(\beta x)^2\delta_{i-1}(\beta x)
\end{align}
along with $\delta_0=1/\beta x$ and $\delta_1=1$.  One of the results of Ref. \citep{Friar:1984wi} is that larger values of $n$ lead to better fits to experimental data.  We use $n=5$ and $\beta/\sqrt{10}=5.687805$ in order to precisely reproduce the binding energy $E_B=2.2246$ MeV.  The wave functions are calculated by integrating in from $r_\text{max}=100$ fm and adding together two linearly independent solutions such that the sum vanishes at $r_\text{min}=0.01$ fm.  As shown in Fig. \ref{fig:opedeuteron}, the results are close to the ``correct" Av18 deuteron.

In Table \ref{tab:deuteron}, we display the quadrupole moment and mean square charge radius of Av18, this OPE potential, and experiment (as quoted in \citep{Friar:1984wi}).
\begin{table}
\caption{\label{tab:deuteron}Deuteron properties.}
\begin{center}
\begin{tabular}{lcc}
\hline\hline
Potential & Q (fm$^2$) &  $\la r^2\ra^{1/2}$ (fm)\\ \hline
Av18 & 0.270 & 1.968\\
OPE ($n=5$) & 0.282 & 1.939\\
Experiment$\quad$ & 0.2859(3) & 1.955(5)\\ \hline\hline
\end{tabular}
\end{center}
\end{table}
It is clear that the form factors in the OPE potential make it difficult to distinguish this construction as less accurate than Av18.  Finally, in Table \ref{tab:rs}, we display the reduced matrix elements for the rescattering pion production diagram evaluated with both the phenomenological potentials and the deuteron of this section.  Since this diagram makes the largest contribution to the cross section we need to verify that neglecting non-OPE parts of the potential does not dramatically change this amplitude.
\begin{table}
\caption{\label{tab:rs}Effect of using OPE deuteron on rescattering diagram.}
\begin{center}
\begin{tabular}{lccc}
\hline\hline
Deuteron & Av18 & Reid '93 & Nijm II\\ \hline
Phenomenological$\quad$ & 67.8 & 69.7 & 71.1\\
OPE ($n=5$) & 69.8 & 72.1 & 74.0\\ \hline\hline
\end{tabular}
\end{center}
\end{table}
Indeed, we observe what should be expected: since the rescattering diagram is not as sensitive to the core of the deuteron, using the OPE wave function in place of the standard one has only a small effect.

\section{\label{sec:isi}Effect of the \texorpdfstring{$\delta$}{delta} Terms: \texorpdfstring{${\cal O}_\text{2B}^a$}{O 2B a} Diagram}

In this section we calculate the correction terms to the first two-body DW amplitude [Eq. (\ref{eq:prodisi})] which is shown in Fig. \ref{fig:iaf}.  Assuming that the $\delta$'s truly are small compared to Eq. (\ref{eq:prodisi2}), we will only worry about calculating them one at a time, numbering the contribution of the $\delta$'s from right to left as 1, 2, and 3.  Note that we will display the results as calculated using the OPE deuteron and the Av18 initial state.

\subsection{\label{sec:isi1}First ${\cal O}_\text{2B}^a$ correction term: $\Delta{\cal O}_1$}
\begin{figure}
\begin{center}
\includegraphics[height=1.5in]{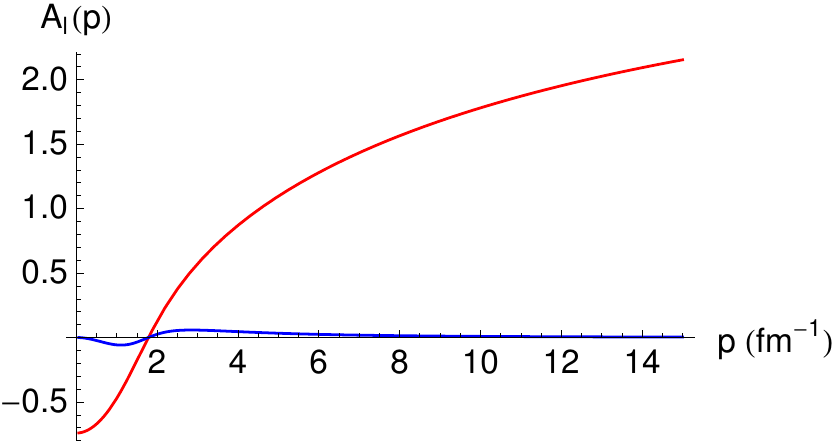}
\caption[Coefficients of the expansion in Eq. (\ref{eq:sphexp})]{\label{fig:Al}(Color online) Coefficients of the expansion in Eq. (\ref{eq:sphexp}).  The red curve shows $l=0$ and the blue shows $l=2$.}
\end{center}
\end{figure}

Consider the rightmost $\delta$ in Eq. (\ref{eq:prodisi}),
\begin{align}\bs
\Delta{\cal O}_1={}&-\frac{\bsig_1\cdot(\bfp-\bfk)\bsig_2\cdot(\bfk-\bfp)}{-(\bfp-\bfk)^2-\mu^2}F^2_{\pi NN}((\bfp-\bfk)^2)\frac{1}{-m_\pi}
\\
{}&\times\frac{\frac{\bfp^2}{2m_N}-\frac{m_\pi}{2}+\frac{\gamma}{2}}{\sqrt{(\bfp-\bfp_i)^2+m_\pi^2}+\frac{\bfp^2}{2m_N}-\frac{m_\pi}{2}+\frac{\gamma}{2}}\bfS\cdot\bfp,\label{eq:blah}
\es\end{align}
where $\mu^2=3m_\pi^2/4$ and $F_{\pi NN}$ is the form factor described in \cref{sec:deuteron}.  The easiest way to evaluate the matrix element of this operator is to let the OPE act to the left on the deuteron in position space.  The resulting expression is then transformed to momentum space.  We can expand the fraction in the integrand of Eq. (\ref{eq:blah}) into spherical harmonics (taking $\hat{\bfp}_i=\hat{{\bf z}}$),
\begin{align}
\frac{\frac{\bfp^2}{2m_N}-\frac{m_\pi}{2}+\frac{\gamma}{2}}{\sqrt{(\bfp-\bfp_i)^2+m_\pi^2}+\frac{\bfp^2}{2m_N}-\frac{m_\pi}{2}+\frac{\gamma}{2}}=\sum_{l}A_{l}(p)Y_{l,0}(\hat{\bfp}),\label{eq:sphexp}
\end{align}
and note that only the $l=0,2$ terms will contribute to $s$-wave production.  The expansion coefficients are shown in Fig. \ref{fig:Al}.
Clearly the $l=2$ term is small and we neglect it here to avoid the extra algebra involved with a $J=2$ operator (resulting in the $A_2$ reduced matrix elements in the notation of \cref{sec:reduced}).  We find
\begin{align}
\frac{\Delta{\cal M}_1}{{\cal M}}=-34\%.
\end{align}

\subsection{\label{sec:isi2}Second ${\cal O}_\text{2B}^a$ correction term: $\Delta{\cal O}_2$}
\begin{figure}
\centering
\parbox{.4\linewidth}{\includegraphics[width=\linewidth]{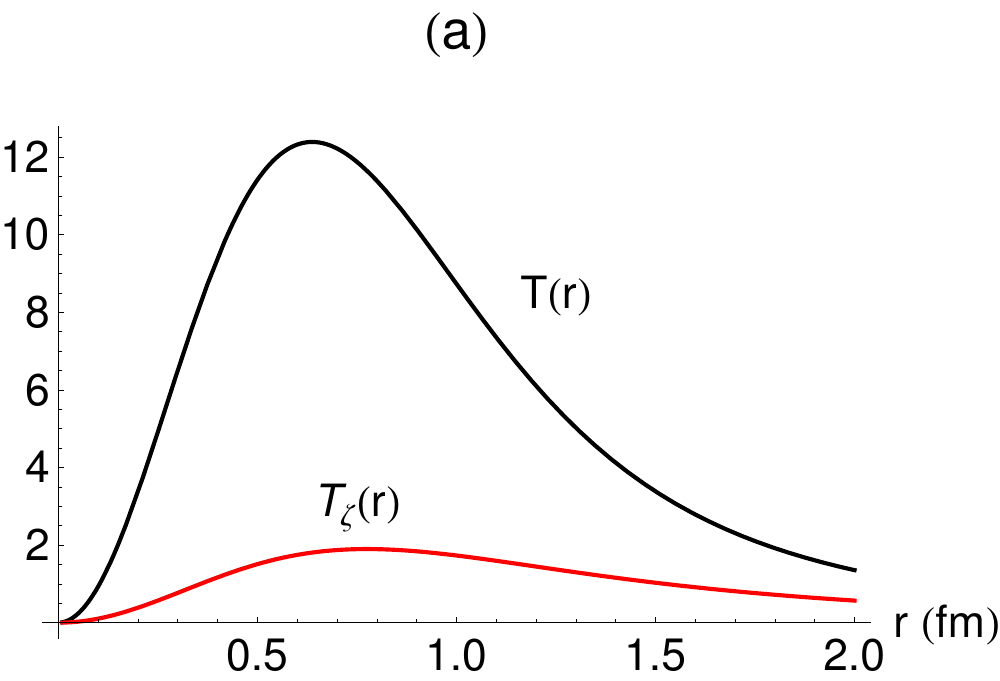}}
\hspace{.05\linewidth}
\parbox{.4\linewidth}{\includegraphics[width=\linewidth]{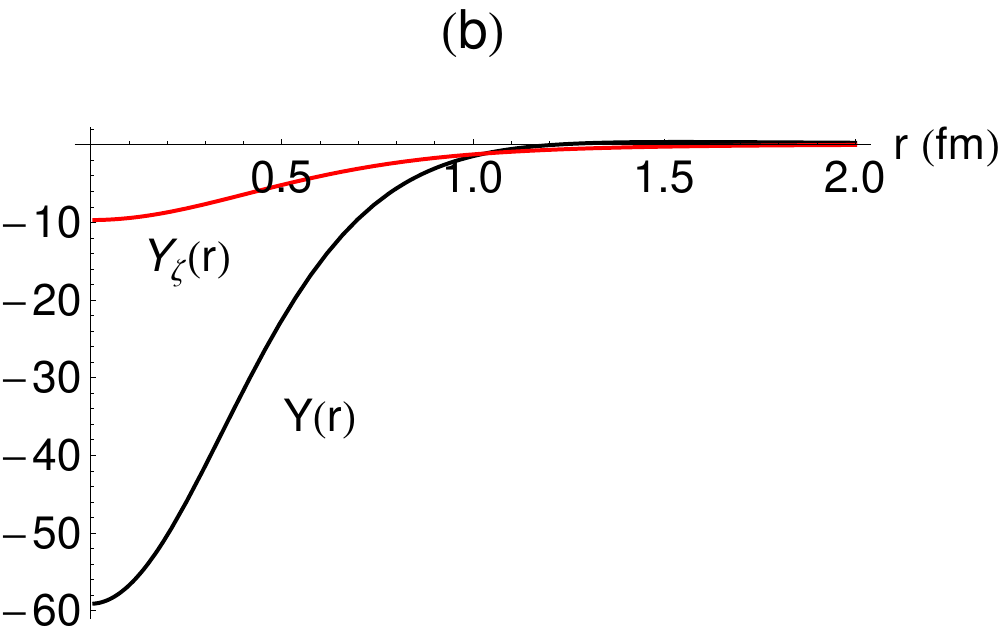}}
\caption[Effect of the square root on the OPE radial functions]{\label{fig:zeta}Effect of the square root on the OPE a) tensor and b) central radial functions.}
\end{figure}

Next consider the term,
\begin{align}\bs
\Delta{\cal O}_2={}&\frac{\bsig_1\cdot(\bfp-\bfk)\bsig_2\cdot(\bfk-\bfp)}{-(\bfp-\bfk)^2-\mu^2}F_{\pi NN}^2((\bfp-\bfk)^2)\frac{1}{\sqrt{(\bfp-\bfk)^2+m_\pi^2}}\frac{1}{-m_\pi}
\\
{}&\times\left(\frac{\bfp^2}{2m_N}-\frac{m_\pi}{2}+\frac{\gamma}{2}\right)\bfS\cdot\bfp.
\es\end{align}
This term has a modified OPE,
\begin{align}\bs
\int\frac{d^3q}{(2\pi)^3}e^{-i\bfq\cdot\bfr}\frac{\bsig_1\cdot\bfq\,\bsig_2\cdot(-\bfq)}{\bfq^2+\mu^2}F_{\pi NN}^2(\bfq^2)\frac{1}{\sqrt{\bfq^2+m_\pi^2}}&\equiv\bsig_1\cdot\nabla\bsig_2\cdot\nabla\frac{\zeta(r)}{4\pi}
\\
&=\frac{\mu^2}{12\pi}\left[S_{12}T_\zeta(r)+\bsig_1\cdot\bsig_2Y_\zeta(r)\right].\label{eq:sqrt}
\es\end{align}
In Fig. \ref{fig:zeta} we compare the functions $T_\zeta(r)$ and $Y_\zeta(r)$ to traditional OPE which has $\mu$ in place of the square root in Eq. (\ref{eq:sqrt}).
We use the Schr\"odinger equation to replace the $\bfp^2/2m_N$ with $V(r)$ and evaluate the matrix element in position space to find
\begin{align}
\frac{\Delta{\cal M}_2}{{\cal M}}=+50\%.
\end{align}

\subsection{\label{sec:isi3}Third ${\cal O}_\text{2B}^a$ correction term: $\Delta{\cal O}_3$}
\begin{figure}
\centering
\parbox{.4\linewidth}{\includegraphics[width=\linewidth]{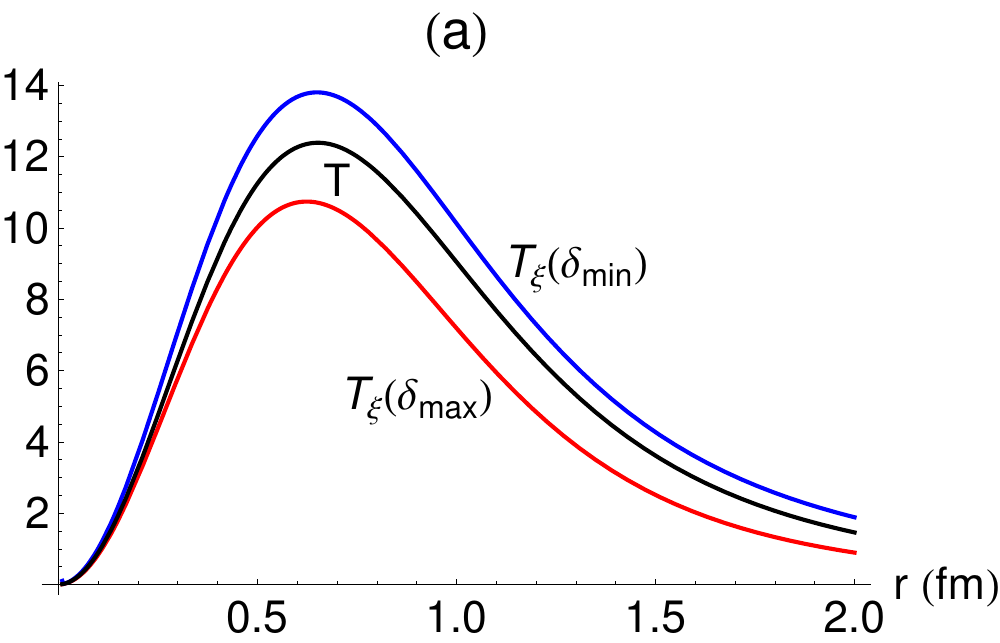}}
\hspace{.05\linewidth}
\parbox{.4\linewidth}{\includegraphics[width=\linewidth]{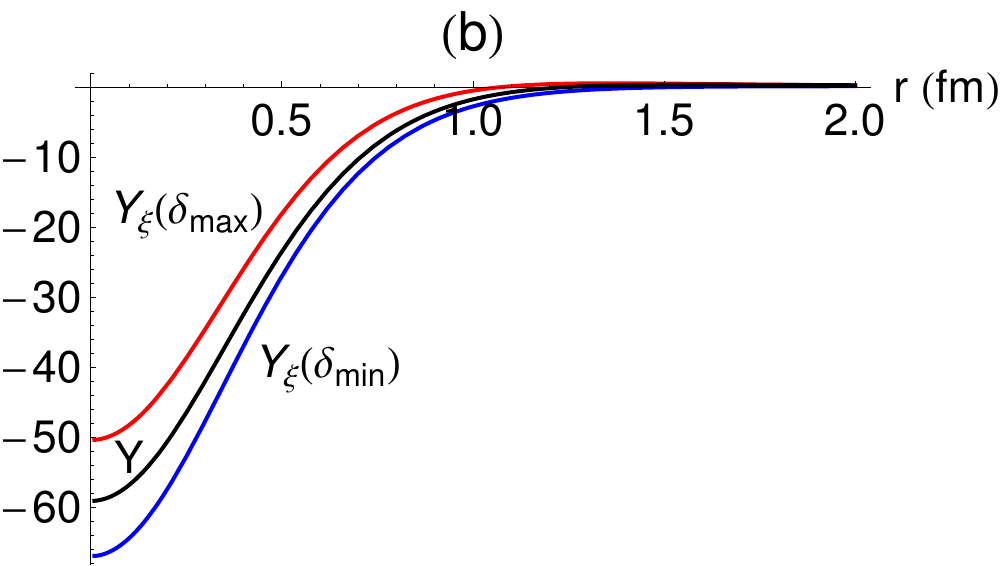}}
\caption[Effect of the $\delta$ on the OPE radial functions]{\label{fig:xiplots}Effect of the $\delta$ on the OPE (a) tensor and (b) central radial functions.}
\end{figure}

Calculating the effects of the $\delta$ in the denominator of the OPE is difficult to do exactly due to the combination of momenta that appear,
\begin{align}
\Delta{\cal O}_3=\frac{\bsig_1\cdot(\bfp-\bfk)\bsig_2\cdot(\bfk-\bfp)}{-\left(\sqrt{(\bfp-\bfk)^2+m_\pi^2}+\delta(\bfp)\right)^2+m_\pi^2/4}F^2_{\pi NN}((\bfp-\bfk)^2)\frac{1}{-m_\pi}\bfS\cdot\bfp,
\end{align}
(recall that $\delta(\bfp)=\bfp^2/2m_N-m_\pi/2+\gamma/2$).  Instead we will evaluate it for fixed values of $\delta$ which represent the deviation of $\bfp$ away from $\bfp_i$,
\begin{align}\bs
\delta_+&=\delta(p_i+m_\pi)=\frac{2p_im_\pi+m_\pi^2}{2m_N}=0.45\,m_\pi
\\
\delta_-&=\delta(p_i-m_\pi)=\frac{-2p_im_\pi+m_\pi^2}{2m_N}=-0.32\,m_\pi.
\es\end{align}
The modified tensor and central functions $T_\xi$ and $Y_\xi$ are shown in Fig \ref{fig:xiplots}.
We define the correction as
\begin{align}
\Delta{\cal M}_3(\delta)={\cal M}(\delta)-{\cal M}(0).
\end{align}
We find,
\begin{align}\bs
\frac{\Delta{\cal M}_3(\delta_+)}{{\cal M}}&=+16\%
\\
\frac{\Delta{\cal M}_3(\delta_-)}{{\cal M}}&=-32\%.
\es\end{align}

\subsection{\label{sec:isisummary}Summary of ${\cal O}_\text{2B}^a$ corrections}
For the purposes of estimating the net result we take the average of the estimates in Sec. \ref{sec:isi3} and find
\begin{align}
\frac{\Delta{\cal M}_\text{tot}}{{\cal M}}\approx-34\%+50\%-8\%=+8\%.
\end{align}
We have successfully shown that the corrections to the first two-body DW amplitude are small, and actually \textit{increase} the amplitude which is already twice as large as the traditional impulse approximation.

\section{\label{sec:fsi}Effect of the \texorpdfstring{$\delta$}{delta} Terms: \texorpdfstring{${\cal O}_\text{2B}^b$}{O 2B b} Diagram}

The second two-body DW amplitude's corrections are evaluated exactly as in the previous sub-sections and we just display the results here,
\begin{align}\bs
\frac{\Delta{\cal M}_1}{{\cal M}}&=-35\%
\\
\frac{\Delta{\cal M}_2}{{\cal M}}&=-39\%
\\
\frac{\Delta{\cal M}_3(\delta_+)}{{\cal M}}=\frac{\Delta{\cal M}_3(\delta_-)}{{\cal M}}&=+3\%,
\es\end{align}
with the net result
\begin{align}
\frac{\Delta{\cal M}_\text{tot}}{{\cal M}}\approx-35\%-39\%+3\%=-71\%.
\end{align}
We see that the corrections to the approximation in \cref{eq:prodfsi2} are fairly large, but this has a negligible effect because the amplitude is already small compared to the first two-body DW amplitude.

\section{\label{sec:sigma}Heavy Meson Exchange}

\begin{figure}
\begin{center}
\includegraphics[height=1.5in]{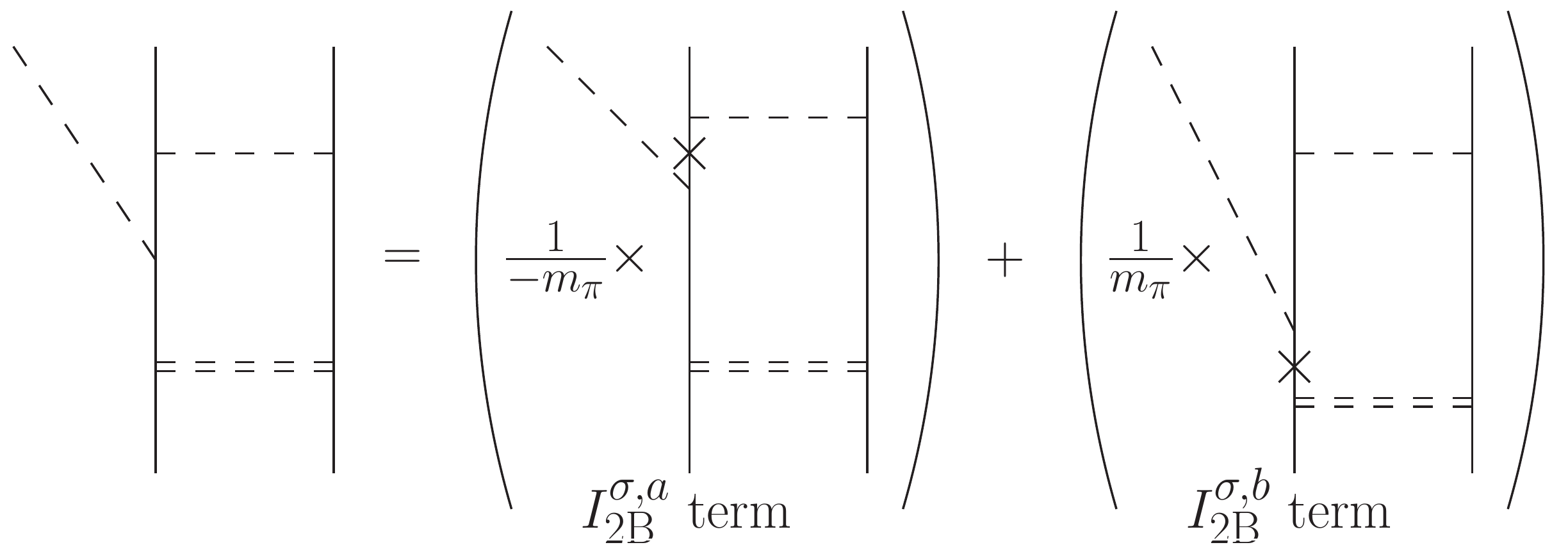}
\caption[Impulse approximation with distorted waves]{\label{fig:sig}Impulse approximation with distorted waves: initial state heavy meson exchange.  Solid lines represent nucleons, dashed lines pions, and the double solid line a $\sigma$ meson.  Crosses represent propagators which are absent due to the partial fractions decomposition.}
\end{center}
\end{figure}

Consider the loop on the left-hand side of Fig. \ref{fig:sig} which is obtained by using OPE for the left $K$ in Eq. (\ref{eq:misi}) and $\sigma$ exchange (the dominant intermediate-range mechanism) for the right $K$.
Note that this loop only differs from Fig. \ref{fig:loop} in two ways: the meson-nucleon vertex (here we consider only scalar-isoscalar) and the meson mass.  We use a typical set of parameters \citep{Ericson:1988gk}, $g^2_\sigma/4\pi=7.1$ and $m_\sigma=550$ MeV.

The result of integrating over energy will proceed exactly as it did with the pion resulting in the two diagrams shown on the RHS of Fig. \ref{fig:sig}.
To interpret the $I_\text{2B}^{\sigma,a}$ term (again, neglecting stretched box diagrams), we absorb the sigma exchange into the initial state and no new term is added.  However, in the $I_\text{2B}^{\sigma,b}$ term, after absorbing the pion exchange into the final state, we are left with a new operator.  The amplitude for this operator can be obtained from that of Fig. \ref{fig:iai} with the following change,
\begin{align}
\left(\frac{g_A}{2f_\pi}\right)^2\frac{\mu^3}{3}\left[2\left(1+\frac{3}{\mu r}+\frac{3}{(\mu r)^2}\right)+1\right]\frac{e^{-\mu r}}{\mu r}\rightarrow g_\sigma^2\mu_\sigma\frac{e^{-\mu_\sigma r}}{\mu_\sigma r},\label{eq:change}
\end{align}
where $\mu^2=3m_\pi^2/4$ and $\mu_\sigma^2=m_\sigma^2-(m_\pi/2)^2$.  We find,
\begin{align}
A_\text{2B}^{\sigma,b}=-7.24,
\end{align}
which is larger in magnitude than the pionic $A_\text{2B}^{\text{DW},b}$ (with the same sign) but smaller than $A_\text{2B}^{\text{DW},a}$ (with the opposite sign).  Since $m_\sigma$ is relatively large, we can safely ignore the two $\delta$ corrections that are competing with $\omega_i$ and only need to evaluate
\begin{align}
\frac{\Delta{\cal M}_1}{{\cal M}}=-27\%.
\end{align}

One natural question is whether the static $\sigma$ exchange already present in the initial-state wave function is a sufficient approximation for the contribution considered in this section.  To answer this question, we can evaluate the traditional impulse approximation with
\begin{align}
|\Psi\ra_i^\sigma=|p_1,p_2\ra+GV_\sigma|\Psi\ra_i,
\end{align}
where here we employ a static $\sigma$ exchange that is present (at least effectively) in the wave function.  Using this initial-state wave function, we calculate
\begin{align}
{\cal M}_\text{1B}^\sigma={}_f\la\phi|{\cal O}_\pi |\Psi\ra_i^\sigma,
\end{align}
and find the reduced matrix element,
\begin{align}
A_\text{1B}^\sigma=-3.3.
\end{align}
Thus we see that the $\sigma$ exchange in the traditional impulse approximation is an underestimate (in magnitude) of the true non-static exchange dictated by the loop integral.

Of course there is no $\sigma$ in traditional B$\chi$PT, so this section is simply telling us that to achieve high accuracy it is indeed important to use more than just simple pion exchange when forming the original box diagram.  Such a calculation is beyond the scope of this work.

\vita{
Daniel Bolton was born in Boulder, CO and grew up twenty miles north in Longmont, CO.  He graduated from Longmont Christian High School in 2002 and the Colorado School of Mines in 2006.  In 2011 he earned his PhD from the University of Washington and will begin his career in the Fall of 2011 as a Lecturer at Baylor University.  He enjoys playing tennis, hiking, running, and reading fiction.
}

\end{document}